\documentclass[a4paper,12pt,twoside]{article}
\usepackage{cite}
\usepackage{a4wide}
\usepackage{amssymb,amsmath,amsthm}
\usepackage[only,llbracket,rrbracket,interleave]{stmaryrd}
\usepackage[pdftex,colorlinks,bookmarksnumbered,bookmarksopen,citecolor=red,urlcolor=red]{hyperref}
\usepackage{color,colordvi,graphicx,xcolor}
\usepackage{subfigure}
\usepackage[makeroom]{cancel}
\usepackage{fancyhdr}\usepackage{lastpage} 
\newenvironment{keywords}{\begin{quote}\emph{\textbf{Keywords:}}}{\end{quote}}
\newtheorem{remark}{Remark}

\usepackage{tikz}
\newcommand{\toh}{\scalebox{0.6}{$\tfrac{1}{2}$}}

\newcommand{\bm}[1]{\text{\boldmath $#1$\unboldmath}}
\newcommand{\abs}[1]{\lvert#1\rvert}
\newcommand{\norm}[1]{\lVert#1\rVert}
\newcommand{\vect}[1]{\mathbf{#1}}
\newcommand{\mat}[1]{\mathbf{#1}}

\newcommand{\Fl}[1]{{#1}_{\!_{\mathcal{F}}}}
\newcommand{\St}[1]{{#1}_{\!_{\mathcal{S}}}}
\newcommand{\Ff}[1]{{#1}_{_{\mathcal{FF}}}}
\newcommand{\Fs}[1]{{#1}_{_{\mathcal{FS}}}}
\newcommand{\Sf}[1]{{#1}_{_{\mathcal{SF}}}}
\newcommand{\Ss}[1]{{#1}_{_{\mathcal{SS}}}}
\newcommand{\FlStintf}[1]{{#1}_{_{\mathcal{I}}}}

\newcommand{\gradS}{\bm{\nabla}_{\!\texttt{S}}}
\newcommand{\gradW}{\bm{\nabla}_{\!\texttt{W}}}

\usepackage{scalerel,stackengine}
\stackMath
\newcommand\reallywidehat[1]{%
\savestack{\tmpbox}{\stretchto{%
  \scaleto{%
    \scalerel*[\widthof{\ensuremath{#1}}]{\kern-.6pt\bigwedge\kern-.6pt}%
    {\rule[-\textheight/2]{1ex}{\textheight}}
  }{\textheight}%
}{0.5ex}}%
\stackon[1pt]{#1}{\tmpbox}%
}

\graphicspath{{figs/}}

\begin{document}
\fancypagestyle{plain}{%
  \renewcommand{\headrulewidth}{0pt}%
  \fancyhead[L]{}%
  \fancyfoot[C]{\footnotesize Page \thepage\ of \pageref{LastPage}}%
}
\title{A weakly compressible \\ hybridizable discontinuous Galerkin formulation \\ for fluid-structure interaction problems}

\author{Andrea La Spina$^{1,2,3}$\footnote{Corresponding author: Andrea La Spina. E-mail: laspina@lnm.mw.tum.de}\hspace{0.5ex}, 
            Martin Kronbichler$^1$,
            Matteo Giacomini$^2$, \\
            Wolfgang A. Wall$^1$
            and 
            Antonio Huerta$^2$}


\date{{\small
         $^1$ Lehrstuhl f\"ur Numerische Mechanik (LNM), Technische Universit\"at M\"unchen (TUM), Garching b. M\"unchen, Germany.\\
         $^2$ Laboratori de Calcul Numeric (LaC\`aN), ETS de Ingenieros de Caminos, Canales y Puertos, Universitat Polit\`ecnica de Catalunya, Barcelona, Spain.\\
         $^3$ Centre Internacional de M\`etodes Num\`erics a l'Enginyeria (CIMNE), Barcelona, Spain.\\[1ex]
         }
         }

\maketitle

\begin{abstract}
A scheme for the solution of fluid-structure interaction (FSI) problems with weakly compressible flows is proposed in this work.
A novel hybridizable discontinuous Galerkin (HDG) method is derived for the discretization of the fluid equations, while the standard continuous Galerkin (CG) approach is adopted for the structural problem. The chosen HDG solver combines robustness of discontinuous Galerkin (DG) approaches in advection-dominated flows with higher order accuracy and efficient implementations. Two coupling strategies are examined in this contribution, namely a partitioned Dirichlet--Neumann scheme in the context of hybrid HDG-CG discretizations and a monolithic approach based on Nitsche's method, exploiting the definition of the numerical flux and the trace of the solution to impose the coupling conditions. Numerical experiments show optimal convergence of the HDG and CG primal and mixed variables and superconvergence of the postprocessed fluid velocity. The robustness and the efficiency of the proposed weakly compressible formulation, in comparison to a fully incompressible one, are also highlighted on a selection of two and three dimensional FSI benchmark problems.
\end{abstract}
\section{Introduction}\label{sec:introduction}

The simulation of the interaction of fluid flows with flexible structures is of great interest in many engineering fields and it has been extensively investigated over the last decades.
The numerical techniques developed for the solution of this challenging multiphysics problem can be catalogued based on many different aspects, for instance, with respect to the spatial discretization, the kinematical description, and the coupling of the fluid and the structure subproblems.

Among many other techniques developed so far, the finite element method is one of the most successful spatial discretization approaches for the solution of the partial differential equations (PDEs) underlying many physical phenomena, including fluid-structure interaction.
The standard continuous Galerkin method provides computationally efficient discretizations with a very limited number of degrees of freedom (DOFs) for the solution of elasticity problems.
For an overview on CG methods for solid mechanics, the interested reader is referred to \cite{Zienkiewicz2000}.
On the other hand, the interest in discontinuous Galerkin methods has increased over the last decades in the computational fluid dynamics community \cite{Persson2009,Krank2017,Fehn2019} because of their distinctive properties, such as the inherited stabilization of the convection terms in conservation laws, the ability to construct high order discretizations on unstructured meshes and the flexibility in performing $p$-adaptivity in addition to the classical $h$-adaptivity.
More recently, hybridizable discontinuous Galerkin methods have gained a lot of attention owing to their reduced computational costs with respect to classical matrix-based DG approaches, thanks to the reduced number of global DOFs in the associated linear systems, especially for high-degree polynomial approximations.
Moreover, the possibility to obtain a superconvergent solution through an efficient element-by-element postprocessing allows to obtain an improved approximation of the solution and to drive efficient degree adaptivity procedures \cite{Giorgiani2013,Giorgiani2014,Sevilla2018a}.
In the context of flow problems, the HDG method has been successfully applied for the discretization of fully compressible flows \cite{Peraire2010} as well as incompressible flows \cite{Giorgiani2014,Nguyen2011,Giacomini2018}.
The strong enforcement of the symmetry of the stress tensor via Voigt notation to retrieve the optimal convergence of the mixed variable and to ensure the superconvergence of the postprocessed solution without additional enrichment of the discrete spaces has been proposed in \cite{Giacomini2018,Sevilla2018} and it is exploited also in the proposed formulation for weakly compressible flows.
The solution of FSI problems with incompressible flows by means of the HDG method for both the fluid and the structure has been formulated in \cite{Sheldon2016,Sheldon2018}.
However, these formulations are computationally much more expensive than the one proposed here and they moreover fail to provide an optimal convergent structural strain field (and therefore a superconvergent displacement field), losing therefore one of the key advantages of the HDG method.

A successful coupling of DG and CG methods for the solution of fluid-structure interaction problems has been proposed in \cite{Froehle2014}, where the high-order accuracy in time given by the implicit-explicit Runge--Kutta method has been demonstrated on a non-trivial test problem.
As opposed to \cite{Froehle2014}, the formulation proposed here aims to couple the HDG method for the discretization of the fluid equations and the CG method for the solution of the structural problem and a high-order accuracy is demonstrated for the hybrid spatial discretization on a problem with manufactured solution.
A first attempt to couple HDG and CG discretizations has been proposed in \cite{Paipuri2019}, while an improved minimally-intrusive HDG-CG coupling has been formulated in \cite{Laspina2020b} for the solution of multi-material structural problems, involving compressible and nearly incompressible solids.

An important aspect in the simulation of multiphysics problems is the choice of an appropriate kinematical description.
Pure fluid problems are commonly solved with an Eulerian description, i.e., the computational mesh is fixed and the fluid moves with respect to the grid.
This approach facilitates the treatment of large distortion in the fluid motion and it is in particular indispensable in case of turbulent flows.
On the contrary, the Lagrangian description is usually used in structural mechanics and with this approach the nodes of the mesh follow the associated material particles during the motion, facilitating the tracking of free surfaces and interfaces between different materials.
Arbitrary Lagrangian--Eulerian (ALE) algorithms \cite{Donea2004} aim to combine the advantages of the classical kinematical descriptions by introducing a computational mesh which can move with a velocity independent of the velocity of the material particles.
This technique is particularly useful for flow problems in the presence of mobile and deforming boundaries, as it happens in fluid-structure interaction \cite{Wall2006,Mayr2015,Schott2018}.

In the formulation derived here, the Arbitrary Lagrangian--Eulerian method is used for the description of the fluid flow, while the total Lagrangian method is adopted for the description of the structural motion.

Regarding the coupling of the single fields, partitioned schemes solve one subproblem per time and exchange the interface information between the fluid and the structure.
The exchange of the interface state is performed just once per time step in the so-called loosely-coupled staggered approaches \cite{Farhat2006}, while the solution of the flow and structural problems are repeated within one time step until a convergence criterion is satisfied in the strongly-coupled staggered approaches \cite{Kuttler2008}.
The partitioned schemes may suffer several stability and convergence issues, but on the other hand they allow the use of well established and optimized single-field solvers.
In particular, these schemes are affected by a detrimental phenomenon, defined in literature {\lq\lq artificial added mass effect\rq\rq} and analyzed in \cite{Causin2005,Forster2007} in the context of incompressible flows.
In a recent work \cite{Laspina2020a}, it is analytically demonstrated how the introduction of a weak compressibility in the fluid formulation alleviates the constraints of the instability condition of the artificial added mass effect, thanks to the reduction of the maximal eigenvalue of the so-called added mass operator.
Moreover, in comparison to a fully incompressible solver, a significant reduction of the coupling iterations and the computational time is observed.
It is worth highlighting that alternative approaches to relax the incompressibility constraint based on the artificial compressibility, as the ones proposed in \cite{Bassi2006,Bassi2007} in the context of DG methods, provide a strongly consistent approximation of the incompressible Navier--Stokes equations.
Embedding such fluid solvers in a partitioned FSI code would not offer any beneficial contribution against the artificial added mass effect.
A general framework for constructing high-order partitioned solvers based on implicit-explicit Runge--Kutta methods for the solution of multiphysics problems (including fluid-structure interaction) has been introduced in \cite{Huang2019}, where four consistent predictors are proposed, leading to different partitioned solvers that preserve the theoretical order of accuracy of the temporal integration scheme.
As opposed to the partitioned strategies, in a monolithic framework \cite{Mayr2015,Heil2004}, a unique solver is in charge of the solution of the complete system of nonlinear equations.
This approach exhibits a higher robustness and it is usually faster compared to partitioned strategies in challenging cases, but it requires an ad hoc implementation.
In addition, the adoption of efficient preconditioners is often needed for the solution of computationally demanding problems.

In the present contribution, a strongly-coupled staggered scheme based on the Dirichlet--Neumann partitioning is revisited for the hybrid HDG-CG coupling and a novel FSI monolithic approach based on Nitsche's method is also introduced.
It is worth noting that the partitioned algorithm adopted here considers a strong Gauss--Seidel-type predictor, which is acknowledged in \cite{Huang2019} to be most stable among the techniques therein analyzed.
A weak compressibility is then considered in the fluid formulation in order to provide an improved robustness and efficiency for the coupled solver.

The present article is organized as follows.
First, the novel HDG formulation for weakly compressible flows is derived in section \ref{sec:hdgformulationweaklycompressibleflows} for pure fluid problems with an Arbitrary Lagrangian--Eulerian description.
In section \ref{sec:hdgcgformulationcoupledFSI} the standard CG formulation for nonlinear elastodynamics is briefly recalled and the two HDG-CG coupling strategies for fluid-structure interaction are presented.
Section \ref{sec:numericalstudies} is devoted to the numerical validation of the pure fluid formulation first and the coupling strategies for fluid-structure interaction afterwards.
Finally, in section \ref{sec:conclusion}, the results of this work are summarized.

\section{HDG formulation for weakly compressible flows}
\label{sec:hdgformulationweaklycompressibleflows}

In this section, the governing equations of unsteady weakly compressible flows are first presented with regards to fixed domains and then formulated according to the Arbitrary Lagrangian--Eulerian description to deal with moving domains.
A brief overview on Voigt notation is provided to handle symmetric tensors and the HDG formulation of the local and global problems is derived together with a postprocessing procedure to construct an improved approximation of the solution.

\subsection{Governing equations}
\label{sec:governingequations}

Let $\Omega_x\in\mathbb{R}^{\texttt{n}_\texttt{sd}}$ be a fixed open bounded domain with boundary $\partial\Omega_x=\Gamma_x^D\cup\Gamma_x^N$ with $\Gamma_x^D\cap\Gamma_x^N=\emptyset$ and $\texttt{n}_\texttt{sd}$ being the number of spatial dimensions and let $\textrm{T}_{\texttt{end}}>0$ be the final time of interest.
The governing equations of time-dependent weakly compressible flows read:
\begin{equation}
\left\{
\begin{aligned}
\dfrac{\partial\rho}{\partial t}+\bm{\nabla}\cdot\left(\rho\bm{\upsilon}\right)&=0
\quad&&\text{in }\Omega_x\times\left(0,\textrm{T}_{\texttt{end}}\right), \\
\dfrac{\partial\rho\bm{\upsilon}}{\partial t}+\bm{\nabla}\cdot\left(\rho\bm{\upsilon}\otimes\bm{\upsilon}\right)-\bm{\nabla}\cdot\boldsymbol{\sigma}&=\rho\mathbf{b}
\quad&&\text{in }\Omega_x\times\left(0,\textrm{T}_{\texttt{end}}\right), \\
p\left(\rho\right)&=0
\quad&&\text{in }\Omega_x\times\left(0,\textrm{T}_{\texttt{end}}\right),
\end{aligned}
\right.
\label{eqn:fluidgoverningequationsfixeddomain}
\end{equation}
where $\rho$ represents the fluid density, $p$ the pressure, $\bm{\upsilon}$ the velocity field, $\boldsymbol{\sigma}$ the Cauchy stress tensor and $\mathbf{b}$ an applied body force.
For a Newtonian fluid, it is assumed that the stress tensor and the augmented strain rate tensor are linearly related, therefore
\begin{equation}
\boldsymbol{\sigma}=-p\mathbf{I}_{\texttt{n}_\texttt{sd}}+2\mu\bm{\nabla}^\texttt{S}\bm{\upsilon}+\lambda\left(\bm{\nabla}\cdot\bm{\upsilon}\right)\mathbf{I}_{\texttt{n}_\texttt{sd}}.
\label{eqn:fluidstresstensor}
\end{equation}
The operator $\bm{\nabla}^\texttt{S}:=\frac{1}{2}\left(\bm{\nabla}+\bm{\nabla}^T\right)$ returns the symmetric part of the gradient while $\mathbf{I}_{\texttt{n}_\texttt{sd}}$ denotes the $\texttt{n}_\texttt{sd}\times \texttt{n}_\texttt{sd}$ identity matrix.
In equation \eqref{eqn:fluidstresstensor}, $\mu$ is the dynamic viscosity and $\lambda$ the so-called second coefficient of viscosity.
Stokes' hypothesis states the following relationship between the two material variables:
\begin{equation}
\lambda=-\dfrac{2}{3}\mu.
\label{eqn:fluidstokeshypothesis}
\end{equation}
The conservation of energy is taken into account by an equation of state $p\left(\rho\right)=0$ defining a relationship between the fluid pressure and the density.
For weakly compressible Newtonian fluids, a linear relationship is used \cite{Venerus2006}:
\begin{equation}
\rho=\rho_0+\varepsilon\left(p-p_0\right),
\label{eqn:fluidequationofstate}
\end{equation}
where $\varepsilon$ is a (small) constant isothermal compressibility coefficient, while $\rho_0$ denotes the mass density evaluated at the reference pressure $p_0$.
Equation \eqref{eqn:fluidequationofstate} has been considered for numerical simulations of weakly compressible flows in long tubes, such as waxy crude oil \cite{Vinay2006} and polymer extrusion \cite{Taliadorou2008}.
Though equation \eqref{eqn:fluidequationofstate} is very attractive for its simplicity, other pressure-density relations have been used in literature, like the Murnaghan--Tait model \cite{Laspina2020a}.

In order to derive the Arbitrary Lagrangian--Eulerian description of the flow, the ALE convective velocity, defined as the velocity of the fluid relative to the moving background mesh (whose velocity is indicated here with $\vect{a}$), is introduced:
\begin{equation}
\mathbf{c}=\bm{\upsilon}-\vect{a}.
\label{eqn:fluidaleconvectivevelocity}
\end{equation}
Using the ALE time derivative (i.e., the time derivative with respect to the reference configuration), the governing equations of the fluid problem under analysis on a deforming domain $\Omega$ can be written as:
\begin{equation}
\left\{
\begin{aligned}
\dfrac{\partial\rho}{\partial t}+\rho\bm{\nabla}\cdot\vect{a}+\bm{\nabla}\cdot\left(\rho\mathbf{c}\right)&=0
\quad&&\text{in }\Omega\times\left(0,\textrm{T}_{\texttt{end}}\right), \\
\begin{aligned}[b]
\dfrac{\partial\rho\bm{\upsilon}}{\partial t}+\rho\bm{\upsilon}\bm{\nabla}\cdot\vect{a}+\bm{\nabla}\cdot\left(\rho\bm{\upsilon}\otimes\mathbf{c}\right)&\\-\bm{\nabla}\cdot\boldsymbol{\sigma}&
\end{aligned}&=\rho\mathbf{b}
\quad&&\text{in }\Omega\times\left(0,\textrm{T}_{\texttt{end}}\right), \\
p\left(\rho\right)&=0
\quad&&\text{in }\Omega\times\left(0,\textrm{T}_{\texttt{end}}\right), \\
\rho&=\rho^0
\quad&&\text{in }\Omega\times\left(0\right), \\
\rho\bm{\upsilon}&=\rho\bm{\upsilon}^0
\quad&&\text{in }\Omega\times\left(0\right), \\
\rho&=\rho^D
\quad&&\text{on }\Gamma^D\times\left(0,\textrm{T}_{\texttt{end}}\right), \\
\rho\bm{\upsilon}&=\rho\bm{\upsilon}^D
\quad&&\text{on }\Gamma^D\times\left(0,\textrm{T}_{\texttt{end}}\right), \\
\boldsymbol{\sigma}\vect{n}&=\mathbf{t}^N
\quad&&\text{on }\Gamma^N\times\left(0,\textrm{T}_{\texttt{end}}\right).
\end{aligned}
\right.
\label{eqn:fluidgoverningequationsmovingdomain}
\end{equation}
The pair $\left(\rho^0,\rho\bm{\upsilon}^0\right)$ defines the initial conditions for the density and the momentum fields, while the quantities $\left(\rho^D,\rho\bm{\upsilon}^D\right)$ and $\mathbf{t}^N$ denote the Dirichlet and the Neumann boundary data applied on $\Gamma^D$ and $\Gamma^N$, respectively.
Finally, $\vect{n}$ denotes the outward-pointing unit normal vector to the corresponding boundary.

Four types of boundary conditions are considered, namely inflow, outflow, no-slip and free-slip conditions.
At the inflow, the momentum profile is imposed via a Dirichlet boundary condition.
At the outflow, the pressure (and therefore the density according to \eqref{eqn:fluidequationofstate}) is set to the given data, while the other quantities are extrapolated.
For the no-slip condition, each velocity component (and therefore each momentum component) is forced to be zero via a Dirichlet boundary condition, while the density is extrapolated.
For the free-slip condition instead, only the normal component of the velocity is set to zero, while the tangential component remains unconstrained.

\subsection{HDG local and global problems}
\label{sec:hdglocalglobalproblems}

In this section, a hybridizable discontinuous Galerkin method is proposed for the solution of weakly compressible flow problems satisfying equations \eqref{eqn:fluidgoverningequationsmovingdomain}.
The so-called broken computational domain is defined by partitioning the fluid domain $\Omega$ in $\texttt{n}^\texttt{el}$ disjoint subdomains $\Omega^e$:
\begin{equation}
\Omega=\bigcup_{e=1}^{\texttt{n}^\texttt{el}}\Omega^e,
\quad \Omega^e\cap\Omega^f=\emptyset\text{ for }e\neq f.
\label{eqn:fluidpartition}
\end{equation}
The internal element boundaries define the internal interface
\begin{equation}
\Gamma:=\left[\bigcup_{e=1}^{\texttt{n}^\texttt{el}}\partial\Omega^e\right]\setminus\partial\Omega
\label{eqn:fluidskeleton}
\end{equation}
and the union of the internal interface with the boundary faces belonging to $\Gamma^N$ constitutes the mesh skeleton, on which the hybrid variables are defined.

The $\mathcal{L}_2$ scalar products for vector-valued functions in the elements and in their boundaries are denoted in the following as:
\begin{equation}
\left(\vect{w},\bm{\upsilon}\right)_{\Omega^e}:=
\int_{\Omega^e}\vect{w}\cdot\bm{\upsilon}d\Omega, \quad\quad
\left<\vect{w},\bm{\upsilon}\right>_{\partial\Omega^e}:=
\sum_{\Gamma_i\subset\partial\Omega^e}\int_{\Gamma_i}\vect{w}\cdot\bm{\upsilon}d\Gamma.
\label{eqn:fluidinternalproducts}
\end{equation}
Given the discontinuous character of the HDG variables, the jump operator $\left\llbracket\cdot\right\rrbracket$ sums values from two adjacent elements $\Omega^e$ and $\Omega^f$ \cite{Montlaur2018}:
\begin{equation}
\left\llbracket\odot\right\rrbracket=\odot^e+\odot^f.
\label{eqn:fluidjumpoperator}
\end{equation}

A key ingredient to preserve the convergence properties of the HDG method and to allow the use of the same polynomial degree for the approximation of the primal and the mixed variables without loss of accuracy is the adoption of the well-known Voigt notation \cite{Giacomini2018,Sevilla2018}.
The Voigt notation allows to strongly enforce the symmetry of the stress tensor in \eqref{eqn:fluidstresstensor} by rearranging its diagonal and off-diagonal terms (according to the ordering in \cite{Fish2007}) and storing only the $\texttt{m}_\texttt{sd}$ independent components, with $\texttt{m}_\texttt{sd}=\texttt{n}_\texttt{sd}(\texttt{n}_\texttt{sd}+1)/2$.
The dynamic viscosity and the second coefficient of viscosity can be embedded in the matrix
\begin{equation}
\mat{D}:=
\begin{bmatrix}
2\mu\mathbf{I}_{\texttt{n}_\texttt{sd}}+\lambda\mathbf{J}_{\texttt{n}_\texttt{sd}} & \mathbf{0}_{\texttt{n}_\texttt{sd}\times \texttt{q}_\texttt{sd}} \\
                  \mathbf{0}_{\texttt{q}_\texttt{sd}\times \texttt{n}_\texttt{sd}} & \mu\mathbf{I}_{\texttt{q}_\texttt{sd}}
\end{bmatrix},
\label{eqn:voigtD}
\end{equation}
with $\texttt{q}_\texttt{sd}=\texttt{m}_\texttt{sd}-\texttt{n}_\texttt{sd}$ and $\mathbf{J}_{\texttt{n}_\texttt{sd}}$ denoting the $\texttt{n}_\texttt{sd}\times \texttt{n}_\texttt{sd}$ matrix of all ones.
Hence, introducing the operator
\begin{equation}
\gradS:=
\left\{
\begin{aligned}
&\begin{bmatrix}
\partial/\partial x &                   0 & \partial/\partial y \\
                  0 & \partial/\partial y & \partial/\partial x
\end{bmatrix}^T
&&\text{in 2D},
\\
&\begin{bmatrix}
\partial/\partial x &                   0 &                   0 & \partial/\partial y & \partial/\partial z &                   0 \\
                  0 & \partial/\partial y &                   0 & \partial/\partial x &                   0 & \partial/\partial z \\
                  0 &                   0 & \partial/\partial z &                   0 & \partial/\partial x & \partial/\partial y  
\end{bmatrix}^T
&&\text{in 3D},
\end{aligned}
\right.
\label{eqn:voigtnablaS}
\end{equation}
and the vector
\begin{equation}
\mat{E}:=
\begin{bmatrix}
\mathbf{1}_{\texttt{n}_\texttt{sd}\times1} \\
\mathbf{0}_{\texttt{q}_\texttt{sd}\times1}
\end{bmatrix},
\label{eqn:voigtE}
\end{equation}
the stress tensor can be expressed in Voigt notation as:
\begin{equation}
\boldsymbol{\sigma}_\texttt{V}=-\mat{E}p\left(\rho\right)+\mat{D}\gradS\bm{\upsilon}.
\label{eqn:voigtstress}
\end{equation}
The normal component of the stress can be computed pre-multiplying $\boldsymbol{\sigma}_\texttt{V}$  by $\vect{n}^T$, with
\begin{equation}
\vect{n}:=
\left\{
\begin{aligned}
&\begin{bmatrix}
n_x &   0 & n_y \\
  0 & n_y & n_x
\end{bmatrix}^T
&&\text{in 2D},
\\
&\begin{bmatrix}
n_x &   0 &   0 & n_y & n_z &   0 \\
  0 & n_y &   0 & n_x &   0 & n_z \\
  0 &   0 & n_z &   0 & n_x & n_y
\end{bmatrix}^T
&&\text{in 3D}.
\end{aligned}
\right.
\label{eqn:voigtN}
\end{equation}

In HDG methods, the solution of the overall problem is split into two phases \cite{Nguyen2011,Cockburn2008,Cockburn2009a,Nguyen2009,Nguyen2009a,Nguyen2010}.
In the first stage, a local element-by-element Dirichlet problem is introduced to compute $\left(\mat{L},\rho,\rho\bm{\upsilon}\right)$ as a function of the unknown hybrid variables $\left(\hat{\rho},\widehat{\rho\bm{\upsilon}}\right)$.
With the notation just introduced, the local problems can be written as:
\begin{equation}
\left\{
\begin{aligned}
\mat{L}+\mat{D}^{\toh}\gradS\bm{\upsilon}&=\mathbf{0}
\quad&&\text{in }\Omega^e\times\left(0,\textrm{T}_{\texttt{end}}\right), \\
\dfrac{\partial\rho}{\partial t}+\rho\bm{\nabla}\cdot\vect{a}+\bm{\nabla}\cdot\left(\rho\mathbf{c}\right)&=0
\quad&&\text{in }\Omega^e\times\left(0,\textrm{T}_{\texttt{end}}\right), \\
\begin{aligned}[b]
\dfrac{\partial\rho\bm{\upsilon}}{\partial t}+\rho\bm{\upsilon}\bm{\nabla}\cdot\vect{a}+\bm{\nabla}\cdot\left(\rho\bm{\upsilon}\otimes\mathbf{c}\right)&\\+\gradS^T\left(\mat{D}^{\toh}\mat{L}+\mat{E}p\left(\rho\right)\right)&
\end{aligned}&=\rho\mathbf{b}
\quad&&\text{in }\Omega^e\times\left(0,\textrm{T}_{\texttt{end}}\right), \\
\rho&=\rho^0
\quad&&\text{in }\Omega^e\times\left(0\right), \\
\rho\bm{\upsilon}&=\rho\bm{\upsilon}^0
\quad&&\text{in }\Omega^e\times\left(0\right), \\
\rho&=\rho^D
\quad&&\text{on }\partial\Omega^e\cap\Gamma^D\times\left(0,\textrm{T}_{\texttt{end}}\right), \\
\rho\bm{\upsilon}&=\rho\bm{\upsilon}^D
\quad&&\text{on }\partial\Omega^e\cap\Gamma^D\times\left(0,\textrm{T}_{\texttt{end}}\right), \\
\rho&=\hat{\rho}
\quad&&\text{on }\partial\Omega^e\setminus\Gamma^D\times\left(0,\textrm{T}_{\texttt{end}}\right), \\
\rho\bm{\upsilon}&=\widehat{\rho\bm{\upsilon}}
\quad&&\text{on }\partial\Omega^e\setminus\Gamma^D\times\left(0,\textrm{T}_{\texttt{end}}\right),
\end{aligned}
\right.
\label{eqn:fluidstrongformlocal}
\end{equation}
for $e=1,\dots,\texttt{n}^\texttt{el}$.
The variable $\mat{L}$ denotes the aforementioned mixed variable, that allows to reduce the second-order problem \eqref{eqn:fluidgoverningequationsmovingdomain} to a system of first-order equations.
In the second stage, the traces of the density $\hat{\rho}$ and the momentum $\widehat{\rho\bm{\upsilon}}$ are computed through the solution of the global problem:
\begin{equation}
\left\{
\begin{aligned}
\left\llbracket\rho\vect{n}\right\rrbracket&=\mathbf{0}
\quad&&\text{on }\Gamma\times\left(0,\textrm{T}_{\texttt{end}}\right), \\
\left\llbracket\rho\bm{\upsilon}\otimes\vect{n}\right\rrbracket&=\mathbf{0}
\quad&&\text{on }\Gamma\times\left(0,\textrm{T}_{\texttt{end}}\right), \\
\left\llbracket\rho\mathbf{c}\otimes\vect{n}\right\rrbracket&=\mathbf{0}
\quad&&\text{on }\Gamma\times\left(0,\textrm{T}_{\texttt{end}}\right), \\
\left\llbracket\left(\rho\bm{\upsilon}\otimes\mathbf{c}\right)\vect{n}\right\rrbracket&=\mathbf{0}
\quad&&\text{on }\Gamma\times\left(0,\textrm{T}_{\texttt{end}}\right), \\
\left\llbracket\vect{n}^T\left(\mat{D}^{\toh}\mat{L}+\mat{E}p\left(\rho\right)\right)\right\rrbracket&=\mathbf{0}
\quad&&\text{on }\Gamma\times\left(0,\textrm{T}_{\texttt{end}}\right), \\
-\vect{n}^T\left(\mat{D}^{\toh}\mat{L}+\mat{E}p\left(\rho\right)\right)&=\mathbf{t}^N
\quad&&\text{on }\Gamma^N\times\left(0,\textrm{T}_{\texttt{end}}\right).
\end{aligned}
\right.
\label{eqn:fluidstrongformglobal}
\end{equation}
These transmission conditions enforce the continuity of the primal variables $\rho$ and $\rho\bm{\upsilon}$ and the normal fluxes across the interface $\Gamma$.
The first two equations in \eqref{eqn:fluidstrongformglobal} are automatically satisfied due to the unique definition of the hybrid variables on each face of the mesh skeleton.
Moreover, it is worth noting that, given the continuous nature of the velocity $\vect{a}$ of the moving background mesh, as opposed to the other HDG variables, no additional conditions have to be enforced in the global problem.

The following discrete functional spaces are introduced to derive the weak form of the problem:
\begin{subequations}
\begin{align}
\mathcal{W}^h\left(\Omega\right):&=
\{w\in\mathcal{L}^2\left(\Omega\right):w\vert_{\Omega^e}\in\mathcal{P}^k\left(\Omega^e\right)\forall\Omega^e\subset\Omega\},
\\
\widehat{\mathcal{W}}^h\left(S\right):&=
\{\hat{w}\in\mathcal{L}^2\left(S\right):\hat{w}\vert_{\Gamma^i}\in\mathcal{P}^k\left(\Gamma^i\right)\forall\Gamma^i\subset S\subseteq\Gamma\cup\partial\Omega\},
\end{align}
\label{eqn:fluidspaces}%
\end{subequations}
where $\mathcal{P}^k\left(\Omega^e\right)$ and $\mathcal{P}^k\left(\Gamma^i\right)$ denote the spaces of polynomials of complete degree at most $k$ in $\Omega^e$ and on $\Gamma^i$, respectively.
The trace of the numerical normal fluxes, arising from the integration by parts of the terms under the divergence operator in \eqref{eqn:fluidstrongformlocal}, are defined as follows:
\begin{subequations}\label{eqn:fluidnumericalflux}%
\begin{align}
\reallywidehat{\rho\mathbf{c}\cdot\vect{n}}:= &\begin{cases}
  \left(\rho\bm{\upsilon}^D-\rho^D\vect{a}\right)\cdot\vect{n}+\tau_\rho\left(\rho-\rho^D\right)
  &\text{on }\partial\Omega^e\cap\Gamma^D, \\
  \left(\widehat{\rho\bm{\upsilon}}-\hat{\rho}\vect{a}\right)\cdot\vect{n}+\tau_\rho\left(\rho-\hat{\rho}\right)
 &\text{on }\partial\Omega^e\setminus\Gamma^D,
\end{cases}
\\
\reallywidehat{\left(\rho\bm{\upsilon}\otimes\mathbf{c}\right)\vect{n}}:= &\begin{cases}
  \Bigl[\rho\bm{\upsilon}^D{\otimes}\bigl((\rho\bm{\upsilon}^D/\rho^D){-}\vect{a}\bigr)\Bigr]\vect{n}
     +\tau_{\rho\upsilon}^c\left(\rho\bm{\upsilon}-\rho\bm{\upsilon}^D\right)
  &\text{on }\partial\Omega^e\cap\Gamma^D, \\[1ex]
  \Bigl[\widehat{\rho\bm{\upsilon}}{\otimes}\bigl((\widehat{\rho\bm{\upsilon}}/\hat{\rho}){-}\vect{a}\bigr)\Bigr]\vect{n}
    +\tau_{\rho\upsilon}^c\left(\rho\bm{\upsilon}-\widehat{\rho\bm{\upsilon}}\right)
  &\text{on }\partial\Omega^e\setminus\Gamma^D,
\end{cases}
\\
\reallywidehat{\vect{n}^T\bigl(\mat{D}^{\toh}\mat{L}{+}\mat{E}p\left(\rho\right)\bigr)}:= &\begin{cases}
  \vect{n}^T\left(\mat{D}^{\toh}\mat{L}+\mat{E}p\left(\rho^D\right)\right)
    +\tau_{\rho\upsilon}^d\left(\rho\bm{\upsilon}-\rho\bm{\upsilon}^D\right)
  &\text{on }\partial\Omega^e\cap\Gamma^D, \\
  \vect{n}^T\left(\mat{D}^{\toh}\mat{L}+\mat{E}p\left(\hat{\rho}\right)\right)
    +\tau_{\rho\upsilon}^d\left(\rho\bm{\upsilon}-\widehat{\rho\bm{\upsilon}}\right)
    &\text{on }\partial\Omega^e\setminus\Gamma^D.
\end{cases}
\end{align}
\end{subequations}
These definitions are rather standard in the context of HDG methods and are closely related to other formulations published in literature.
More precisely, the definition of the trace in the continuity equation is analogous to the one adopted for the solution of convection-diffusion problems in \cite{Nguyen2009}, whereas the definition of the traces in the momentum equation follows the one proposed for the solution of flow problems in \cite{Giacomini2018,Giacomini2020}.
However, the specific form of the definitions adopted in the context of the density-momentum formulation of the governing equations on deforming domains is original.
The stabilization parameters $\tau_\rho$, $\tau_{\rho\upsilon}^c$ and $\tau_{\rho\upsilon}^d$ account for the compressibility, the convection and the diffusion effects, respectively, and they play a crucial role on the stability and the convergence of the HDG method \cite{Cockburn2008,Cockburn2009a,Soon2009}.
Dimensional analysis provides a practical choice for the stabilization parameters:
\begin{equation}
\tau_\rho=C_\rho\dfrac{1}{\varepsilon\left|\bm{\upsilon}\right|},
\quad\quad\quad
\tau_{\rho\upsilon}^c=C_{\rho\upsilon}^c\left|\bm{\upsilon}\right|,
\quad\quad\quad
\tau_{\rho\upsilon}^d=C_{\rho\upsilon}^d\dfrac{\mu}{\rho_0l},
\label{eqn:fluidstabilizationparameters}
\end{equation}
with $\left|\bm{\upsilon}\right|$ and $l$ being a representative flow velocity and length scale, respectively, and $C_\rho$, $C_{\rho\upsilon}^c$ and $C_{\rho\upsilon}^d$ denoting suitable positive scaling factors.
It is empirically observed that choosing the scaling factors in the range $(1,10)$ provides a good balance for the quality of the approximation of the primal, the mixed and the postprocessed variables, regardless of the polynomial degree, the type of element and the dimensionality of the problem.
These considerations are in agreement with the established results in the HDG literature \cite{Giacomini2018,Sevilla2018,Cockburn2008,Kirby2012}.
Without loss of generality, a unique parameter $\tau_{\rho\upsilon}=\tau_{\rho\upsilon}^c+\tau_{\rho\upsilon}^d$, taking into account both the convection and the diffusion effects, will be considered in the following.

With this definition of the numerical fluxes and expliciting all the unknowns, the discrete weak form of the local problems \eqref{eqn:fluidstrongformlocal} reads: given $\left(\rho^0,\rho\bm{\upsilon}^0\right)$ in $\Omega^e\times\left(0\right)$, $\left(\rho^D,\rho\bm{\upsilon}^D\right)$ on $\Gamma^D$ and $\left(\hat{\rho}^h,\widehat{\rho\bm{\upsilon}}^h\right)$ on $\Gamma\cup\Gamma^N$, find $\left(\mat{L}^h,\rho^h,\rho\bm{\upsilon}^h\right)\in\left[\mathcal{W}^h\left(\Omega^e\right)\right]^{\texttt{m}_\texttt{sd}}\times\mathcal{W}^h\left(\Omega^e\right)\times\left[\mathcal{W}^h\left(\Omega^e\right)\right]^{\texttt{n}_\texttt{sd}}$ for $e=1,\dots,\texttt{n}^\texttt{el}$ such that
\begin{subequations}\label{eqn:fluidweakformlocal}%
\begin{multline}
  {-}\left(\mat{L},\mat{L}^h\right)_{\Omega^e}
  +\Bigl(\gradS^T\mat{D}^{\toh}\mat{L},\frac{\rho\bm{\upsilon}^h}{\rho^h}\Bigr)_{\Omega^e} \\
  = 
  \Bigl<\vect{n}^T\mat{D}^{\toh}\mat{L},\dfrac{\rho\bm{\upsilon}^D}{\rho^D}\Bigr>_{\partial\Omega^e\cap\Gamma^D}
  +\Bigl<\vect{n}^T\mat{D}^{\toh}\mat{L},\dfrac{\widehat{\rho\bm{\upsilon}}^h}{\hat{\rho}^h}\Bigr>_{\partial\Omega^e\setminus\Gamma^D},
\end{multline}
\begin{multline}
  \Bigl(w,\frac{\partial\rho^h}{\partial t}\Bigr)_{\Omega^e}
  +\left(w,\rho^h\bm{\nabla}\cdot\vect{a}\right)_{\Omega^e} 
  -\left(\bm{\nabla}w,\rho\bm{\upsilon}^h-\rho^h\vect{a}\right)_{\Omega^e}
  +\left<w,\tau_\rho\rho^h\right>_{\partial\Omega^e} \\
  = 
  -\left<w,\left(\rho\bm{\upsilon}^D-\rho^D\vect{a}\right)\cdot\vect{n}-\tau_\rho\rho^D\right>_{\partial\Omega^e\cap\Gamma^D} \\
  -\left<w,\left(\widehat{\rho\bm{\upsilon}}^h-\hat{\rho}^h\vect{a}\right)\cdot\vect{n}-\tau_\rho\hat{\rho}^h\right>_{\partial\Omega^e\setminus\Gamma^D},
\end{multline}
\begin{multline}
  \Bigl(\vect{w},\frac{\partial\rho\bm{\upsilon}^h}{\partial t}\Bigr)_{\Omega^e}
  +\left(\vect{w},\rho\bm{\upsilon}^h\bm{\nabla}\cdot\vect{a}\right)_{\Omega^e}
  -\left(\bm{\nabla}\vect{w},\rho\bm{\upsilon}^h\otimes\left(\dfrac{\rho\bm{\upsilon}^h}{\rho^h}-\vect{a}\right)\right)_{\Omega^e} \\
  +\bigl(\vect{w},\gradS^T\bigl[\mat{D}^{\toh}\mat{L}^h+\mat{E}p(\rho^h)\bigr]\bigr)_{\Omega^e} 
  +\left<\vect{w},\tau_{\rho\upsilon}\rho\bm{\upsilon}^h\right>_{\partial\Omega^e}
  -\left(\vect{w},\rho^h\mathbf{b}\right)_{\Omega^e} \\
  = 
  \Bigl<\vect{w},\bigl[\rho\bm{\upsilon}^D\otimes\bigl(\frac{\rho\bm{\upsilon}^D}{\rho^D}-\vect{a}\bigr)\bigr]\vect{n}-\tau_{\rho\upsilon}\rho\bm{\upsilon}^D\Bigr>_{\partial\Omega^e\cap\Gamma^D} \\
 \Bigl<\vect{w},\bigl[\widehat{\rho\bm{\upsilon}}^h\otimes\bigl(\frac{\widehat{\rho\bm{\upsilon}}^h}{\hat{\rho}^h}-\vect{a}\bigr)\bigr]\vect{n}-\tau_{\rho\upsilon}\widehat{\rho\bm{\upsilon}}^h\Bigr>_{\partial\Omega^e\setminus\Gamma^D},
\end{multline}
\end{subequations}
for all $(\mat{L},w,\vect{w})\in[\mathcal{W}^h(\Omega^e)]^{\texttt{m}_\texttt{sd}}\times\mathcal{W}^h(\Omega^e)\times[\mathcal{W}^h(\Omega^e)]^{\texttt{n}_\texttt{sd}}$.

The discrete weak form of the global problem \eqref{eqn:fluidstrongformglobal} instead reads: find $(\hat{\rho}^h,\widehat{\rho\bm{\upsilon}}^h)\in\widehat{\mathcal{W}}^h(\Gamma\cup\Gamma^N)\times[\widehat{\mathcal{W}}^h(\Gamma\cup\Gamma^N)]^{\texttt{n}_\texttt{sd}}$ such that
\begin{subequations}\label{eqn:fluidweakformglobal}%
\begin{gather}
  \sum_{e=1}^{\texttt{n}^\texttt{el}}\bigl<\hat{w},\tau_\rho (\rho^h-\hat{\rho}^h) \bigr>_{\partial\Omega^e\setminus\Gamma^D} = 0,
\\
 -\sum_{e=1}^{\texttt{n}^\texttt{el}}\bigl<\hat{\vect{w}},\vect{n}^T\bigl(\mat{D}^{\toh}\mat{L}^h+\mat{E}p(\hat{\rho}^h)\bigr)+\tau_{\rho\upsilon}(\rho\bm{\upsilon}^h-\widehat{\rho\bm{\upsilon}}^h)\bigr>_{\partial\Omega^e\setminus\Gamma^D}
  = \sum_{e=1}^{\texttt{n}^\texttt{el}}\left<\hat{\vect{w}},\mathbf{t}^N\right>_{\partial\Omega^e\cap\Gamma^N},
\end{gather}
\end{subequations}
for all $(\hat{w},\hat{\vect{w}})\in\widehat{\mathcal{W}}^h(\Gamma\cup\Gamma^N)\times[\widehat{\mathcal{W}}^h(\Gamma\cup\Gamma^N)]^{\texttt{n}_\texttt{sd}}$.

\subsection{Local postprocessing}
\label{sec:localpostprocessing}

A key feature of the HDG method is the possibility to exploit the optimal convergence of the mixed variable in order to construct a better approximation of the solution, converging in a superoptimal fashion.
In HDG methods, the postprocessed variable is usually the same physical quantity as the primal variable given by the solution of the local problems, for instance the temperature in thermal problems, the displacement in elasticity problems \cite{Sevilla2018,Laspina2020b,Soon2009,Cockburn2012} and the velocity in incompressible flow problems \cite{Nguyen2011,Giacomini2018,Nguyen2010}.
The local postprocessing proposed in this contribution allows to construct a superconvergent velocity field $\bm{\upsilon}^\star$, although the primal variables are represented by the density $\rho$ and the momentum $\rho\bm{\upsilon}$.
To the best of the authors' knowledge, such feature is not present in any HDG formulation presented so far.
The additional functional spaces are introduced:
\begin{subequations}
\begin{align}
{\mathcal{W}^\star}^h(\Omega):&=
\{w^\star\in\mathcal{L}^2(\Omega):w^\star\vert_{\Omega^e}\in\mathcal{P}^{k+1}(\Omega^e)\forall\Omega^e\subset\Omega\},
\\
\mathcal{U}^h(\Omega):&=
\{u\in\mathcal{L}^2(\Omega):u\vert_{\Omega^e}\in\mathcal{P}^{0}(\Omega^e)\forall\Omega^e\subset\Omega\}.
\end{align}
\label{eqn:fluidspacespostprocessing}%
\end{subequations}

The discrete weak form of the local postprocessing then reads: given $(\mat{L}^h,\rho^h,\rho\bm{\upsilon}^h)$ in $\Omega^e$, $(\rho^D,\rho\bm{\upsilon}^D)$ on $\Gamma^D$ and $(\hat{\rho}^h,\widehat{\rho\bm{\upsilon}}^h)$ on $\Gamma\cup\Gamma^N$, find ${\bm{\upsilon}^\star}^h\in[{\mathcal{W}^\star}^h(\Omega^e)]^{\texttt{n}_\texttt{sd}}$ for $e=1,\dots,\texttt{n}^\texttt{el}$ such that
\begin{subequations}\label{eqn:fluidweakformpostprocessing}%
\begin{align}
  -\Bigl(\gradS\vect{w}^\star,\mat{D}^{\toh}\gradS{\bm{\upsilon}^\star}^h\Bigr)_{\Omega^e}
  &= \bigl(\gradS\vect{w}^\star,\mat{L}^h\bigr)_{\Omega^e},
\\
  \bigl(\mathbf{u}_\texttt{T},{\bm{\upsilon}^\star}^h\bigr)_{\Omega^e}
  &=\Bigl(\mathbf{u}_\texttt{T},\dfrac{\rho\bm{\upsilon}^h}{\rho^h}\Bigr)_{\Omega^e},
\\
  \bigl(\mathbf{u}_\texttt{R},\gradW{\bm{\upsilon}^\star}^h\bigr)_{\Omega^e}
  &= \Bigl\langle\mathbf{u}_\texttt{R},\mathbf{T}\dfrac{\rho\bm{\upsilon}^D}{\rho^D}\Bigl\rangle_{\partial\Omega^e\cap\Gamma^D} 
    + \Bigl\langle\mathbf{u}_\texttt{R},\mathbf{T}\dfrac{\widehat{\rho\bm{\upsilon}}^h}{\hat{\rho}^h}\Bigl\rangle_{\partial\Omega^e\setminus\Gamma^D},
\end{align}
\end{subequations}
for all $(\vect{w}^\star,\mathbf{u}_\texttt{T},\mathbf{u}_\texttt{R})\in[{\mathcal{W}^\star}^h(\Omega^e)]^{\texttt{n}_\texttt{sd}}\times[{\mathcal{U}}^h(\Omega^e)]^{\texttt{n}_\texttt{sd}}\times[{\mathcal{U}}^h(\Omega^e)]^{\texttt{q}_\texttt{sd}}$.
The vorticity operator in \eqref{eqn:fluidweakformpostprocessing} is defined in Voigt notations as
\begin{equation}
\gradW:=
\begin{cases}
 \begin{bmatrix}-\partial/\partial y &  \partial/\partial x \end{bmatrix}  &\text{in 2D},
\\
 \begin{bmatrix}                0 & -\partial/\partial z &  \partial/\partial y \\
                 \partial/\partial z &                    0 & -\partial/\partial x \\
                -\partial/\partial y &  \partial/\partial x &                    0 \end{bmatrix} &\text{in 3D},
\end{cases}
\label{eqn:voigtnablaW}
\end{equation}
while the matrix $\mathbf{T}$ accounts for the tangential direction to the boundary and it is defined as
\begin{equation}
\mathbf{T}:=
\begin{cases}
  \begin{bmatrix} -n_y &  n_x \end{bmatrix} &\text{in 2D},
\\
  \begin{bmatrix}   0 & -n_z &  n_y \\
                         n_z &      0 & -n_x \\
                        -n_y &  n_x &    0   \end{bmatrix} &\text{in 3D}.
\end{cases}
\label{eqn:voigtT}
\end{equation}
The postprocessing presented here is formally similar to the one proposed in \cite{Giacomini2018}, but with a different definition of the matrix $\mat{D}$ (including here the second coefficient of viscosity $\lambda$) and with the presence of the ratio between the momentum and the density instead of the velocity itself.
The first equation in \eqref{eqn:fluidweakformpostprocessing} directly follows from the definition of the mixed variable in \eqref{eqn:fluidstrongformlocal} and is a least-squares fit to the accurate variable $\mat{L}^h$, while the last two equations remove the underdetermination of the problem by constraining the rigid motions.
It is worth recalling that the space $\mathcal{U}^h$, containing the functions of all ones in the elements, in the vector case selectively goes through the various components, hence providing $\texttt{n}_\texttt{sd}$ constraints for the translations and $\texttt{q}_\texttt{sd}$ constraints for the rotations in \eqref{eqn:fluidweakformpostprocessing}.

\section{HDG-CG formulation for the coupled FSI problem}
\label{sec:hdgcgformulationcoupledFSI}

In this section, the governing equations of nonlinear elastodynamics are first presented together with the associated standard CG formulation.
Then, the conditions to couple the fluid and the structural fields are briefly presented and two different coupling strategies, namely the Dirichlet--Neumann coupling and the Nitsche-based coupling, are proposed and their distinctive properties discussed.
No specific indices have been used in section \ref{sec:hdgformulationweaklycompressibleflows} to refer to the fluid quantities in order ease the comprehension of the proposed HDG formulation for weakly compressible flows.
In the following, however, the fluid and the structural quantities are distinguished by means of the subscripts $\Fl{(\cdot)}$ and $\St{(\cdot)}$.

\subsection{CG formulation for nonlinear elastodynamics}
\label{sec:cgformulationnonlinearelastodynamics}

The strong form of the time-dependent nonlinear elastic problem can be written with respect to the undeformed structural domain $\St{\Omega}$ as:
\begin{equation}
\left\{
\begin{aligned}
\St{\rho}\dfrac{d^2\St{\mathbf{u}}}{dt^2}-\bm{\nabla}\cdot\St{\mathbf{P}}&=\St{\rho}\St{\mathbf{b}}
\quad&&\text{in }\St{\Omega}\times\left(0,\textrm{T}_{\texttt{end}}\right), \\
\St{\mathbf{u}}&=\St{\mathbf{u}}^0
\quad&&\text{in }\St{\Omega}\times\left(0\right), \\
\dfrac{d\St{\mathbf{u}}}{dt}&=\St{\dot{\mathbf{u}}}^0
\quad&&\text{in }\St{\Omega}\times\left(0\right), \\
\St{\mathbf{u}}&=\St{\mathbf{u}}^D
\quad&&\text{on }\St{\Gamma}^D\times\left(0,\textrm{T}_{\texttt{end}}\right), \\
\St{\mathbf{P}}\St{\vect{n}}&=\St{\mathbf{t}}^N
\quad&&\text{on }\St{\Gamma}^N\times\left(0,\textrm{T}_{\texttt{end}}\right),
\end{aligned}
\right.
\label{eqn:structurestrongform}
\end{equation}
where $\St{\mathbf{u}}$ represents the unknown displacement field, $\St{\rho}$ the structural density, $\St{\mathbf{P}}$ the first Piola--Kirchhoff stress tensor and $\St{\mathbf{b}}$ an external body force per unit undeformed volume.
The pair ($\St{\mathbf{u}}^0$, $\St{\dot{\mathbf{u}}}^0$) defines the initial conditions for the displacement and the velocity, while the quantities $\St{\mathbf{u}}^D$ and $\St{\mathbf{t}}^N$ denote the Dirichlet and the Neumann boundary data applied on $\St{\Gamma}^D$ and $\St{\Gamma}^N$, respectively.
For hyperelastic materials, the first Piola--Kirchhoff stress tensor is defined as
\begin{equation}
\St{\mathbf{P}}=\frac{\partial\St{\psi}}{\partial\St{\mathbf{F}}},
\label{eqn:structurefirstpiolakirchhoff}
\end{equation}
with $\St{\mathbf{F}}$ being the deformation gradient and $\St{\psi}$ the strain energy density function.
The former is derived from the displacement field as
\begin{equation}
\St{\mathbf{F}}=\bm{\nabla}\St{\mathbf{u}}+\mathbf{I}_{\texttt{n}_\texttt{sd}},
\label{eqn:structuredeformationgradient}
\end{equation}
while the latter is defined as
\begin{equation}
  \St{\psi}= \begin{cases}
  \frac{\St{\mu}}{2}\bigl[\text{tr}(\St{\mathbf{F}}^T\St{\mathbf{F}}){-}\texttt{n}_\texttt{sd}{-}2\ln(|\St{\mathbf{F}}|)\bigl]
                       {+}\frac{\St{\lambda}}{2}\left[\ln\left(\left|\St{\mathbf{F}}\right|\right)\right]^2
  & \text{for Neo-Hooke},
\\
  \St{\mu}\St{\mat{E}}{:}\St{\mat{E}}{+}\frac{\St{\lambda}}{2}\left[\text{tr}\left(\St{\mat{E}}\right)\right]^2
  \,\text{with}\,\St{\mat{E}}{=}\tfrac{1}{2}(\St{\mathbf{F}}^T\St{\mathbf{F}}{-}\mathbf{I}_{\texttt{n}_\texttt{sd}}) 
  & \text{for St.\ Venant--Kirchhoff},
\end{cases}
\label{eqn:structurestrainenergyfunction}
\end{equation}
for the two popular material models used here.
The Lam\'e parameters $\St{\mu}$ and $\St{\lambda}$ can be evaluated as functions of the Young modulus $\St{E}$ and the Poisson ratio $\St{\nu}$ of the material through the following relations:
\begin{equation}
  \St{\mu}=\frac{\St{E}}{2\left(1+\St{\nu}\right)} \quad\text{and}\quad
 \St{\lambda}=\frac{\St{\nu}\St{E}}{\left(1+\St{\nu}\right)\left(1-2\St{\nu}\right)}.
\label{eqn:structurelamecoefficients}
\end{equation}

The following discrete functional spaces are introduced:
\begin{subequations}
\begin{align}
\mathcal{V}^h\left(\Omega\right):&=
\{v\in\mathcal{H}^1\left(\Omega\right):v\vert_{\Omega^e}\in\mathcal{P}^k\left(\Omega^e\right)\forall\Omega^e\subset\Omega,\;v\vert_{\Gamma^D}=\St{u}^D\},
\\
\mathcal{V}_0^h\left(\Omega\right):&=
\{v\in\mathcal{H}^1\left(\Omega\right):v\vert_{\Omega^e}\in\mathcal{P}^k\left(\Omega^e\right)\forall\Omega^e\subset\Omega,\;v\vert_{\Gamma^D}=0\}.
\end{align}
\label{eqn:structurespaces}%
\end{subequations}
As usual in CG methods, the weak form of the problem is obtained by multiplying the governing equation with the test functions and integrating by parts the term with second order derivatives.
The discrete weak form of the structural problem then reads: given $(\St{\mathbf{u}}^0,\St{\dot{\mathbf{u}}}^0)$ in $\St{\Omega}\times(0)$, find $\St{\mathbf{u}}^h\in[\mathcal{V}^h(\St{\Omega})]^{\texttt{n}_\texttt{sd}}$ such that
\begin{equation}
  \Bigl(\mathbf{v},\St{\rho}\frac{d^2\St{\mathbf{u}}^h}{dt^2}\Bigr)_{\St{\Omega}} + \left(\bm{\nabla}\mathbf{v},\St{\mathbf{P}}^h\right)_{\St{\Omega}}
  = \left(\mathbf{v},\St{\rho}\St{\mathbf{b}}\right)_{\St{\Omega}} +\left<\mathbf{v},\St{\mathbf{t}}^N\right>_{\St{\Gamma}^N},
\label{eqn:structureweakform}
\end{equation}
for all $\mathbf{v}\in[\mathcal{V}_0^h(\St{\Omega})]^{\texttt{n}_\texttt{sd}}$.
Of course, $\St{\mathbf{P}}^h$ features a nonlinear dependence on the displacement through \eqref{eqn:structurestrainenergyfunction} and a Newton--Raphson procedure is utilized for the solution of \eqref{eqn:structureweakform}.

\subsection{The fluid-structure coupling}
\label{sec:fluidstructurecoupling}

In order to couple the fluid problem, whose weak form has been derived section in \ref{sec:hdgformulationweaklycompressibleflows}, and the structural problem, whose weak form has been presented in section \ref{sec:cgformulationnonlinearelastodynamics}, kinematic and dynamic continuity conditions have to be enforced at the fluid-structure interface $\Gamma^I=\Fl{\Omega}\cap\St{\Omega}$.
First, the no-slip condition
\begin{equation}
\Fl{\bm{\upsilon}}-\St{\bm{\upsilon}}=\mathbf{0}
\label{eqn:fsivelocitycompatibility}
\end{equation}
prohibits a fluid flow across the interface and a relative tangential movement of fluid and structure at the interface.
Here, the fluid velocity is evaluated as the ratio of the momentum and the density, while the structural velocity is computed as the time derivative of the structural unknown displacement.
Second, the traction equilibrium
\begin{equation}
\Fl{\mathbf{t}}+\St{\mathbf{t}}=\mathbf{0}
\label{eqn:fsitractionequilibrium}
\end{equation}
states the equilibrium of the fluid and the structural forces at the interface.
The way in which the coupling conditions \eqref{eqn:fsivelocitycompatibility} and \eqref{eqn:fsitractionequilibrium} are imposed differs depending on the coupling strategy adopted.

It is worth mentioning that the deformation of the fluid computational mesh is evaluated as a function of the structural displacement at the interface, by means of a unique ALE mapping
\begin{equation}
\mathbf{d}=\boldsymbol{\varphi}(\St{\mathbf{u}}).
\label{eqn:alemapping}
\end{equation}
The grid motion strategy is an artificial problem and does not affect the physics of the coupled problem.
Its sole purpose is to generate a proper mesh for the solution of the fluid problem.
The velocity of the computational mesh, which is independent of the velocity of the material particles, is computed as the time derivative of the mesh displacement.

\begin{figure}
\begin{center}
\begin{tikzpicture}
\def \Lx{5.0}
\def \Ly{5.0}
\def \nelxF{3}
\def \nelyF{3}
\def \degF{2}
\def \intgap{0.5}
\def \gapFS{\intgap/2}
\def \nelxS{3}
\def \nelyS{3}
\def \degS{2}
\def \dofradius{0.1}
\def \dofedge{0.20}

\def \eledgexF{(\Lx-\intgap*(\nelxF-1))/\nelxF}
\def \eledgeyF{(\Ly-\intgap*(\nelyF-1))/\nelyF}
\def \nlinex{\numexpr(\nelxF)+1}
\def \nliney{\numexpr(\nelyF)+1}
\def \extgap{\intgap/2.0}
\def \ndofF{\numexpr(\degF)+1}

\foreach \i in {1,...,\nelyF}
{
  \foreach \j in {1,...,\nelxF}
  {
    \def \xi{\extgap+(\j-1)*(\eledgexF+\intgap)}
    \def \yi{\extgap+(\i-1)*(\eledgeyF+\intgap)}
    \def \xf{\xi+\eledgexF}
    \def \yf{\yi+\eledgeyF}
    \draw[thick] ({\xi},{\yi}) rectangle ({\xf},{\yf});
  }
}

\foreach \i in {1,...,\nliney}
{
  \foreach \j in {1,...,\nelxF}
  {
    \def \xi{\extgap+(\j-1)*(\eledgexF+\intgap)}
    \def \yi{(\i-1)*(\eledgeyF+\intgap)}
    \def \xf{\xi+\eledgexF}
    \def \yf{\yi}
    \draw[dashed,thick] ({\xi},{\yi}) -- ({\xf},{\yf});
  }
}

\foreach \i in {1,...,\nelyF}
{
  \foreach \j in {1,...,\nlinex}
  {
    \def \xi{(\j-1)*(\eledgexF+\intgap)}
    \def \yi{\extgap+(\i-1)*(\eledgeyF+\intgap)}
    \def \xf{\xi}
    \def \yf{\extgap+(\i-1)*(\eledgeyF+\intgap)+\eledgeyF}
    \draw[dashed,thick] ({\xi},{\yi}) -- ({\xf},{\yf});
  }
}

\foreach \i in {1,...,\nelyF}
{
  \foreach \j in {1,...,\nelxF}
  {
    \foreach \idof in {1,...,\ndofF}
    {
      \foreach \jdof in {1,...,\ndofF}
      {
        \def \xdof{(\jdof-1)*\eledgexF/\degF}
        \def \ydof{(\idof-1)*\eledgeyF/\degF}
        \def \xc{\extgap+\xdof+(\j-1)*(\eledgexF+\intgap)}
        \def \yc{\extgap+\ydof+(\i-1)*(\eledgeyF+\intgap)}
        \filldraw[fill=blue,draw=black,thick] ({\xc},{\yc}) circle (\dofradius);
      }
    }
  } 
}

\foreach \i in {1,...,\nliney}
{
  \foreach \j in {1,...,\nelxF}
  {
    \foreach \jdof in {1,...,\ndofF}
    {
      \def \xdof{(\jdof-1)*\eledgexF/\degF}
      \def \xc{\extgap+\xdof+(\j-1)*(\eledgexF+\intgap)}
      \def \yc{(\i-1)*(\eledgeyF+\intgap)}
      \def \xi{\xc-\dofedge/2}
      \def \yi{\yc-\dofedge/2}
      \def \xf{\xc+\dofedge/2}
      \def \yf{\yc+\dofedge/2}
      \filldraw[fill=blue,draw=black,thick] ({\xi},{\yi}) rectangle ({\xf},{\yf});
    }
  } 
}

\foreach \i in {1,...,\nelyF}
{
  \foreach \j in {1,...,\nlinex}
  {
    \foreach \idof in {1,...,\ndofF}
    {
      \def \ydof{(\idof-1)*\eledgeyF/\degF}
      \def \xc{(\j-1)*(\eledgexF+\intgap)}
      \def \yc{\extgap+\ydof+(\i-1)*(\eledgeyF+\intgap)}
      \def \xi{\xc-\dofedge/2}
      \def \yi{\yc-\dofedge/2}
      \def \xf{\xc+\dofedge/2}
      \def \yf{\yc+\dofedge/2}
      \filldraw[fill=blue,draw=black,thick] ({\xi},{\yi}) rectangle ({\xf},{\yf});
    }
  } 
}

\def \eledgexS{(\nelxF*\eledgexF+(\nelxF-1)*\intgap)/\nelxS}
\def \eledgeyS{(\nelyF*\eledgeyF+(\nelyF-1)*\intgap)/\nelyS}
\def \xstart{\nelxF*(\eledgexF+2*\extgap)+\gapFS}
\def \ystart{\extgap}
\def \xend{\xstart+\nelxS*\eledgexS}
\def \yend{\ystart+\nelyS*\eledgeyS}
\def \ndofS{\numexpr(\degS)+1}

\foreach \i in {1,...,\nelyS}
{
  \foreach \j in {1,...,\nelxS}
  {
    \def \xi{\xstart+(\j-1)*\eledgexS}
    \def \yi{\ystart+(\i-1)*\eledgeyS}
    \def \xf{\xi+\eledgexS}
    \def \yf{\yi+\eledgeyS}
    \draw[thick] ({\xi},{\yi}) rectangle ({\xf},{\yf});
  }
}

\foreach \i in {1,...,\nelyS}
{
  \foreach \j in {1,...,\nelxS}
  {
    \foreach \idof in {1,...,\ndofS}
    {
      \foreach \jdof in {1,...,\ndofS}
      {
        \def \xdof{(\jdof-1)*\eledgexS/\degS}
        \def \ydof{(\idof-1)*\eledgeyS/\degS}
        \def \xc{\xstart+\xdof+(\j-1)*\eledgexS}
        \def \yc{\ystart+\ydof+(\i-1)*\eledgeyS}
        \filldraw[fill=red,draw=black,thick] ({\xc},{\yc}) circle (\dofradius);
      }
    }
  } 
}
\end{tikzpicture}
\end{center}
\caption{Degrees of freedom of the coupled HDG-CG discretization using polynomial approximation of degree $k=2$ in the HDG fluid subdomain $\Fl{\Omega}$ (in blue) and in the CG structural subdomain $\St{\Omega}$ (in red).}
\label{fig:discretization}
\end{figure}
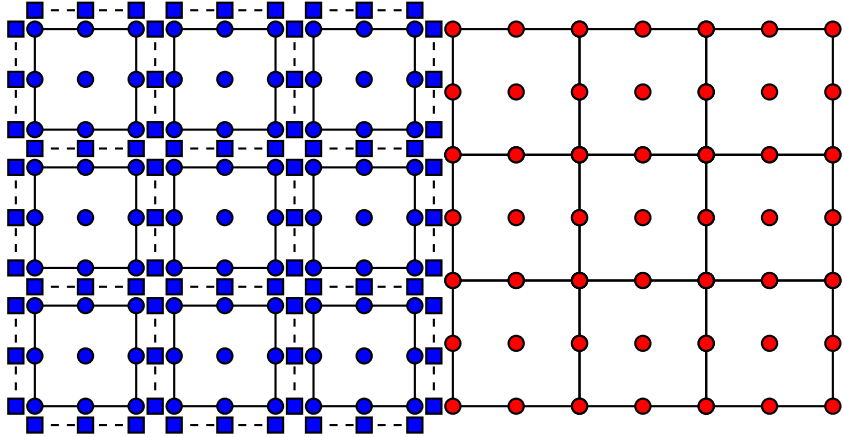

In this contribution, special attention is devoted to the spatial discretization of the weakly compressible flow problem by means of the HDG method and to the coupling of the fluid field with the structural one, discretized by means of the CG method.
A sketch of the heterogeneous HDG-CG discretization is exemplarily shown in Figure \ref{fig:discretization}.
The temporal discretization does not represent the focus of this work and the implicit backward differentiation formulas (BDF) are adopted here for the sake of simplicity.
Better time integration schemes have been developed in literature, such as the generalized-$\alpha$ method for the fluid \cite{Jansen2000} and the structure \cite{Chung1993}, and the possibility of independently choosing them in order to meet the needs of the individual fields has been exploited in \cite{Mayr2015}.

For the sake of readability, the superscript $(\cdot)^h$ associated with the numerical approximation of the unknowns will be henceforth omitted.
Moreover, it is assumed that the fluid quantities refer to the deformed domain $\Fl{\Omega}$ and the structural quantities refer to the undeformed domain $\St{\Omega}$.

\subsection{A partitioned Dirichlet--Neumann algorithm for the HDG-CG coupling}
\label{sec:partitioneddirichletneumannalgorithm}

The first coupling strategy is a partitioned Dirichlet--Neumann coupling, since it builds a Dirichlet-to-Neumann map by taking the structural solution at the interface as a Dirichlet boundary condition for the fluid problem and imposing the fluid normal flux as a Neumann boundary condition for the structural problem.
This represents a popular partitioned scheme for the solution of FSI problems and it has been used and analyzed for instance in \cite{Kuttler2008,Degroote2009}.
In \cite{Laspina2020a} it has moreover been shown that the introduction of a weak compressibility in the fluid field alleviates the constraints of the instability condition of the artificial added mass effect and, in comparison to a fully incompressible solver, it reduces the number of coupling iterations required and it leads to an increase of the dynamic relaxation parameter.

In this contribution, the partitioned Dirichlet--Neumann scheme is revisited for the coupling of weakly compressible flows and elastic structures in which the fluid and the structural problems are discretized by means of the HDG and the CG method, respectively.

The solver coupling at each time steps can be schematized in the following steps:
\begin{enumerate}
\item Predict the structural displacement $\St{\mathbf{u}}$ on $\Gamma^I$, assuming for instance a constant displacement, velocity or acceleration field.
\item Update the fluid mesh configuration by means of the ALE mapping \eqref{eqn:alemapping}.
\item Solve the fluid problem on the newly deformed domain by imposing the velocity compatibility \eqref{eqn:fsivelocitycompatibility} as a Dirichlet-type boundary condition on the local problems as
\begin{equation}
\frac{\Fl{\rho\bm{\upsilon}}}{\Fl{\rho}}-\frac{d\St{\mathbf{u}}}{dt}=\mathbf{0}\quad\text{on }\Gamma^I.
\label{eqn:fsivelocitycompatibilityDN}
\end{equation}

The weak form of the fluid local problems then reads: given $(\Fl{\rho}^0,\Fl{\rho\bm{\upsilon}}^0)$ in $\Fl{\Omega}^e\times(0)$, $(\Fl{\rho}^D,\Fl{\rho\bm{\upsilon}}^D)$ on $\Fl{\Gamma}^D$, $(\Fl{\hat{\rho}},\Fl{\widehat{\rho\bm{\upsilon}}})$ on $\Fl{\Gamma}\cup\Fl{\Gamma}^N$ and $(\Fl{\hat{\rho}},\St{\bm{\upsilon}})$ on $\Gamma^I$, find $(\Fl{\mat{L}},\Fl{\rho},\Fl{\rho\bm{\upsilon}})\in[\mathcal{W}(\Fl{\Omega}^e)]^{\texttt{m}_\texttt{sd}}\times\mathcal{W}(\Fl{\Omega}^e)\times[\mathcal{W}(\Fl{\Omega}^e)]^{\texttt{n}_\texttt{sd}}$ for $e=1,\dots,\Fl{\texttt{n}}^\texttt{el}$ such that
\begin{subequations}\label{eqn:fsifluidweakformlocalDN}%
\begin{multline}
  -\left(\mat{L},\Fl{\mat{L}}\right)_{\Fl{\Omega}^e}
  +\Bigl(\gradS^T\Fl{\mat{D}}^{\toh}\mat{L},\frac{\Fl{\rho\bm{\upsilon}}}{\Fl{\rho}}\Bigr)_{\Fl{\Omega}^e} 
  = \Bigl<\Fl{\vect{n}}^T\Fl{\mat{D}}^{\toh}\mat{L},\frac{\Fl{\rho\bm{\upsilon}}^D}{\Fl{\rho}^D}\Bigr>_{\partial\Fl{\Omega}^e\cap\Fl{\Gamma}^D} \\
    {+}\Bigl<\Fl{\vect{n}}^T\Fl{\mat{D}}^{\toh}\mat{L},\frac{\Fl{\widehat{\rho\bm{\upsilon}}}}{\Fl{\hat{\rho}}}\Bigr>_{\partial\Fl{\Omega}^e\setminus\Fl{\Gamma}^D\setminus\Gamma^I}
    {+}\Bigl<\Fl{\vect{n}}^T\Fl{\mat{D}}^{\toh}\mat{L},\frac{d\St{\mathbf{u}}}{dt}\Bigr>_{\partial\Fl{\Omega}^e\cap\Gamma^I},
\end{multline}
\begin{multline}
    \Bigl(w,\frac{\partial\Fl{\rho}}{\partial t}\Bigr)_{\Fl{\Omega}^e}
  {+}(w,\Fl{\rho}\bm{\nabla}\cdot\Fl{\vect{a}})_{\Fl{\Omega}^e} 
   {-}(\bm{\nabla}w,\Fl{\rho\bm{\upsilon}}{-}\Fl{\rho}\Fl{\vect{a}})_{\Fl{\Omega}^e}
   {+}\bigl<w,\tau_\rho\Fl{\rho}\bigl>_{\partial\Fl{\Omega}^e} \\
  = 
  {-}\bigl<w,(\Fl{\rho\bm{\upsilon}}^D{-}\Fl{\rho}^D\Fl{\vect{a}})\cdot\Fl{\vect{n}}{-}\tau_\rho\Fl{\rho}^D\bigr>_{\partial\Fl{\Omega}^e\cap\Fl{\Gamma}^D} 
  {-}\bigl<w,(\Fl{\widehat{\rho\bm{\upsilon}}}{-}\Fl{\hat{\rho}}\Fl{\vect{a}})\cdot\Fl{\vect{n}}{-}\tau_\rho\Fl{\hat{\rho}}\bigr>_{\partial\Fl{\Omega}^e\setminus\Fl{\Gamma}^D\setminus\Gamma^I} \\
  -\bigl<w,\Fl{\hat{\rho}}\bigl(\frac{d\St{\mathbf{u}}}{dt}-\Fl{\vect{a}}\bigl)\cdot\Fl{\vect{n}}-\tau_\rho\Fl{\hat{\rho}}\bigr>_{\partial\Fl{\Omega}^e\cap\Gamma^I},
\end{multline}
\begin{multline}
    \Bigl(\vect{w},\frac{\partial\Fl{\rho\bm{\upsilon}}}{\partial t}\Bigr)_{\Fl{\Omega}^e}
  {+}(\vect{w},\Fl{\rho\bm{\upsilon}}\bm{\nabla}\cdot\Fl{\vect{a}})_{\Fl{\Omega}^e}
  {-}\bigl(\bm{\nabla}\vect{w},\Fl{\rho\bm{\upsilon}}{\otimes}\bigl(\frac{\Fl{\rho\bm{\upsilon}}}{\Fl{\rho}}{-}\Fl{\vect{a}}\bigr)\bigr)_{\Fl{\Omega}^e} \\
  {+}\Bigl(\vect{w},\gradS^T\bigl(\Fl{\mat{D}}^{\toh}\Fl{\mat{L}}+\mat{E}\Fl{p}(\Fl{\rho})\bigr)\Bigr)_{\Fl{\Omega}^e} 
  {+}\bigl<\vect{w},\tau_{\rho\upsilon}\Fl{\rho\bm{\upsilon}}\bigr>_{\partial\Fl{\Omega}^e}
  {-}(\vect{w},\Fl{\rho}\Fl{\mathbf{b}})_{\Fl{\Omega}^e} \\
  = 
  {-}\Bigl<\vect{w},\bigl[\Fl{\rho\bm{\upsilon}}^D{\otimes}\bigl(\frac{\Fl{\rho\bm{\upsilon}}^D}{\Fl{\rho}^D}{-}\Fl{\vect{a}}\bigr)\bigr]\Fl{\vect{n}}{-}\tau_{\rho\upsilon}\Fl{\rho\bm{\upsilon}}^D\Bigr>_{\partial\Fl{\Omega}^e\cap\Fl{\Gamma}^D} \\
  {-}\Bigl<\vect{w},\bigl[\Fl{\widehat{\rho\bm{\upsilon}}}{\otimes}\bigl(\frac{\Fl{\widehat{\rho\bm{\upsilon}}}}{\Fl{\hat{\rho}}}{-}\Fl{\vect{a}}\bigr)\bigr]\Fl{\vect{n}}{-}\tau_{\rho\upsilon}\Fl{\widehat{\rho\bm{\upsilon}}}\Bigr>_{\partial\Fl{\Omega}^e\setminus\Fl{\Gamma}^D\setminus\Gamma^I} \\
  {-}\Bigl<\vect{w},\bigl[\Fl{\hat{\rho}}\frac{d\St{\mathbf{u}}}{dt}{\otimes}\bigl(\frac{d\St{\mathbf{u}}}{dt}{-}\Fl{\vect{a}}\bigr)\bigr]\Fl{\vect{n}}{-}\tau_{\rho\upsilon}\Fl{\hat{\rho}}\frac{d\St{\mathbf{u}}}{dt}\Bigr>_{\partial\Fl{\Omega}^e\cap\Gamma^I},
\end{multline}
\end{subequations}
for all $(\mat{L},w,\vect{w})\in[\mathcal{W}(\Fl{\Omega}^e)]^{\texttt{m}_\texttt{sd}}\times\mathcal{W}(\Fl{\Omega}^e)\times[\mathcal{W}(\Fl{\Omega}^e)]^{\texttt{n}_\texttt{sd}}$.
From a practical point of view, equations \eqref{eqn:fsifluidweakformlocalDN} are obtained from \eqref{eqn:fluidweakformlocal} by replacing $\Fl{\widehat{\rho\bm{\upsilon}}}$ with $\Fl{\hat{\rho}}\frac{d\St{\mathbf{u}}}{dt}$ at the fluid-structure interface $\Gamma^I$.

The weak form of the fluid global problem instead reads: find $(\Fl{\hat{\rho}},\Fl{\widehat{\rho\bm{\upsilon}}})\in\widehat{\mathcal{W}}(\Fl{\Gamma}\cup\Fl{\Gamma}^N\cup\Gamma^I)\times[\widehat{\mathcal{W}}(\Fl{\Gamma}\cup\Fl{\Gamma}^N)]^{\texttt{n}_\texttt{sd}}$ such that
\begin{subequations}\label{eqn:fsifluidweakformglobalDN}%
\begin{gather}
  \sum_{e=1}^{\Fl{\texttt{n}}^\texttt{el}}\bigl<\hat{w},\tau_\rho (\Fl{\rho}-\Fl{\hat{\rho}}) \bigr>_{\partial\Fl{\Omega}^e\setminus\Fl{\Gamma}^D} = 0,
\\
  {-}\sum_{e=1}^{\Fl{\texttt{n}}^\texttt{el}}
  \!
  \bigl<\hat{\vect{w}},\Fl{\vect{n}}^T\bigl(\Fl{\mat{D}}^{\toh}\Fl{\mat{L}}{+}\mat{E}\Fl{p}(\Fl{\hat{\rho}})\bigr){+}\tau_{\rho\upsilon} (\Fl{\rho\bm{\upsilon}}{-}\Fl{\widehat{\rho\bm{\upsilon}}})\bigr>_{\partial\Fl{\Omega}^e\setminus\Fl{\Gamma}^D\setminus\Gamma^I}
  {=} \sum_{e=1}^{\Fl{\texttt{n}}^\texttt{el}}
  \!
  \langle\hat{\vect{w}},\Fl{\mathbf{t}}^N\rangle_{\partial\Fl{\Omega}^e\cap\Fl{\Gamma}^N},
\end{gather}
\end{subequations}
for all $(\hat{w},\hat{\vect{w}})\in\widehat{\mathcal{W}}(\Fl{\Gamma}\cup\Fl{\Gamma}^N\cup\Gamma^I)\times[\widehat{\mathcal{W}}(\Fl{\Gamma}\cup\Fl{\Gamma}^N)]^{\texttt{n}_\texttt{sd}}$.
\item Solve the structural problem by imposing the coupling condition \eqref{eqn:fsitractionequilibrium} as a Neumann-type boundary condition, consistently expressed in terms of the first Piola--Kirchhoff stress tensor as
\begin{equation}
(\Fl{\mathbf{P}}-\St{\mathbf{P}})\St{\vect{n}}=\mathbf{0}\quad\text{on }\Gamma^I,
\label{eqn:fsitractionequilibriumDN}
\end{equation}
where the fluid Cauchy stress is transformed by means of the pull-back operation
\begin{equation}
\Fl{\mathbf{P}}=
-|\Fl{\mathbf{F}}|\text{V}^{-1}(\Fl{\mat{D}}^{\toh}\Fl{\mat{L}}+\mat{E}\Fl{p})\Fl{\mathbf{F}}^{-T}.
\label{eqn:fsistresstransformationDN}
\end{equation}
The deformation gradient $\Fl{\mathbf{F}}$ is evaluated as in \eqref{eqn:structuredeformationgradient} but with respect to the fluid mesh displacement.
Given a $\texttt{m}_\texttt{sd}\times 1$ vector in Voigt notation, the operator $\text{V}^{-1}$ returns the associated $\texttt{n}_\texttt{sd}\times \texttt{n}_\texttt{sd}$ symmetric tensor:
\begin{equation}\label{eqn:voigtV}
  \text{V}^{-1}:=
  \begin{cases} \begin{bmatrix} \sigma_{xx} & \sigma_{yy} & \sigma_{xy} \end{bmatrix}^T
                        \longrightarrow
                        \begin{bmatrix}  \sigma_{xx} & \sigma_{xy} \\
                                                  \sigma_{xy} & \sigma_{yy} \end{bmatrix}  &\text{in 2D},
  \\
                        \begin{bmatrix} \sigma_{xx} & \sigma_{yy} & \sigma_{zz} & \sigma_{xy} & \sigma_{xz} & \sigma_{yz} \end{bmatrix}^T
                        \longrightarrow
                        \begin{bmatrix} \sigma_{xx} & \sigma_{xy} & \sigma_{xz} \\
                                                 \sigma_{xy} & \sigma_{yy} & \sigma_{yz} \\
                                                 \sigma_{xz} & \sigma_{yz} & \sigma_{zz}  \end{bmatrix} &\text{in 3D}.
  \end{cases}
\end{equation}

Analogously to the viscous stress in \eqref{eqn:fluidnumericalflux}, the following trace of the numerical normal flux is introduced:
\begin{equation}
  \reallywidehat{\Fl{\mathbf{P}}\Fl{\vect{n}}}
  := {-} \abs{\Fl{\mathbf{F}}} \text{V}^{-1} (\Fl{\mat{D}}^{\toh}\Fl{\mat{L}}{+}\mat{E}\Fl{p} (\Fl{\hat{\rho}}) ) \Fl{\mathbf{F}}^{-T}\Fl{\vect{n}}
      {-}\tau_{\rho\upsilon} \bigl(\Fl{\rho\bm{\upsilon}}-\Fl{\hat{\rho}}\frac{d\St{\mathbf{u}}}{dt} \bigr)
      \;\text{on }\Gamma^I.
\label{eqn:fsinumericalfluxhdg}
\end{equation}

The weak form of the structural problem then reads: given $(\St{\mathbf{u}}^0,\St{\dot{\mathbf{u}}}^0)$ in $\St{\Omega}\times(0)$ and $\Fl{\mathbf{t}}$ on $\Gamma^I$, find $\St{\mathbf{u}}\in[\mathcal{V}(\St{\Omega})]^{\texttt{n}_\texttt{sd}}$ such that
\begin{multline}\label{eqn:fsistructureweakformDN}
  \Bigl(\mathbf{v},\St{\rho}\dfrac{d^2\St{\mathbf{u}}}{dt^2}\Bigr)_{\St{\Omega}} {+} \bigl(\bm{\nabla}\mathbf{v},\St{\mathbf{P}}\bigl)_{\St{\Omega}} \\
  +\bigl<\mathbf{v},\abs{\Fl{\mathbf{F}}} \text{V}^{-1} \bigl(\Fl{\mat{D}}^{\toh}\Fl{\mat{L}}{+}\mat{E}\Fl{p}(\Fl{\hat{\rho}})\bigr)\Fl{\mathbf{F}}^{-T}\St{\vect{n}}{-}\tau_{\rho\upsilon}\bigl(\Fl{\rho\bm{\upsilon}}{-}\Fl{\hat{\rho}}\frac{d\St{\mathbf{u}}}{dt}\bigr)\bigr>_{\Gamma^I} \\
  = \bigl(\mathbf{v},\St{\rho}\St{\mathbf{b}}\bigr)_{\St{\Omega}}
   {+}\bigl<\mathbf{v},\St{\mathbf{t}}^N\bigr>_{\St{\Gamma}^N},
\end{multline}
for all $\mathbf{v}\in[\mathcal{V}_0(\St{\Omega})]^{\texttt{n}_\texttt{sd}}$.
\item Check for convergence: continue with next time step if the algorithm is converged, otherwise return to step 2.
The convergence is considered satisfied if
\begin{equation}
\St{\mathbf{r}}^{i+1}=\norm{\St{\tilde{\mathbf{u}}}^{i+1}-\St{\mathbf{u}}^{i}} < \eta\quad\text{on }\Gamma^I,
\label{eqn:fsiconvergencecheckDN}
\end{equation}
with $\St{\mathbf{u}}^{i}$ denoting the structural interface displacement at the $i$-th coupling iteration and $\St{\tilde{\mathbf{u}}}^{i+1}$ the newly computed one by solving \eqref{eqn:fsistructureweakformDN}.
The parameter $\eta$ represents instead a user-defined convergence tolerance.

To accelerate the convergence of the fixed-point scheme, a relaxation of the structural interface displacement is performed
\begin{equation}
\St{\mathbf{u}}^{i+1}=\omega^{i}\St{\tilde{\mathbf{u}}}^{i+1}+(1-\omega^{i})\St{\mathbf{u}}^{i},
\label{eqn:fsirelaxationDN}
\end{equation}
where the relaxation parameter $\omega$ is evaluated at each coupling iteration by means of the Aitken $\Delta^2$ method \cite{Kuttler2008}:
\begin{equation}
\omega^{i+1}=-\omega^{i}\frac{(\St{\mathbf{r}}^{i+1})^T(\St{\mathbf{r}}^{i+2}-\St{\mathbf{r}}^{i+1})}{\|\St{\mathbf{r}}^{i+2}-\St{\mathbf{r}}^{i+1}\|}.
\label{eqn:fsiaitkendelta2methodDN}
\end{equation}
\end{enumerate}

\begin{remark}
The strategy presented in this section to couple the fluid and the structure could also be implemented in a monolithic fashion, by simultaneously solving the problems \eqref{eqn:fsifluidweakformlocalDN}-\eqref{eqn:fsifluidweakformglobalDN}-\eqref{eqn:fsistructureweakformDN} in a large linear system.
An analogous method to couple HDG and CG discretization has been presented in \cite{Paipuri2019} in the context of conjugate heat transfer problems.
However, such a method results in a coupling of local and global degrees of freedom of the HDG problem with the ones of the CG discretization, making the implementation of this strategy in existing HDG and CG libraries rather intrusive.
\end{remark}

\subsection{A monolithic algorithm for the HDG-CG coupling based on Nitsche's method}
\label{sec:monolithicalgorithmnitsche}

The second coupling strategy imposes the structural numerical normal flux as a Neumann boundary condition for the fluid problem and takes the fluid hybrid variables at the interface as a Dirichlet-type boundary condition for the structural problem.
The resulting hybrid HDG-CG coupling does not affect the structure of the core CG and HDG matrices, thus leading to a minimally-intrusive implementation of this technique in existing finite element codes.
This approach to couple HDG and CG discretization has been recently presented in \cite{Laspina2020b} for the solution of elasticity problems involving compressible and nearly incompressible solids.

The key feature of the Nitsche-based coupling is that it allows to impose the coupling conditions solely in the global problem and, as a consequence, the local problems remain the same as in the pure HDG case (equations \eqref{eqn:fluidweakformlocal}).

With the notation introduced in section \ref{sec:partitioneddirichletneumannalgorithm} for the coupled problem, the HDG local problems read: given $(\Fl{\rho}^0,\Fl{\rho\bm{\upsilon}}^0)$ in $\Fl{\Omega}^e\times(0)$, $(\Fl{\rho}^D,\Fl{\rho\bm{\upsilon}}^D)$ on $\Fl{\Gamma}^D$ and $(\Fl{\hat{\rho}},\Fl{\widehat{\rho\bm{\upsilon}}})$ on $\Fl{\Gamma}\cup\Fl{\Gamma}^N\cup\Gamma^I$, find $(\Fl{\mat{L}},\Fl{\rho},\Fl{\rho\bm{\upsilon}})\in[\mathcal{W}(\Fl{\Omega}^e)]^{\texttt{m}_\texttt{sd}}\times\mathcal{W}(\Fl{\Omega}^e)\times[\mathcal{W}(\Fl{\Omega}^e)]^{\texttt{n}_\texttt{sd}}$ for $e=1,\dots,\Fl{\texttt{n}}^\texttt{el}$ such that
\begin{subequations}\label{eqn:fsiweakformlocalNitsche}%
\begin{equation}
  {-}\bigl(\mat{L},\Fl{\mat{L}}\bigr)_{\Fl{\Omega}^e}
  {+}\bigl(\gradS^T\Fl{\mat{D}}^{\toh}\mat{L},\dfrac{\Fl{\rho\bm{\upsilon}}}{\Fl{\rho}}\bigr)_{\Fl{\Omega}^e} 
  {=}\bigl<\Fl{\vect{n}}^T\Fl{\mat{D}}^{\toh}\mat{L},\dfrac{\Fl{\rho\bm{\upsilon}}^D}{\Fl{\rho}^D}\bigr>_{\partial\Fl{\Omega}^e\cap\Fl{\Gamma}^D}
  {+}\bigl<\Fl{\vect{n}}^T\Fl{\mat{D}}^{\toh}\mat{L},\dfrac{\Fl{\widehat{\rho\bm{\upsilon}}}}{\Fl{\hat{\rho}}}\bigr>_{\partial\Fl{\Omega}^e\setminus\Fl{\Gamma}^D},
\end{equation}
\begin{multline}
    \Bigl(w,\dfrac{\partial\Fl{\rho}}{\partial t}\Bigr)_{\Fl{\Omega}^e}
  {+}\bigl(w,\Fl{\rho}\bm{\nabla}{\cdot}\Fl{\vect{a}}\bigr)_{\Fl{\Omega}^e}
   {-}\bigl(\bm{\nabla}w,\Fl{\rho\bm{\upsilon}}-\Fl{\rho}\Fl{\vect{a}}\bigr)_{\Fl{\Omega}^e}
  {+}\bigl<w,\tau_\rho\Fl{\rho}\bigr>_{\partial\Fl{\Omega}^e} \\
  = {-}\bigl<w, (\Fl{\rho\bm{\upsilon}}^D{-}\Fl{\rho}^D\Fl{\vect{a}}) {\cdot}\Fl{\vect{n}}{-}\tau_\rho\Fl{\rho}^D\bigr>_{\partial\Fl{\Omega}^e\cap\Fl{\Gamma}^D} \\
     {-}\bigl<w, (\Fl{\widehat{\rho\bm{\upsilon}}}{-}\Fl{\hat{\rho}}\Fl{\vect{a}}){\cdot}\Fl{\vect{n}}{-}\tau_\rho\Fl{\hat{\rho}}\bigr>_{\partial\Fl{\Omega}^e\setminus\Fl{\Gamma}^D},
\end{multline}
\begin{multline}
    \Bigl(\vect{w},\dfrac{\partial\Fl{\rho\bm{\upsilon}}}{\partial t}\Bigr)_{\Fl{\Omega}^e}
  {+}\bigl(\vect{w},\Fl{\rho\bm{\upsilon}}\bm{\nabla}\cdot\Fl{\vect{a}}\bigr)_{\Fl{\Omega}^e} 
   {-}\bigl(\bm{\nabla}\vect{w},\Fl{\rho\bm{\upsilon}}{\otimes}\bigl(\frac{\Fl{\rho\bm{\upsilon}}}{\Fl{\rho}}-\Fl{\vect{a}}\bigr)\bigr)_{\Fl{\Omega}^e} \\
  {+}\bigl(\vect{w},\gradS^T\bigl(\Fl{\mat{D}}^{\toh}\Fl{\mat{L}}{+}\mat{E}\Fl{p}(\Fl{\rho})\bigr)\bigr)_{\Fl{\Omega}^e} 
  {+}\bigl<\vect{w},\tau_{\rho\upsilon}\Fl{\rho\bm{\upsilon}}\bigr>_{\partial\Fl{\Omega}^e}
  {-}\bigl(\vect{w},\Fl{\rho}\Fl{\mathbf{b}}\bigr)_{\Fl{\Omega}^e} \\
  ={-}\bigl<\vect{w},\bigl[\Fl{\rho\bm{\upsilon}}^D{\otimes}\bigl(\frac{\Fl{\rho\bm{\upsilon}}^D}{\Fl{\rho}^D}{-}\Fl{\vect{a}}\bigr)\bigr] \Fl{\vect{n}}
                              {-}\tau_{\rho\upsilon}\Fl{\rho\bm{\upsilon}}^D\bigr>_{\partial\Fl{\Omega}^e\cap\Fl{\Gamma}^D} \\
    {-}\bigl<\vect{w},\bigl[\Fl{\widehat{\rho\bm{\upsilon}}}{\otimes}\bigl(\frac{\Fl{\widehat{\rho\bm{\upsilon}}}}{\Fl{\hat{\rho}}}{-}\Fl{\vect{a}}\bigr)\bigr] \Fl{\vect{n}}
                              {-}\tau_{\rho\upsilon}\Fl{\widehat{\rho\bm{\upsilon}}}\bigr>_{\partial\Fl{\Omega}^e\setminus\Fl{\Gamma}^D},
\end{multline}
\end{subequations}
for all $(\mat{L},w,\vect{w})\in[\mathcal{W}(\Fl{\Omega}^e)]^{\texttt{m}_\texttt{sd}}\times\mathcal{W}(\Fl{\Omega}^e)\times[\mathcal{W}(\Fl{\Omega}^e)]^{\texttt{n}_\texttt{sd}}$.
It is worth recalling that equations \eqref{eqn:fsiweakformlocalNitsche} are identical to \eqref{eqn:fluidweakformlocal} and they are rewritten here for completeness with the specific notation adopted for fluid-structure interaction problems.

The coupling condition \eqref{eqn:fsitractionequilibrium} is imposed as a Neumann-type boundary condition on the global fluid problem and it is consistently expressed in terms of the Cauchy stress tensor.
The dynamic equilibrium can therefore be written as
\begin{equation}
(\Fl{\boldsymbol{\sigma}}-\St{\boldsymbol{\sigma}})\Fl{\vect{n}}=\mathbf{0}\quad\text{on }\Gamma^I,
\label{eqn:fsitractionequilibriumNitsche}
\end{equation}
where the first Piola--Kirchhoff stress is transformed by means of the push-forward operation
\begin{equation} 
  \St{\boldsymbol{\sigma}} = \abs{\St{\mathbf{F}}}^{-1}\St{\mathbf{P}}{\St{\mathbf{F}}}^T.
\label{eqn:fsistresstransformationNitsche}
\end{equation}

In order to impose the coupling conditions on the interface, the following definition of the trace of the CG numerical normal flux is adopted:
\begin{equation}
\reallywidehat{\St{\mathbf{P}}\St{\vect{n}}} := \St{\mathbf{P}}\St{\vect{n}} 
                                     -\frac{\gamma}{h}\bigl(\dfrac{d\St{\mathbf{u}}}{dt}-\frac{\Fl{\widehat{\rho\bm{\upsilon}}}}{\Fl{\hat{\rho}}}\bigr)
                                    \quad\text{on }\Gamma^I,
\label{eqn:fsinumericalfluxcg}
\end{equation}
where $h$ denotes a characteristic element size on $\Gamma^I$ and $\gamma$ is a sufficiently large positive value, called in the following {\lq\lq Nitsche's parameter\rq\rq}, that is commonly used to enforce coercivity of the discrete bilinear form in CG discretizations with Nitsche's imposition of essential boundary conditions.
The influence of the Nitsche parameter on the accuracy of the hybrid HDG-CG coupling has been investigated in a previous work \cite{Laspina2020b} on a computationally cheap scalar problem.
The analysis revealed that for low values of $\gamma$ an insufficient stabilization produces unreliable results, whereas values of $\gamma$ above a certain lower bound ensure the stability of the scheme, in agreement with established results in literature \cite{Hansbo2005}.
However, although an estimation of such a lower bound can be obtained by solving an auxiliary generalized eigenvalue problem as suggested in \cite{Griebel2003}, $\gamma$ is problem-dependent and affected both by the equation under analysis and the material parameters.
In terms of the scaling of the penalty parameter with the polynomial degree, a $k^2$ scaling can be expected from the results established by the symmetric interior penalty community \cite{Epshteyn2007}, a topic not further investigated in the present work.

It is worth recalling that in the standard CG approach the numerical normal fluxes are naturally equilibrated on the internal faces of the triangulation.
The velocity compatibility \eqref{eqn:fsivelocitycompatibility} is weakly imposed as a Dirichlet-type boundary condition on the structural problem, exploiting the definition \eqref{eqn:fsinumericalfluxcg}.

The weak form of the global problem then reads: given $(\St{\mathbf{u}}^0,\St{\dot{\mathbf{u}}}^0)$ in $\St{\Omega}\times(0)$, find $(\Fl{\hat{\rho}},\Fl{\widehat{\rho\bm{\upsilon}}},\St{\mathbf{u}})\in\widehat{\mathcal{W}}(\Fl{\Gamma}\cup\Fl{\Gamma}^N\cup\Gamma^I)\times[\widehat{\mathcal{W}}(\Fl{\Gamma}\cup\Fl{\Gamma}^N\cup\Gamma^I)]^{\texttt{n}_\texttt{sd}}\times[\mathcal{V}(\St{\Omega})]^{\texttt{n}_\texttt{sd}}$ such that
\begin{subequations}\label{eqn:fsiweakformglobalNitsche}%
\begin{equation}
  \sum_{e=1}^{\Fl{\texttt{n}}^\texttt{el}} \bigl<\hat{w},\tau_\rho (\Fl{\rho}-\Fl{\hat{\rho}}) \bigr>_{\partial\Fl{\Omega}^e\setminus\Fl{\Gamma}^D} = 0,
\end{equation}
\begin{multline}
  -\sum_{e=1}^{\Fl{\texttt{n}}^\texttt{el}}
                                                  \bigg\{
   \bigl<\hat{\vect{w}},\Fl{\vect{n}}^T\bigl(\Fl{\mat{D}}^{\toh}\Fl{\mat{L}}{+}\mat{E}\Fl{p}(\Fl{\hat{\rho}})\bigr)
                                 {+}\tau_{\rho\upsilon}(\Fl{\rho\bm{\upsilon}}{-}\Fl{\widehat{\rho\bm{\upsilon}}})\bigr>_{\partial\Fl{\Omega}^e\setminus\Fl{\Gamma}^D}
                                                  \biggr. \\ 
                                                  \biggl.
  +\Bigl<\hat{\vect{w}},\abs{\St{\mathbf{F}}}^{-1}\St{\mathbf{P}}{\St{\mathbf{F}}}^T\Fl{\vect{n}}
              {+}\frac{\gamma}{h}\Bigl(\frac{d\St{\mathbf{u}}}{dt}-\frac{\Fl{\widehat{\rho\bm{\upsilon}}}}{\Fl{\hat{\rho}}}\Bigl)\Bigr>_{\partial\Fl{\Omega}^e\cap\Gamma^I}
                                                  \biggr\} 
  = \sum_{e=1}^{\Fl{\texttt{n}}^\texttt{el}}\bigl<\hat{\vect{w}},\Fl{\mathbf{t}}^N\bigr>_{\partial\Fl{\Omega}^e\cap\Fl{\Gamma}^N},
\end{multline}
\begin{multline}
  \Bigl(\mathbf{v},\St{\rho}\dfrac{d^2\St{\mathbf{u}}}{dt^2}\Bigr)_{\St{\Omega}}
  +\bigl(\bm{\nabla}\mathbf{v},\St{\mathbf{P}}\bigr)_{\St{\Omega}}
  -\Bigl<\mathbf{v},\St{\mathbf{P}}\St{\vect{n}}-\dfrac{\gamma}{h}\Bigl(\frac{d\St{\mathbf{u}}}{dt}{-}\frac{\Fl{\widehat{\rho\bm{\upsilon}}}}{\Fl{\hat{\rho}}}\Bigr)\Bigr>_{\Gamma^I} \\
  -\Bigl<\frac{\partial\St{\mathbf{P}}}{\partial\bm{\nabla}\St{\mathbf{u}}}\bm{\nabla}\mathbf{v}\,\St{\vect{n}},\frac{d\St{\mathbf{u}}}{dt}-\frac{\Fl{\widehat{\rho\bm{\upsilon}}}}{\Fl{\hat{\rho}}}\Bigl>_{\Gamma^I}
  =
  \bigl(\mathbf{v},\St{\rho}\St{\mathbf{b}}\bigr)_{\St{\Omega}}
  +\bigl<\mathbf{v},\St{\mathbf{t}}^N\bigr>_{\St{\Gamma}^N},
\end{multline}
\end{subequations}
for all $(\hat{w},\hat{\vect{w}},\mathbf{v})\in\widehat{\mathcal{W}}(\Fl{\Gamma}\cup\Fl{\Gamma}^N\cup\Gamma^I)\times[\widehat{\mathcal{W}}(\Fl{\Gamma}\cup\Fl{\Gamma}^N\cup\Gamma^I)]^{\texttt{n}_\texttt{sd}}\times[\mathcal{V}_0(\St{\Omega})]^{\texttt{n}_\texttt{sd}}$.

After standard finite element discretization and assembly, the following linear system is obtained in terms of the increments of the global unknowns $\Fl{\vect{\widehat{U}}}=\begin{bmatrix}\Fl{\vect{\hat{\rho}}} & \Fl{\vect{\widehat{\rho\bm{\upsilon}}}}\end{bmatrix}^T$ and $\St{\vect{u}}$:
\begin{equation}
\begin{bmatrix}
\Ff{\mat{K}} & \Fs{\mat{K}} \\
\Sf{\mat{K}} & \Ss{\mat{K}} \\
\end{bmatrix}
\begin{bmatrix}
\delta\Fl{\vect{\widehat{U}}} \\
\delta\St{\vect{u}}
\end{bmatrix}
=
\begin{bmatrix}
\Fl{\vect{f}} \\
\St{\vect{f}}
\end{bmatrix},
\label{eqn:fsisystemglobalNitsche}
\end{equation}
with the left hand side matrices computed as
\begin{subequations}
\begin{align}
\begin{bmatrix}
\Ff{\mat{K}}
\end{bmatrix}
&=
\sum_{e=1}^{\Fl{\texttt{n}}^\texttt{el}}
\Bigl\{
\begin{bmatrix}
\mat{K}_{\widehat{U}\widehat{U}}
\end{bmatrix}_e
-
\begin{bmatrix}
\mat{K}_{\widehat{U}L} & \mat{K}_{\widehat{U}U}
\end{bmatrix}_e
\begin{bmatrix}
\mat{K}_{LL} & \mat{K}_{LU} \\
\mat{K}_{UL} & \mat{K}_{UU}
\end{bmatrix}_e^{-1}
\begin{bmatrix}
\mat{K}_{L\widehat{U}} \\
\mat{K}_{U\widehat{U}}
\end{bmatrix}_e
\Bigr\},
\\
\begin{bmatrix}
\Fs{\mat{K}}
\end{bmatrix}
&=
\sum_{e=1}^{\FlStintf{\texttt{n}}^\texttt{el}}
\Bigl\{
\begin{bmatrix}
\mat{K}_{\widehat{U}u}
\end{bmatrix}_e
\vphantom{
\begin{bmatrix}
\mat{K}_{LL} & \mat{K}_{LU} \\
\mat{K}_{UL} & \mat{K}_{UU}
\end{bmatrix}_e^{-1}}
\Bigr\},
\\
\begin{bmatrix}
\Sf{\mat{K}}
\end{bmatrix}
&=
\sum_{e=1}^{\FlStintf{\texttt{n}}^\texttt{el}}
\Bigl\{
\begin{bmatrix}
\mat{K}_{u\widehat{U}}
\end{bmatrix}_e
\vphantom{
\begin{bmatrix}
\mat{K}_{LL} & \mat{K}_{LU} \\
\mat{K}_{UL} & \mat{K}_{UU}
\end{bmatrix}_e^{-1}}
\Bigr\},
\\
\begin{bmatrix}
\Ss{\mat{K}}
\end{bmatrix}
&=
\sum_{e=1}^{\St{\texttt{n}}^\texttt{el}}
\Bigl\{
\begin{bmatrix}
\mat{K}_{uu}
\end{bmatrix}_e
\vphantom{
\begin{bmatrix}
\mat{K}_{LL} & \mat{K}_{LU} \\
\mat{K}_{UL} & \mat{K}_{UU}
\end{bmatrix}_e^{-1}}
\Bigr\},
\end{align}
\label{eqn:fsisystemgloballhsNitsche}%
\end{subequations}
and the right hand side vectors computed as
\begin{subequations}
\begin{align}
\begin{bmatrix}
\Fl{\vect{f}}
\end{bmatrix}
&=
\sum_{e=1}^{\Fl{\texttt{n}}^\texttt{el}}
\Bigl\{
\begin{bmatrix}
\vect{f}_{\widehat{U}}
\end{bmatrix}_e
-
\begin{bmatrix}
\mat{K}_{\widehat{U}L} & \mat{K}_{\widehat{U}U}
\end{bmatrix}_e
\begin{bmatrix}
\mat{K}_{LL} & \mat{K}_{LU} \\
\mat{K}_{UL} & \mat{K}_{UU}
\end{bmatrix}_e^{-1}
\begin{bmatrix}
\vect{f}_{L} \\
\vect{f}_{U}
\end{bmatrix}_e
\Bigr\},
\\
\begin{bmatrix}
\St{\vect{f}}
\end{bmatrix}
&=
\sum_{e=1}^{\St{\texttt{n}}^\texttt{el}}
\Bigl\{
\begin{bmatrix}
\vect{f}_{u}
\end{bmatrix}_e
\vphantom{
\begin{bmatrix}
\mat{K}_{LL} & \mat{K}_{LU} \\
\mat{K}_{UL} & \mat{K}_{UU}
\end{bmatrix}_e^{-1}}
\Bigr\}.
\end{align}
\label{eqn:fsisystemglobalrhsNitsche}%
\end{subequations}
The terms $\Fl{\texttt{n}}^\texttt{el}$ and $\St{\texttt{n}}^\texttt{el}$ denote the number of elements in the fluid and structural discretization, respectively, whereas $\FlStintf{\texttt{n}}^\texttt{el}$ refers to the number of elements adjacent to the interface (belonging either to the fluid or the structural subdomain).
In the formulas \eqref{eqn:fsisystemgloballhsNitsche}--\eqref{eqn:fsisystemglobalrhsNitsche}, the summation over elements is understood as the usual assembly process, adding the local matrices and vectors into the associated positions of the global matrices and vectors.
It can be observed from \eqref{eqn:fsisystemgloballhsNitsche} how $\Ff{\mat{K}}$ and $\Ss{\mat{K}}$ feature the usual structure of the matrices of the HDG and CG global problem, respectively, and they differ from the single-field matrices only for the inclusion of a small number of terms in $\mat{K}_{\widehat{U}\widehat{U}}$ and $\mat{K}_{uu}$ arising from the definition \eqref{eqn:fsinumericalfluxcg}.
The blocks $\Fs{\mat{K}}$ and $\Sf{\mat{K}}$ are responsible for the coupling and they simply stem from the linearization and discretization of $-\bigl<\hat{\vect{w}},\abs{\St{\mathbf{F}}}^{-1}\St{\mathbf{P}}{\St{\mathbf{F}}}^T\Fl{\vect{n}}+\gamma h^{-1}(d\St{\mathbf{u}}/dt)\bigr>$ and $\bigl<(\partial\St{\mathbf{P}}/\partial\bm{\nabla}\St{\mathbf{u}})\bm{\nabla}\mathbf{v}\St{\vect{n}},\Fl{\widehat{\rho\bm{\upsilon}}}\Fl{\hat{\rho}}^{-1}\bigr>
-\bigl<\mathbf{v},\gamma h^{-1}\Fl{\widehat{\rho\bm{\upsilon}}}\Fl{\hat{\rho}}^{-1}\bigr>$ along the interface, respectively.
The vectors $\Fl{\vect{f}}$ and $\St{\vect{f}}$ represent the residuals of the fluid and the structural global problems.
In the spirit of the Nitsche-based coupling of HDG and CG discretizations proposed in \cite{Laspina2020b}, the fluid HDG local problems remain unchanged and they require, at each Newton iteration, the solution of the linear systems
\begin{equation}
\begin{bmatrix}
\mat{K}_{LL} & \mat{K}_{LU} \\
\mat{K}_{UL} & \mat{K}_{UU}
\end{bmatrix}_e
\begin{bmatrix}
\delta\Fl{\vect{L}} \\
\delta\Fl{\vect{U}}
\end{bmatrix}_e
=
\begin{bmatrix}
\vect{f}_{L} \\
\vect{f}_{U}
\end{bmatrix}_e
-
\begin{bmatrix}
\mat{K}_{L\widehat{U}} \\
\mat{K}_{U\widehat{U}}
\end{bmatrix}_e
\begin{bmatrix}
\delta\Fl{\vect{\widehat{U}}}
\end{bmatrix}_e,
\label{eqn:fsisystemlocalNitsche}
\end{equation}
for $e=1,\dots,\Fl{\texttt{n}}^\texttt{el}$.

Since here the coupling takes place only at a global level, unlike the monolithic version of the Dirichlet--Neumann coupling mentioned in section \ref{sec:partitioneddirichletneumannalgorithm}, the communication of the HDG local matrices is not required outside the fluid block of the matrix and the right hand side in \eqref{eqn:fsisystemglobalNitsche}.
This segregation between the HDG local DOFs and and the CG DOFs leads to the minimally-intrusive computer implementation already mentioned.
Moreover, the HDG local problems and the HDG local postprocessing remain the same as in the pure HDG case and no special treatment of the interface elements is required.
Last but not least, since the Nitsche-based coupling solely relies on the hybrid variables to impose the coupling conditions, the treatment of non-matching grids and/or non-uniform polynomial degrees is easily handled without the need of introducing special projection operators.

\begin{remark}
The strategy presented in this section to couple the fluid and the structure could in theory be implemented in a partitioned fashion, by alternating the solution of pure fluid and structural problems and exchanging the interface information among the fields.
This method can be referred to as {\lq\lq partitioned Neumann--Dirichlet coupling\rq\rq} and it has been proposed for instance in \cite{Kuttler2006} as a possible remedy for the so-called incompressibility dilemma.
However, as stated by the same authors of \cite{Kuttler2006}, such a method fails to solve real world problems, since the response of stiff structures to varying interface displacements will be too sensitive for any numerical approach to find the equilibrium.
\end{remark}

\section{Numerical studies}
\label{sec:numericalstudies}

In this section several numerical studies are presented to assess the performance of the proposed HDG-CG formulation for weakly compressible fluid-structure interaction.
The convergence properties of the HDG formulation for weakly compressible flows introduced in section \ref{sec:hdgformulationweaklycompressibleflows} are analyzed first with respect to a simple steady state Poiseuille flow equipped with analytical solution and then with respect to an unsteady flow on a moving mesh which exercises all the terms present in the fluid PDEs.
The third example verifies the optimal convergence of the HDG-CG coupling schemes on a problem with manufactured solution and the following examples solve weakly compressible fluid-structure interaction problems on two and three dimensions.

\subsection{Weakly compressible Poiseuille flow}
\label{sec:weaklycompressiblepoiseuilleflow}

The first numerical example considers a steady state isothermal Poiseuille flow of a weakly compressible Newtonian fluid in a straight channel.
The goal of this study is to show on a simple and physically meaningful example the convergence properties of the proposed HDG formulation, as well as its robustness with respect to the compressibility level.
In the work \cite{Housiadas2016}, the authors derive an analytical solution by representing the primary flow variables as asymptotic expansions of the compressibility coefficient, which is assumed to be a small parameter, and perturbing them with respect to the same coefficient.
The solution is then found up to the first order in $\varepsilon$.
The study \cite{Housiadas2016} considers also a pressure-dependent viscosity, but this feature has been neglected because unimportant in the context of the present contribution.
Since this example concerns a pure flow problem, the subscript $\Fl{(\cdot)}$ is omitted for brevity.

The analytical solution in terms of velocity and pressure can be written as
\begin{equation}
\begin{split}
  \upsilon_x(x,y)& = \frac{3}{2}U\Bigl[1-\Bigl(\frac{y}{R}\Bigr)^2\Bigr]-\frac{9}{2}\frac{\mu LU^2}{\rho_0 R^2}\Bigl(1-\frac{x}{L}\Bigr)\Bigl[1-\Bigl(\frac{y}{R}\Bigr)^2\Bigr]\varepsilon, 
  \\
  \upsilon_y(x,y)& = 0, 
  \\
  p(x,y)& = p_0+3\frac{\mu LU}{R^2}\Bigl(1-\frac{x}{L}\Bigr) 
                 -\frac{3}{2} \frac{\mu^2U^2}{\rho_0R^2} \Bigl\{3\Bigl(\frac{L}{R}\Bigr)^2\Bigl(1-\frac{x}{L}\Bigr)^2-\Bigl[1-\Bigl(\frac{y}{R}\Bigr)^2\Bigr]\Bigr\}\varepsilon,
\end{split}
\label{eqn:analyticalsolutionweaklycompressiblepoiseuilleflow}
\end{equation}
where $L$ and $R$ represent the length and the half-height of the channel, respectively, while $U$ denotes the mean velocity at the channel exit.
The solution of the density is derived from the equation of state \eqref{eqn:fluidequationofstate} using the expression of the pressure in \eqref{eqn:analyticalsolutionweaklycompressiblepoiseuilleflow} and the solution of the momentum is then obtained by multiplying the density just derived with the velocity field in \eqref{eqn:analyticalsolutionweaklycompressiblepoiseuilleflow}.
No body forces in the momentum equation appear in this expansion.
Due to the asymptotic expansion of the solution in terms of $\varepsilon$ up to the first order, a residual $\mathcal{O}\left(\varepsilon^2\right)$ is added to the right hand side of the continuity equation in \eqref{eqn:fluidgoverningequationsfixeddomain} in the spirit of manufactured solutions:
\begin{equation}
  \mathcal{R}(x,y) = \frac{27\mu^2U^3}{4\rho_0 R^8}\left(R^2{-}y^2\right)
                               \left\{6R^2\left(L{-}x\right)\varepsilon^2 {-}\mu U\rho_0^{-1} \left[9\left(L{-}x\right)^2{-}\left(R^2{-}y^2\right)\right]\varepsilon^3\right\}.
\label{eqn:residualcontinuityweaklycompressiblepoiseuilleflow}
\end{equation}
From the expressions \eqref{eqn:analyticalsolutionweaklycompressiblepoiseuilleflow} it can be easily observed that the solution of a classic incompressible Poiseuille flow in a straight channel is fully recovered when $\varepsilon\rightarrow0$.
In this situation the axial velocity assumes a simple parabolic profile, while the pressure varies linearly along the channel.

\begin{figure}
\centering
\subfigure[$\upsilon_x(x,0)$]
{\includegraphics[width=0.49\textwidth]{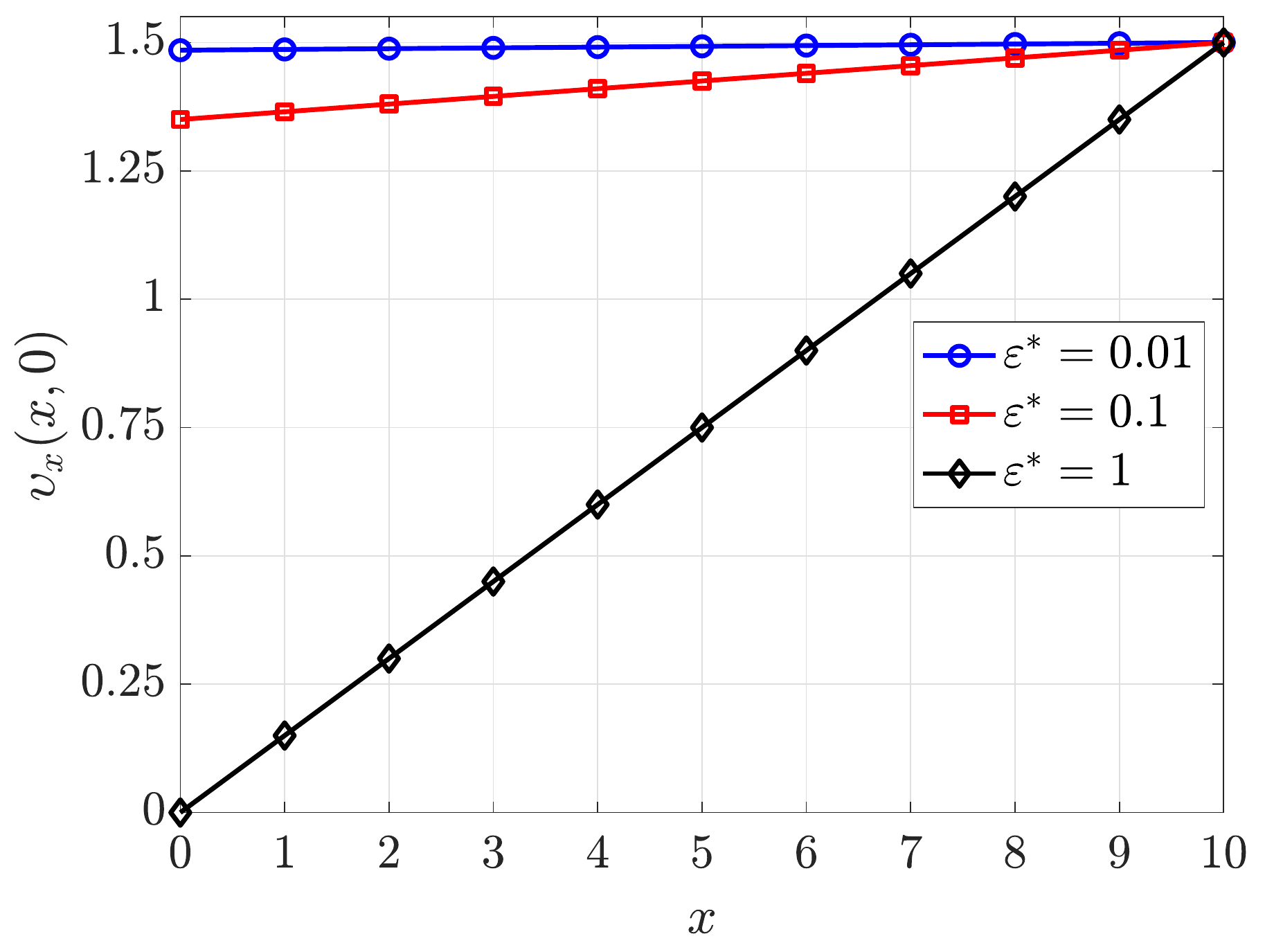}}
\hfill
\subfigure[$p(x,0)$]
{\includegraphics[width=0.49\textwidth]{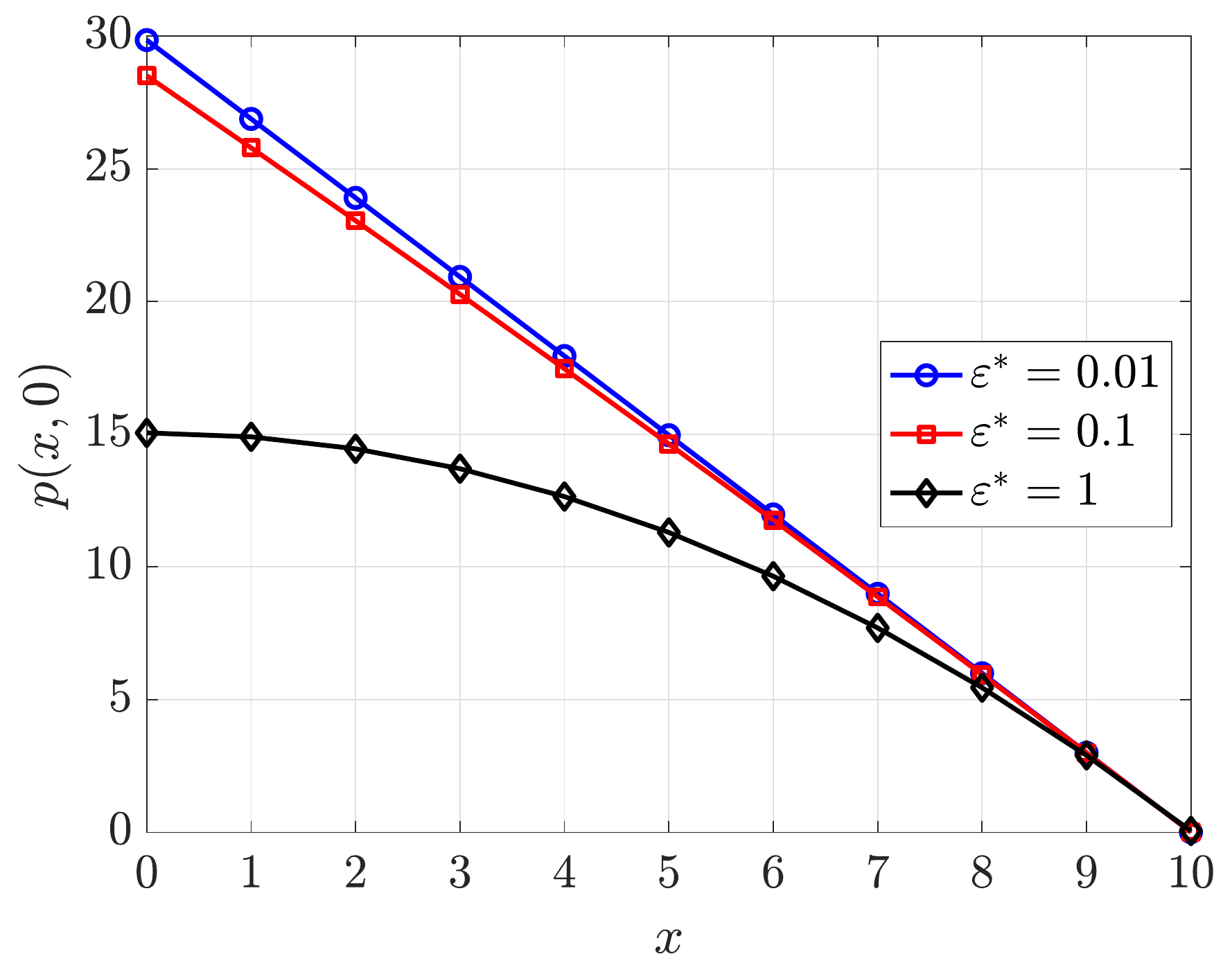}}
\caption{Axial velocity (left) and pressure (right) along the horizontal axis of the weakly compressible Poiseuille flow with different values of the dimensionless compressibility coefficient.}
\label{fig:solutionaxisweaklycompressiblepoiseuilleflow}
\end{figure}

The fluid domain is the rectangle $\Omega=[0,L]\times[-R,R]$, with $L=10$ and $R=1$.
The viscosity $\mu$ is considered equal to $1$ as well as the mean velocity at the channel exit $U$ and the reference density $\rho_0$, evaluated at the reference pressure $p_0=0$.
A dimensionless compressibility number can be defined as
\begin{equation}
\varepsilon^*=\frac{3\mu LU}{\rho_0 R^2}\varepsilon
\label{eqn:dimensionlesscompressibilityweaklycompressiblepoiseuilleflow}
\end{equation}
and three different orders of magnitude are considered in the following studies, i.e., $\varepsilon^*=[0.01,0.1,1]$.
In Figure \ref{fig:solutionaxisweaklycompressiblepoiseuilleflow} the analytical axial velocity and pressure along the horizontal axis are plotted for the three different dimensionless compressibility numbers considered.
With regards to the velocity, it varies linearly along the horizontal axis and it always reaches the value $1.5$ (corresponding to a mean velocity $U=1$ at the channel exit).
Clearly, the solution has a physical meaning only when the volumetric flow rate at the entrance of the channel is positive, therefore $\varepsilon^*=1$ constitutes an upper limit of validity of the analytical solution in terms of the compressibility coefficient.
Although this last case lacks physical meaningfulness, it is interesting to investigate the performance of the numerical method proposed also in this extreme case.
With regards to the pressure, it varies quadratically along the horizontal axis and it always reaches the reference value ($p_0=0$) at the channel exit corners.
Moreover, the average pressure drop required to drive the flow decreases with the compressibility.

\begin{figure}
\centering
\subfigure[Mesh with $m=1$]
{\includegraphics[width=0.32\textwidth]{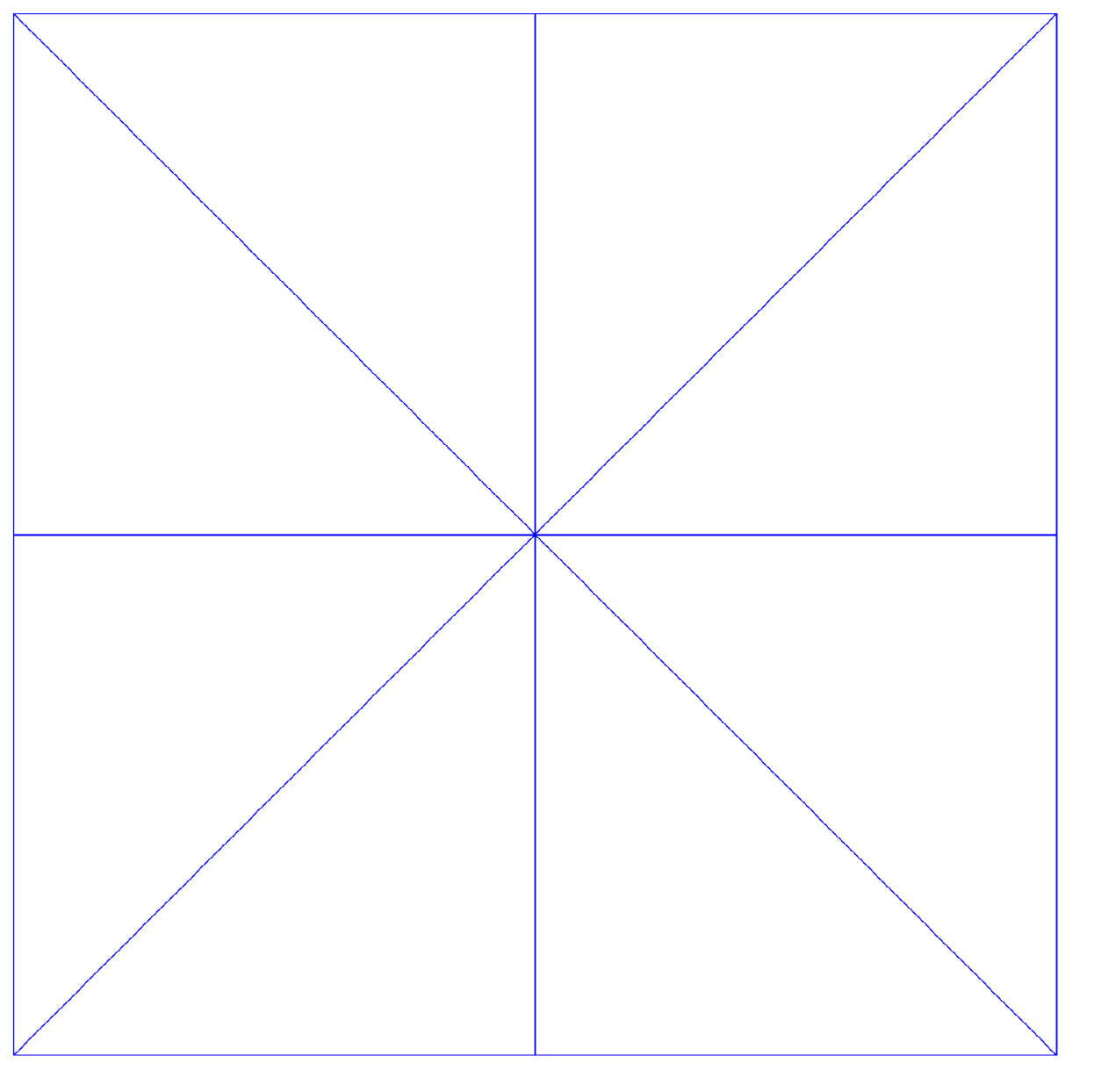}}
\hfill
\subfigure[Mesh with $m=2$]
{\includegraphics[width=0.32\textwidth]{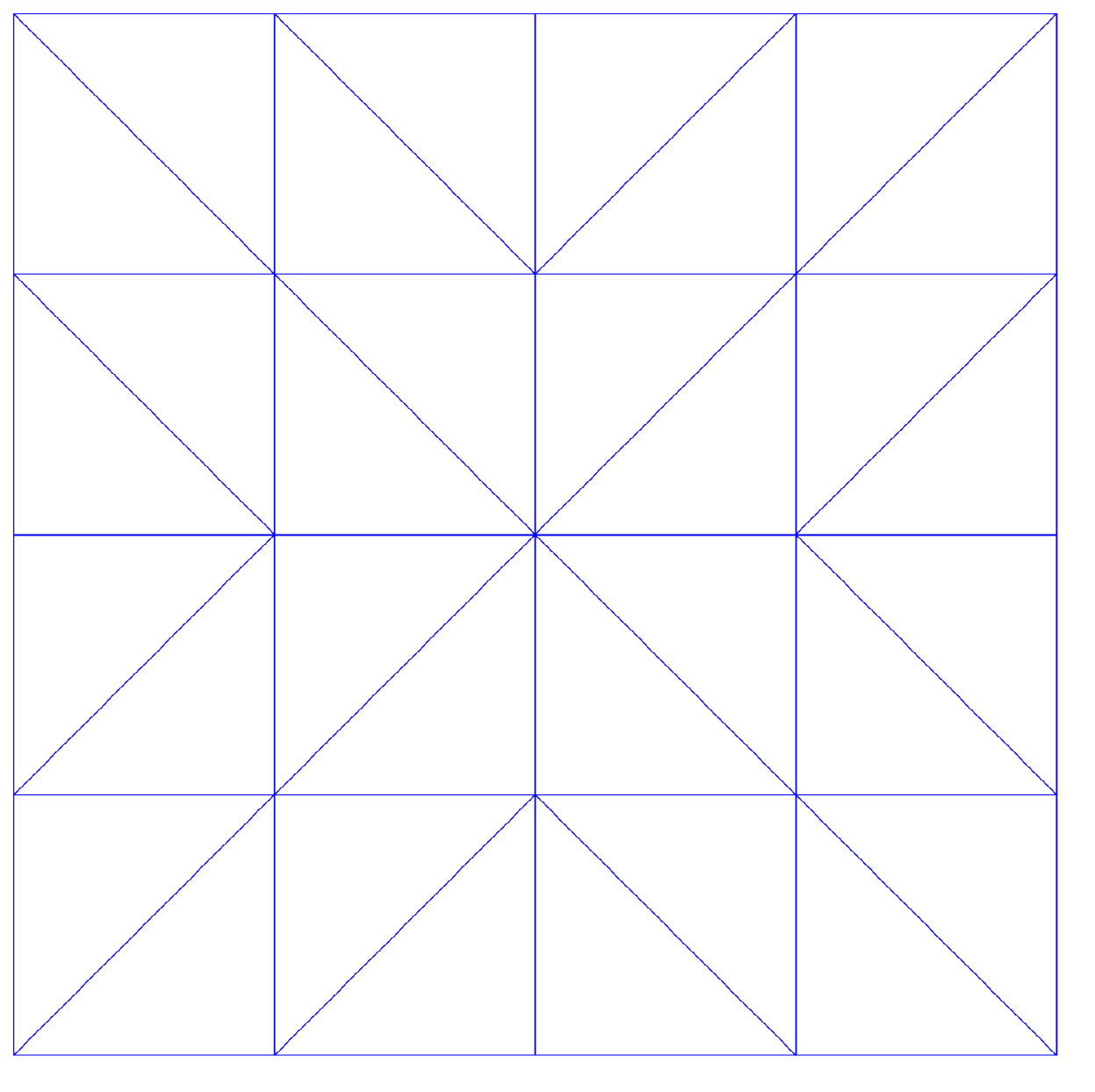}}
\hfill
\subfigure[Mesh with $m=3$]
{\includegraphics[width=0.32\textwidth]{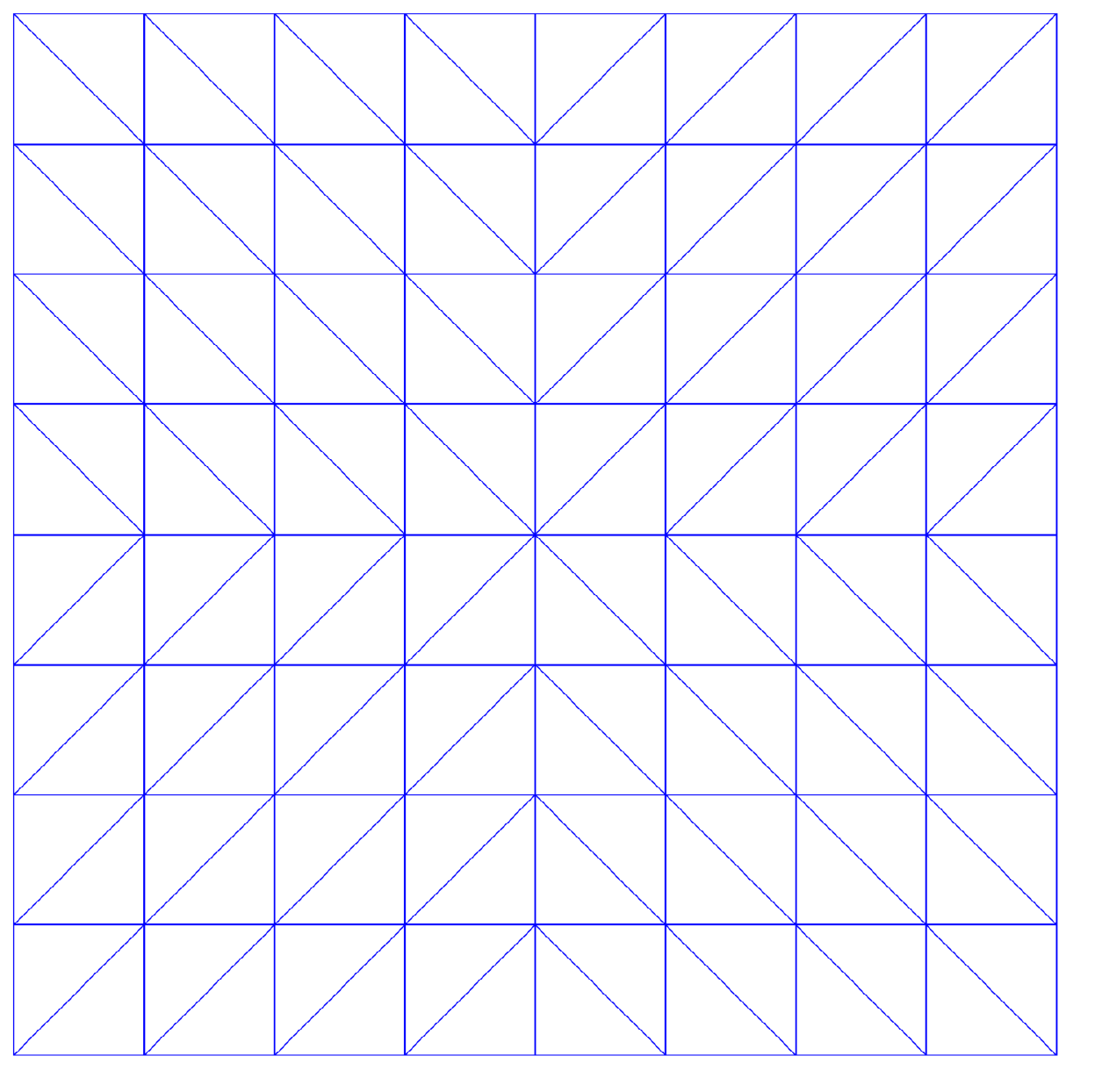}}
\caption{First three levels of refinement of the mesh used for the convergence studies of the weakly compressible Poiseuille flow.}
\label{fig:meshweaklycompressiblepoiseuilleflow}
\end{figure}

\begin{figure}
\centering
\subfigure[$\rho$ with $\varepsilon^*=0.01$]
{\includegraphics[width=0.32\textwidth]{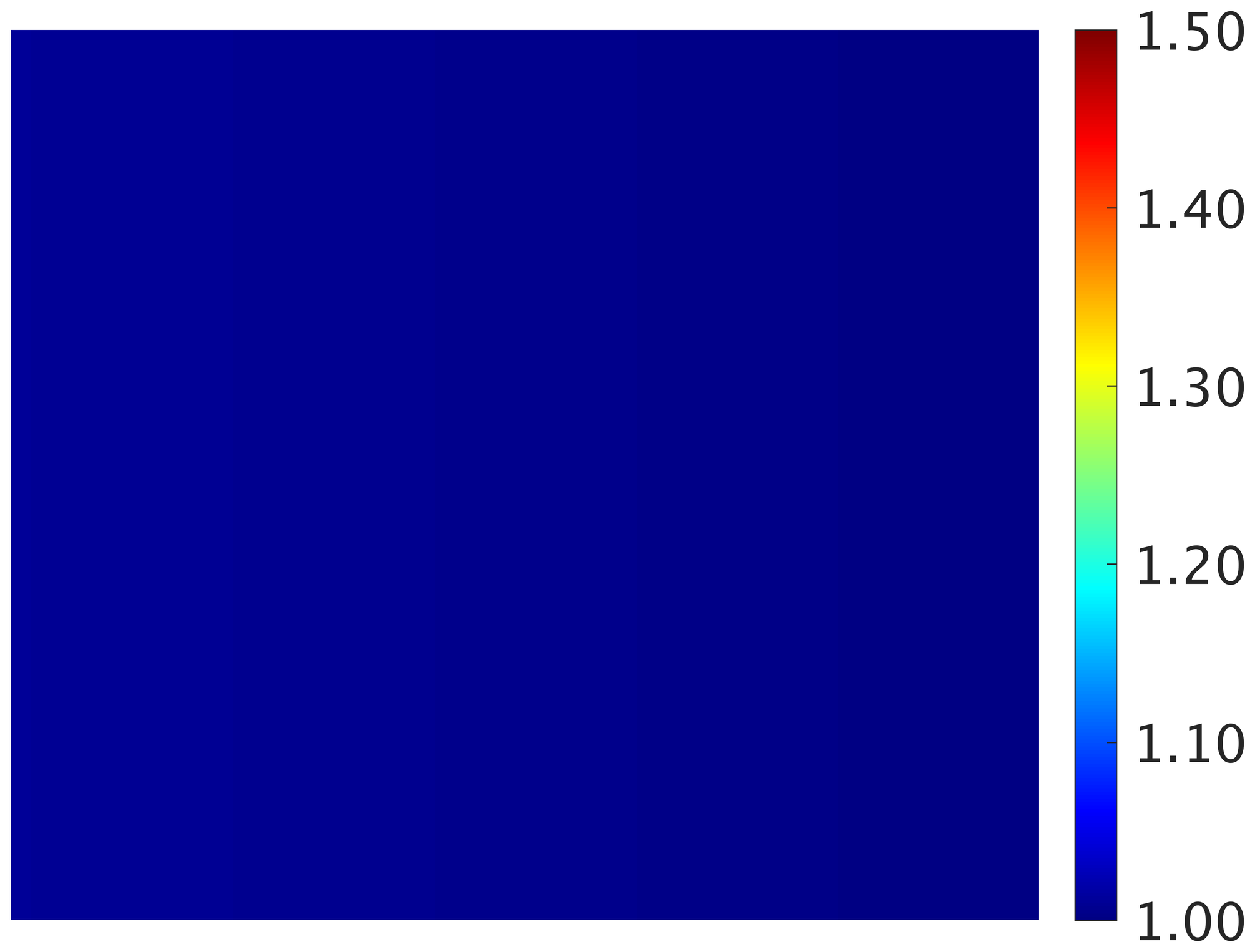}}
\hfill
\subfigure[$\rho$ with $\varepsilon^*=0.1$]
{\includegraphics[width=0.32\textwidth]{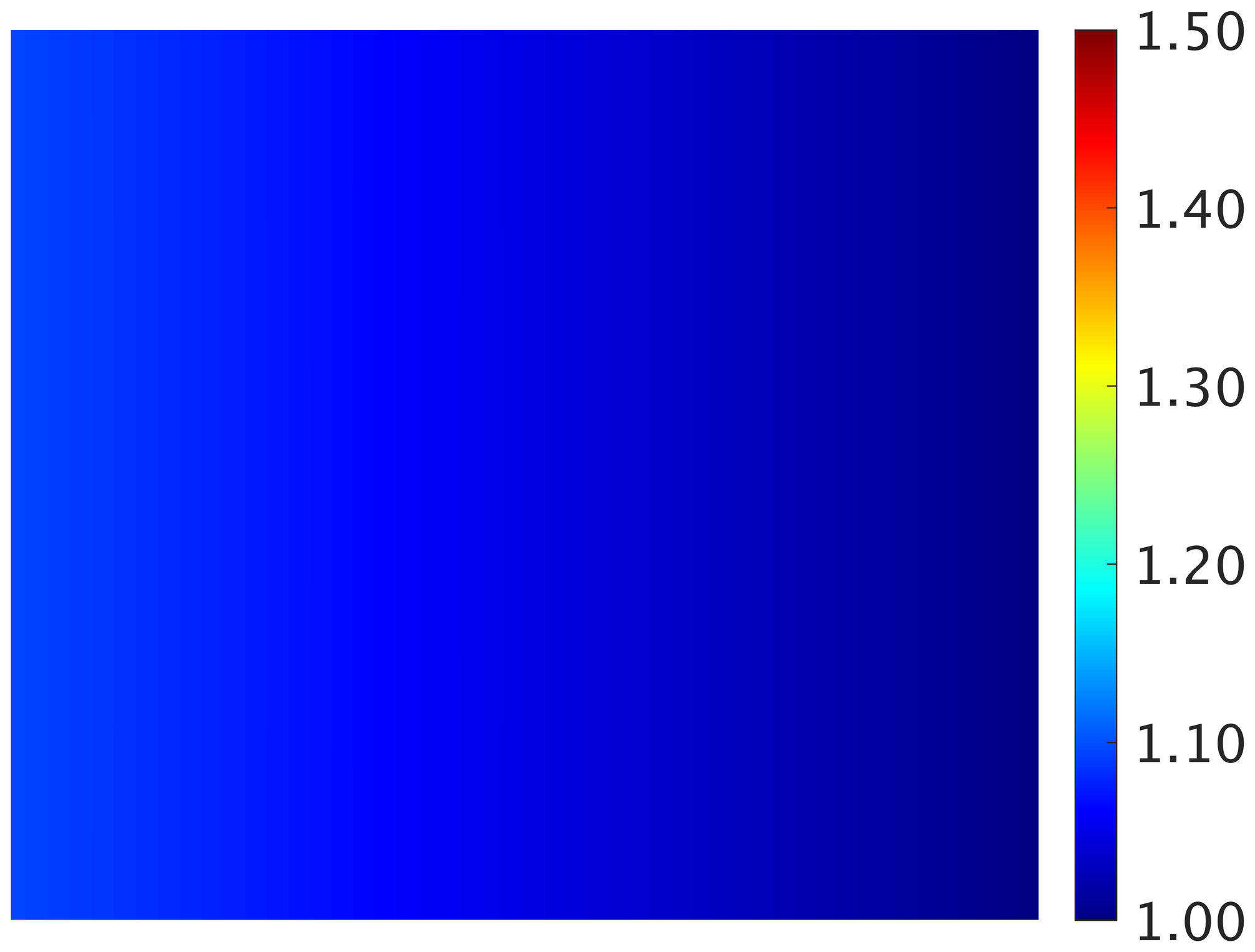}}
\hfill
\subfigure[$\rho$ with $\varepsilon^*=1$]
{\includegraphics[width=0.32\textwidth]{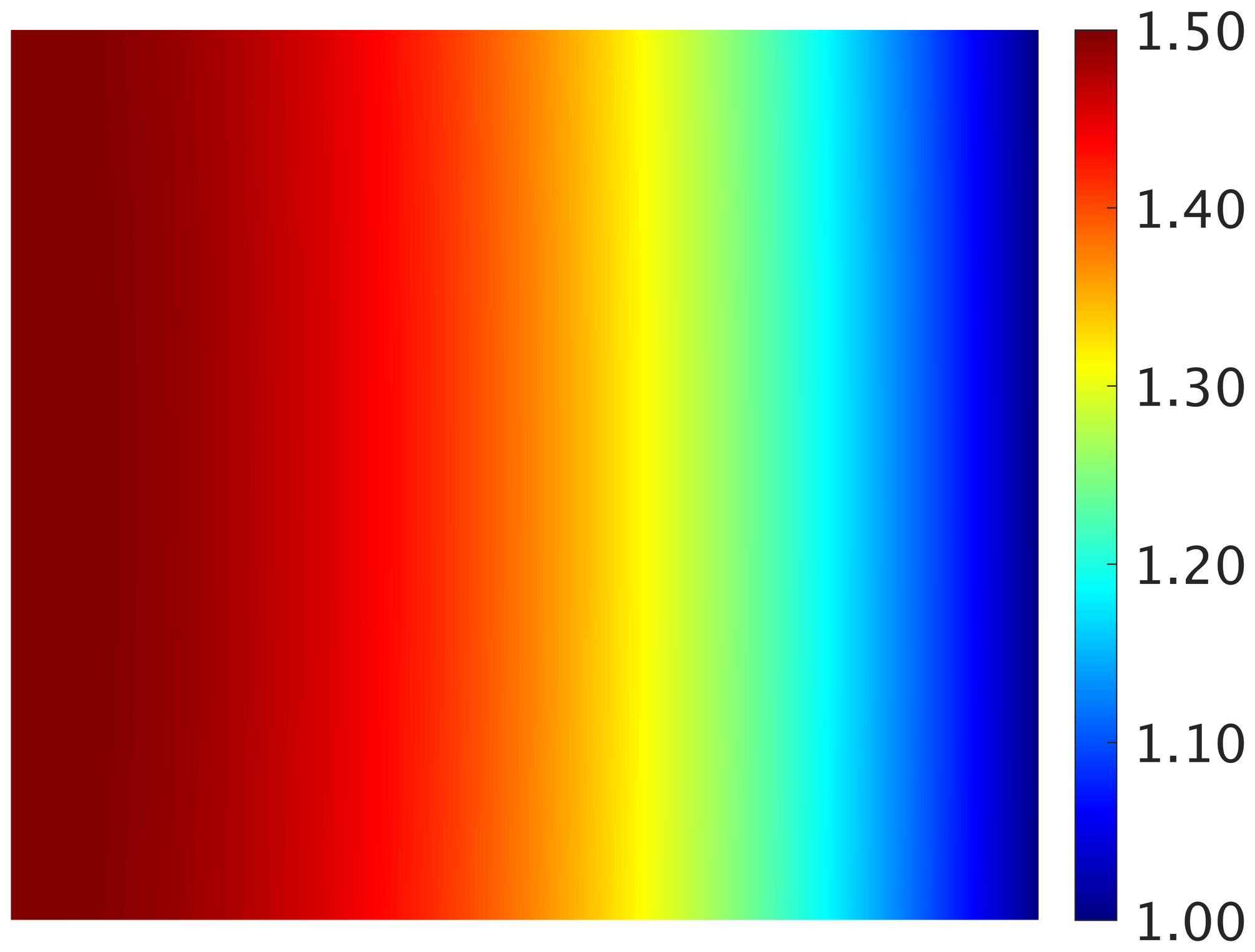}} \\
\subfigure[$\rho\upsilon_x$ with $\varepsilon^*=0.01$]
{\includegraphics[width=0.32\textwidth]{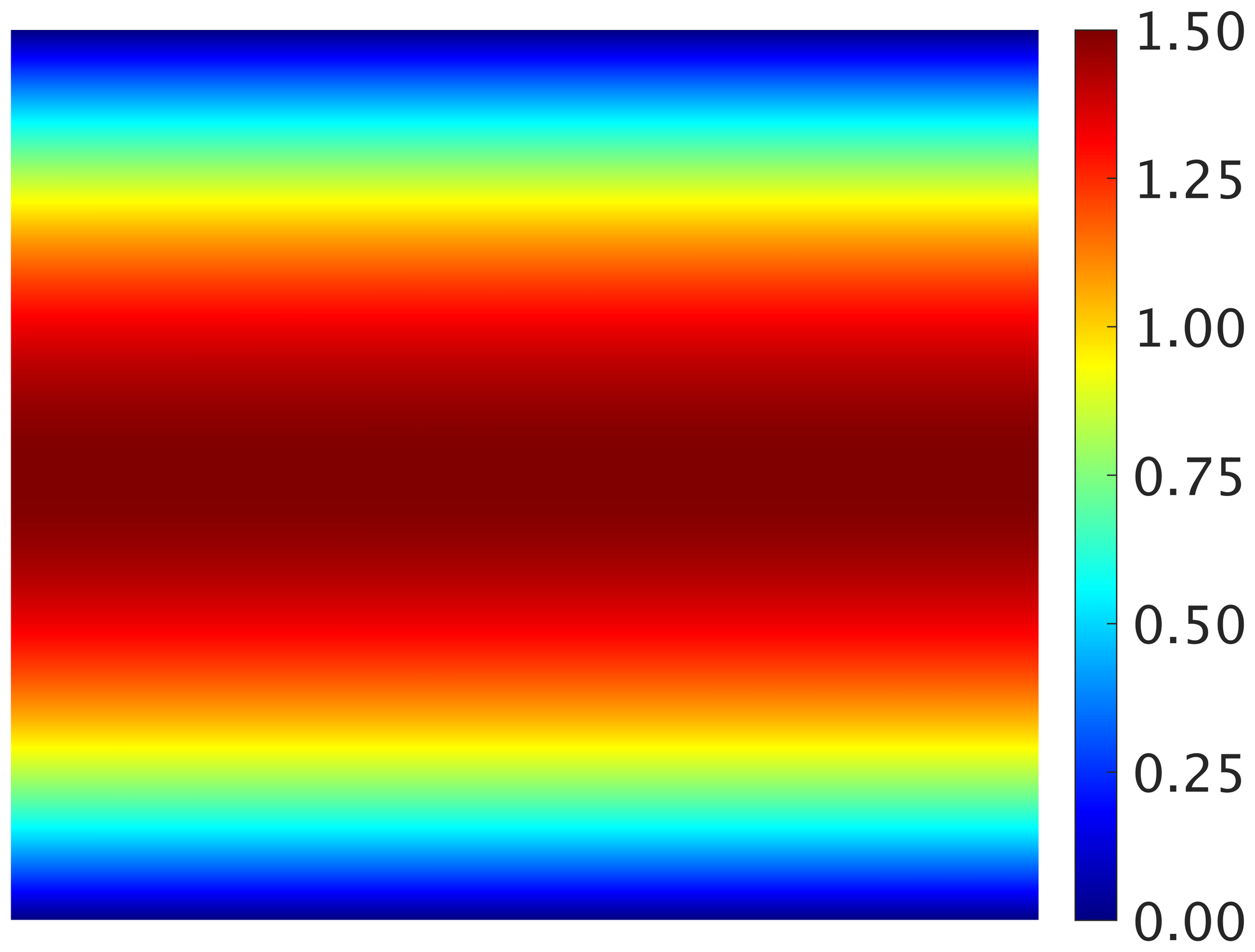}}
\hfill
\subfigure[$\rho\upsilon_x$ with $\varepsilon^*=0.1$]
{\includegraphics[width=0.32\textwidth]{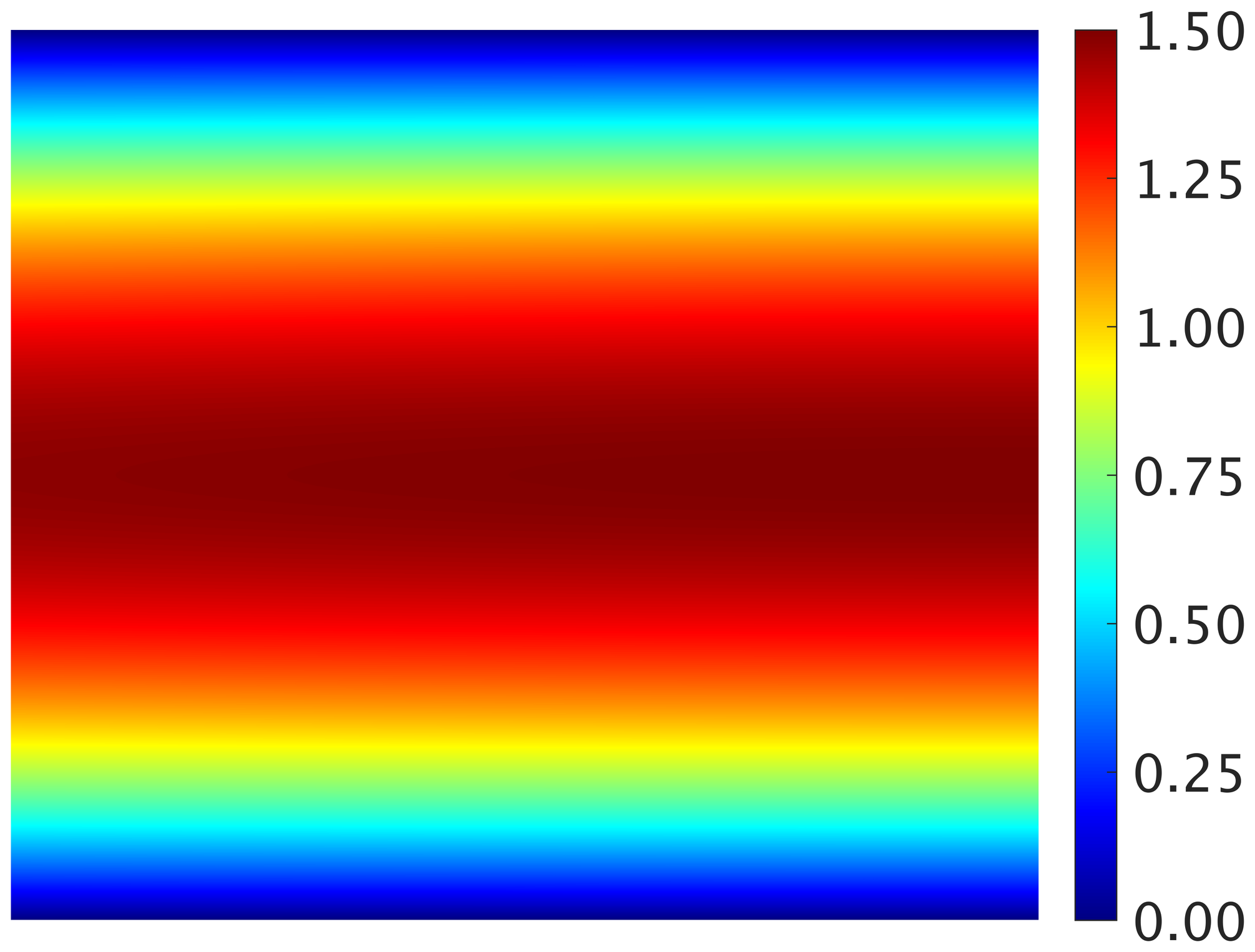}}
\hfill
\subfigure[$\rho\upsilon_x$ with $\varepsilon^*=1$]
{\includegraphics[width=0.32\textwidth]{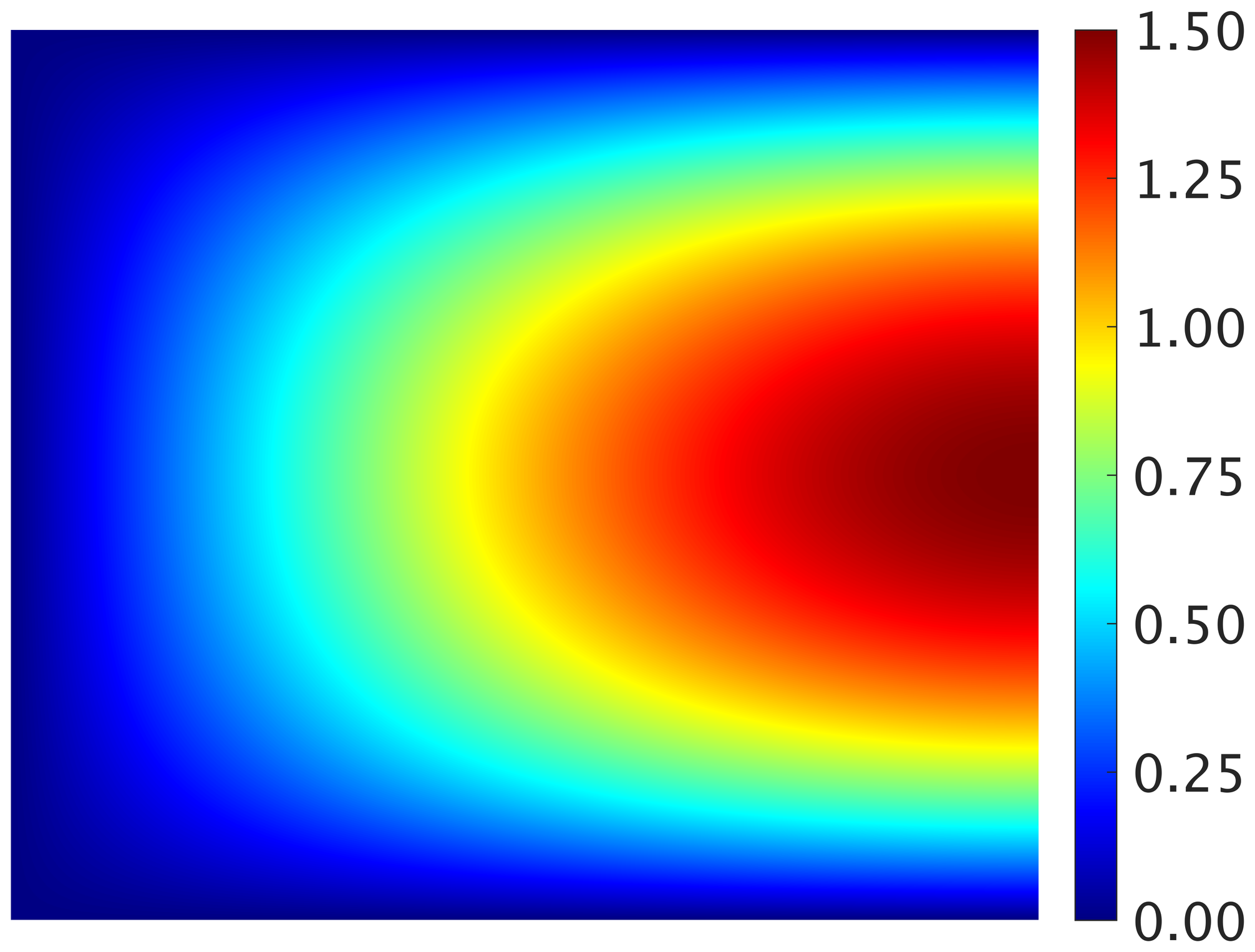}}
\caption{Approximation of the density and the momentum field of the weakly compressible Poiseuille flow with different values of the dimensionless compressibility coefficient.}
\label{fig:solutionweaklycompressiblepoiseuilleflow}
\end{figure}

\begin{figure}
\centering
\subfigure[$\upsilon_x$ with $\varepsilon^*=0.1$]
{\includegraphics[width=0.40\textwidth]{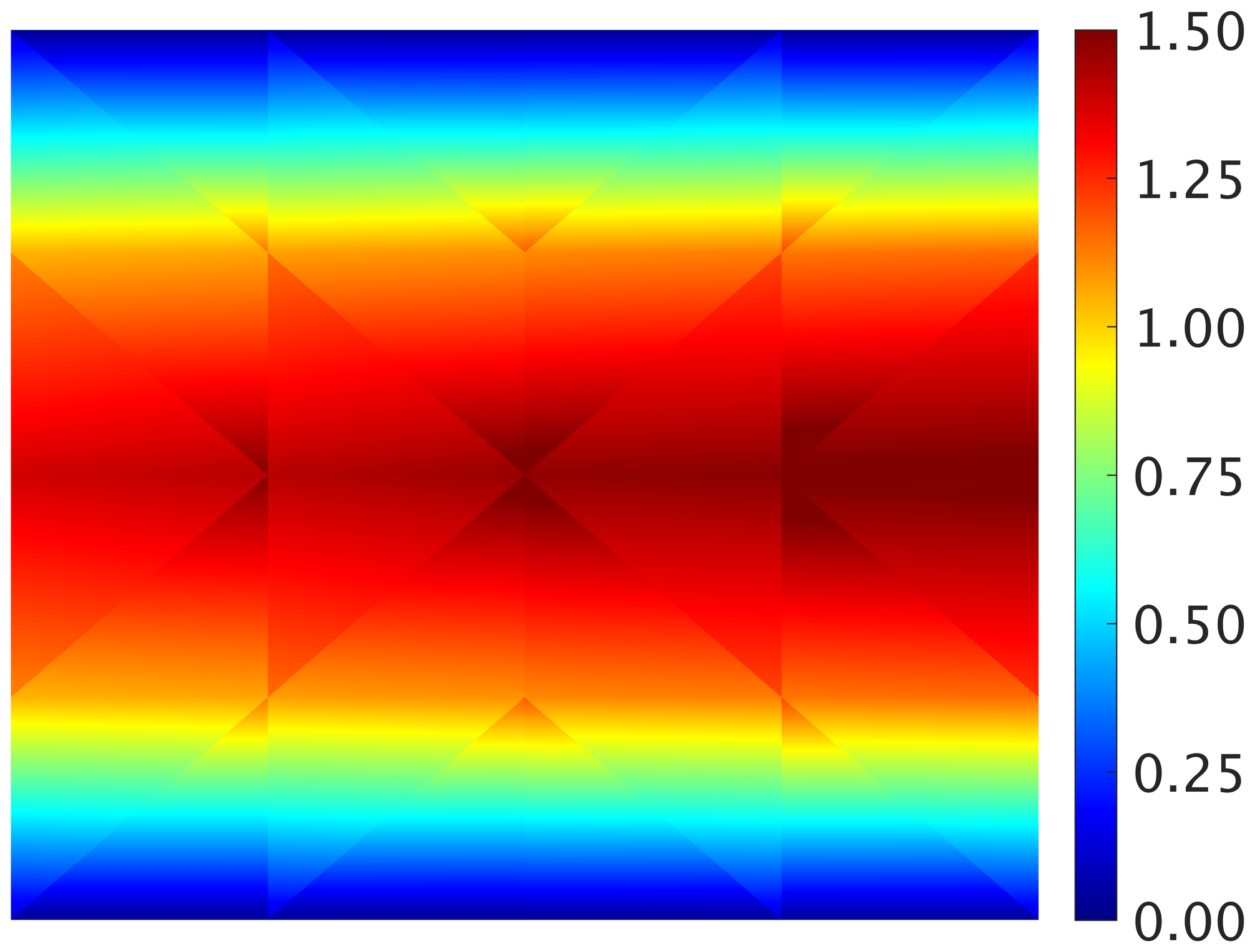}}
{\qquad}
\subfigure[$\upsilon^\star_x$ with $\varepsilon^*=0.1$]
{\includegraphics[width=0.40\textwidth]{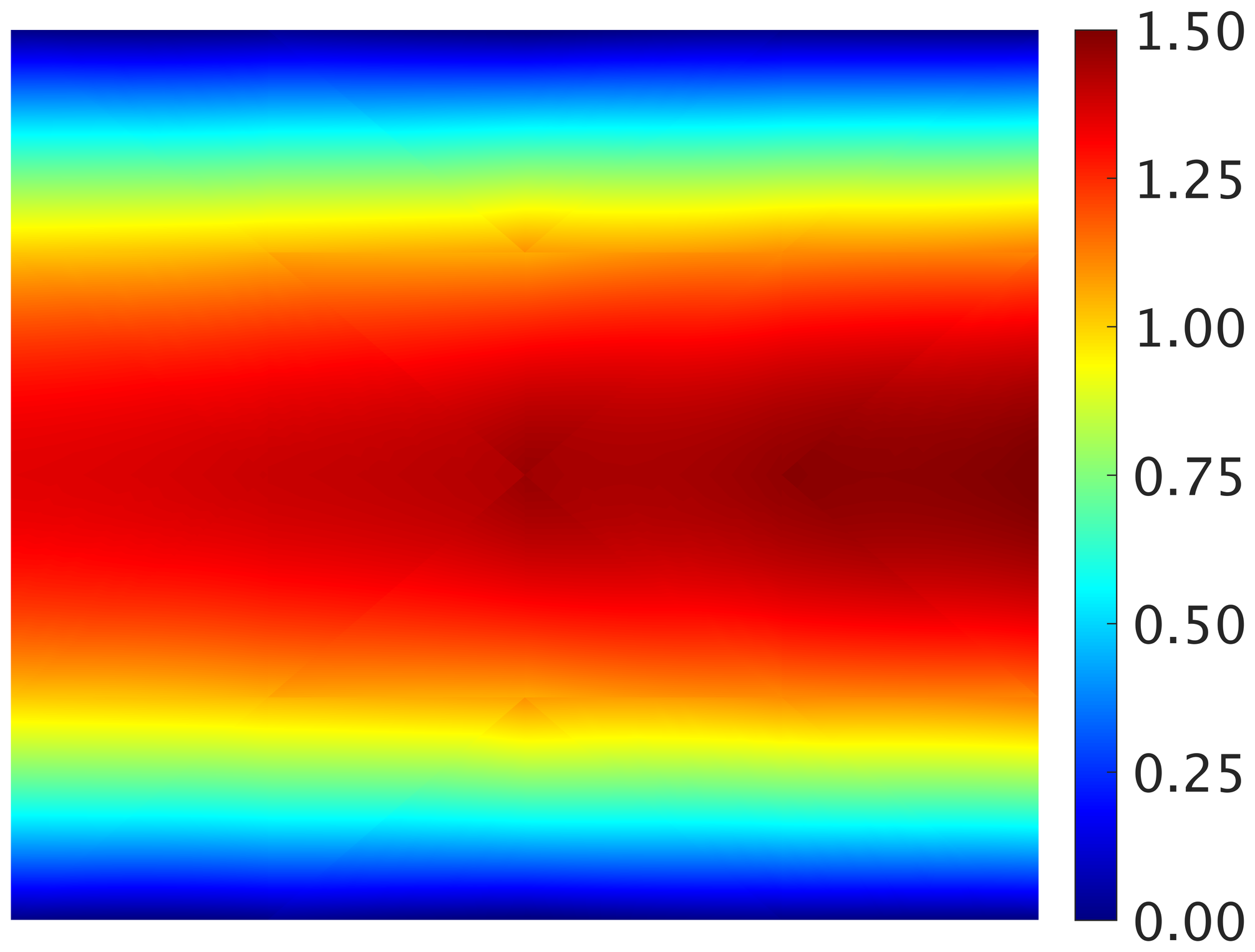}}
\caption{Approximation of the velocity and the postprocessed velocity field of the weakly compressible Poiseuille flow with $\varepsilon^*=0.1$.}
\label{fig:solutionpostweaklycompressiblepoiseuilleflow}
\end{figure}

In the following experiments, Dirichlet boundary conditions, corresponding to the restriction of the analytical solution to the domain boundary, are imposed on $\Gamma^D=\partial\Omega$.
Uniform meshes of triangular elements are considered for the whole domain by splitting a regular $2^m\times 2^m$ Cartesian grid (with $m$ ranging from $1$ to $5$) into a total of $2^{2m+1}$ triangles, giving element sizes of $h=L/2^m$.
The first three levels of refinement of the mesh used for the convergence studies are shown in Figure \ref{fig:meshweaklycompressiblepoiseuilleflow}.
The degree of approximation $k$ used ranges from $1$ to $2$ for $\varepsilon^*=0.01$, from $1$ to $3$ for $\varepsilon^*=0.1$ and from $1$ to $4$ for $\varepsilon^*=1$.
The stabilization parameters are computed according to \eqref{eqn:fluidstabilizationparameters}, considering $\left|\bm{\upsilon}\right|=U$ and $l=R$ as representative velocity and length, respectively.
The scaling factors are chosen as $C_\rho=3.33$ and $C_{\rho\upsilon}^d=1$ (no convective effects are included in this flow configuration), returning the stabilization parameters $\tau_\rho=3.33/\varepsilon$ and $\tau_{\rho\upsilon}=1$ for the density and the momentum, respectively.
In Figure \ref{fig:solutionweaklycompressiblepoiseuilleflow} the solution of the density and the momentum field obtained with the proposed HDG formulation using $m=5$ and $k=2$ is shown.
With regards to the density, its maximum variation from the reference value is about $1\%$ for $\varepsilon^*=0.01$, $10\%$ for $\varepsilon^*=0.1$ and $50\%$ for $\varepsilon^*=1$, for which the maximum value of the density reaches $1.50$.
In Figure \ref{fig:solutionpostweaklycompressiblepoiseuilleflow} the improvement of the approximation of the velocity field given by the local postprocessing described in section \ref{sec:localpostprocessing} is exemplarily shown for the intermediate compressibility coefficient ($\varepsilon^*=0.1$) using $m=2$ and $k=1$.

\begin{figure}
\centering
\subfigure[$\mat{L}$ with $\varepsilon^*=0.01$]
{\includegraphics[width=0.32\textwidth]{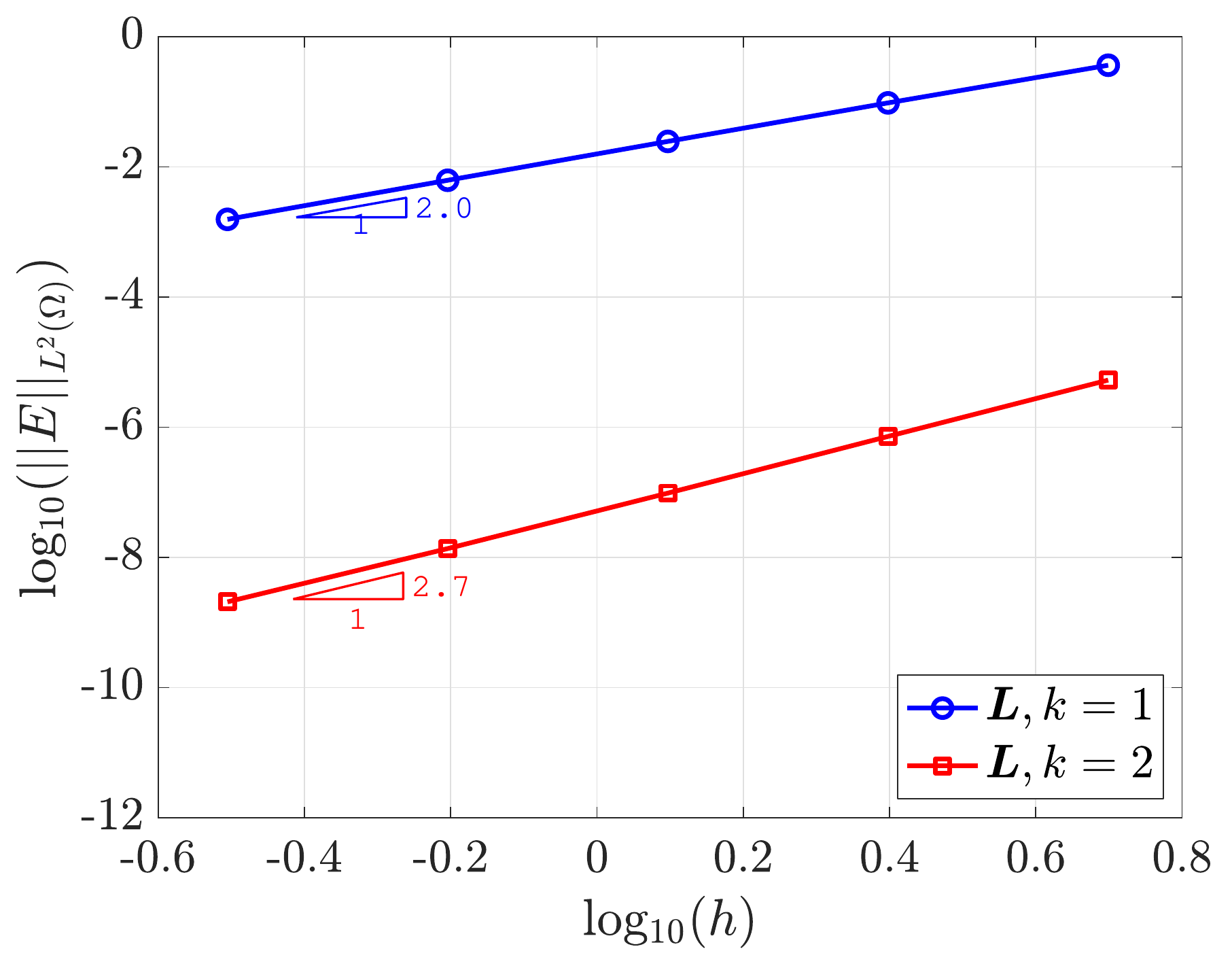}}
\subfigure[$\mat{L}$ with $\varepsilon^*=0.1$]
{\includegraphics[width=0.32\textwidth]{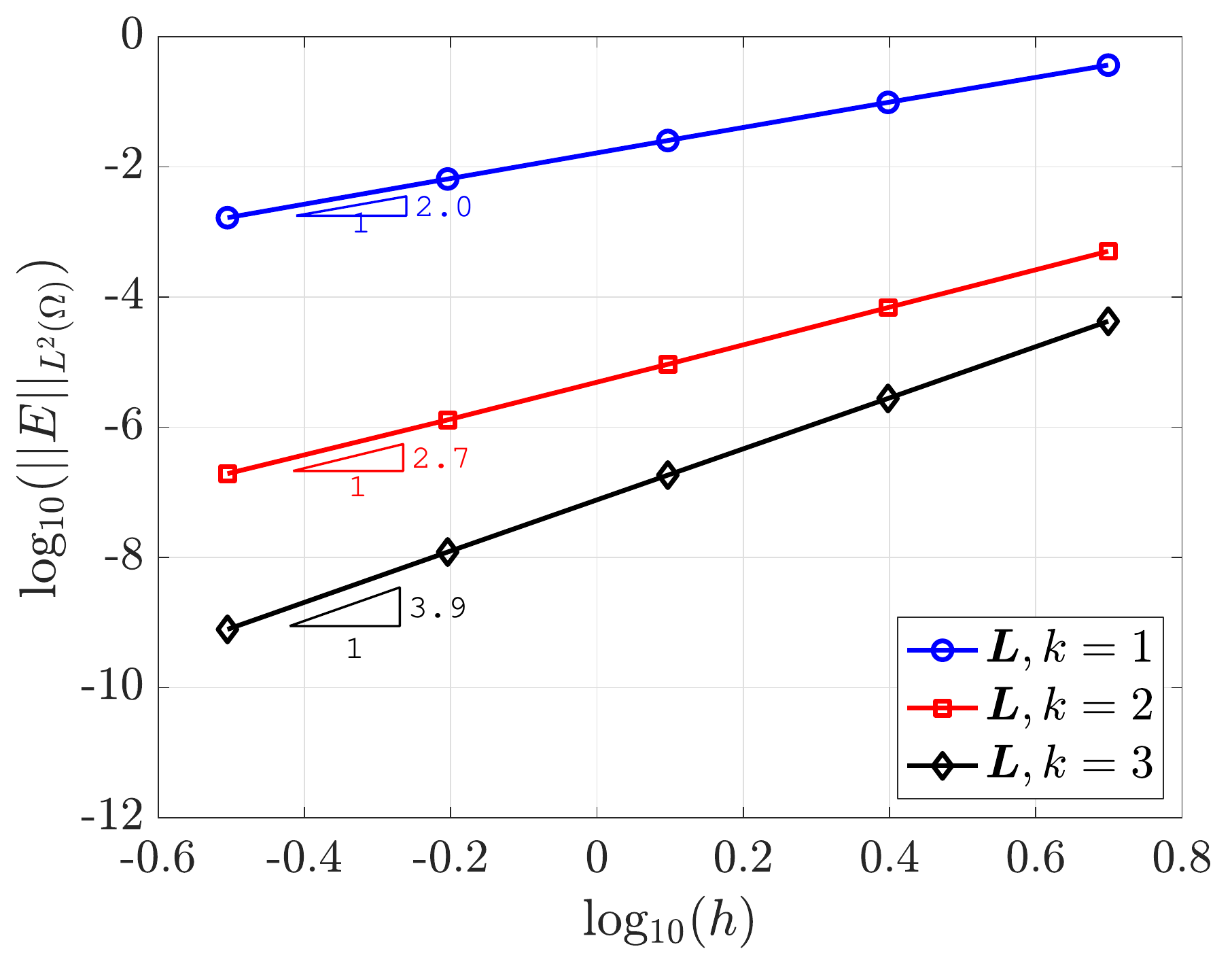}}
\subfigure[$\mat{L}$ with $\varepsilon^*=1$]
{\includegraphics[width=0.32\textwidth]{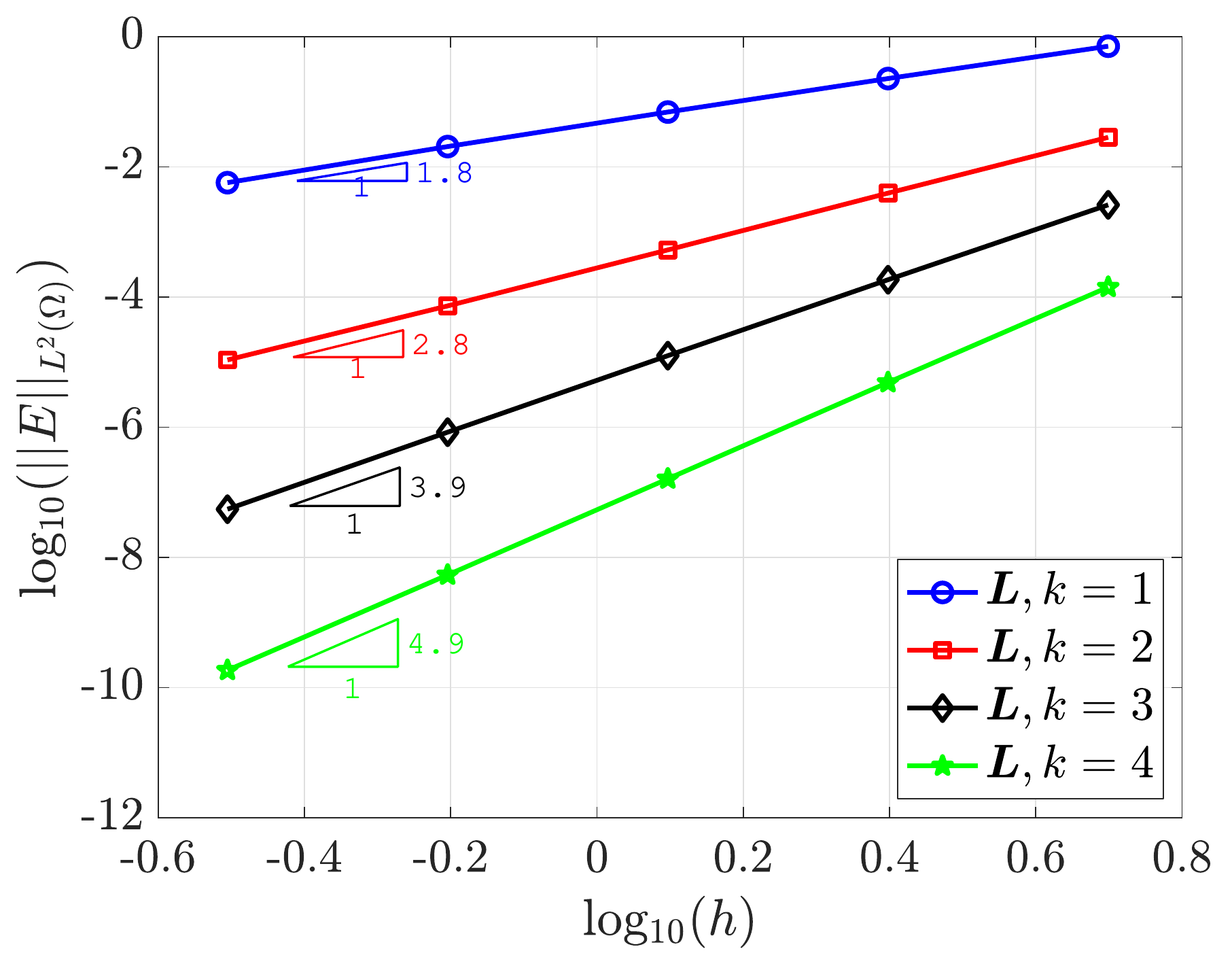}} \\
\subfigure[$\rho$ with $\varepsilon^*=0.01$]
{\includegraphics[width=0.32\textwidth]{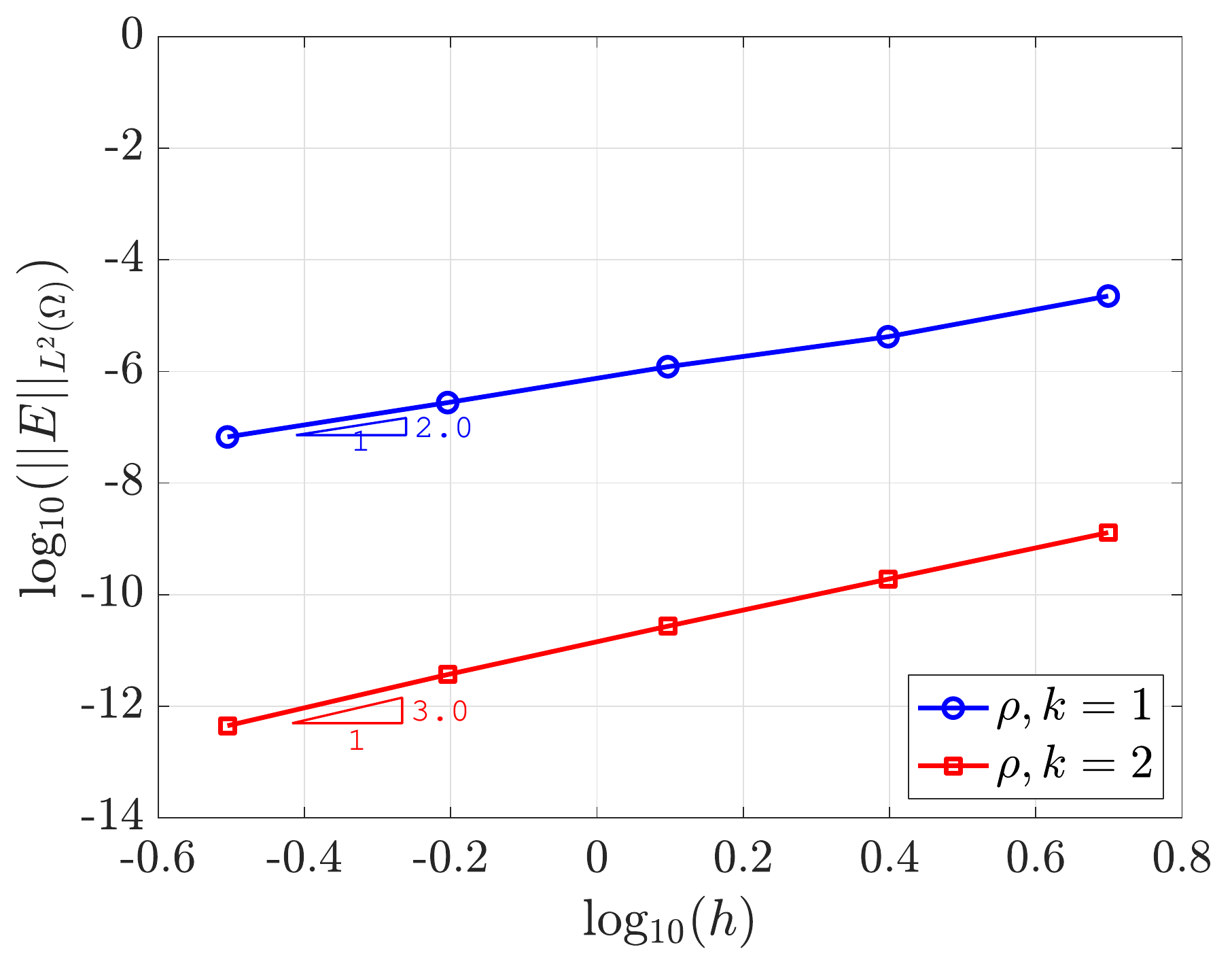}}
\subfigure[$\rho$ with $\varepsilon^*=0.1$]
{\includegraphics[width=0.32\textwidth]{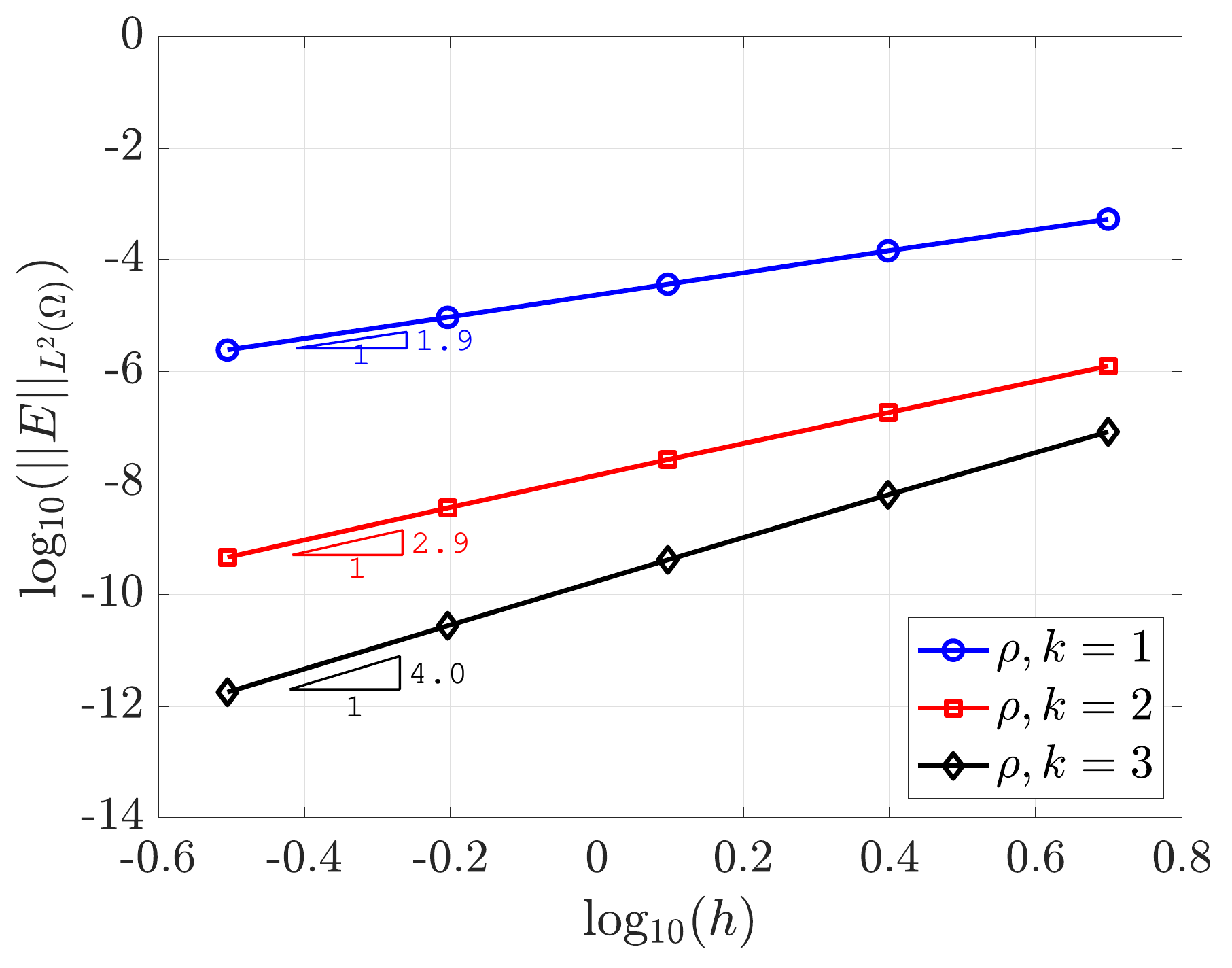}}
\subfigure[$\rho$ with $\varepsilon^*=1$]
{\includegraphics[width=0.32\textwidth]{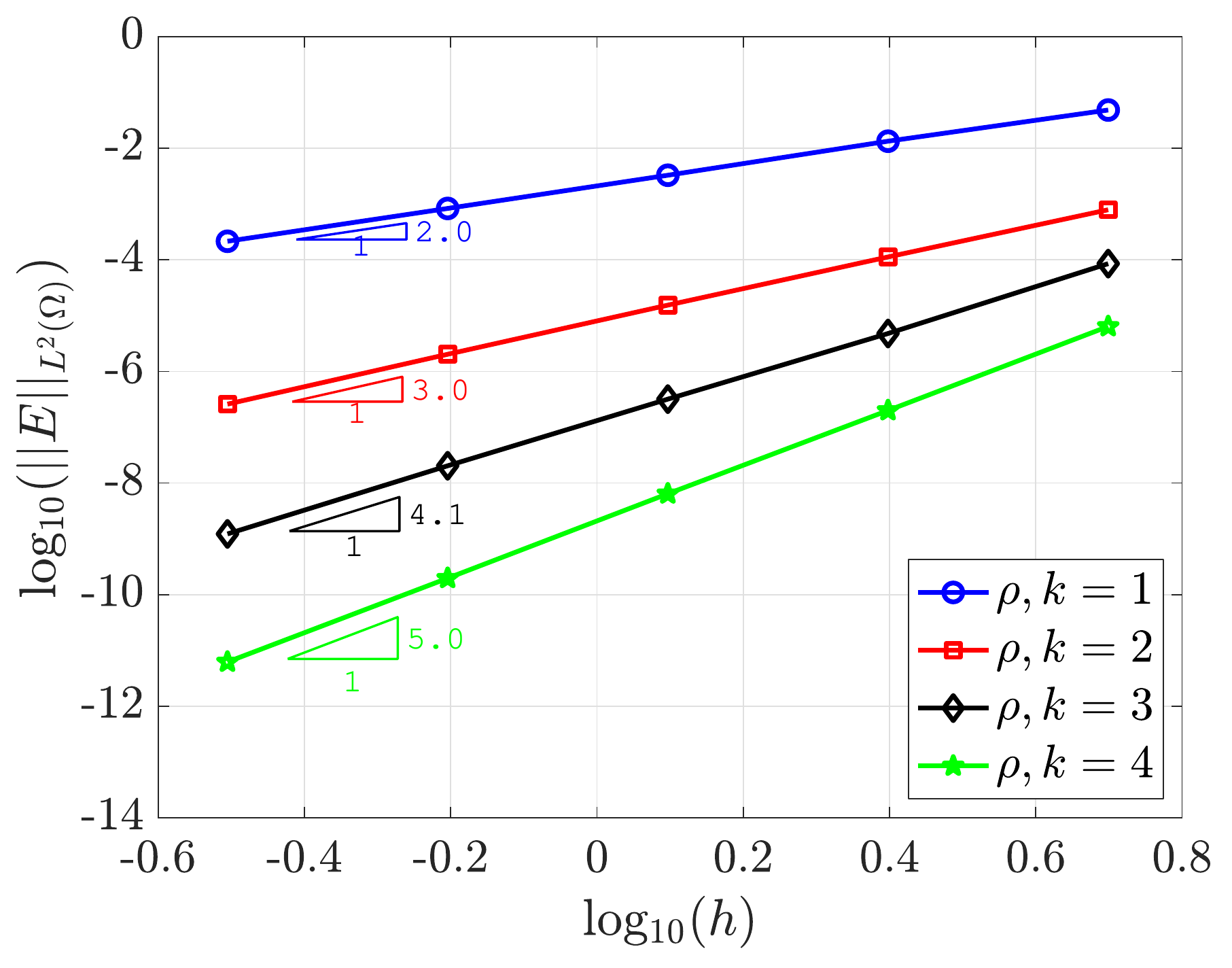}} \\
\subfigure[$\rho\bm{\upsilon}$ with $\varepsilon^*=0.01$]
{\includegraphics[width=0.32\textwidth]{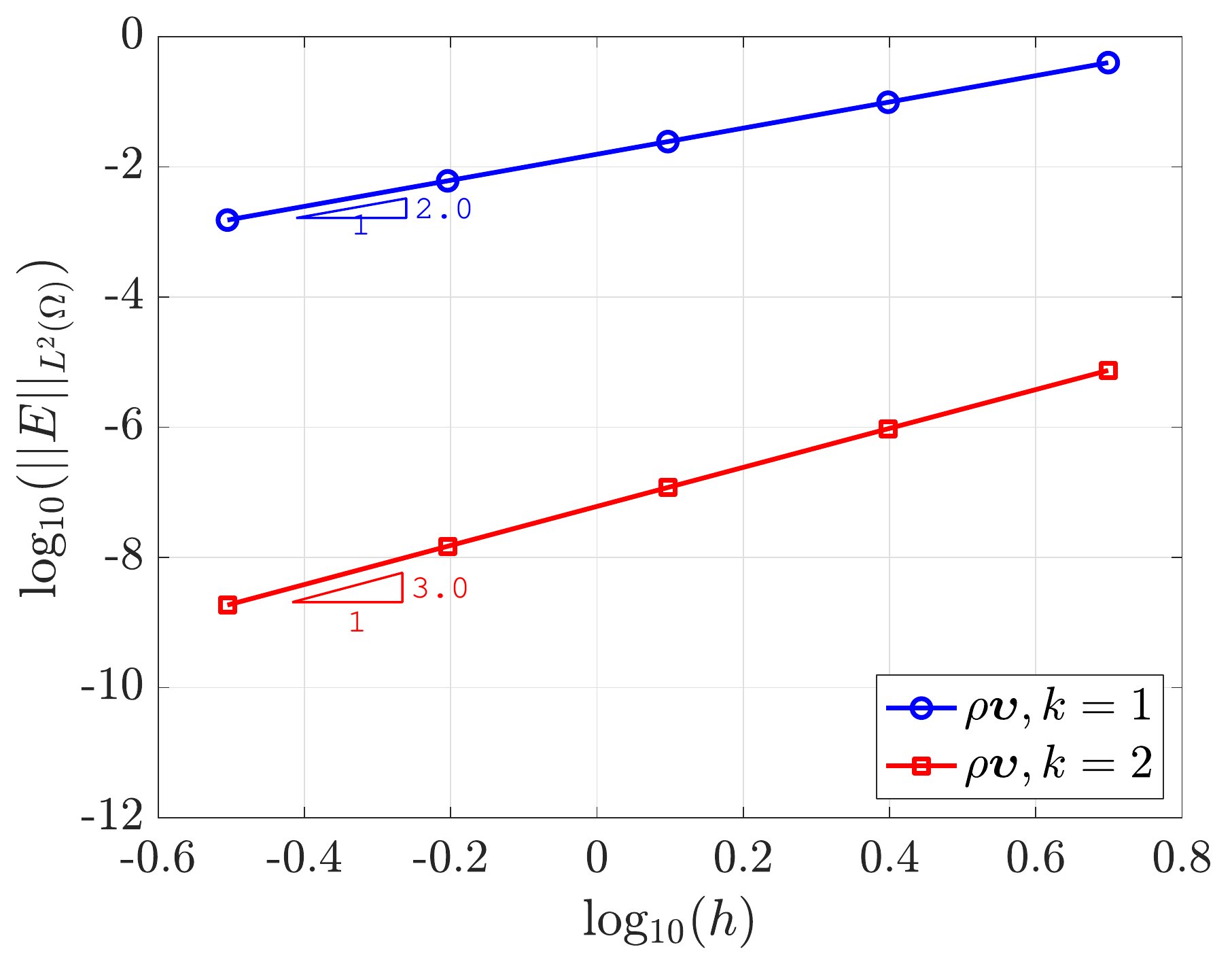}}
\subfigure[$\rho\bm{\upsilon}$ with $\varepsilon^*=0.1$]
{\includegraphics[width=0.32\textwidth]{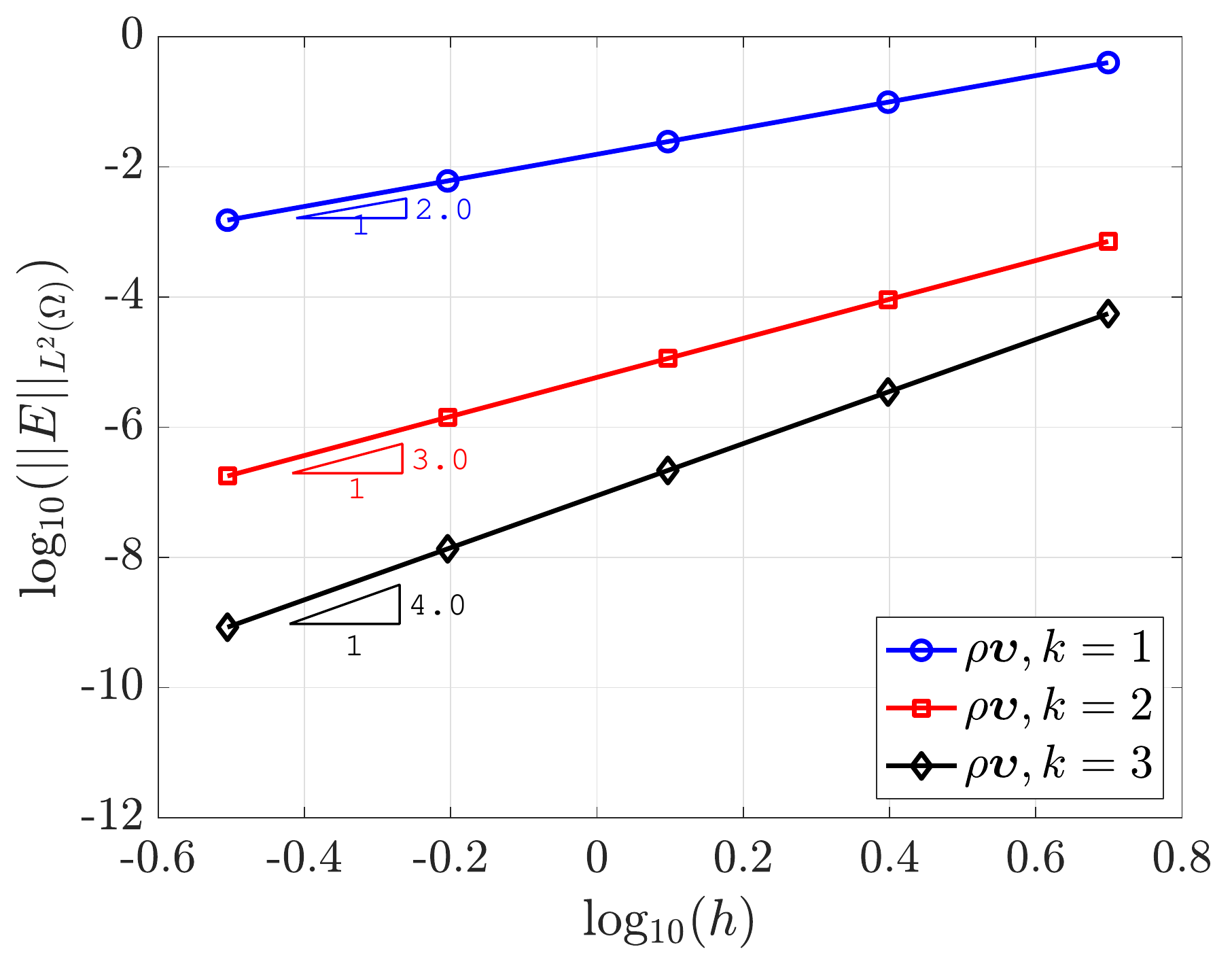}}
\subfigure[$\rho\bm{\upsilon}$ with $\varepsilon^*=1$]
{\includegraphics[width=0.32\textwidth]{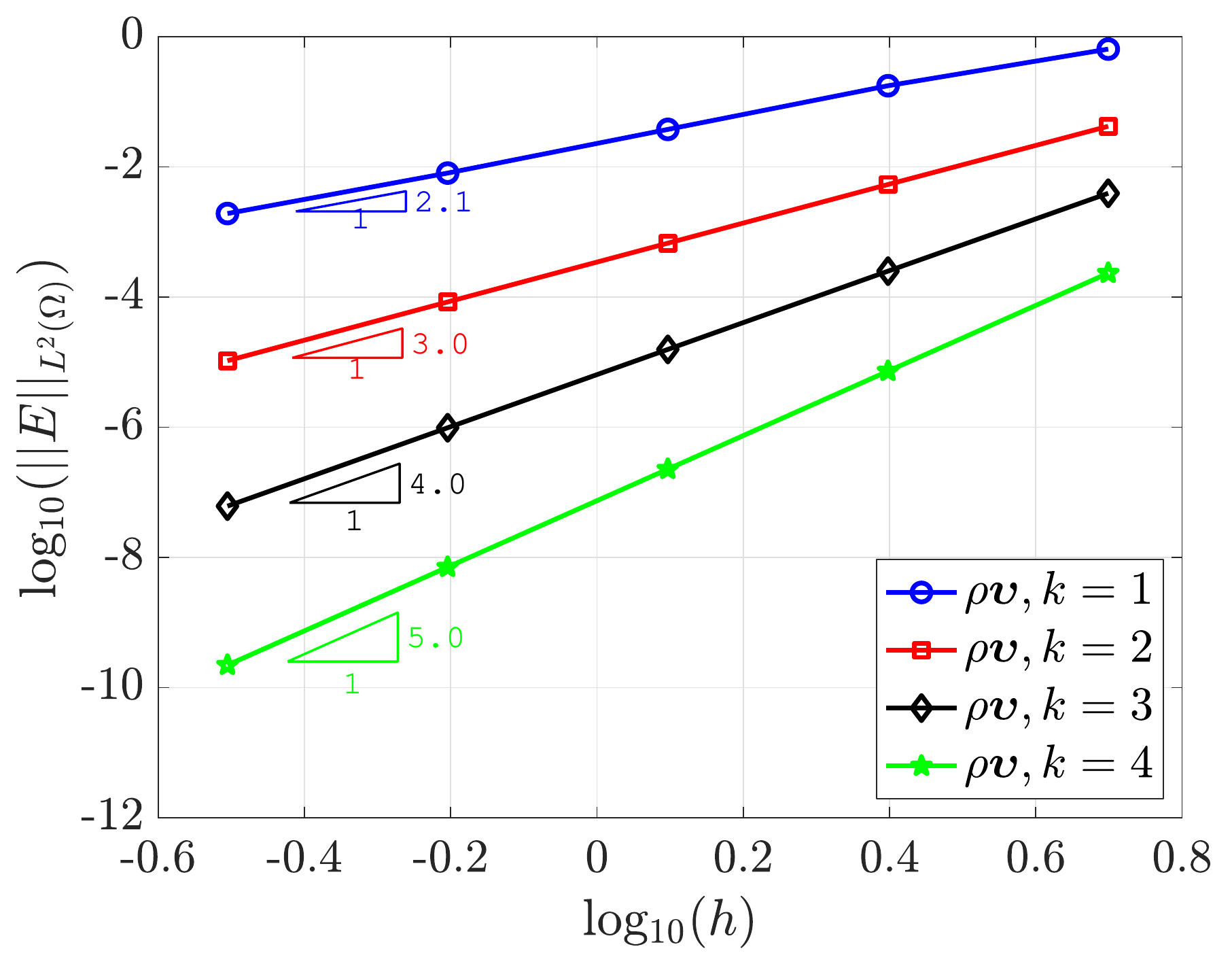}} \\
\subfigure[$\bm{\upsilon}^\star$ with $\varepsilon^*=0.01$]
{\includegraphics[width=0.32\textwidth]{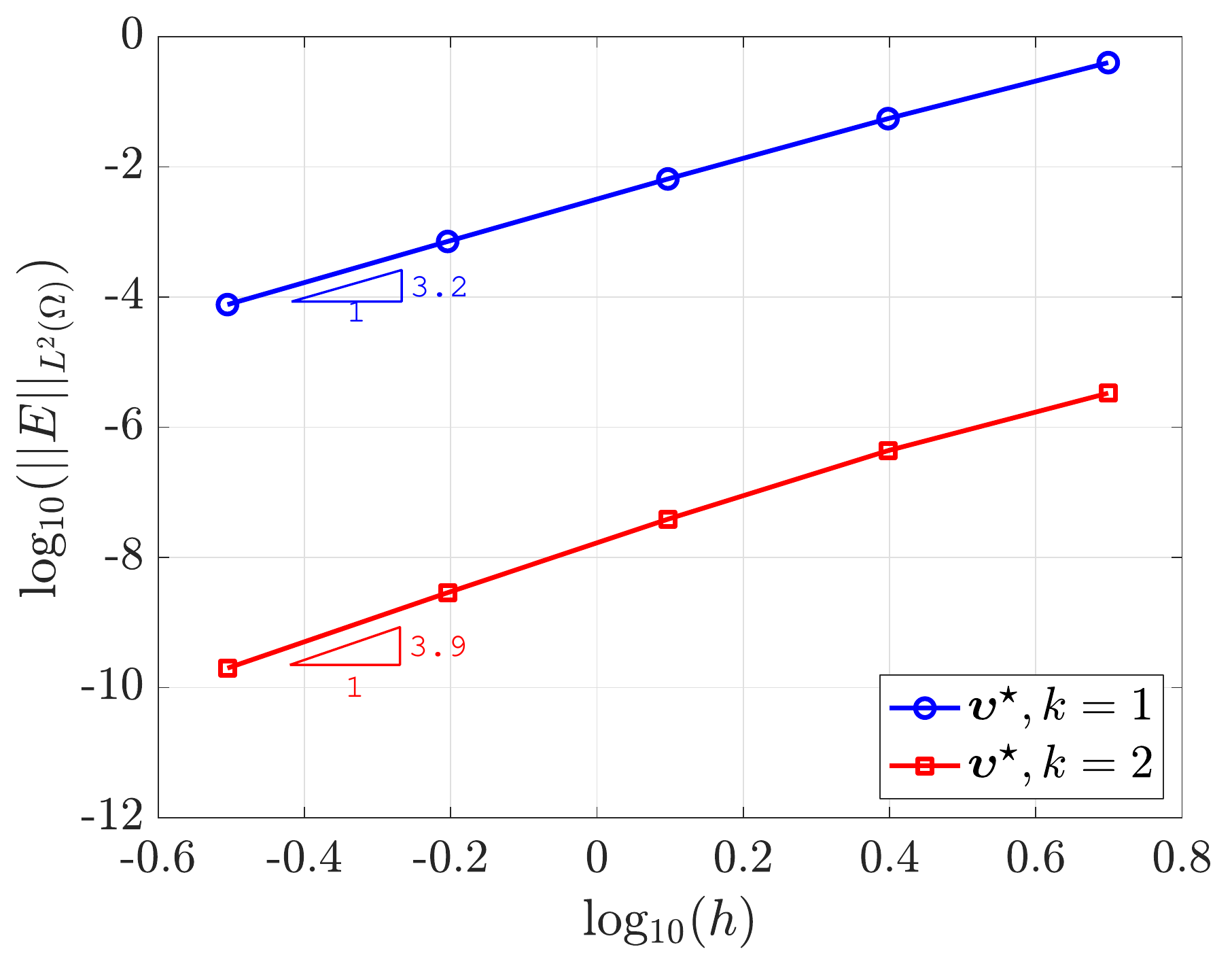}}
\subfigure[$\bm{\upsilon}^\star$ with $\varepsilon^*=0.1$]
{\includegraphics[width=0.32\textwidth]{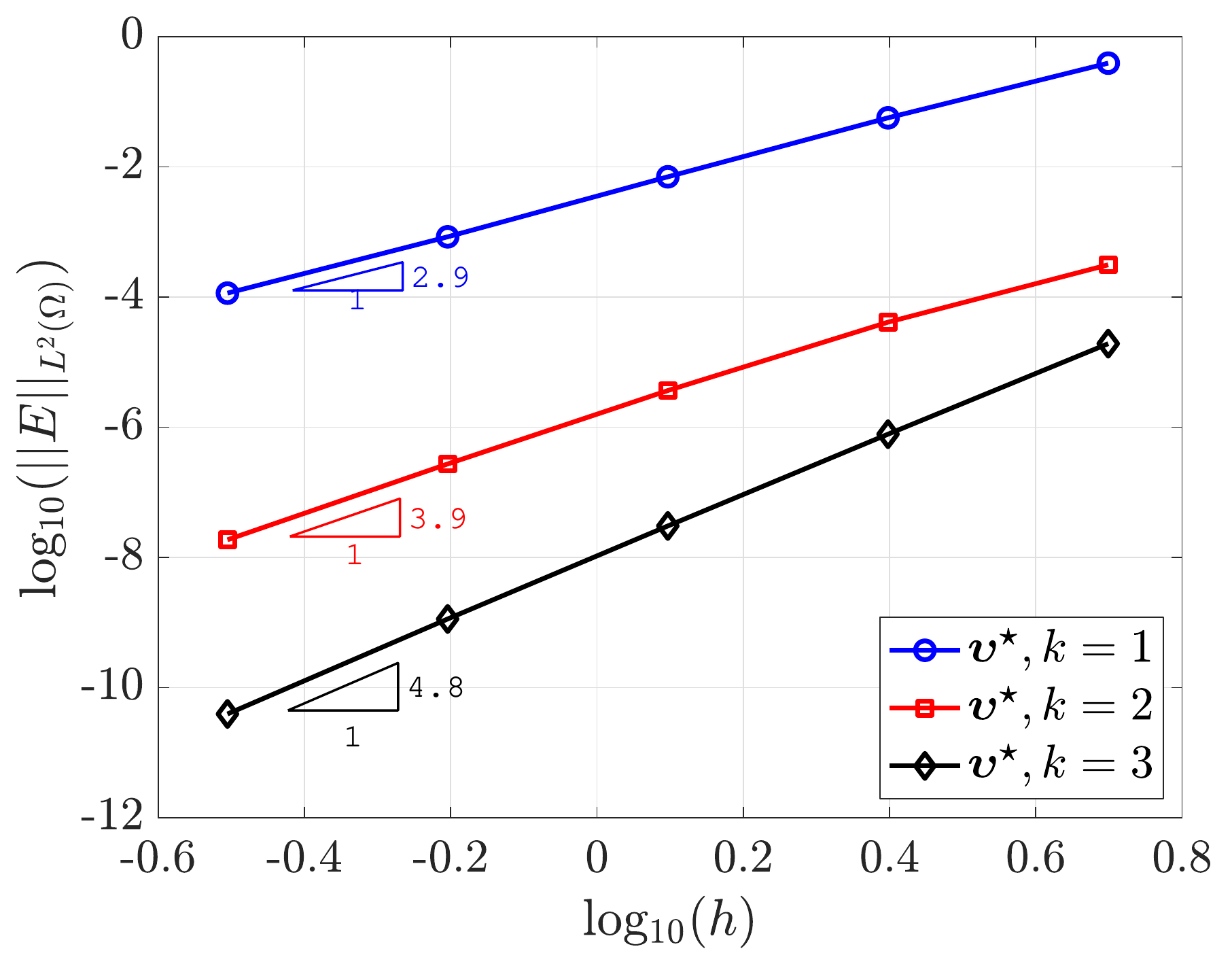}}
\subfigure[$\bm{\upsilon}^\star$ with $\varepsilon^*=1$]
{\includegraphics[width=0.32\textwidth]{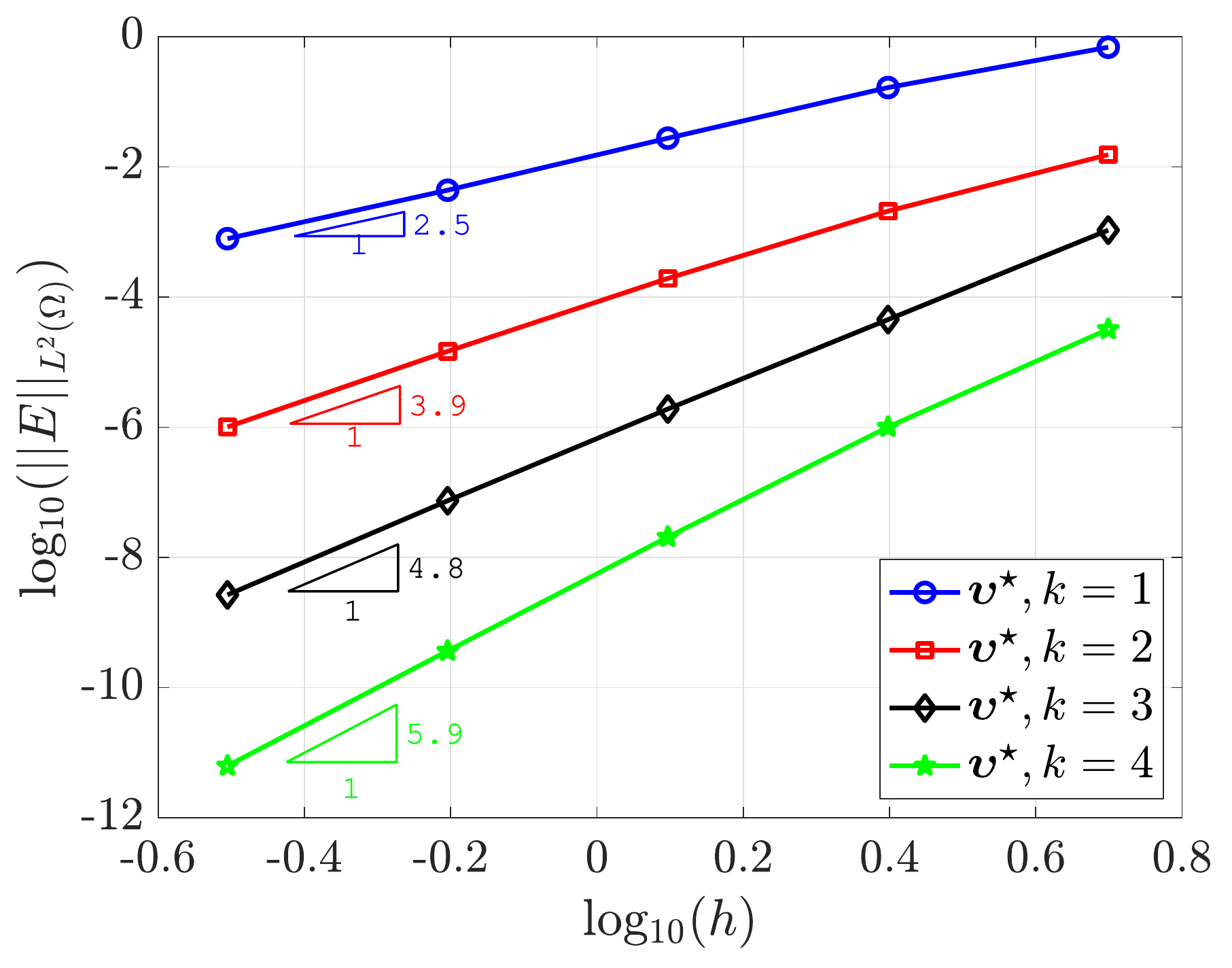}}
\caption{Spatial convergence of the $\mathcal{L}_2$-error of the mixed, primal and postprocessed variables for the weakly compressible Poiseuille flow with different values of the dimensionless compressibility coefficient.}
\label{fig:convergenceweaklycompressiblepoiseuilleflow}
\end{figure}

The convergence of the error measured in the $\mathcal{L}_2$ norm as a function of the characteristic element size $h$ is represented in Figure \ref{fig:convergenceweaklycompressiblepoiseuilleflow}, for the different compressibility coefficients considered.
It is worth noting that the error decreases by orders of magnitude when decreasing the compressibility coefficient because in the limit case of $\varepsilon=0$ the solution belongs to the space of quadratic polynomials $\mathcal{P}^2\left(\Omega\right)$.
On the other hand, the analytical solution is fully recovered up to machine precision with $k\geq5$ regardless of the compressibility coefficient $\varepsilon$, since the pressure (and therefore the density according to \eqref{eqn:fluidequationofstate}) belongs to $\mathcal{P}^2\left(\Omega\right)$ and the velocity belongs to $\mathcal{P}^3\left(\Omega\right)$, thus the momentum belongs to $\mathcal{P}^5\left(\Omega\right)$.
The convergence studies are therefore performed with variable ranges of polynomial degrees for the different compressibility coefficients, in order to clearly visualize the convergence rates in the asymptotic regime by avoiding the errors to reach the machine precision for excessively coarse meshes.
Optimal convergence rates (with order $k+1$) are obtained for the mixed variable $\mat{L}$ thanks to the adoption of Voigt notation, strongly enforcing the symmetry of the stress tensor, and for the primary variables $\rho$ and $\rho\bm{\upsilon}$, regardless of the compressibility coefficient $\varepsilon$.
The optimal convergence of the mixed variable and the procedure presented in section \ref{sec:localpostprocessing} to resolve the underdetermination of the rigid body motions allow the construction of a superconvergent velocity field $\bm{\upsilon}^\star$ (converging therefore with order $k+2$).
A reduction of $0.5$ is observed only for $k=1$ and $\varepsilon^*=1$, which represents however a limit case, not meaningful from a physical point of view.

\subsection{Fluid problem with manufactured solution on a moving mesh}
\label{sec:fluidproblemmanufacturedsolutionmovingmesh}

The second numerical experiment considers a two dimensional fluid problem with manufactured solution.
The solution aims to exercise all the terms present in the fluid partial differential equations, including the time-dependent terms, and to tackle all the nonlinearities, i.e., the weak compressibility and the convection.
Again, the subscript $\Fl{(\cdot)}$ is omitted here for brevity.
The analytical solution of the problem reads
\begin{subequations}
\begin{multline}
  \upsilon_x(x,y,t)  = \sin(\pi x)\sin(\pi y)\sin(\pi t ) \\ {-}\frac{\pi}{8\rho_0} \bigl[4\sin(\pi x)\sin(\pi y)\cos(\pi t)
                                    {+}\bigl(\sin(2\pi x)+{+}2\pi x\cos(2\pi y)\bigr)\sin^2(\pi t)\bigr]\varepsilon, 
\end{multline}
\begin{multline}
  \upsilon_y(x,y,t) = \cos(\pi x)\cos(\pi y)\sin(\pi t) \\ {+}\frac{\pi}{8\rho_0} \bigl[4\cos(\pi x)\cos(\pi y)\cos(\pi t)
                                  {-}\bigl(\sin(2\pi y){+}2\pi y\cos(2\pi x)\bigr)\sin^2(\pi t)\bigr]\varepsilon,
\end{multline}
\begin{equation}
  p(x,y,t) = \pi\cos(\pi x)\sin(\pi y)\sin(\pi t),
\end{equation}
\label{eqn:analyticalsolutionfluidproblemmanufacturedsolutionmovingmesh}
\end{subequations}
and it has been obtained by adding to a divergence-free velocity field specific $\mathcal{O}(\varepsilon)$ terms, such that the residual of the continuity equation in \eqref{eqn:fluidgoverningequationsmovingdomain}
\begin{multline}
\mathcal{R}(x,y,t)= 
 \bigg\{ \frac{\pi^3}{8\rho_0}\sin(\pi x)\sin(\pi y)\sin(\pi t) \\
 \bigl[4\sin(\pi x)\sin(\pi y)\cos(\pi t){+}\bigl(\sin(2\pi x){+}2\pi x\cos(2\pi y)\bigr)\sin^2(\pi t)\bigr] \\
 {+}\frac{\pi^3}{8\rho_0}\cos(\pi x)\cos(\pi y)\sin(\pi t)
\\ 
  \bigl[4\cos(\pi x)\cos(\pi y)\cos(\pi t){-}\bigl(\sin(2\pi y){+}2\pi y\cos(2\pi x)\bigr)\sin^2(\pi t)\bigr] \\
  +\frac{\pi^2}{\rho_0}\bigl(p_0{-}\pi\cos(\pi x)\sin(\pi y)\sin(\pi t)\bigr)  \\ 
  \bigl[\cos(\pi x)\sin(\pi y)\cos(\pi t){+}\bigl(\cos^2(\pi x){+}\cos^2(\pi y){-}1\bigr)\sin^2(\pi t)\bigr]  \biggr\}\varepsilon^2
\label{eqn:residualcontinuityfluidproblemmanufacturedsolutionmovingmesh}
\end{multline}
is of second order in $\varepsilon$.
A body force to cancel out any imbalance is then added to the right hand side of the momentum equation in \eqref{eqn:fluidgoverningequationsmovingdomain}.
If $\varepsilon=0$, the solution satisfies the fully incompressible Navier--Stokes equations with no need of adding the term \eqref{eqn:residualcontinuityfluidproblemmanufacturedsolutionmovingmesh}.

\begin{figure}
\centering
\subfigure[Fixed mesh]
{\includegraphics[width=0.49\textwidth]{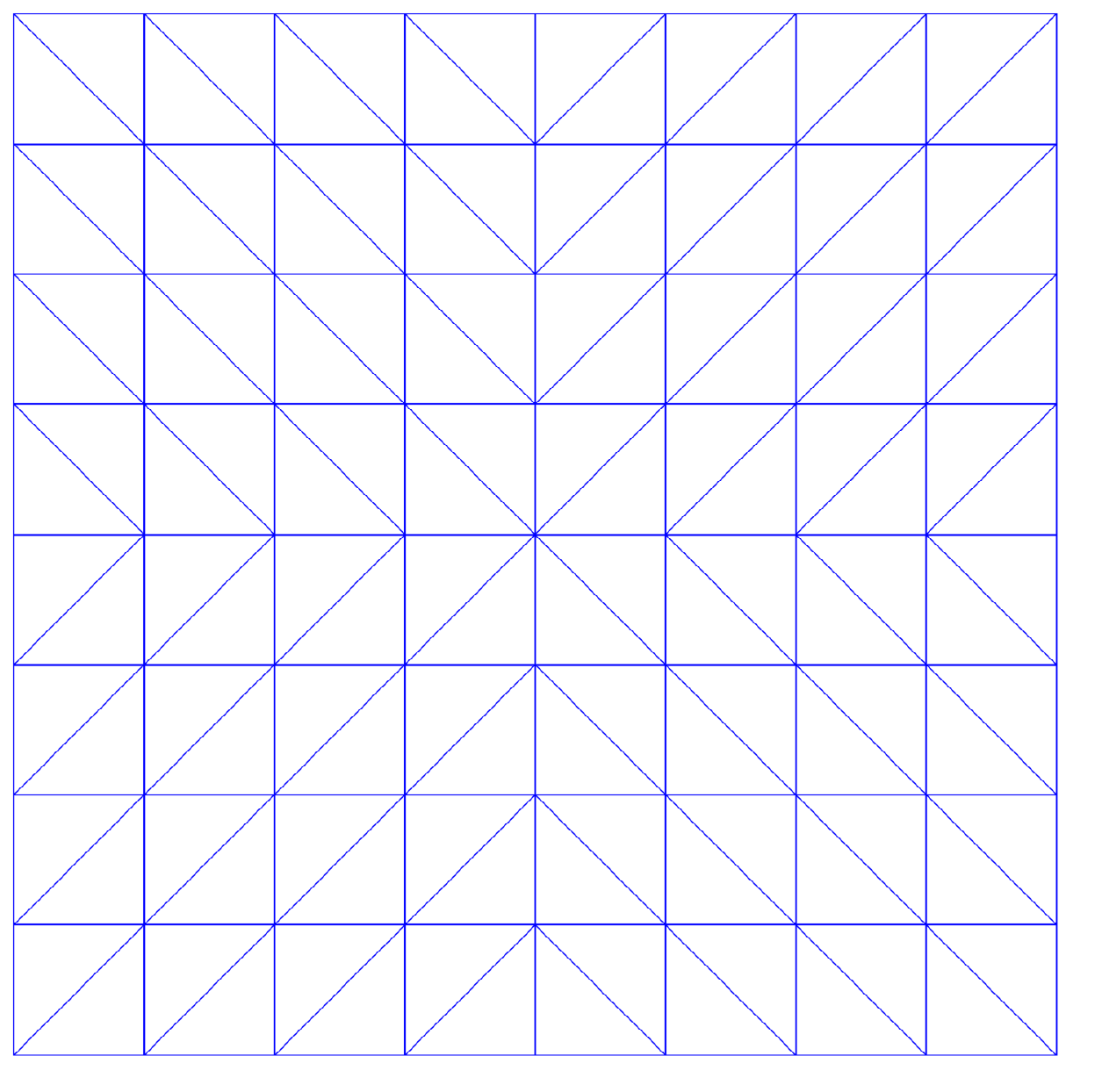}}
\subfigure[Moving mesh at $t=0.5$]
{\includegraphics[width=0.49\textwidth]{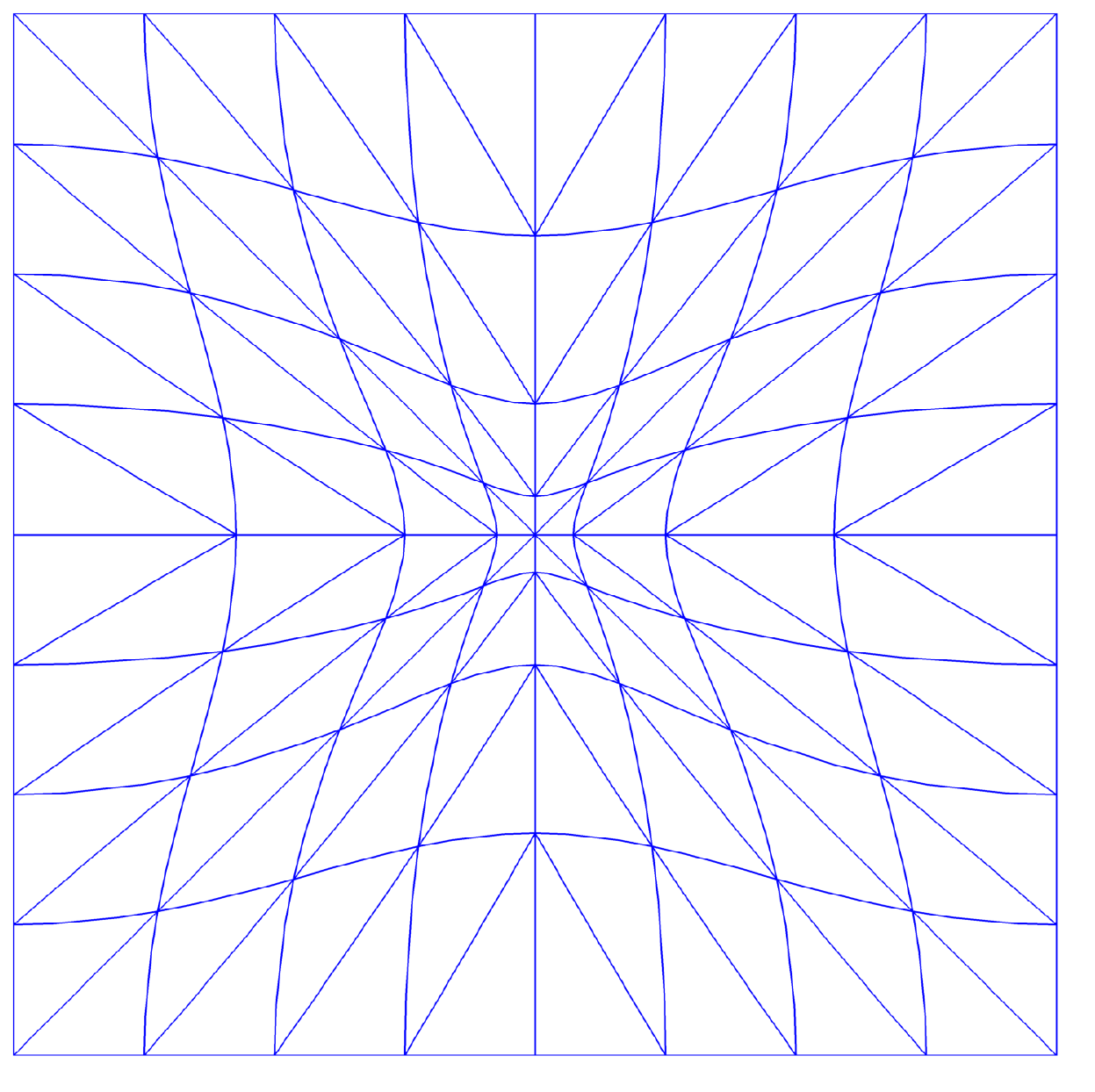}}
\caption{Third level of refinement of the fixed (left) and moving (right) mesh at $t=0.5$ used for the convergence studies of the fluid problem with manufactured solution.}
\label{fig:meshrfluidproblemmanufacturedsolutionmovingmesh}
\end{figure}

In order to validate the proposed HDG method formulated in the ALE form in section \ref{sec:hdgformulationweaklycompressibleflows}, two cases are considered:
\begin{enumerate}
\item the problem is solved on a fixed mesh,
\item the problem is solved on a moving mesh, whose displacement is described by a predefined function $\mathbf{d}\left(x,y,t\right)$, with
\end{enumerate}
\begin{equation}
\begin{split}
  d_x(x,y,t)&= \frac{1}{4}\sin(2\pi x)\bigl[1-\cos(2\pi y)\bigr]\bigl[1-\cos(2\pi t)\bigr]\bar{d}, \\
  d_y(x,y,t)&= \frac{1}{4}\bigl[1-\cos(2\pi x)\bigr]\sin(2\pi y)\bigl[1-\cos(2\pi t)\bigr]\bar{d},
\end{split}
\label{eqn:analyticaldisplacementfluidproblemmanufacturedsolutionmovingmesh}
\end{equation}
with $\bar{d}=0.125$, while the corresponding velocity $\vect{a}\left(x,y,t\right)$ is evaluated at the elemental level through standard finite differentiation techniques.
The fluid domain is the unit square $\Omega=[0,1]\times[0,1]$ and the same triangular pattern of the previous example is considered for the meshes used to perform the convergence studies.
Figure \ref{fig:meshrfluidproblemmanufacturedsolutionmovingmesh} shows the third level of refinement of the fixed (left) and moving (right) mesh at $t=0.5$.
The degrees of approximation considered for the convergence studies are $k=[1,3,5]$.
The viscosity $\mu$ is considered equal to $0.1$, while the reference density $\rho_0$ is taken equal to $1$ and evaluated at the reference pressure $p_0=0$, with a compressibility coefficient $\varepsilon$ equal to $0.1$.
The stabilization parameters are set as $\tau_\rho=10/\varepsilon$ and $\tau_{\rho\upsilon}=1$.
The final time of the simulation is $t=0.5$ and the fourth-order backward difference formula (BDF4) is deployed for the temporal discretization.
In order keep the temporal error sufficiently small to perform the spatial convergence studies, the time steps considered are $\Delta t=[2^{-5},2^{-7},2^{-9}]$ for $k=[1,3,5]$, respectively.
The initial conditions and the boundary conditions imposed on $\Gamma^D=\partial\Omega$ are computed from the analytical solution \eqref{eqn:analyticalsolutionfluidproblemmanufacturedsolutionmovingmesh}.

\begin{figure}
\centering
\subfigure[$\rho$ with $k=1$ on fixed mesh]
{\includegraphics[width=0.35\textwidth]{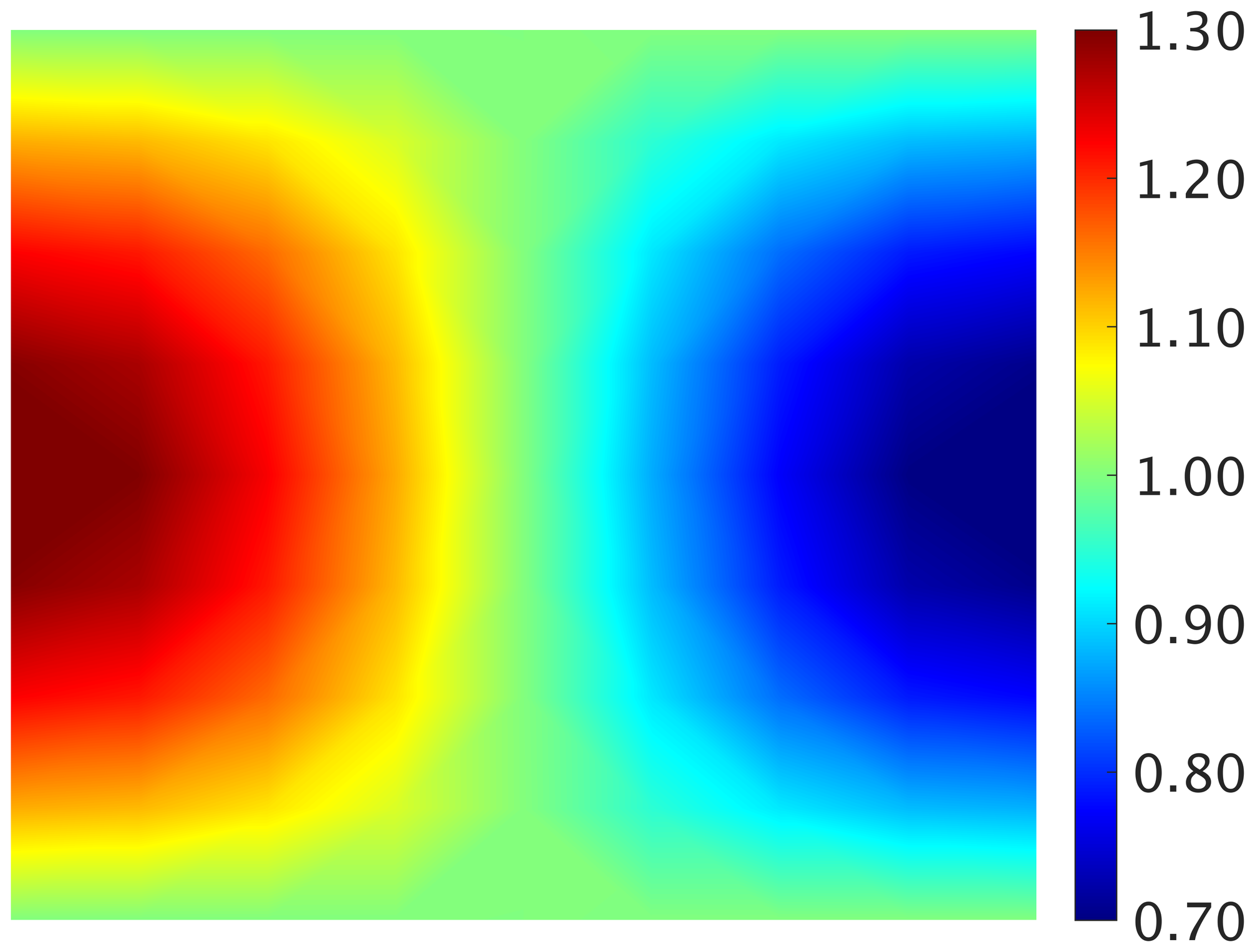}}
{\qquad}
\subfigure[$\rho$ with $k=1$ on moving mesh]
{\includegraphics[width=0.35\textwidth]{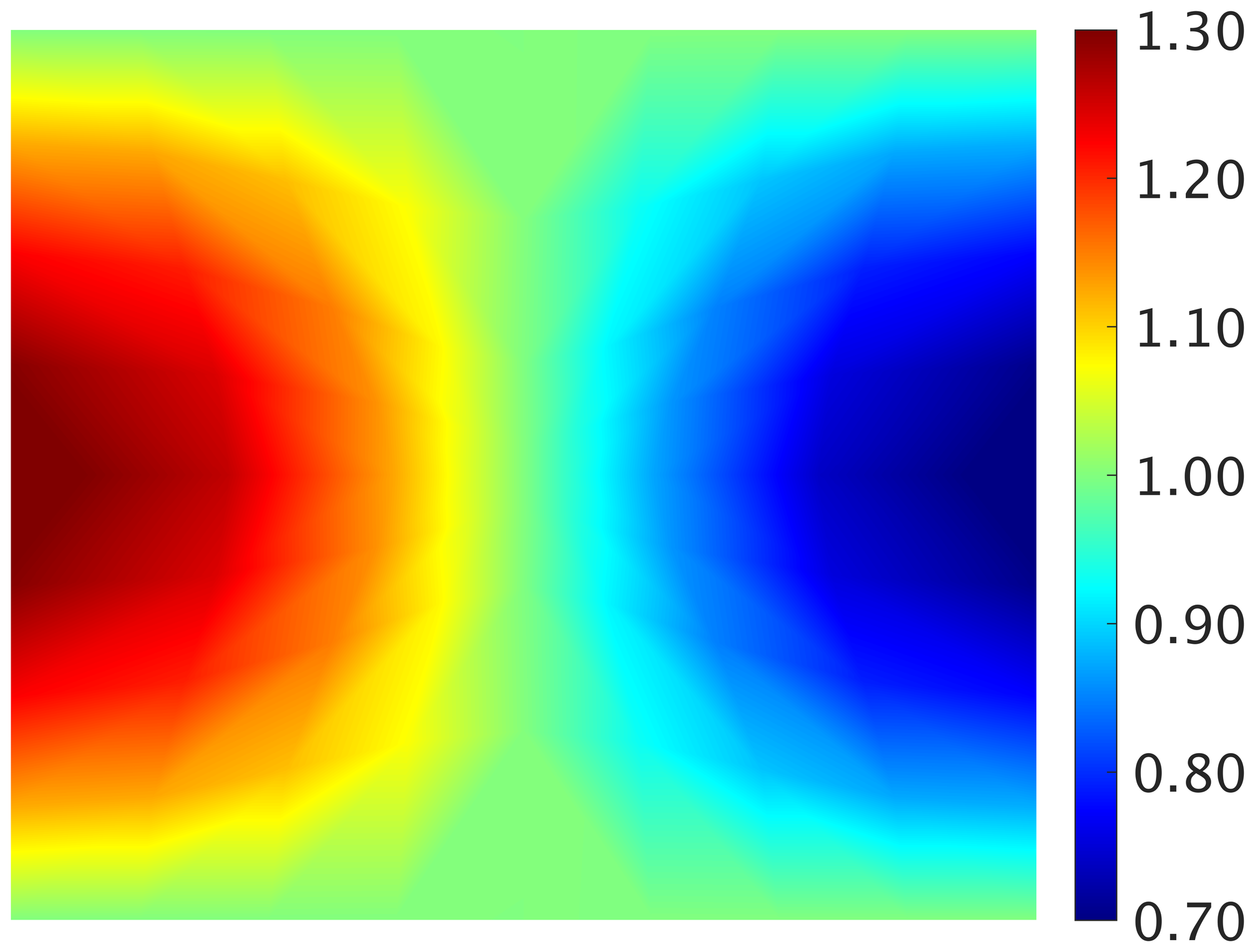}} \\
\subfigure[$\left|\rho\bm{\upsilon}\right|$ with $k=1$ on fixed mesh]
{\includegraphics[width=0.35\textwidth]{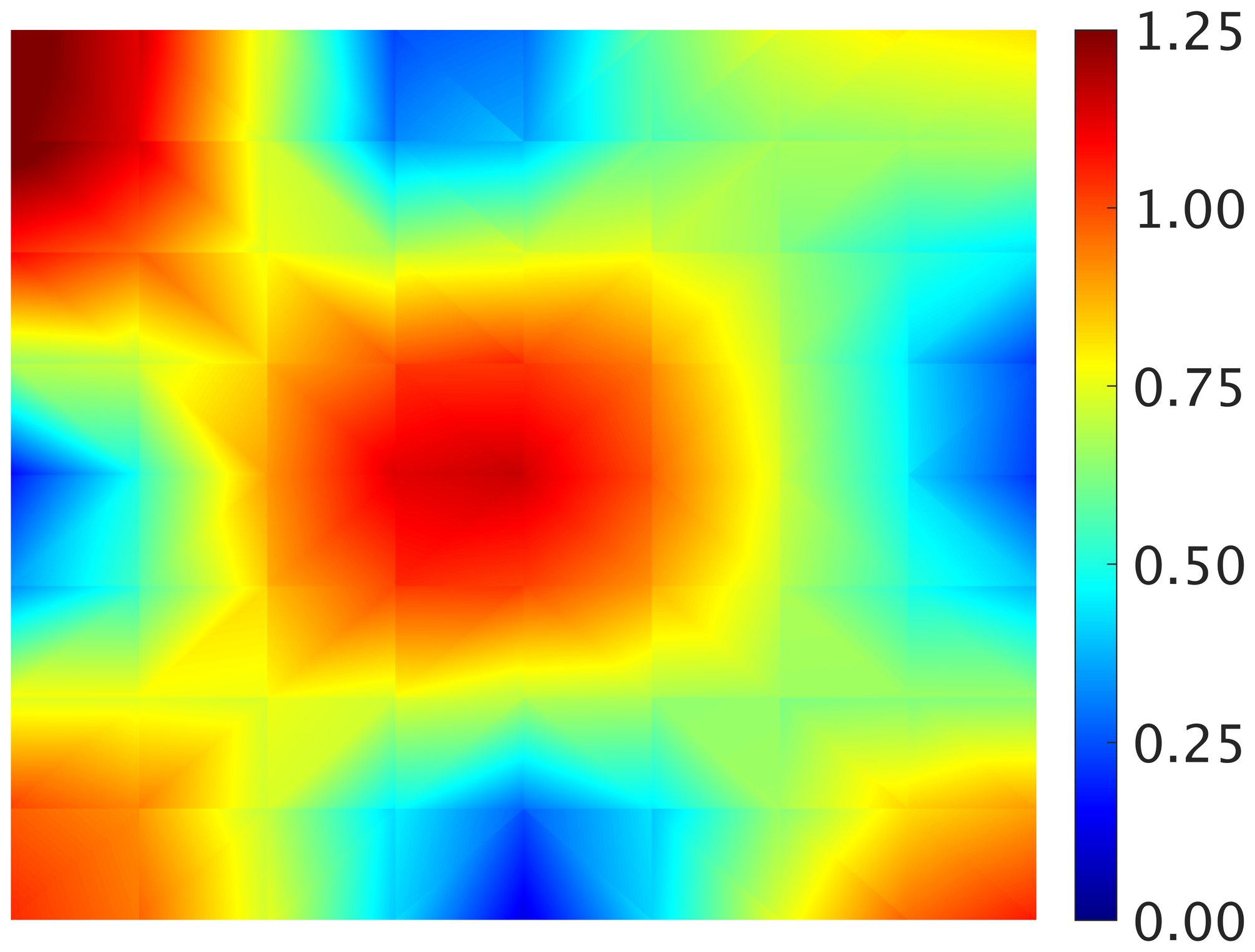}}
{\qquad}
\subfigure[$\left|\rho\bm{\upsilon}\right|$ with $k=1$ on moving mesh]
{\includegraphics[width=0.35\textwidth]{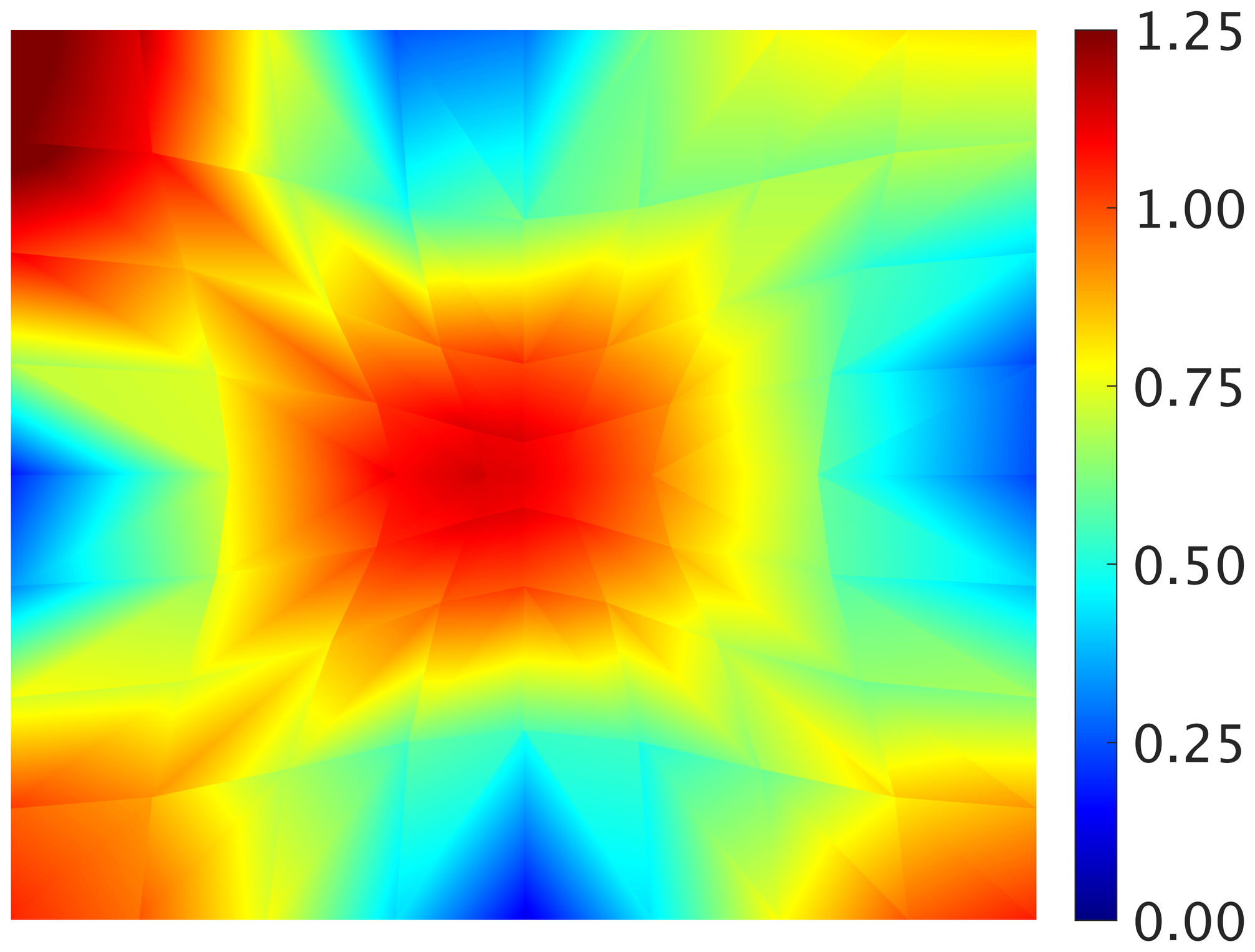}} \\
\subfigure[$\rho$ with $k=3$ on fixed mesh]
{\includegraphics[width=0.35\textwidth]{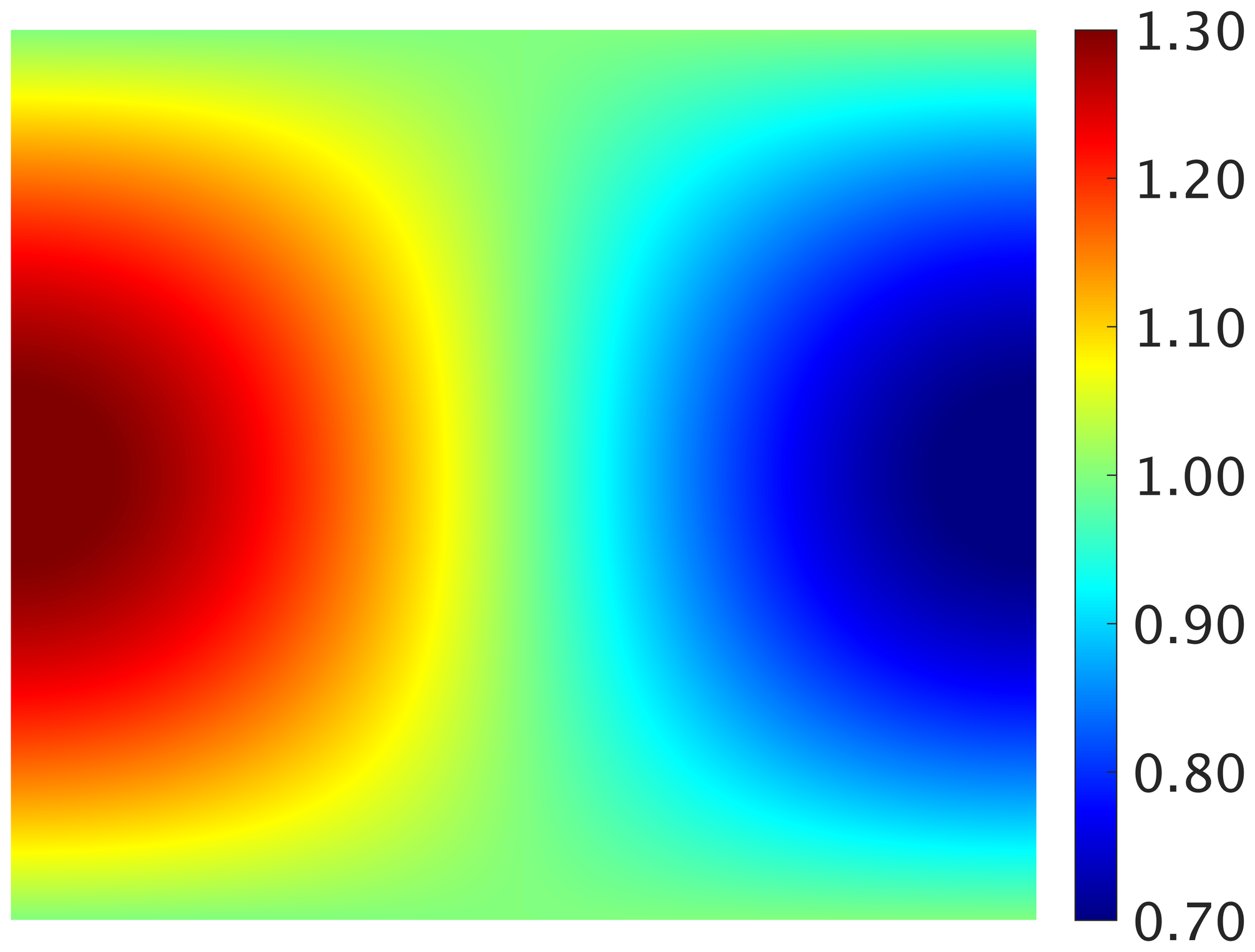}}
{\qquad}
\subfigure[$\rho$ with $k=3$ on moving mesh]
{\includegraphics[width=0.35\textwidth]{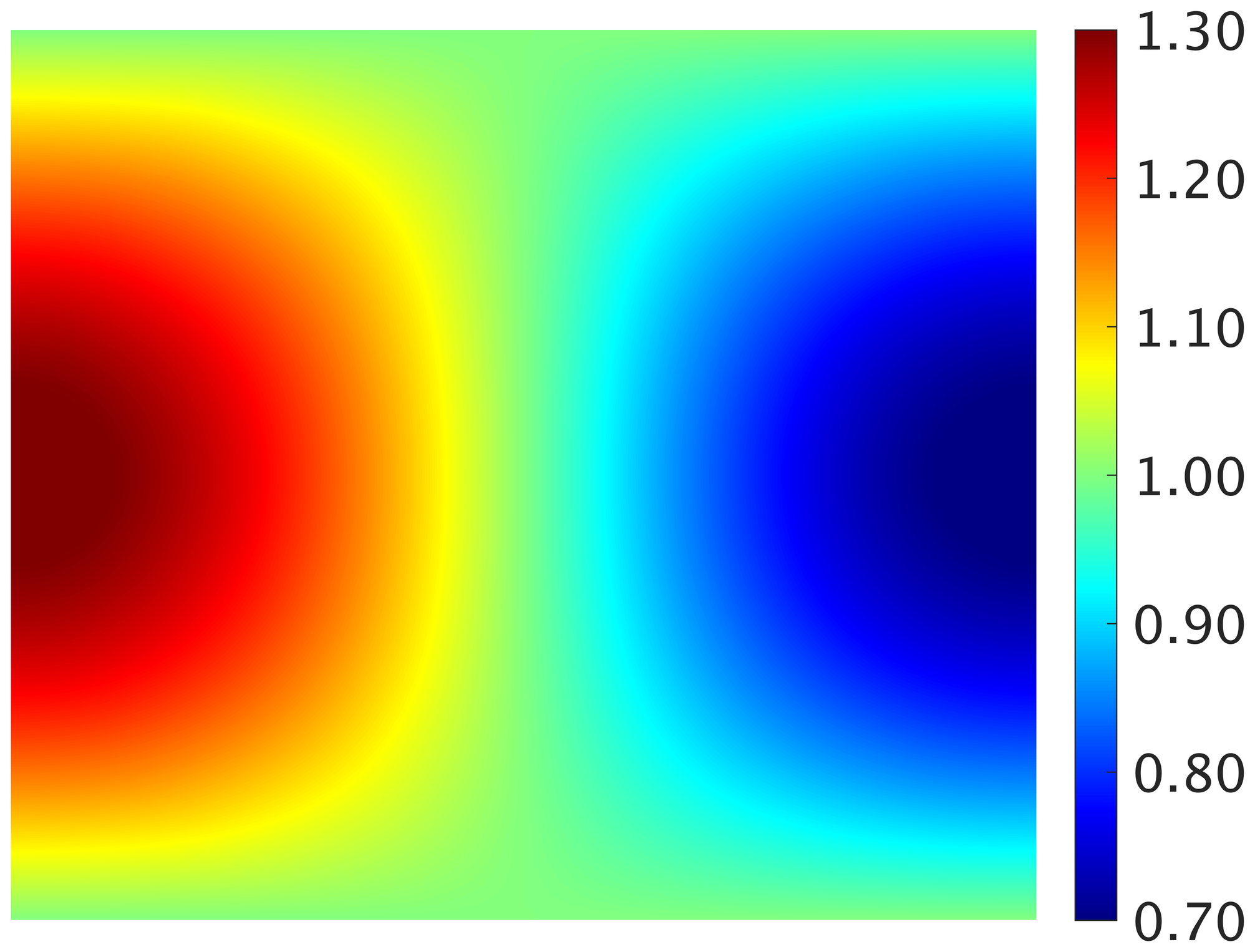}} \\
\subfigure[$\left|\rho\bm{\upsilon}\right|$ with $k=3$ on fixed mesh]
{\includegraphics[width=0.35\textwidth]{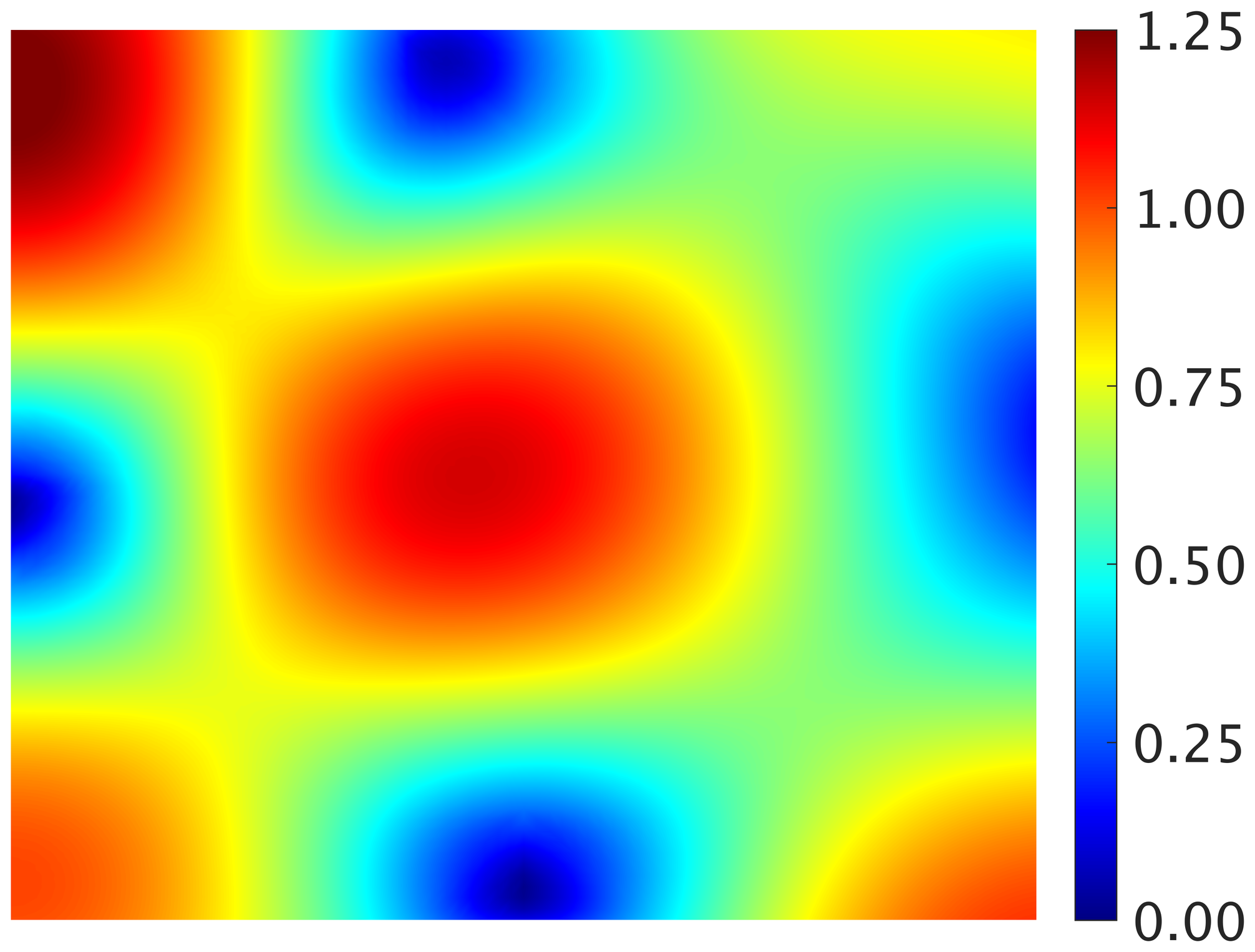}}
{\qquad}
\subfigure[$\left|\rho\bm{\upsilon}\right|$ with $k=3$ on moving mesh]
{\includegraphics[width=0.35\textwidth]{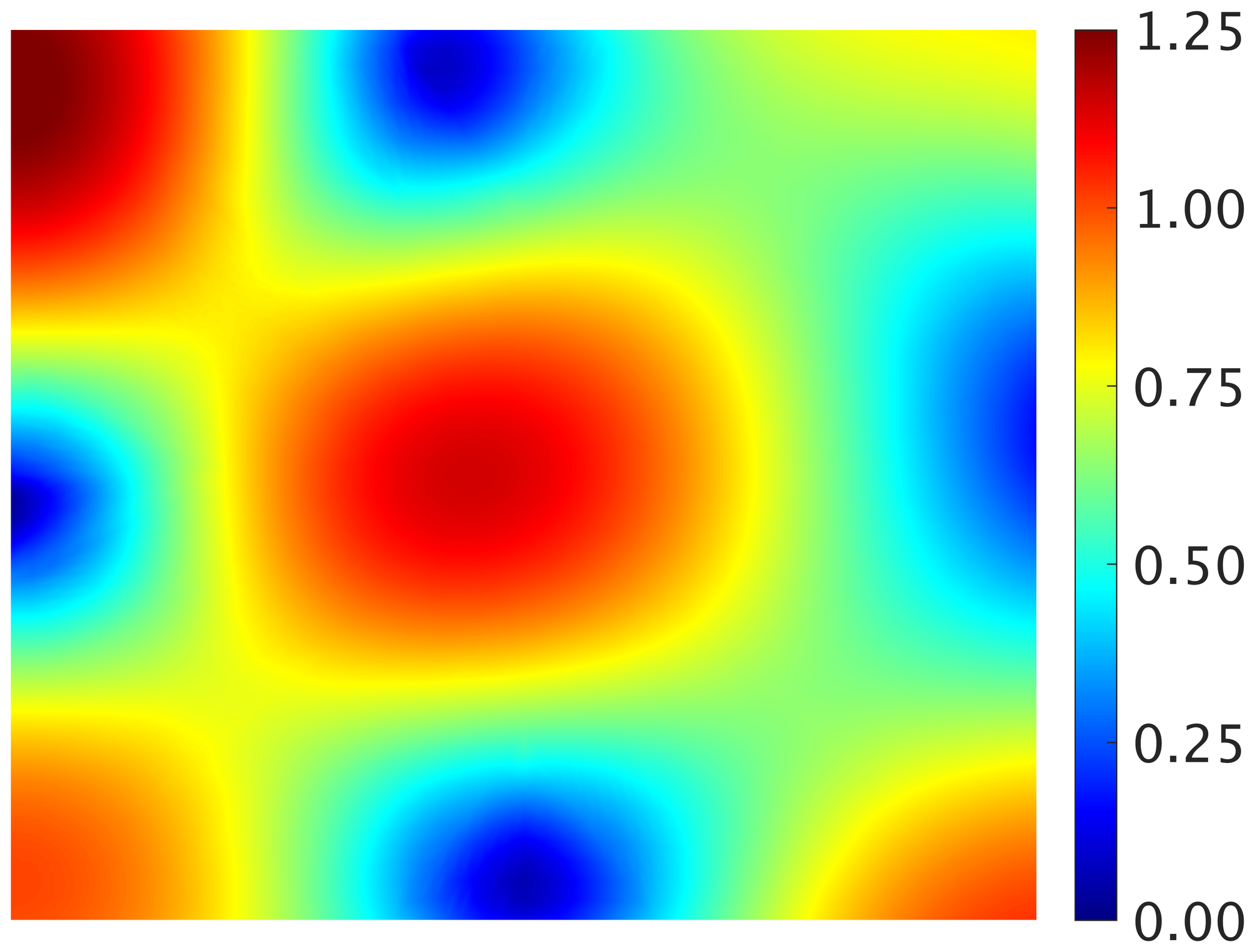}}
\caption{Approximation of the density and the momentum field of the fluid problem with manufactured solution at $t=0.5$ on the fixed (left) and moving (right) mesh.}
\label{fig:solutionfluidproblemmanufacturedsolutionmovingmesh}
\end{figure}

In Figure \ref{fig:solutionfluidproblemmanufacturedsolutionmovingmesh} the solution of the density and the momentum field obtained with the proposed HDG formulation using $m=3$ and $k=1$ (top) and $k=3$ (bottom) on the fixed (left) and moving (right) mesh at the final time is shown.
With this choice of the compressibility coefficient, the maximum variation of the density from the reference value is about $\pm30\%$.
On the one hand, the improvement of the approximation of the solution when increasing the polynomial degree is clearly observed by comparing the plots on the top with the plots on the bottom.
On the other hand, no differences can be captured by comparing the plots on the left with the plots on the right for a sufficiently accurate solution, confirming the correct implementation of the ALE framework.

\begin{figure}
\centering
\subfigure[$\mat{L}$ on fixed mesh]
{\includegraphics[width=0.45\textwidth]{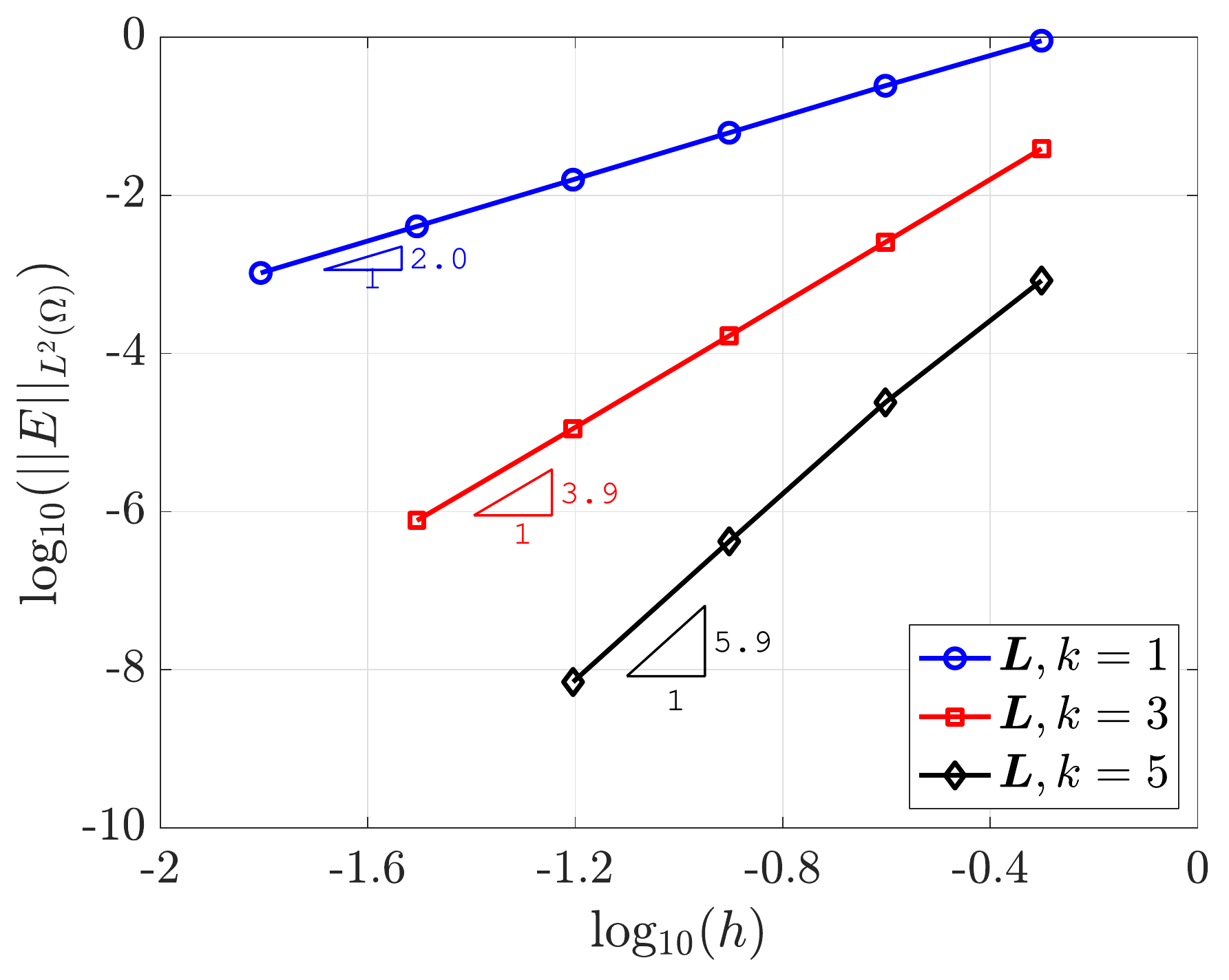}}
\subfigure[$\mat{L}$ on moving mesh]
{\includegraphics[width=0.45\textwidth]{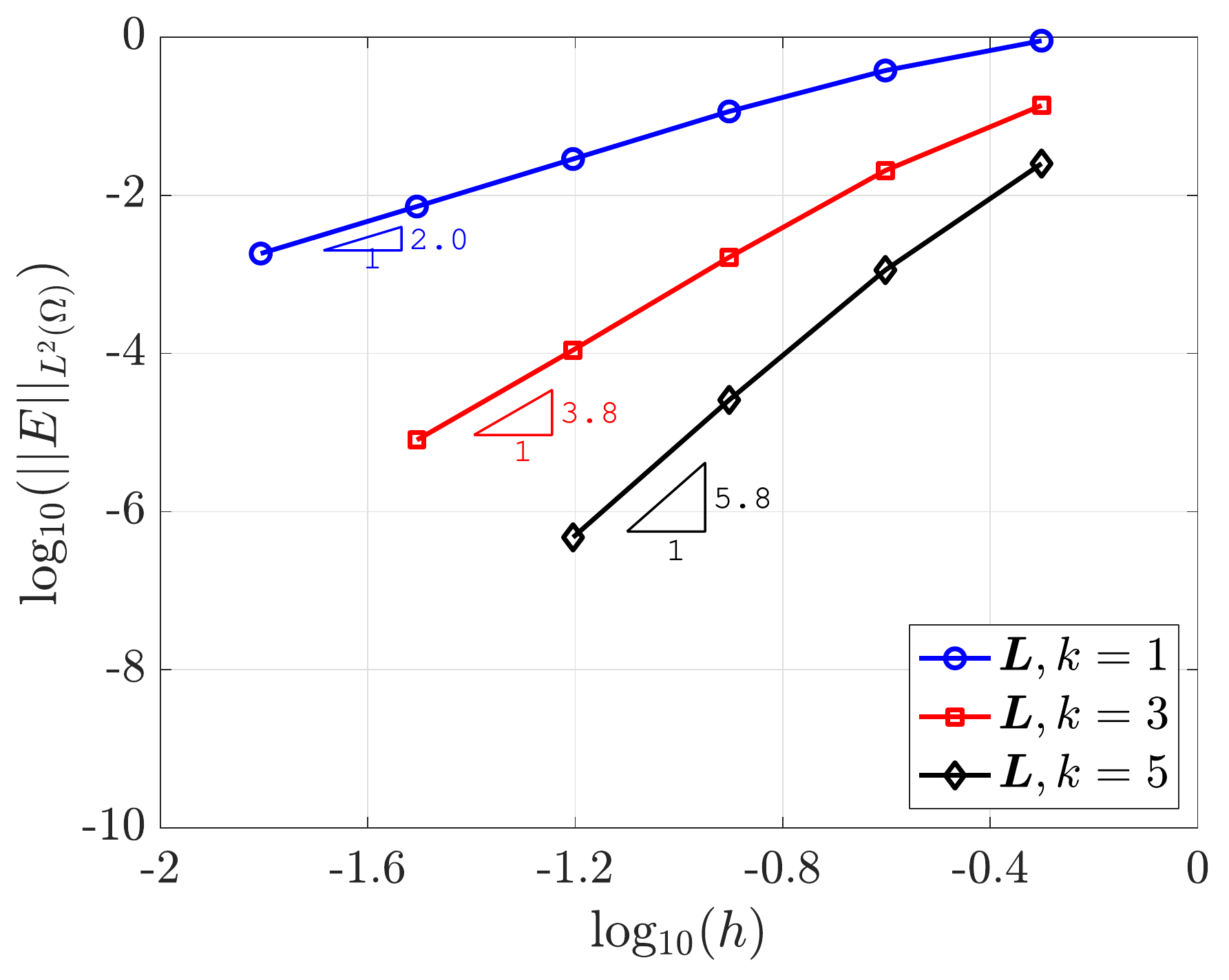}} \\
\subfigure[$\rho$ on fixed mesh]
{\includegraphics[width=0.45\textwidth]{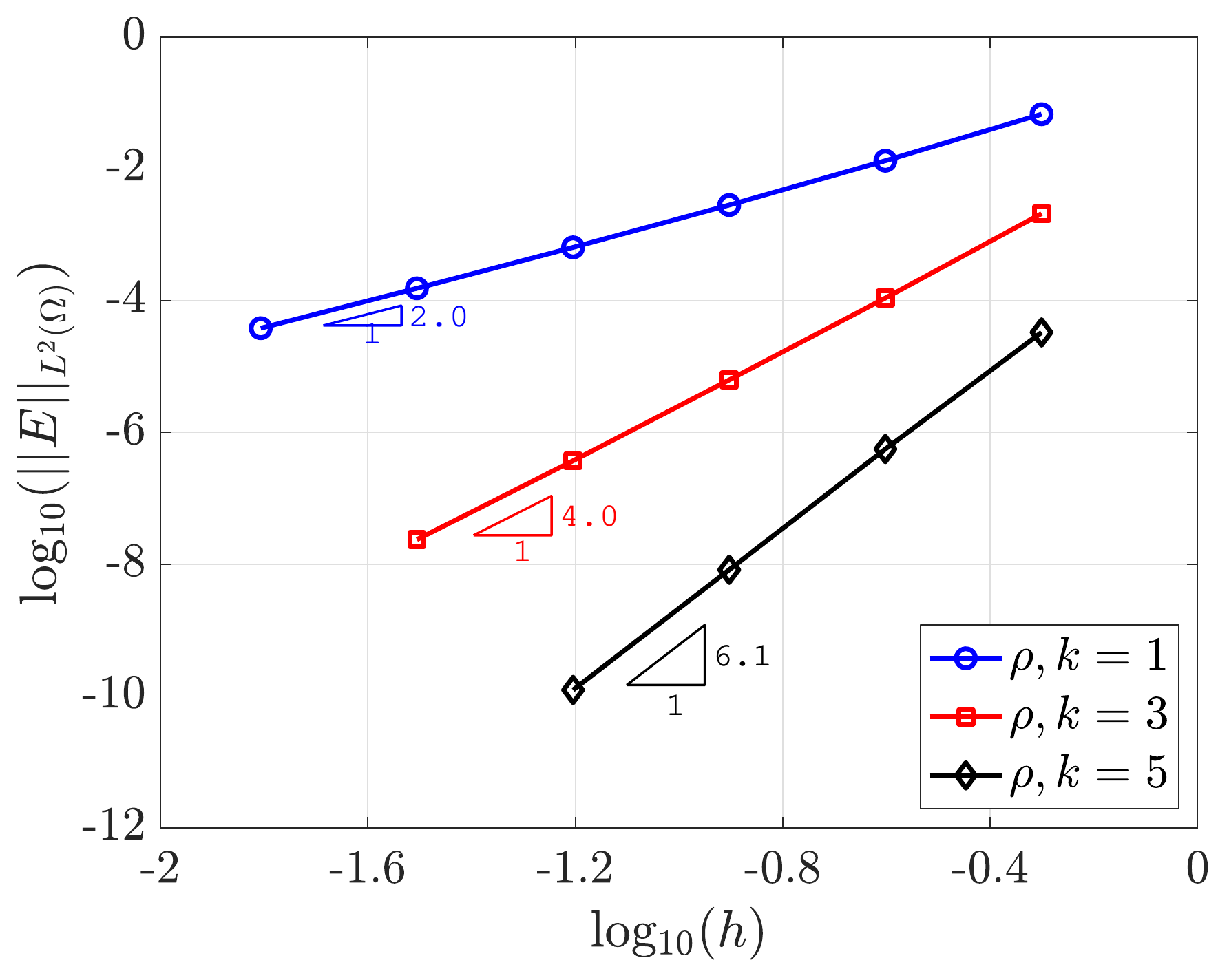}}
\subfigure[$\rho$ on moving mesh]
{\includegraphics[width=0.45\textwidth]{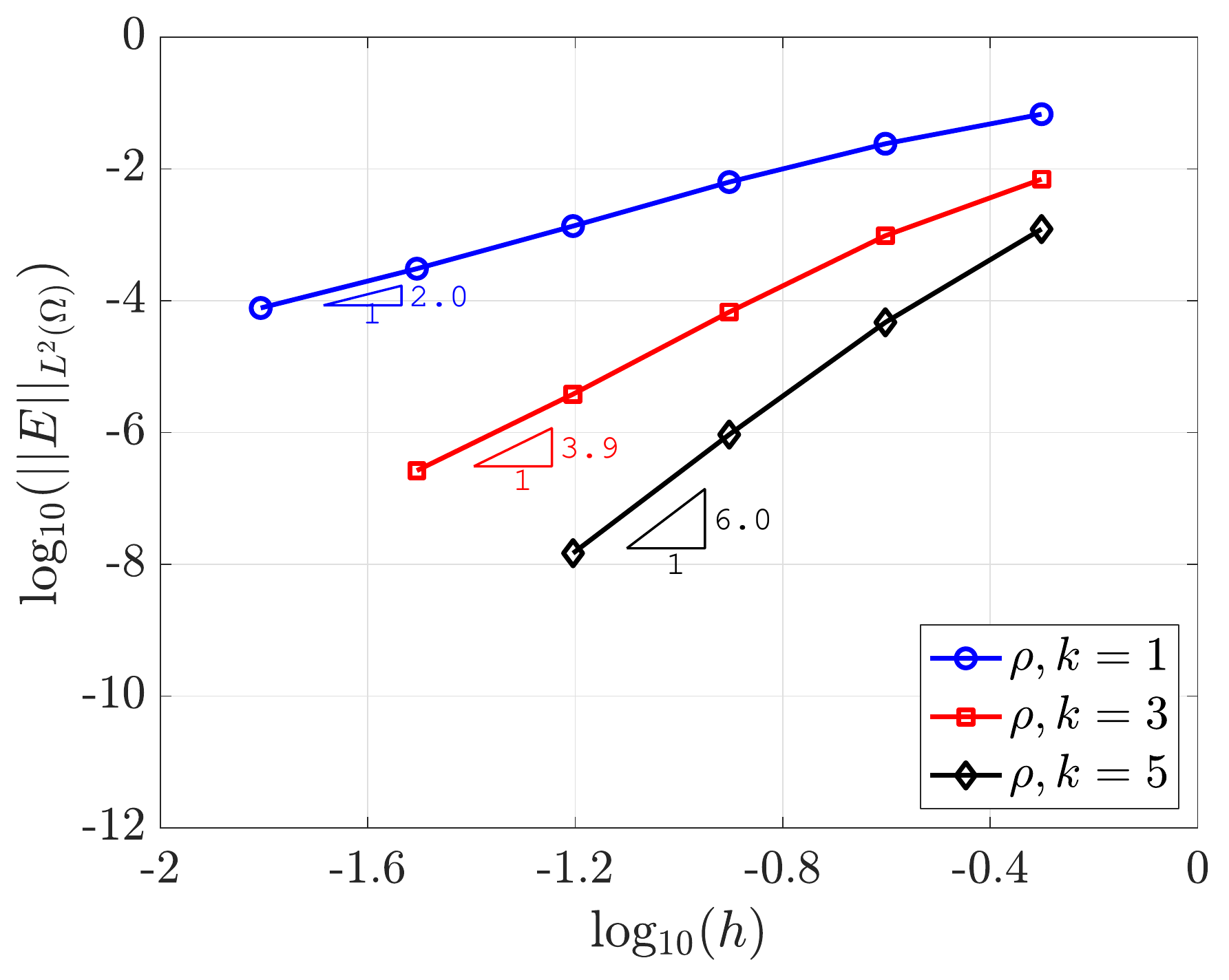}} \\
\subfigure[$\rho\bm{\upsilon}$ on fixed mesh]
{\includegraphics[width=0.45\textwidth]{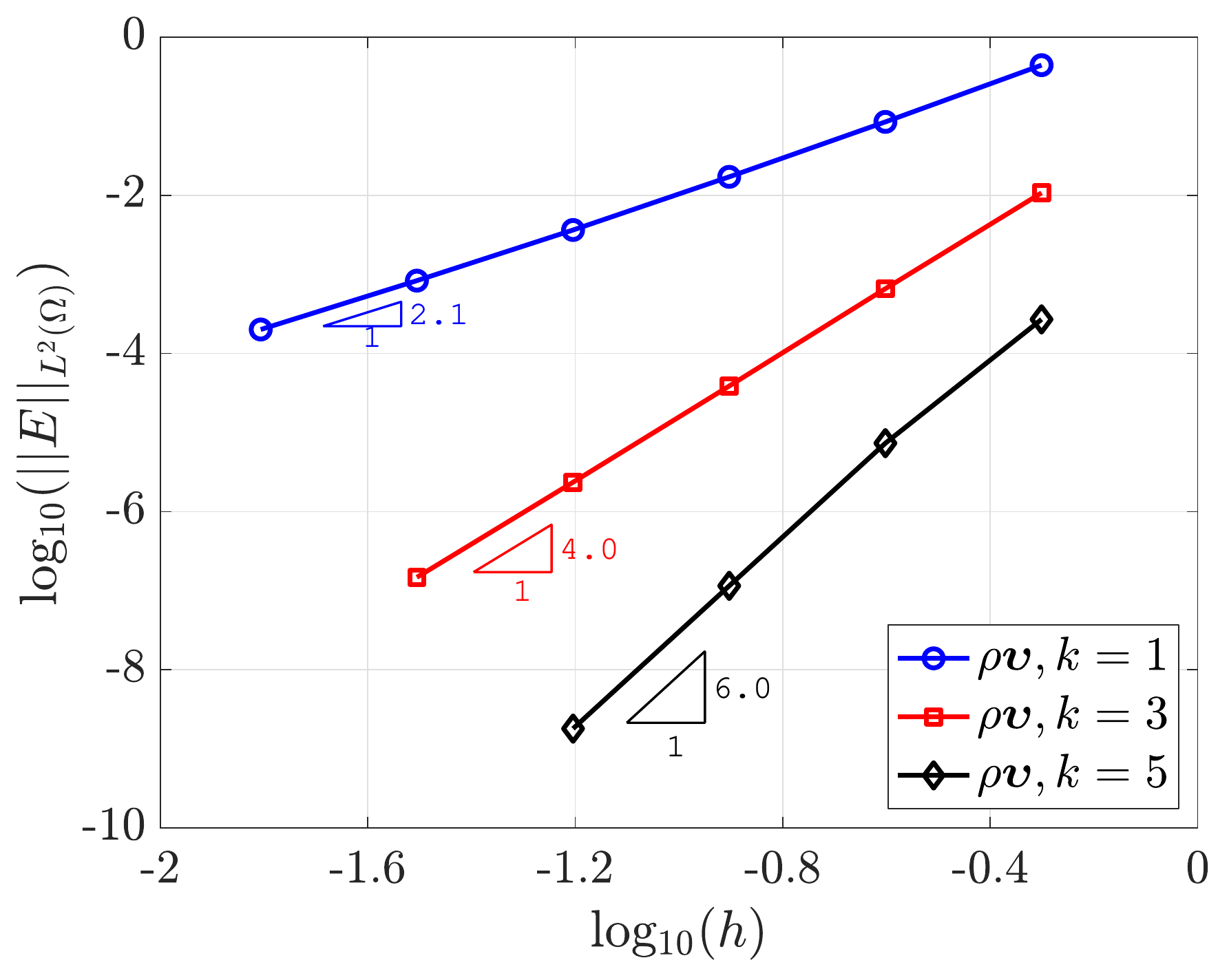}}
\subfigure[$\rho\bm{\upsilon}$ on moving mesh]
{\includegraphics[width=0.45\textwidth]{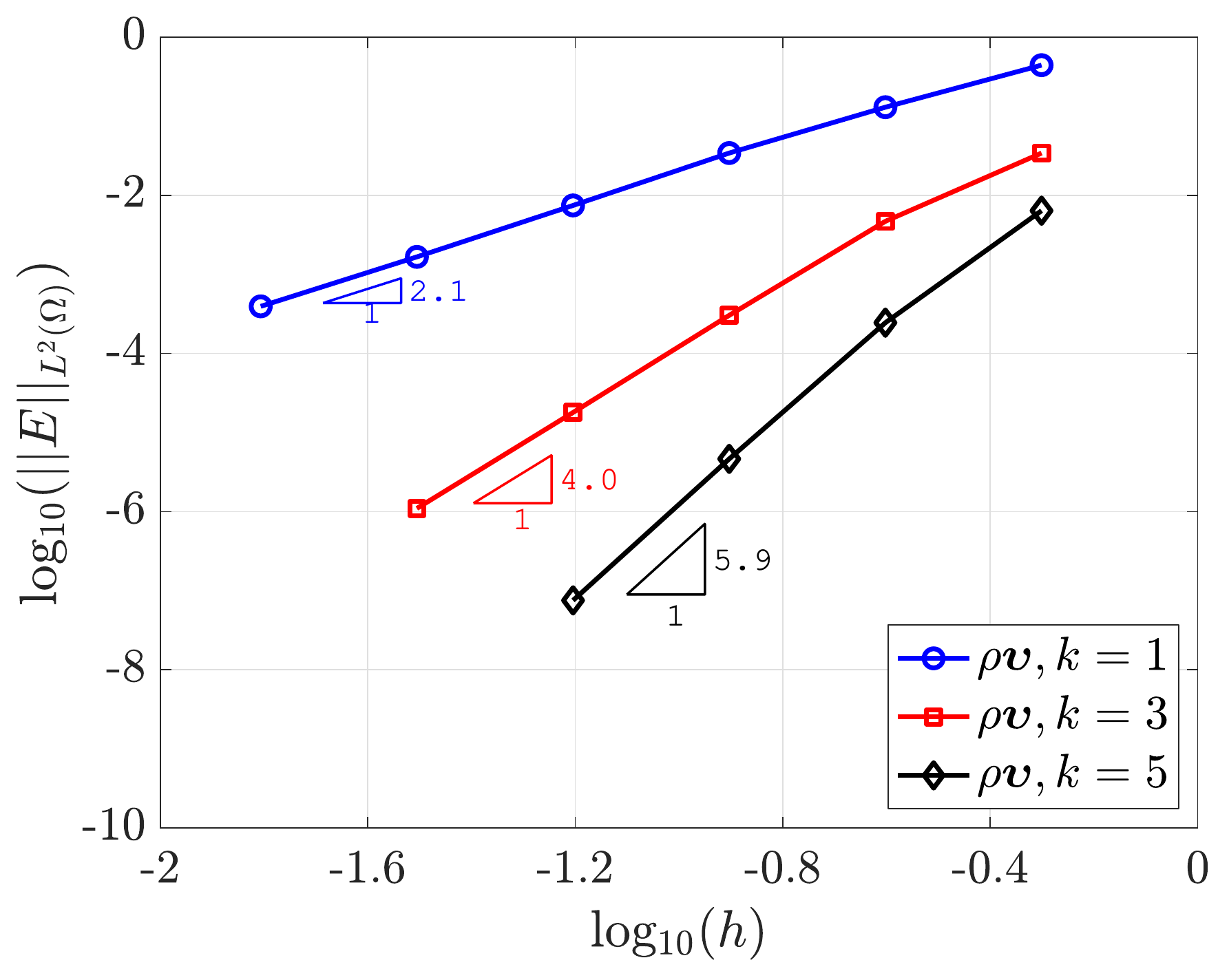}}
\caption{Spatial convergence of the $\mathcal{L}_2$-error of the mixed and primal variables evaluated at $t=0.5$ for the fluid problem with manufactured solution on the fixed (left) and moving (right) mesh.}
\label{fig:convergencefluidproblemmanufacturedsolutionmovingmesh}
\end{figure}

The convergence of the error of the mixed and primal variables measured in the $\mathcal{L}_2$ norm as a function of the characteristic element size $h$ is presented in Figure \ref{fig:convergencefluidproblemmanufacturedsolutionmovingmesh}, for the case $1$ (left) and the case $2$ (right).
Although the errors in case $2$ (evaluated on a distorted mesh) are systematically larger than the corresponding ones in case $1$ by about $0.5$, $1$ and $2$ orders of magnitude for the finest meshes considered for $k$ equal to $1$, $3$ and $5$, respectively, the optimal convergence rates are nicely preserved.
The capability of the method to preserve the optimal convergence on arbitrarily moving meshes is a crucial feature in order to accurately solve multiphysics problems like fluid-structure interaction.

\subsection{FSI problem with manufactured solution}
\label{sec:fsimanufacturedsolution}

The third numerical example aims to verify the convergence properties of the two coupling strategies presented in section \ref{sec:hdgcgformulationcoupledFSI} for the solution of fluid-structure interaction problems.
The generation of manufactured solutions for the verification of FSI formulations can be a quite complex task and a rigorous procedure to generate non-trivial solutions has been developed in \cite{Etienne2012}.
The problem under analysis considers a two dimensional unsteady incompressible flow interacting with a flexible structure and this flow configuration is simulated by choosing a sufficiently small compressibility coefficient.
The compressibility level however does not play an important role for the coupling techniques we aim to test.

\begin{figure}
\centering
\subfigure[Initial configuration]
{\includegraphics[width=0.49\textwidth]{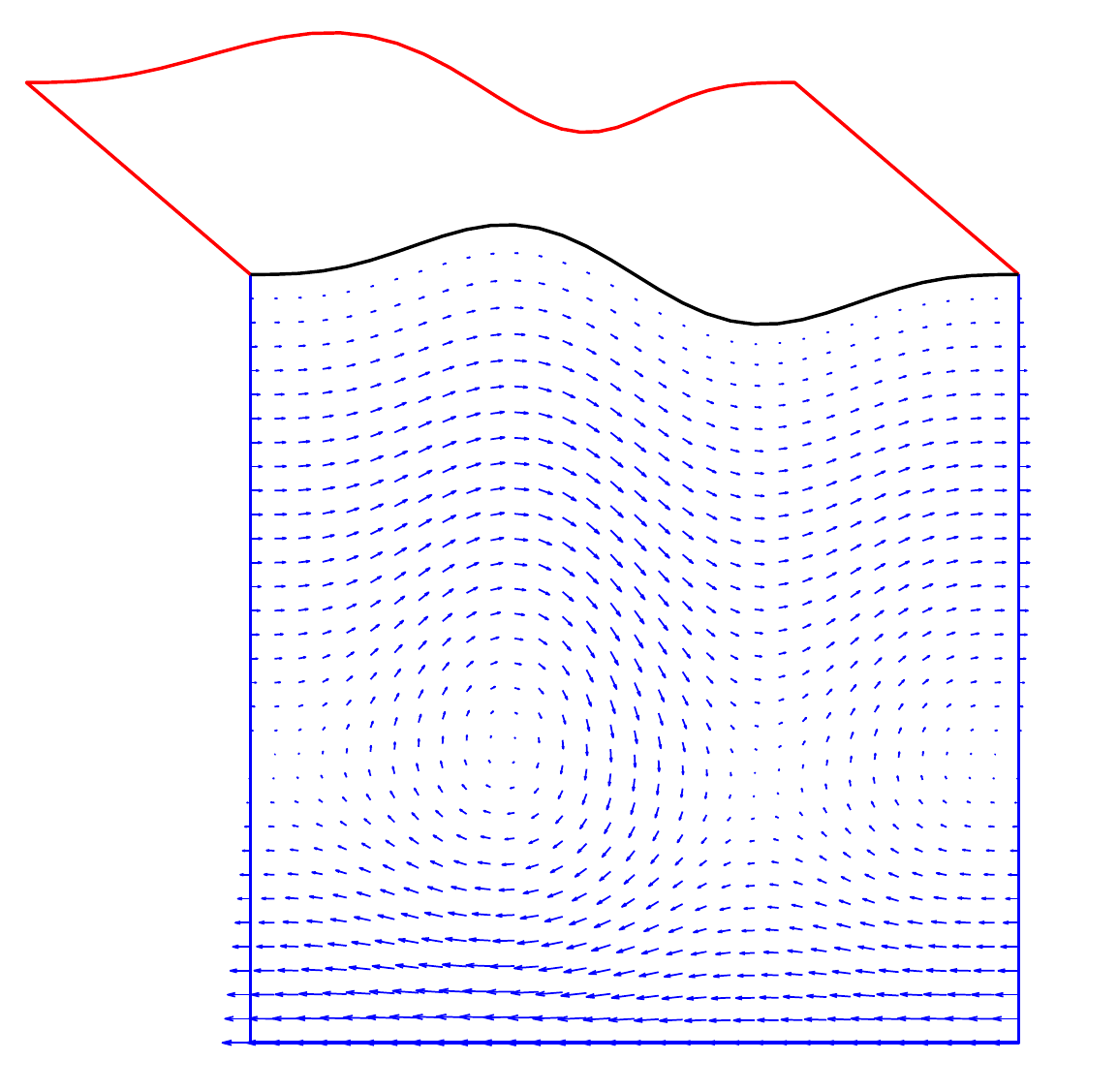}}
\subfigure[Final configurationh]
{\includegraphics[width=0.49\textwidth]{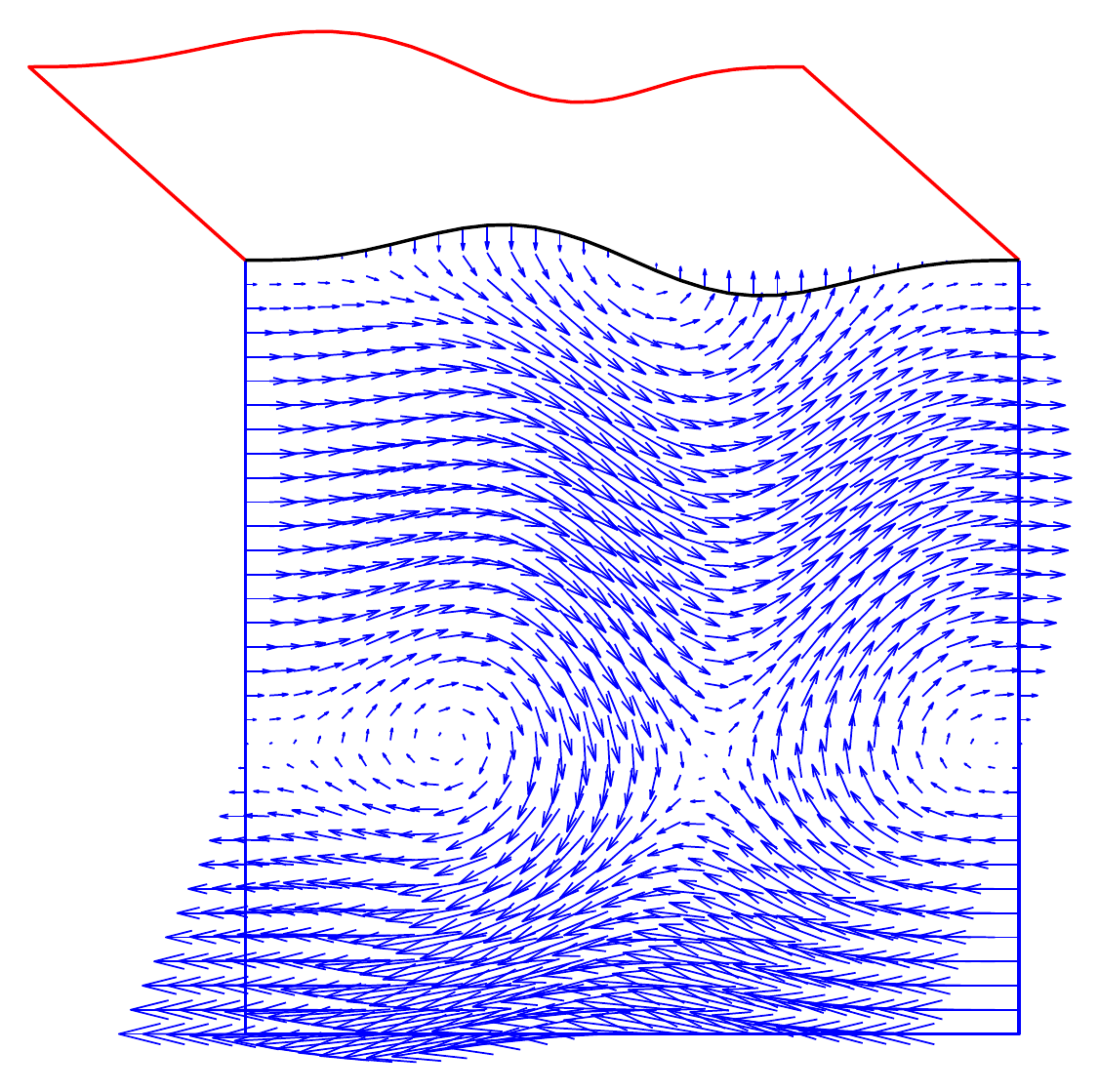}}
\caption{Velocity field of the FSI problem with manufactured solution on the initial (left) and final (right) deformed configuration.}
\label{fig:solutionvectorialFSIproblemmanufacturedsolution}
\end{figure}

The fluid domain is the square $\Fl{\Omega}=[0,1]\times[0,1]$ while the structural domain is defined as $\St{\Omega}=[0,1]\times[1,1.25]$, such that the interface (in the undeformed configuration) can be identified as $\Gamma^I=\left\{\left(x,y\right)\in\mathbb{R}^2 \mid y=1\right\}$.
The parameters considered to generate the solution according to the procedure in \cite{Etienne2012} are:
\begin{equation}
\begin{split}
\delta\left(t\right)&=\delta_0\sin\left[2\pi\left(t+t_0\right)\right], \\
f(x,t)&=1+\delta\left(t\right)\left[1-\cos\left(2\pi x\right)\right]\sin\left(2\pi x\right), \\
K(x,t)&=x\left[8-7\left(h\left(t\right)/\delta_0\right)^2\right], \\
j&=1, \\
a\left(x,y,t\right)&=\left[1+10\delta\left(t\right)\cos\left(2\pi x\right)/3\right]\left(1-y\right), \\
b\left(x,y,t\right)&=1,
\end{split}
\label{eqn:analyticalsolutionFSIproblemmanufacturedsolutioninput}
\end{equation}
with $\delta_0=0.05$ and $t_0=0.25$.
The function $f(x,t)$ describes the deformed fluid-structure interface, $K(x,t)$ is a user supplied data characterizing the fluid flow, while the chosen parameter $j=1$ provides a non zero fluid velocity profile along the bottom boundary.
Moreover, $a\left(x,y,t\right)$ and $b\left(x,y,t\right)$ define the structural solution.
Two auxiliary functions are then derived to ease the notation:
\begin{equation}
M(x,t)=\int_{0}^{x}K\left(z,t\right)dz, \quad\quad
L(x,t)=\int_{0}^{x}K\left(z,t\right)zdz.
\label{eqn:analyticalsolutionFSIproblemmanufacturedsolutionfunctionals}
\end{equation}
The structural displacement is defined as follows:
\begin{equation}
\begin{split}
{\St{u}}_x\left(x,y,t\right)&=a\left(x,y,t\right), \\
{\St{u}}_y\left(x,y,t\right)&=b\left(x,y,t\right)\left(f(x,t)-1\right).
\end{split}
\label{eqn:analyticalsolutionFSIproblemmanufacturedsolutionstructure}
\end{equation}
An appropriate body force canceling out the imbalance in the structural PDEs is added in the right hand side of the equation in \eqref{eqn:structurestrongform}.
The fluid velocity is then constructed as:
\begin{equation}
\begin{split}
  {\Fl{\upsilon}}_x(x,y,t)& = (j+1)y^j\bigl(M[f(x,t),t]-M(y,t)\bigr) - jy^{j-1}\bigl(L[f(x,t),t]-L(y,t)\bigr), \\
   {\Fl{\upsilon}}_y(x,y,t)& = y^j\bigl(f(x,t)-y\bigr)K\bigl[f(x,t),t\bigr]\frac{\partial f}{\partial x}(x,t)+\frac{\partial f}{\partial t}(x,t).
\end{split}
\label{eqn:analyticalsolutionFSIproblemmanufacturedsolutionfluid}
\end{equation}
The parameters chosen induce a sufficiently complex flow which is shown in Figure \ref{fig:solutionvectorialFSIproblemmanufacturedsolution} on the initial (left) and final (right) deformed configuration.
The velocity field presents two pronounced vortices circulating in opposite directions and it equals the structural velocity on the fluid-structure interface.
The structural forces on the deformed interface are evaluated through the push-forward operation
\begin{equation}
\mathbf{S}(x,t)= \bigl[\abs{\St{\mathbf{F}}}^{-1}\St{\mathbf{P}}\St{\mathbf{F}}^T\vect{n}\bigr]\Big|_{y=f(x,t)},
\label{eqn:analyticalsolutionFSIproblemmanufacturedsolutionstructureforceinterface}
\end{equation}
where $\vect{n}$ is the unit normal vector to the interface whose Cartesian components can be evaluated from the definition of the deformed interface $f(x,t)$ as
\begin{equation}
\begin{split}
n_x(x,t)& = -\frac{\partial f}{\partial x}(x,t)\Bigl[1+\Bigl(\frac{\partial f}{\partial x}(x,t)\Bigr)^2\Bigr]^{-1/2}, \\
n_y(x,t)& = \Bigl[1+\Bigl(\frac{\partial f}{\partial x}(x,t)\Bigr)^2\Bigr]^{-1/2}.
\end{split}
\label{eqn:analyticalsolutionFSIproblemmanufacturedsolutionnormalinterface}
\end{equation}
The fluid pressure is defined as
\begin{equation}
\Fl{p}(x,t)=\dfrac{-AS_yn_x-BS_yn_y+BS_xn_x+CS_xn_y}{-Bn_x^2-Cn_xn_y+An_xn_y+Bn_y^2},
\label{eqn:analyticalsolutionFSIproblemmanufacturedsolutionpressure}
\end{equation}
where
\begin{equation}
 A(x,t)=2\frac{\partial{\Fl{\upsilon}}_x}{\partial x}\Big|_{y=f(x,t)}, \;
 B(x,t)=\Bigl(\frac{\partial{\Fl{\upsilon}}_x}{\partial y}+\frac{\partial{\Fl{\upsilon}}_y}{\partial x}\Bigr)\Big|_{y=f(x,t)}, \;
 C(x,t)=2\frac{\partial{\Fl{\upsilon}}_y}{\partial y}\Big|_{y=f(x,t)}.
\label{eqn:analyticalsolutionFSIproblemmanufacturedsolutionfluidderivatives}
\end{equation}
In order to ensure the continuity of the fluid and solid forces without introducing additional terms at the interface (that would require substantial modification for most codes), an appropriate spatially varying viscosity is taken into account:
\begin{equation}
\Fl{\mu}(x,t)=\frac{S_xn_y-S_yn_x}{-Bn_x^2-Cn_xn_y+An_xn_y+Bn_y^2}.
\label{eqn:analyticalsolutionFSIproblemmanufacturedsolutionfluidviscosity}
\end{equation}
The last ingredient is the introduction of a fluid body force satisfying the momentum equation in \eqref{eqn:fluidgoverningequationsmovingdomain}.
The complexity of algebraic manipulation needed to compute this term is quite challenging and the symbolic math toolbox of MATLAB has been used for this purpose.

\begin{figure}
\centering
\subfigure[Undeformed mesh]
{\includegraphics[width=0.32\textwidth]{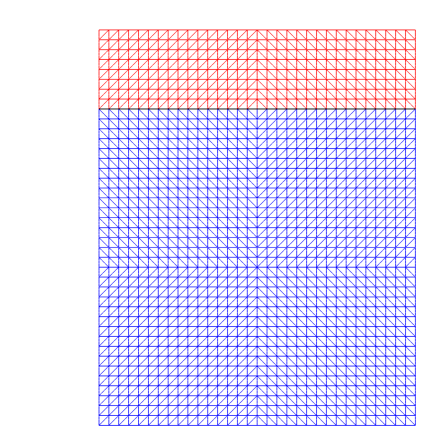}}
\subfigure[Initial deformed mesh]
{\includegraphics[width=0.32\textwidth]{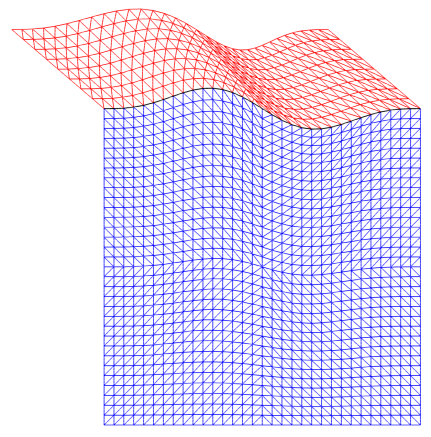}}
\subfigure[Final deformed mesh]
{\includegraphics[width=0.32\textwidth]{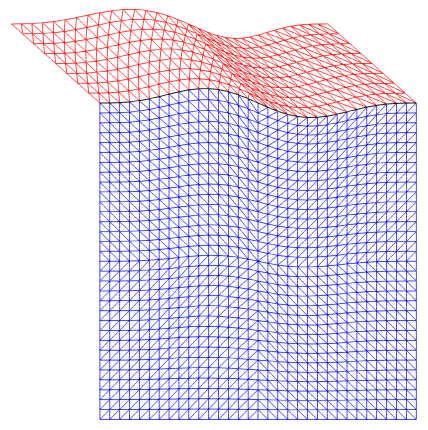}}
\caption{Fourth level of refinement of the undeformed (left), initial deformed (center) and final deformed (right) mesh used for the convergence studies of the FSI problem with manufactured solution.
In blue is represented the fluid domain, in red the structural domain and in black the fluid-structure interface.}
\label{fig:meshFSIproblemmanufacturedsolution}
\end{figure}

In order to test the coupling strategies formulated in section \ref{sec:hdgcgformulationcoupledFSI}, two cases are considered:
\begin{enumerate}
\item the problem is solved with the partitioned Dirichlet--Neumann coupling presented in section \ref{sec:partitioneddirichletneumannalgorithm},
\item the problem is solved with the monolithic Nitsche-based coupling presented in section \ref{sec:monolithicalgorithmnitsche}.
\end{enumerate}
The convergence studies are performed through uniform mesh refinement of the fluid and solid domains with triangular elements and considering as degree of approximation $k=[1,2,3]$.
In Figure \ref{fig:meshFSIproblemmanufacturedsolution} the fourth level of refinement of the undeformed (left), initial deformed (center) and final deformed (right) mesh is shown.
To mimic incompressible flow, the fluid parameters are chosen as $\Fl{\rho_0}=1$, $\Fl{p_0}=0$ and $\Fl{\varepsilon}=10^{-6}$.
The Young modulus and the Poisson ratio of the elastic structure are $\St{E}=1$ and $\St{\nu}=0.49$.
The stabilization parameters are $\tau_\rho=1/\Fl{\varepsilon}$ and $\tau_{\rho\upsilon}=10$ and the Nitsche parameter is set as $\gamma=10^5$ for the case 2.
The time span considered is $\left(0,0.125\right)$ and in order to keep the temporal error sufficiently small to perform the spatial convergence studies, the BDF2 method is adopted with a time step $\Delta t=2^{-12}$.
The initial conditions and the boundary conditions imposed on $\Gamma^D=\partial\left(\Fl{\Omega}\cup\St{\Omega}\right)$ are computed from the analytical solution developed here.

\begin{figure}
\centering
\subfigure[Initial configuration]
{\includegraphics[width=0.49\textwidth]{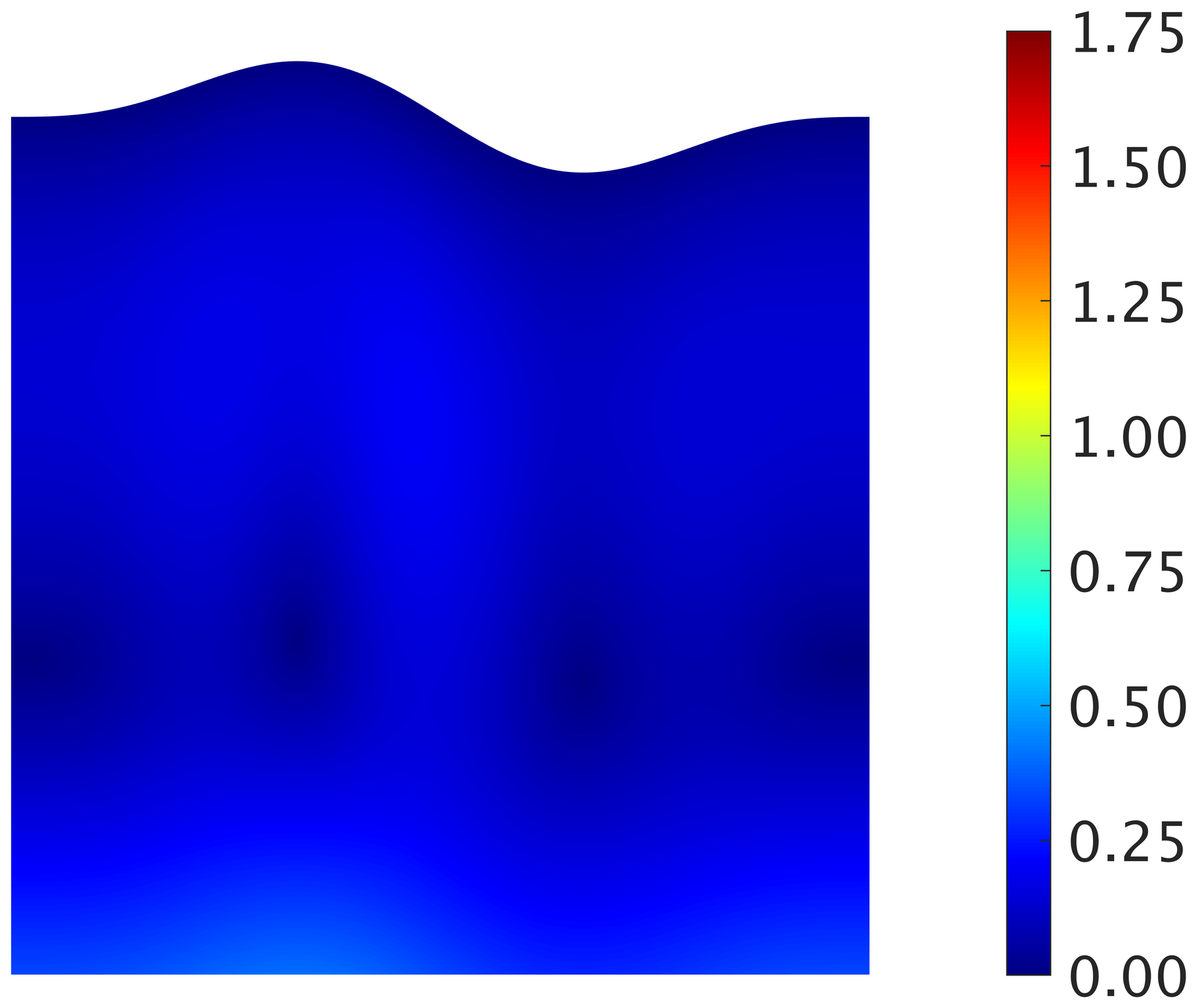}}
\subfigure[Final configurationh]
{\includegraphics[width=0.49\textwidth]{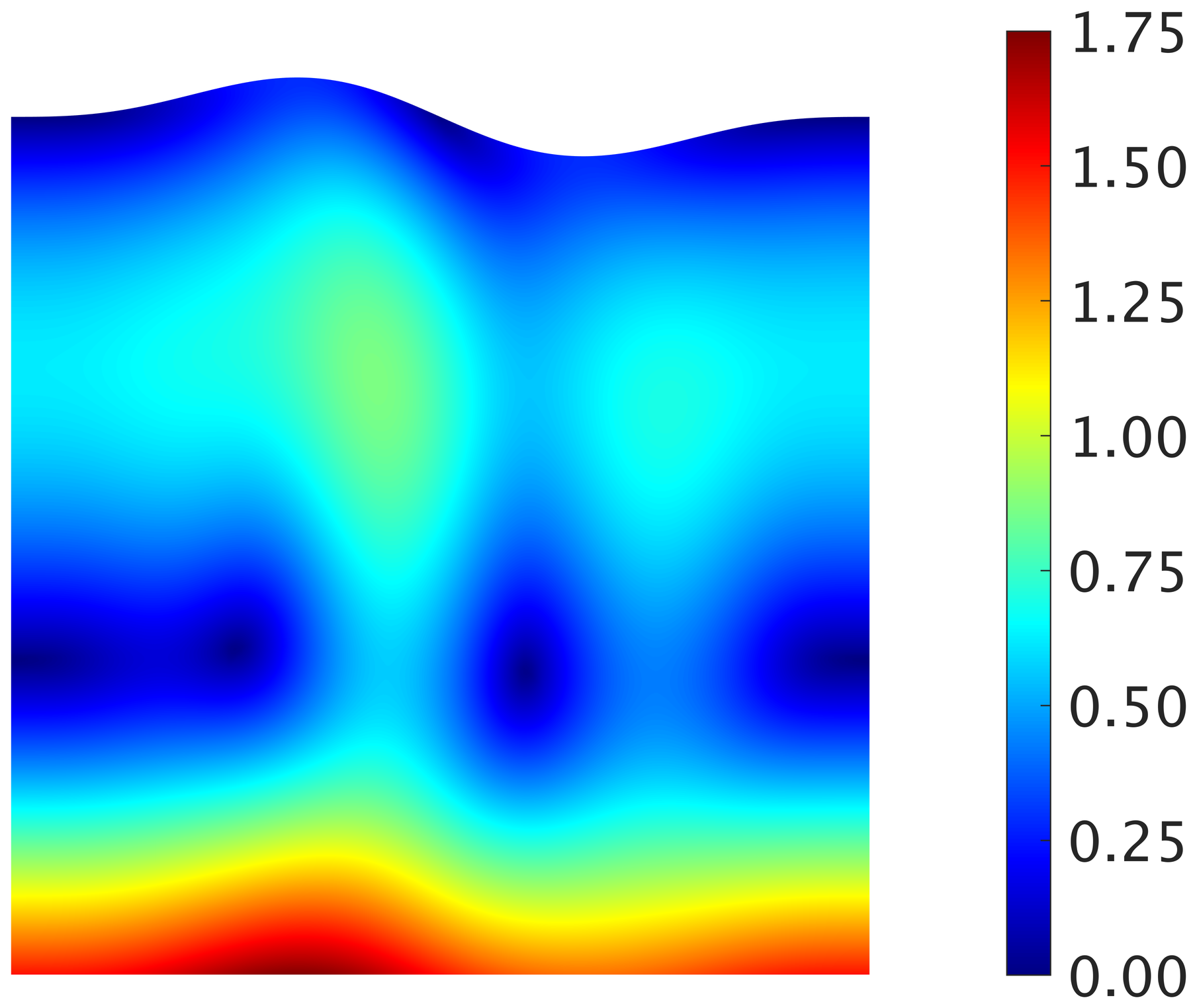}}
\caption{Approximation of the velocity field of the FSI problem with manufactured solution on the initial (left) and final (right) deformed configuration.}
\label{fig:solutionFSIproblemmanufacturedsolution}
\end{figure}

In Figure \ref{fig:solutionFSIproblemmanufacturedsolution} the solution of the fluid velocity field obtained with the monolithic Nitsche-based coupling using $m=4$ and $k=3$ on the initial (left) and final (right) deformed configuration is shown.

\begin{figure}
\centering
\subfigure[$\Fl{\mat{L}}$ with partitioned Dirichlet--Neumann coupling]
{\includegraphics[width=0.23\textwidth]{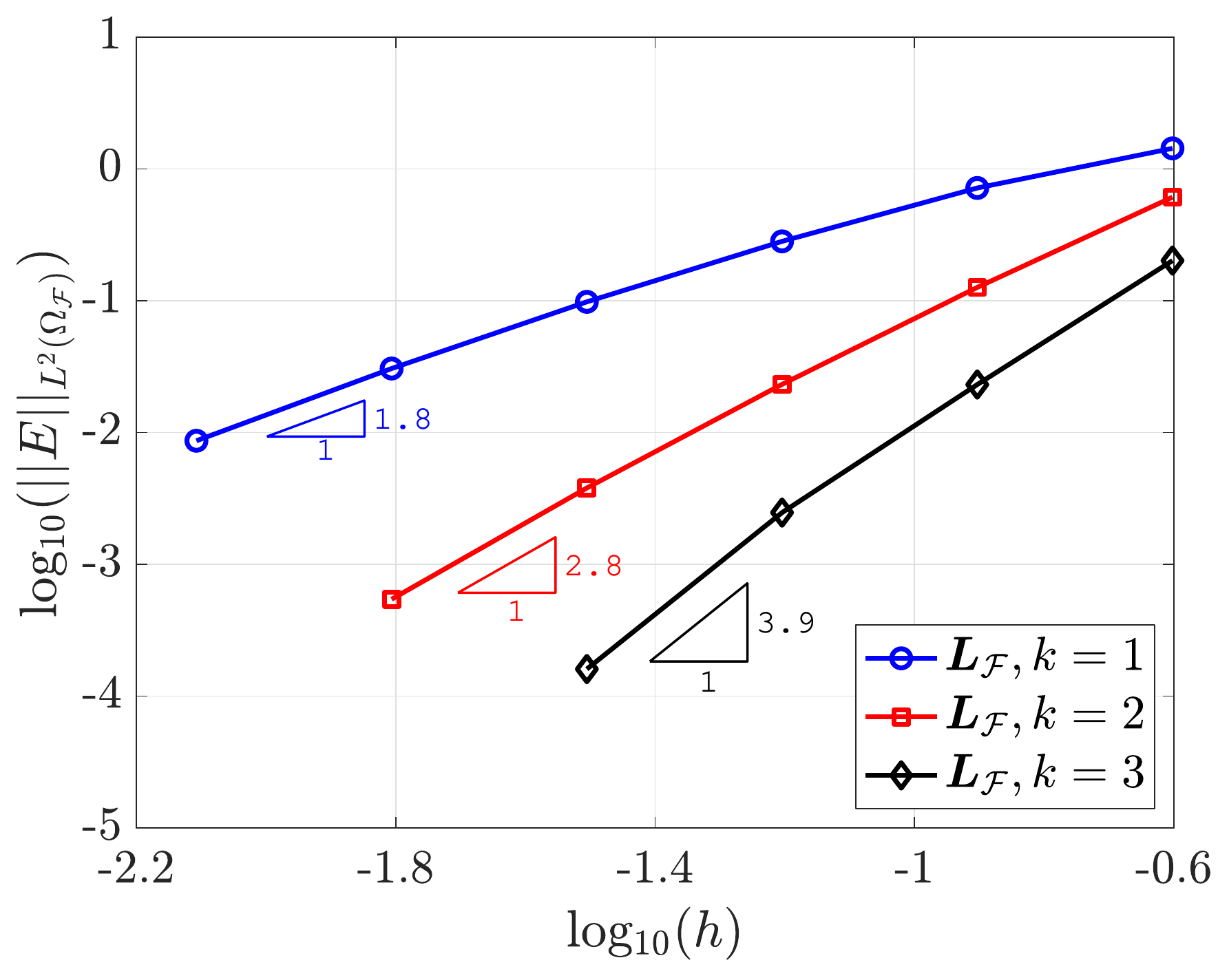}}
\subfigure[$\Fl{\mat{L}}$ with monolithic Nitsche-based coupling]
{\includegraphics[width=0.23\textwidth]{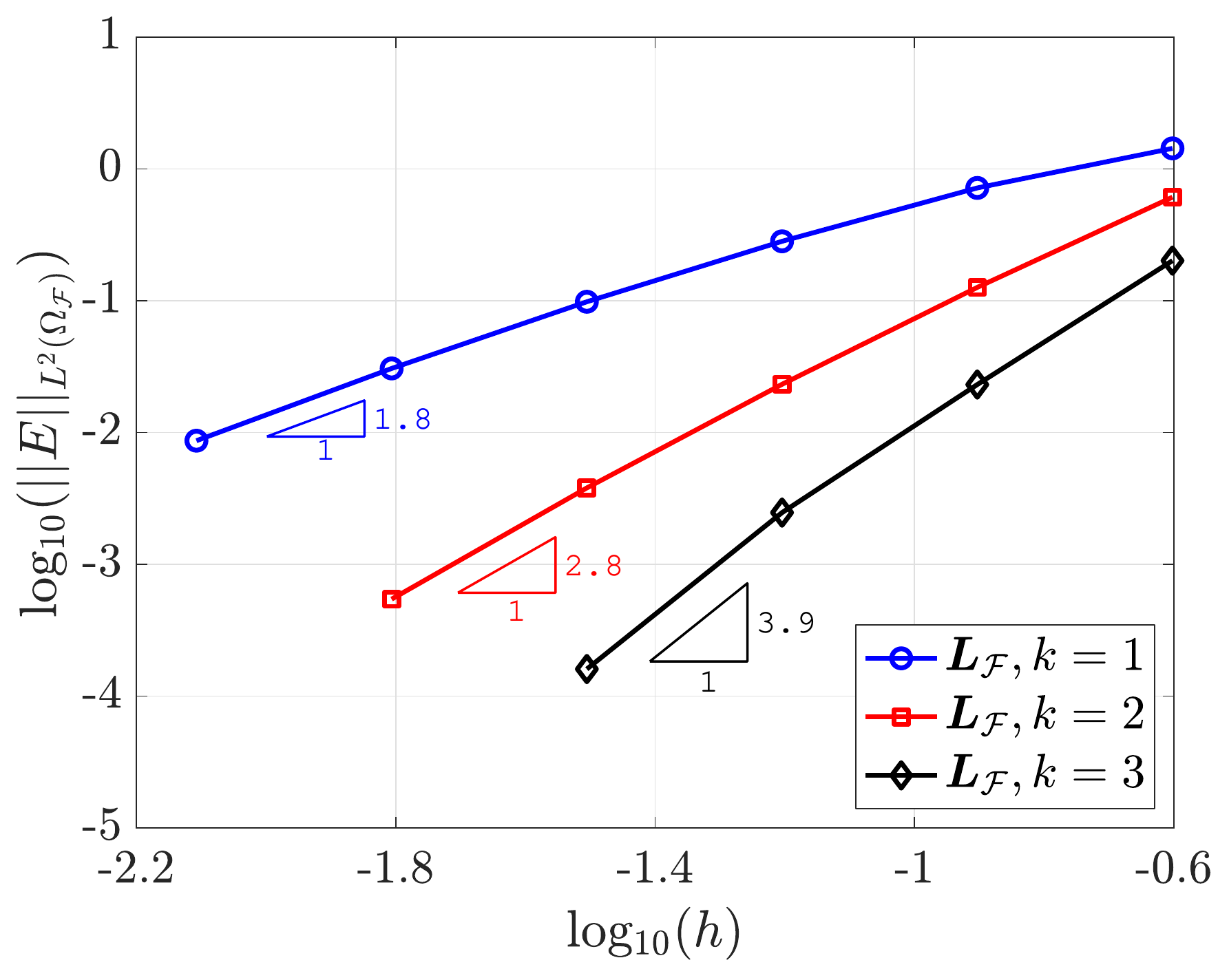}} 
\subfigure[$\Fl{\rho}$ with partitioned Dirichlet--Neumann coupling]
{\includegraphics[width=0.23\textwidth]{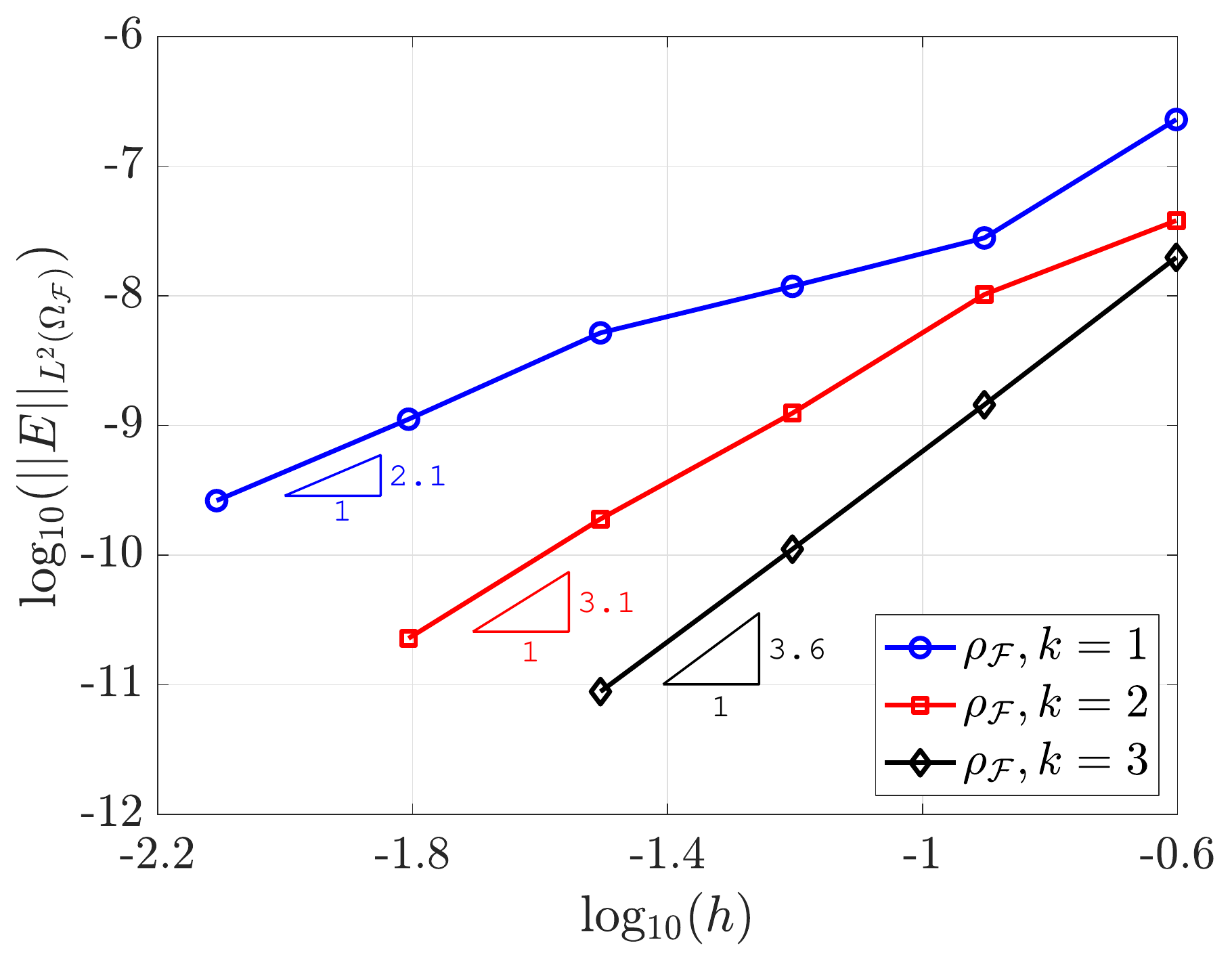}}
\subfigure[$\Fl{\rho}$ with monolithic Nitsche-based coupling]
{\includegraphics[width=0.23\textwidth]{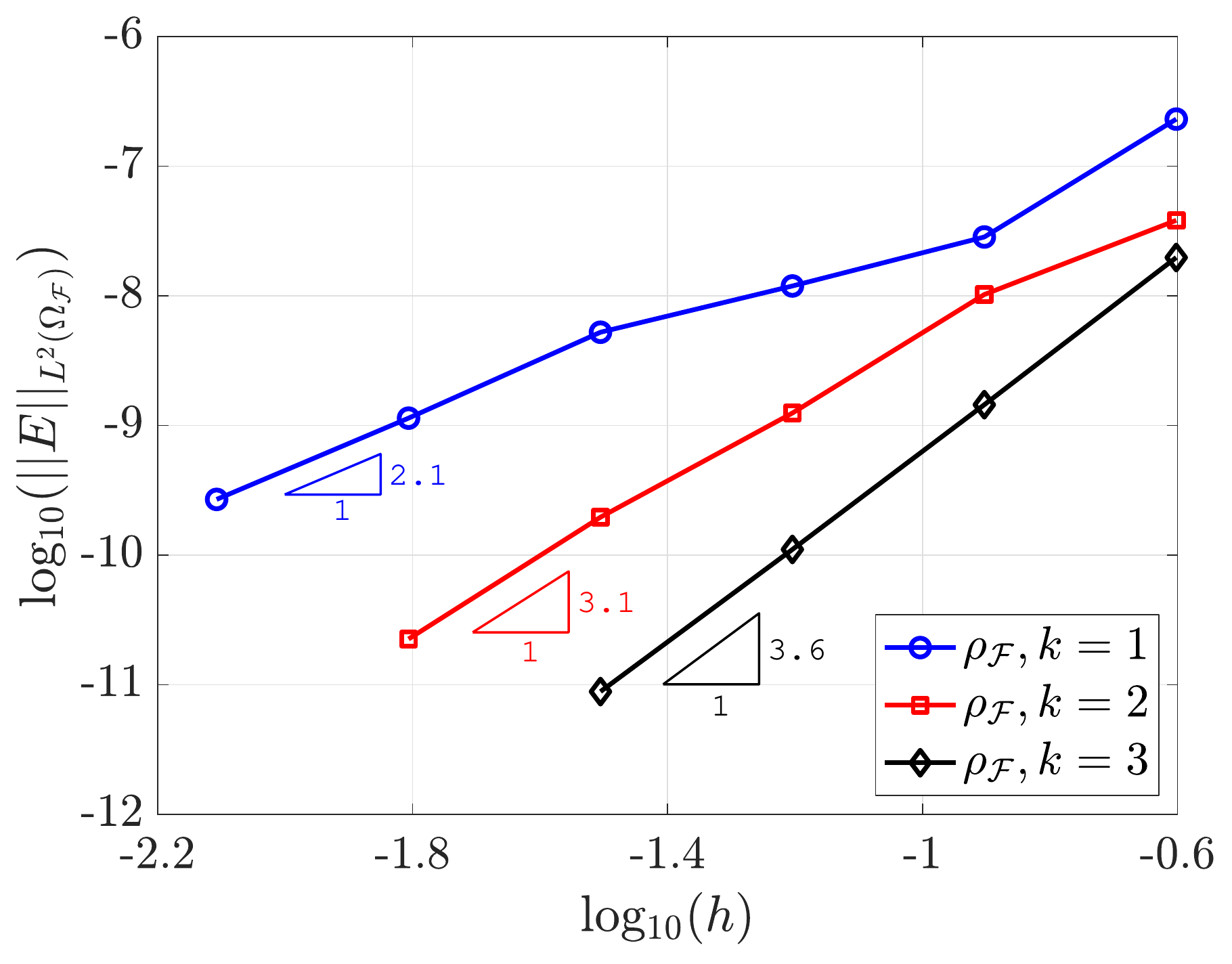}} \\
\subfigure[$\Fl{\rho\bm{\upsilon}}$ with partitioned Dirichlet--Neumann coupling]
{\includegraphics[width=0.23\textwidth]{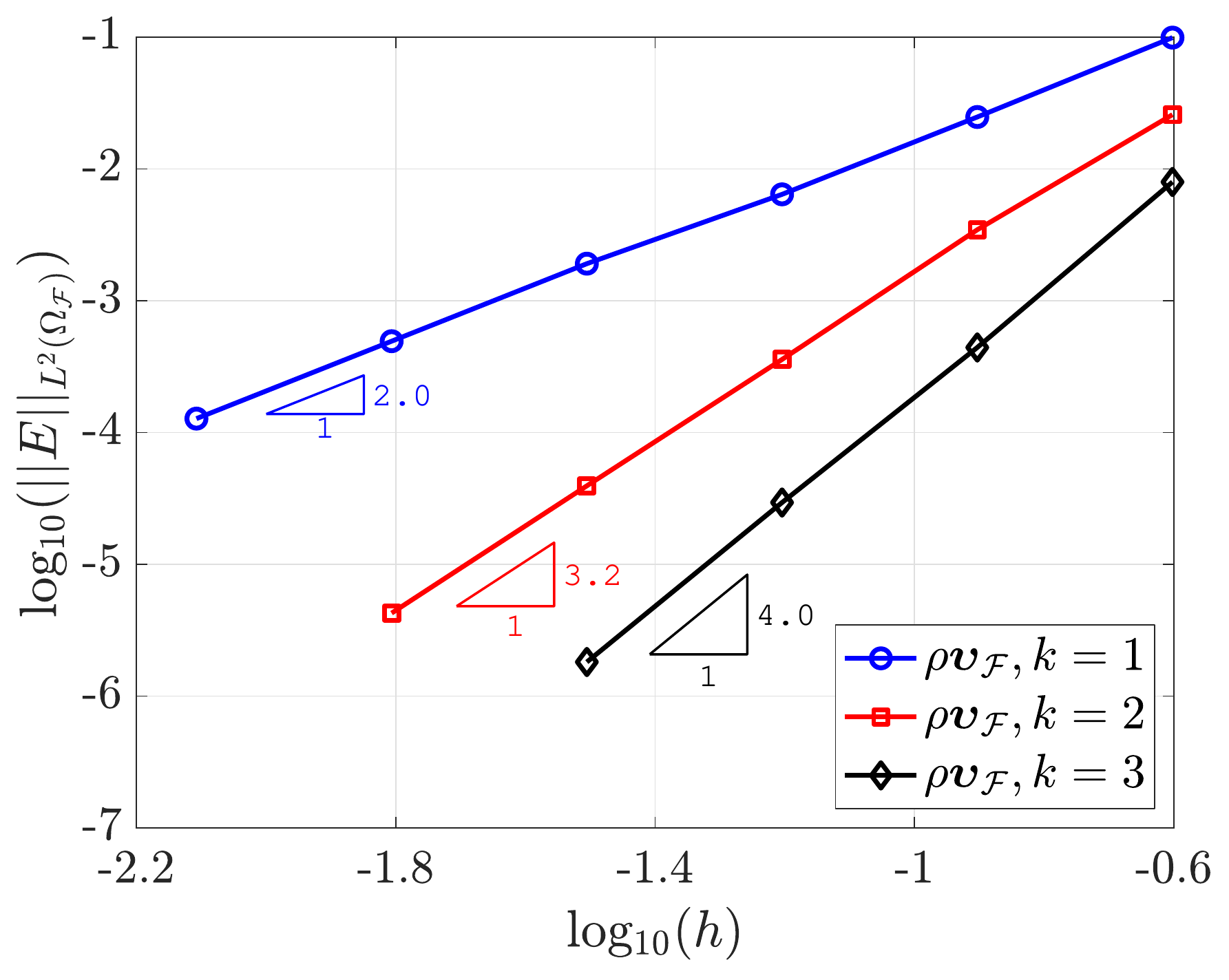}}
\subfigure[$\Fl{\rho\bm{\upsilon}}$ with monolithic Nitsche-based coupling]
{\includegraphics[width=0.23\textwidth]{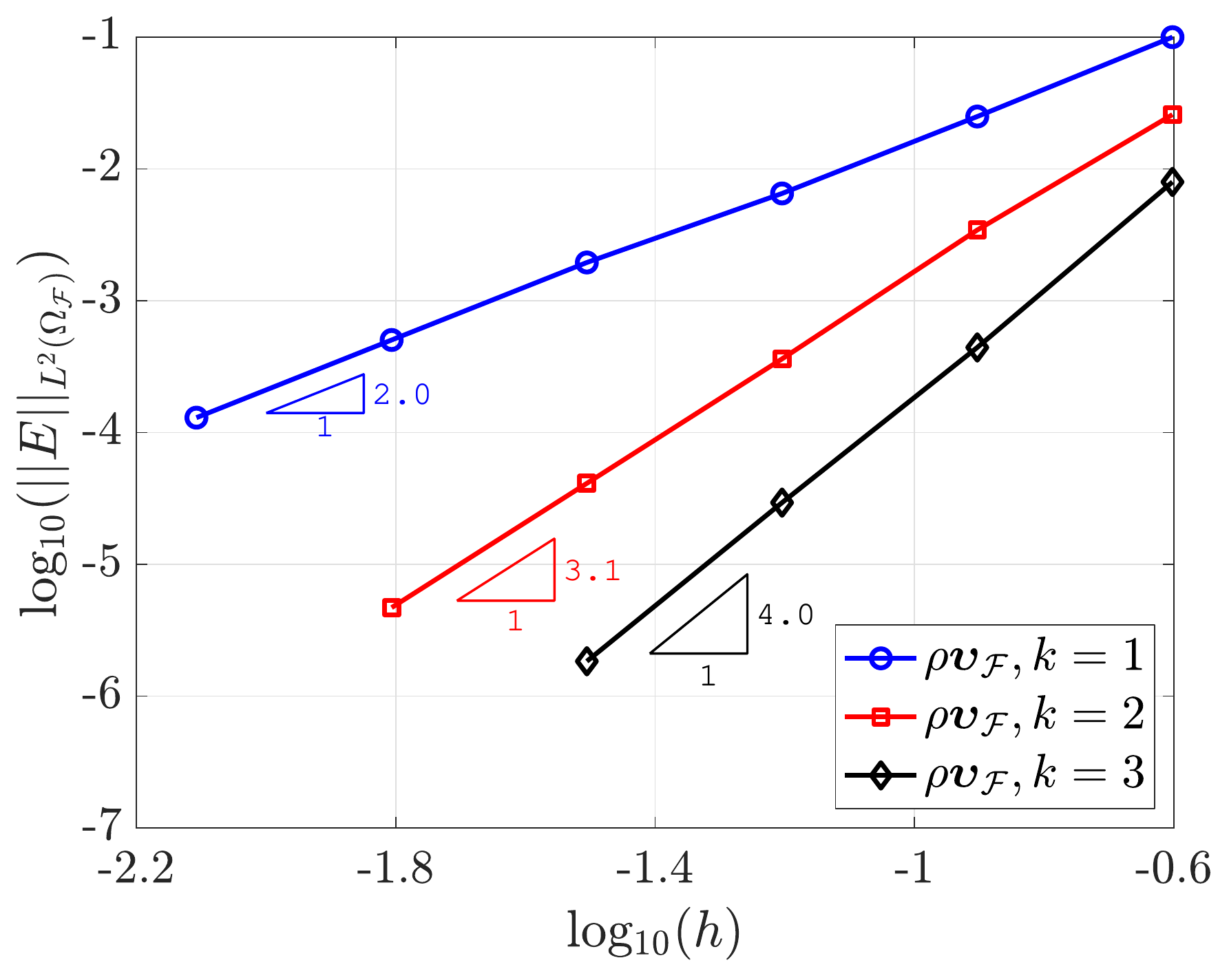}} 
\subfigure[$\St{\mathbf{u}}$ with partitioned Dirichlet--Neumann coupling]
{\includegraphics[width=0.23\textwidth]{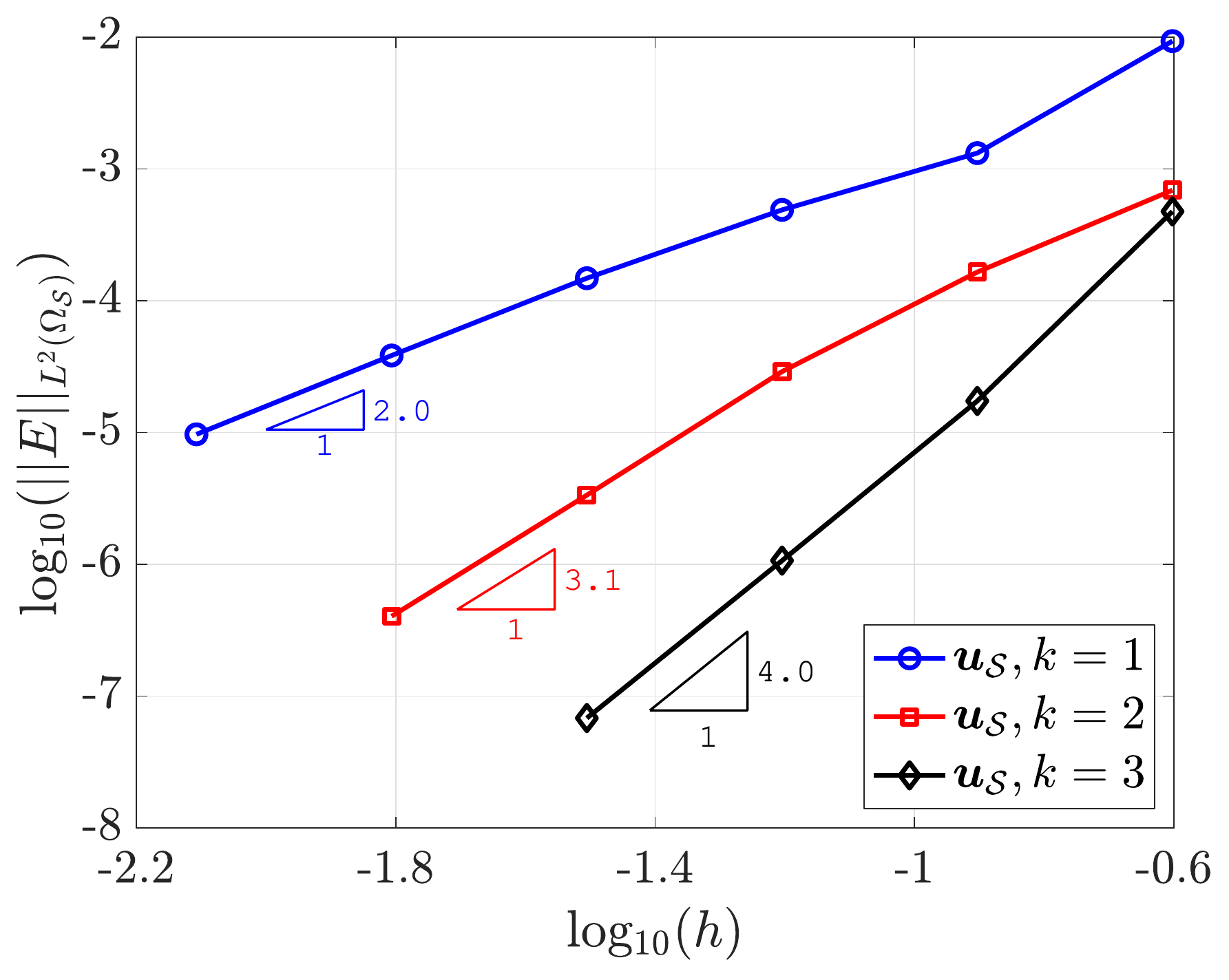}}
\subfigure[$\St{\mathbf{u}}$ with monolithic Nitsche-based coupling]
{\includegraphics[width=0.23\textwidth]{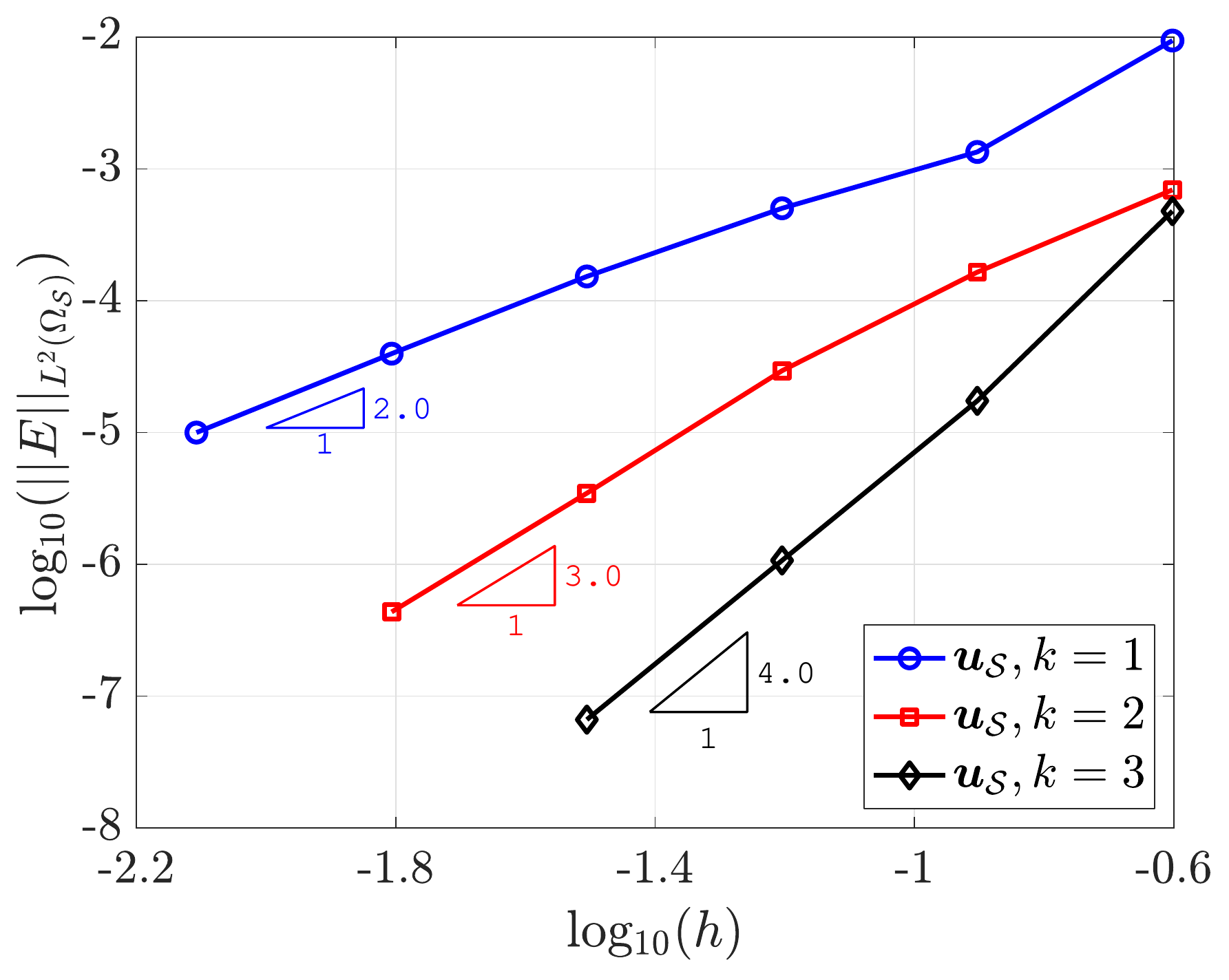}}
\caption{Spatial convergence of the $\mathcal{L}_2$-error of the mixed and primal variables evaluated at $t=0.125$ for the FSI problem with manufactured solution solved with the partitioned Dirichlet--Neumann coupling (left) and the monolithic Nitsche-based coupling (right).}
\label{fig:convergenceFSIproblemmanufacturedsolution}
\end{figure}

The convergence of the error of the HDG fluid solution and the CG structural solution measured in the $\mathcal{L}_2$ norm as a function of the characteristic element size $h$ is represented in Figure \ref{fig:convergenceFSIproblemmanufacturedsolution}, for the case $1$ (left) and the case $2$ (right).
Optimal convergence rates are observed for all the variables in both subdomains.
By comparing the plots on the left and the ones on the right, it can be observed that the two coupling strategies provide almost identical results.

\subsection{Channel with flexible wall}
\label{sec:channelflexiblewall}

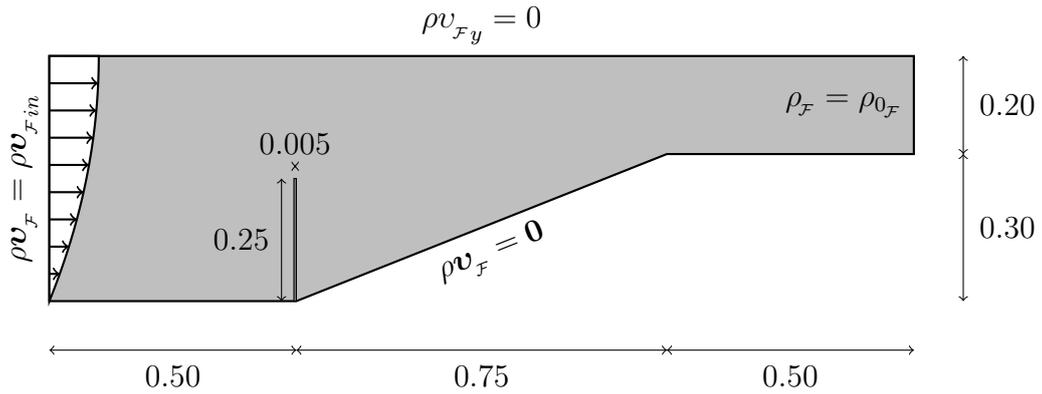
\begin{figure}
\begin{center}
\begin{tikzpicture}
\def \fac{6.5}
\def \Launs{0.50}
\def \Lbuns{0.75}
\def \Lcuns{0.50}
\def \Hauns{0.30}
\def \Hbuns{0.20}
\def \tuns{0.005}
\def \huns{0.25}
\def \umaxuns{0.1}
\def \dsizeuns{0.1}
\def \dtextuns{0.3}
\def \dbcuns{0.02}
\def \arrows{10}
\def \nconst{3}
\def \dconstuns{0.01}

\def \La{\Launs*\fac}
\def \Lb{\Lbuns*\fac}
\def \Lc{\Lcuns*\fac}
\def \Ha{\Hauns*\fac}
\def \Hb{\Hbuns*\fac}
\def \t{\tuns*\fac}
\def \h{\huns*\fac}
\def \umax{\umaxuns*\fac}
\def \dsize{\dsizeuns*\fac}
\def \dtext{\dtextuns*\fac}
\def \dbc{\dbcuns*\fac}
\def \dconst{\dconstuns*\fac}

\draw[fill=lightgray,draw=black,thick]
(0,0)                 --
(\La,0)               -- node[below=\dbc,sloped]            {$\Fl{\rho\bm{\upsilon}}=\mathbf{0}$}
(\La+\Lb,\Ha)         --
(\La+\Lb+\Lc,\Ha)     -- node[left=\dbc]                    {$\Fl{\rho}=\Fl{\rho_0}$}
(\La+\Lb+\Lc,\Ha+\Hb) -- node[above=\dbc]                   {${\Fl{\rho\upsilon}}_y=0$}
(0,\Ha+\Hb)           -- node[above=\dbc,sloped,rotate=180] {$\Fl{\rho\bm{\upsilon}}={\Fl{\rho\bm{\upsilon}}}_{in}$}
cycle;

\draw[fill=gray,draw=black,thin]
(\La-\t,0)  -- 
(\La,0)     -- 
(\La,\h)    -- 
(\La-\t,\h) -- 
cycle;

\draw[fill=white,draw=black,thick,domain={0}:{\Ha+\Hb},samples=100] plot[smooth] ({4*(\umax)*(\x/\fac)*(1-\x/\fac)},{\x}) --
(0,\Ha+\Hb) --
cycle;
\foreach \i in {2,...,\numexpr(\arrows)-1}
{
  \def \xi{0}
  \def \yi{(\i-1)*(\Ha+\Hb)/(\arrows-1)}
  \def \xf{{4*(\umax)*(\yi/\fac)*(1-\yi/\fac)}}
  \def \yf{(\i-1)*(\Ha+\Hb)/(\arrows-1)}
  \draw [->,draw=black,thick] ({\xi},{\yi}) -- ({\xf},{\yf});
}

\draw [<->,draw=black,thin] ({0},{-\dsize}) -- node[below=\dtext] {$0.50$} ({\La},{-\dsize});

\draw [<->,draw=black,thin] ({\La},{-\dsize}) -- node[below=\dtext] {$0.75$} ({\La+\Lb},{-\dsize});

\draw [<->,draw=black,thin] ({\La+\Lb},{-\dsize}) -- node[below=\dtext] {$0.50$} ({\La+\Lb+\Lc},{-\dsize});

\draw [<->,draw=black,thin] ({\La+\Lb+\Lc+\dsize},{0}) -- node[right=\dtext] {$0.30$} ({\La+\Lb+\Lc+\dsize},{\Ha});

\draw [<->,draw=black,thin] ({\La+\Lb+\Lc+\dsize},{\Ha}) -- node[right=\dtext] {$0.20$} ({\La+\Lb+\Lc+\dsize},{\Ha+\Hb});

\draw [<->,draw=black,thin] ({\La-\t},{\h+\dsize/4}) -- node[above=\dtext/4] {$0.005$} ({\La},{\h+\dsize/4});

\draw [<->,draw=black,thin] ({\La-\t-\dsize/4},{0}) -- node[left=\dtext/4] {$0.25$} ({\La-\t-\dsize/4},{\h});
\end{tikzpicture}
\end{center}
\caption{Geometry (dimensions in meters) and boundary conditions of the channel with flexible wall.}
\label{fig:channelflexiblewallgeometrybcs}
\end{figure}

The fourth numerical experiment considers a convergent fluid channel containing a flexible wall structure attached to its bottom.
This example is inspired by \cite{Mok2001} and its main feature is the strong coupling between the fluid and structural fields, given by their similar densities.
The geometry and the boundary conditions of the problem are depicted in Figure \ref{fig:channelflexiblewallgeometrybcs}.

The dynamic viscosity of the fluid is $\Fl{\mu}=0.145$ $kg/(m \cdot s)$ and the reference density $\Fl{\rho_0}=956$ $kg/m^3$ evaluated at the reference pressure $\Fl{p_0}=0$ $N/m^2$.
Three different orders of magnitude are considered for the compressibility coefficient, i.e., $\Fl{\varepsilon}=[0.001,0.01,0.1]$ $kg/(N \cdot m)$.
The structure is modelled as a St. Venant--Kirchhoff material with density $\St{\rho}=1500$ $kg/m^3$, Young's modulus $\St{E}=2.3\times10^6$ $N/m^2$ and Poisson's ratio $\St{\nu}=0.45$.
The following parabolic momentum profile is imposed at the inlet:
\begin{equation}
\begin{split}
{\Fl{\rho\upsilon}}_x\left(0,y,t\right)&=
\left\{
\begin{aligned}
&4y\left(1-y\right)\dfrac{1-\cos\left(\pi t/10\right)}{2}\Fl{\rho_0}\Fl{\bar{\upsilon}}
\quad&&\text{if }t\leq 10\text{ }s, \\
&4y\left(1-y\right)\Fl{\rho_0}\Fl{\bar{\upsilon}}
\quad&&\text{if }t  >  10\text{ }s,
\end{aligned}
\right. \\
{\Fl{\rho\upsilon}}_y\left(0,y,t\right)&=0,
\end{split}
\label{eqn:inflowchannelflexiblewall}
\end{equation}
with $\Fl{\bar{\upsilon}}=0.06067$ $m/s$, resulting in a Reynolds number of about $100$ after the flow has been completely formed.
No-slip and free-slip boundary conditions are considered on the bottom and on the top sides of the channel, respectively.
At the channel exit, the fluid density is imposed equal to its reference value.

The fluid domain is discretized with the HDG method and contains $772$ triangular elements, while the structural domain is discretized with the CG method and contains $200$ triangular elements.
A boundary layer mesh is constructed near the physical walls.
The time interval studied is $25$ $s$ and the temporal integration is performed by means of the BDF2 method.
Three different orders of magnitude of the time step size are considered, i.e., $\Delta t=[0.1,0.01,0.001]$ $s$.
The stabilization parameters are $\tau_\rho=10/\Fl{\varepsilon}$ and $\tau_{\rho\upsilon}=1$.

Two cases are considered:
\begin{enumerate}
\item the problem is solved with the partitioned Dirichlet--Neumann coupling presented in section \ref{sec:partitioneddirichletneumannalgorithm},
\item the problem is solved with the monolithic Nitsche-based coupling presented in section \ref{sec:monolithicalgorithmnitsche}.
\end{enumerate}
For the first case, the convergence tolerance for the coupling iterations is set to a very small value, i.e., $\eta=10^{-9}$.
For the second case, the Nitsche parameter is set to $\gamma=100$.

\begin{figure}
\centering
{\includegraphics[width=0.49\textwidth]{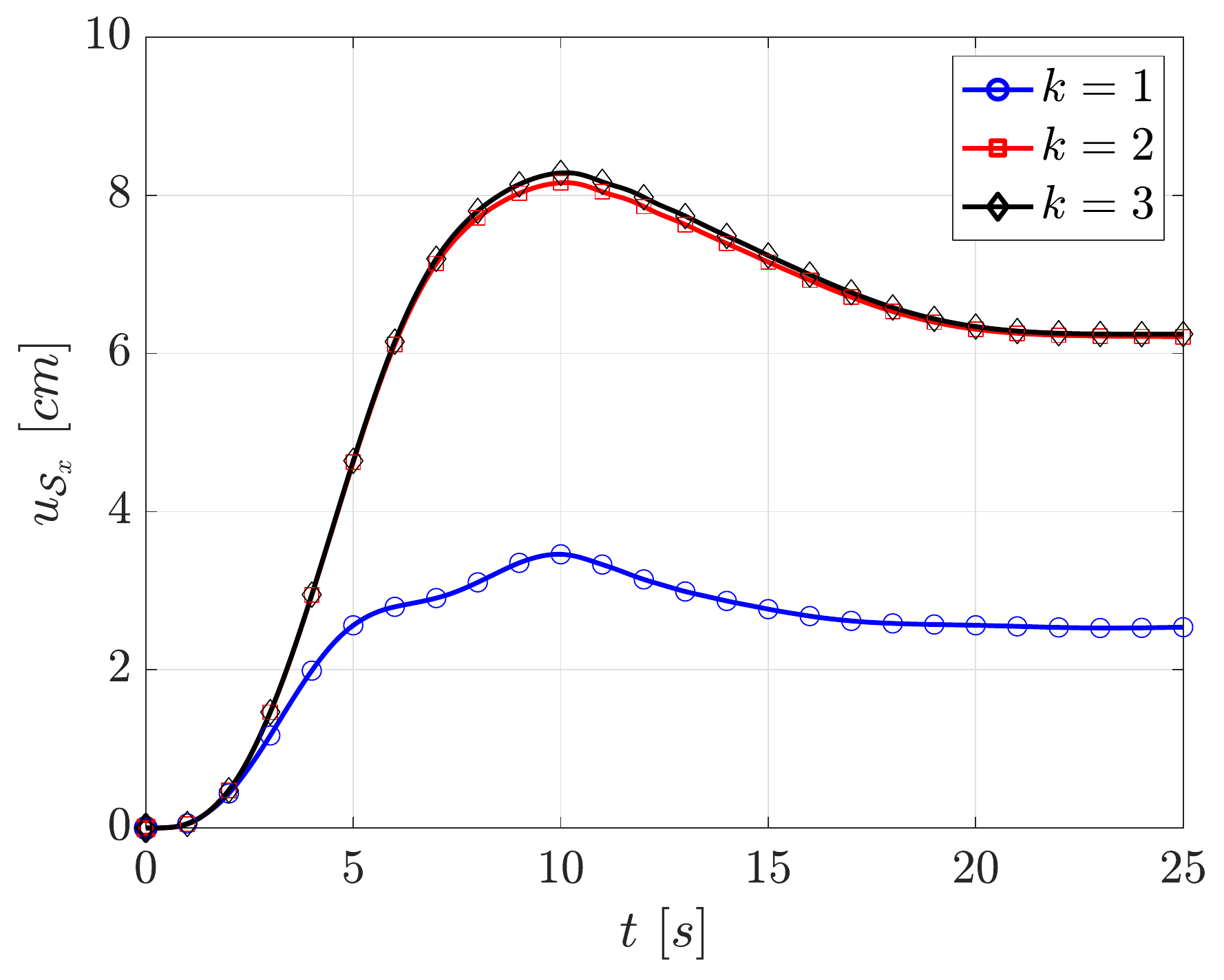}}
\caption{$x$-displacement component of the center top point of the flexible wall with different polynomial degrees of approximation computed with the monolithic Nitsche-based coupling with $\Fl{\varepsilon}=0.01$ $kg/(N \cdot m)$ and $\Delta t=0.1$ $s$.}
\label{fig:displacementchannelflexiblewall_k_1_2_3}
\end{figure}

A preliminary study, conducted with the monolithic scheme for the intermediate compressibility coefficient ($\Fl{\varepsilon}=0.01$ $kg/(N \cdot m)$) and the largest time step size ($\Delta t=0.1$ $s$), has highlighted the need for adopting a higher-order polynomial degree of approximation.
Figure \ref{fig:displacementchannelflexiblewall_k_1_2_3} shows how the resulting horizontal displacement of the center top point of the flexible wall is severely underestimated when computed with linear elements ($k=1$), whereas $k=2$ and $k=3$ produce a reliable response with a final structural displacement differing by less than $1\%$.
Hence, a degree of approximation $k=2$ is considered in the following for both subdomains.

\begin{figure}
\centering
\subfigure[${\St{u}}_x$ with partitioned Dirichlet--Neumann coupling]
{\includegraphics[width=0.49\textwidth]{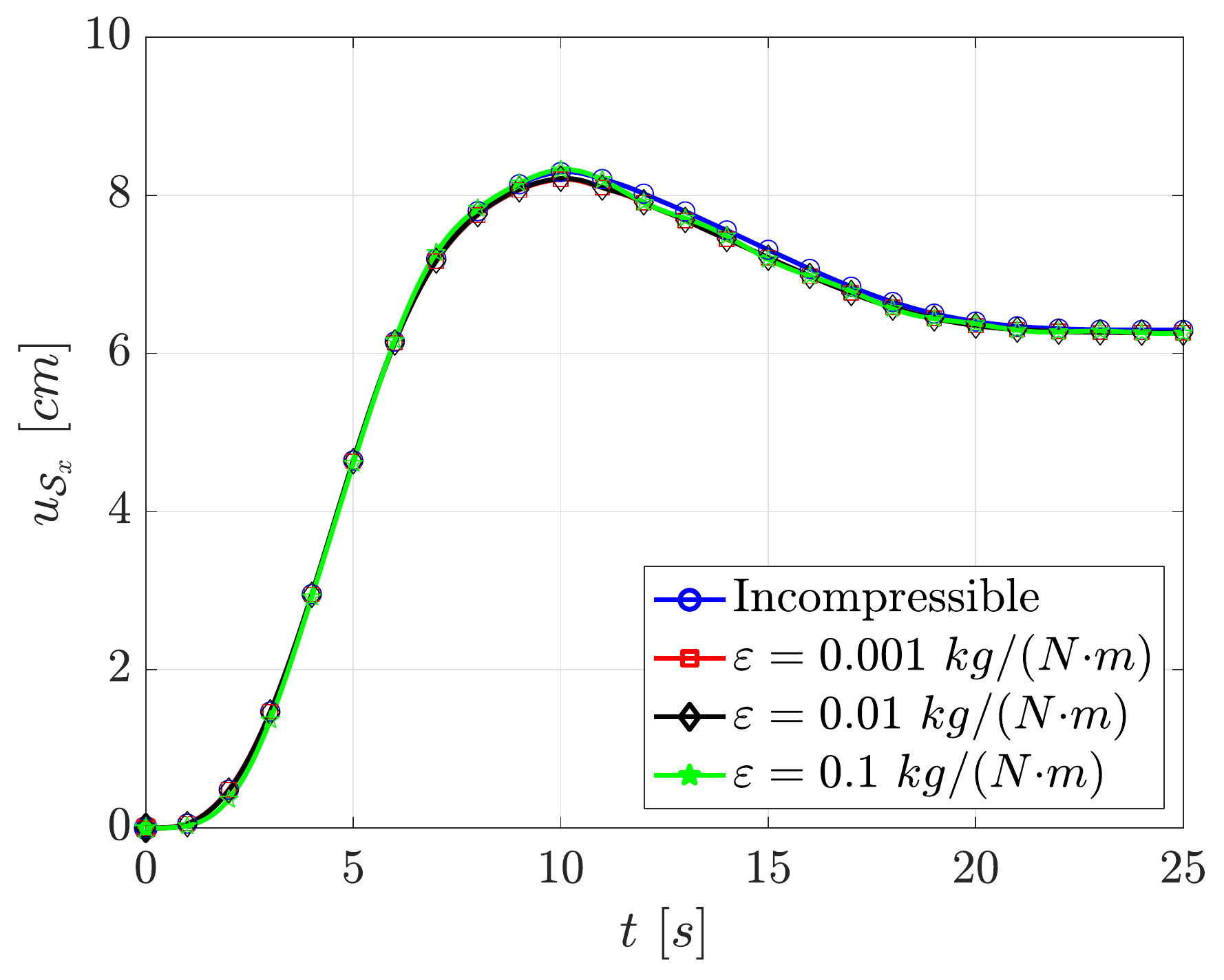}}
\subfigure[${\St{u}}_x$ with monolithic Nitsche-based coupling]
{\includegraphics[width=0.49\textwidth]{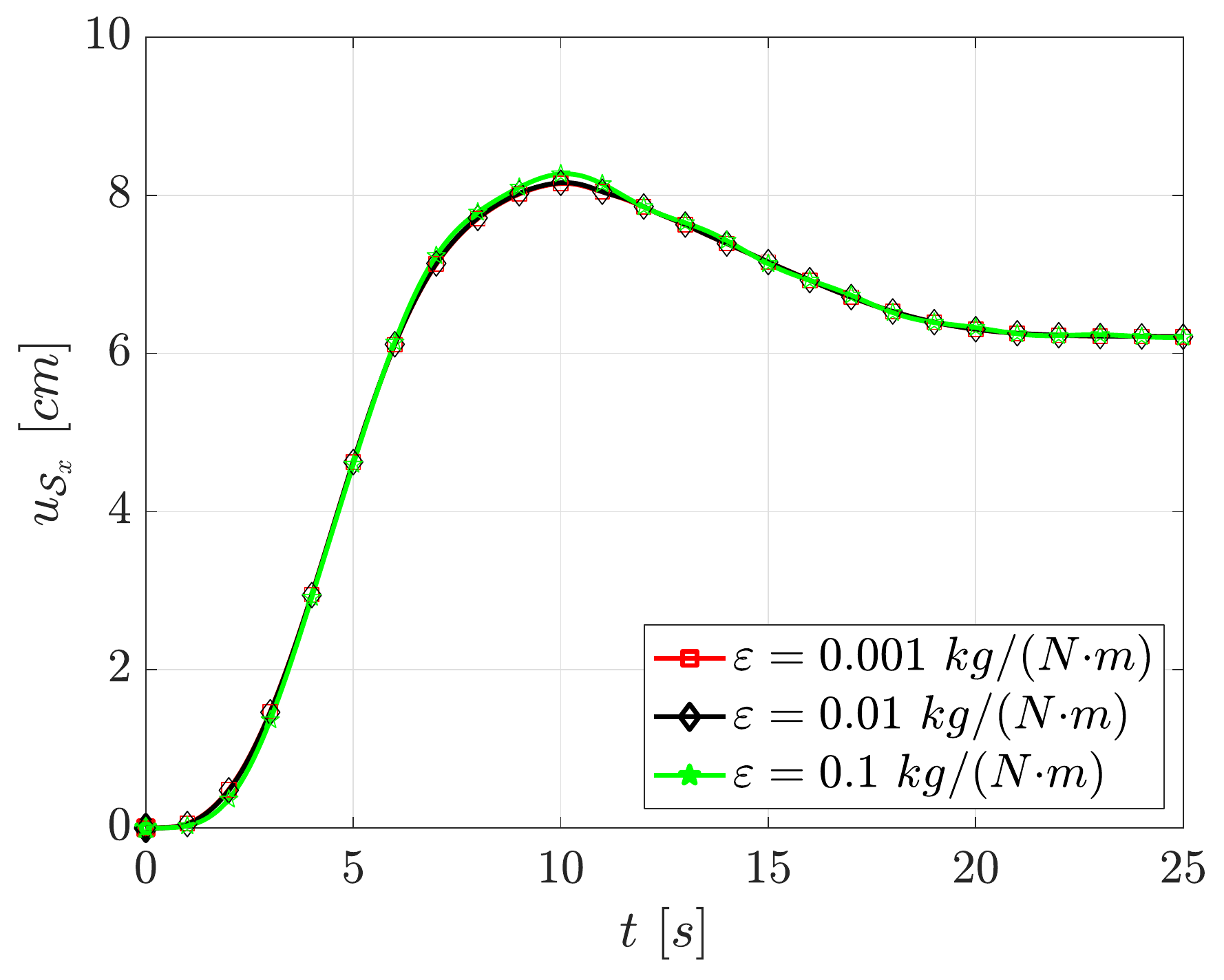}}
\caption{$x$-displacement component of the center top point of the flexible wall with different values of the compressibility coefficient computed with the partitioned Dirichlet--Neumann coupling (left) and the monolithic Nitsche-based coupling (right) with $\Delta t=0.1$ $s$.}
\label{fig:displacementchannelflexiblewall}
\end{figure}

The structural displacement obtained with both the partitioned and the monolithic scheme is shown in Figure \ref{fig:displacementchannelflexiblewall} for the different compressibility coefficients.
For the sake of validation and comparison, the displacement is also computed on the same mesh with a fully incompressible solver, based on an equal order stabilized finite element formulation, with the same settings used for the partitioned Dirichlet--Neumann coupling.
The results are added in the left panel of Figure \ref{fig:displacementchannelflexiblewall}.
By comparing the left and the right panels, it can be observed that the two coupling strategies provide almost identical results.
Moreover, for the compressibility levels considered, the physical results are sufficiently close to those obtained by considering the flow fully incompressible.

\begin{figure}
\centering
\subfigure[$\Fl{\rho}$ at $t=5$ $s$  with $\Fl{\varepsilon}=0.1$ $kg/(N \cdot m)$]
{\includegraphics[width=0.49\textwidth]{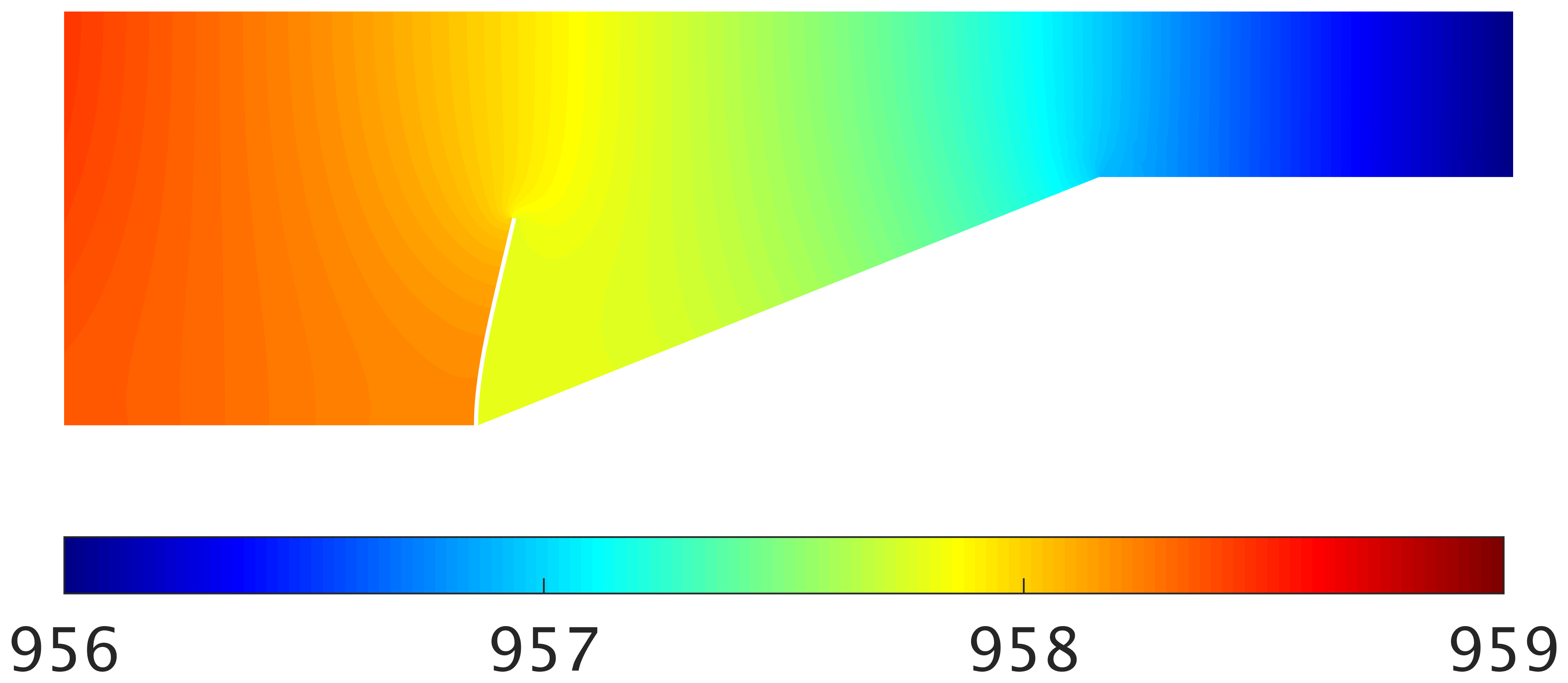}}
\subfigure[$\abs{\Fl{\rho\bm{\upsilon}}}$ at $t=5$ $s$ with $\Fl{\varepsilon}=0.1$ $kg/(N \cdot m)$]
{\includegraphics[width=0.49\textwidth]{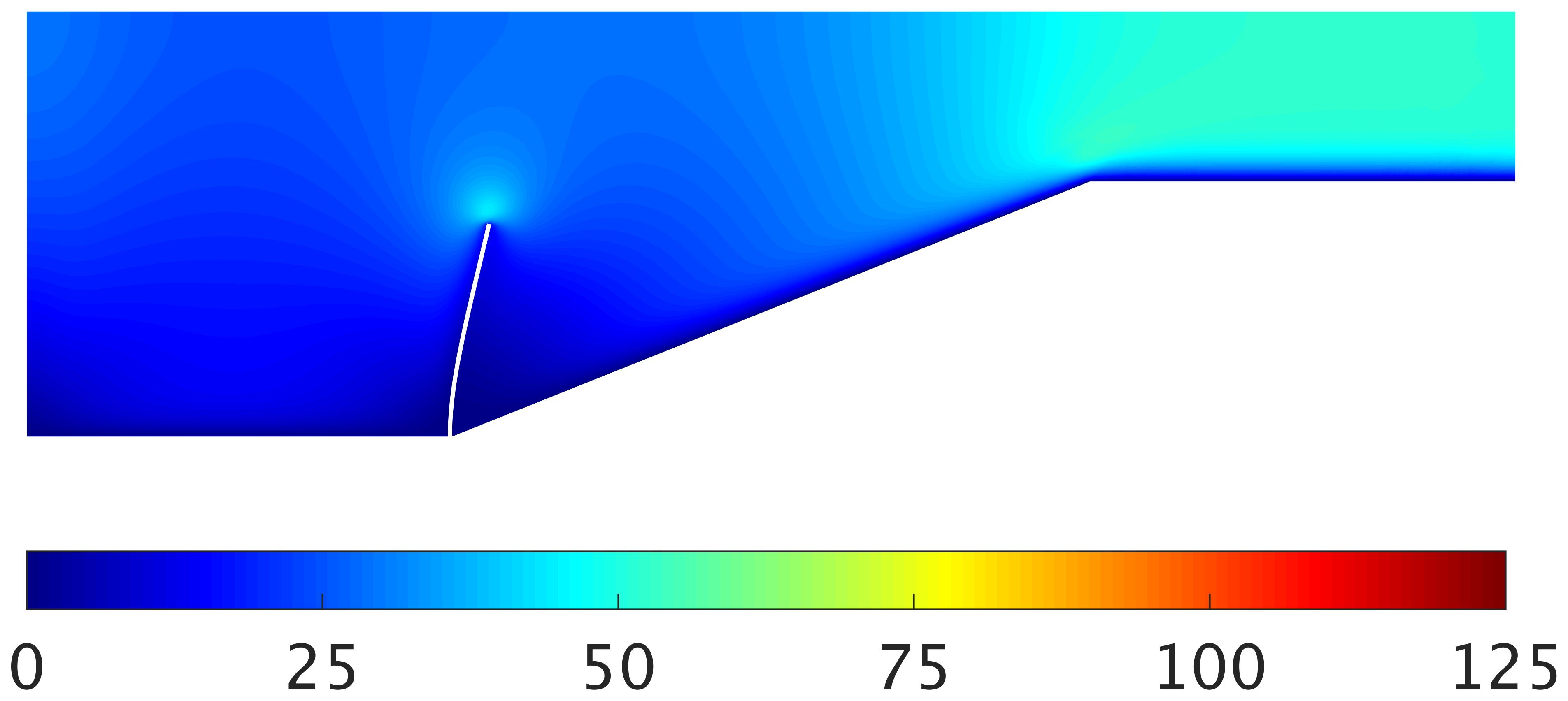}} \\
\subfigure[$\Fl{\rho}$ at $t=10$ $s$ with $\Fl{\varepsilon}=0.1$ $kg/(N \cdot m)$]
{\includegraphics[width=0.49\textwidth]{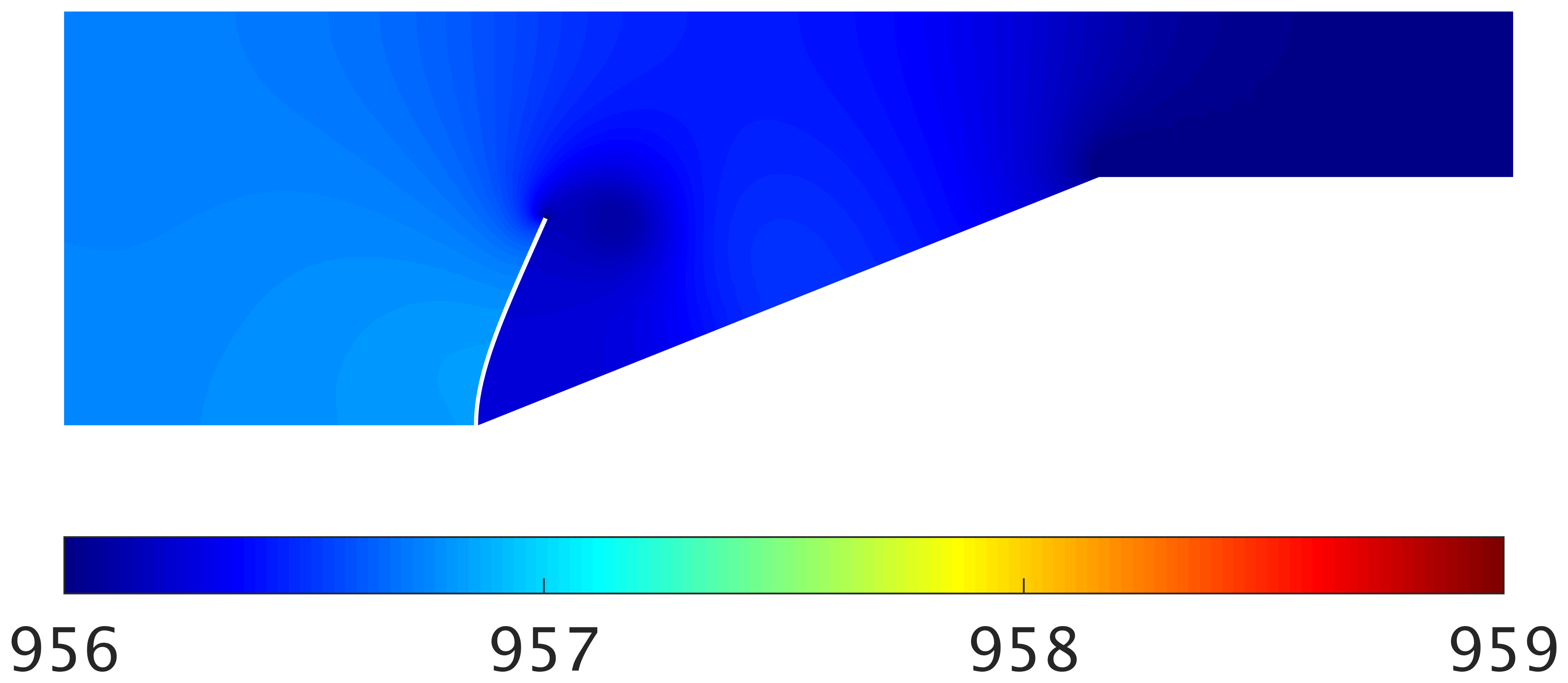}}
\subfigure[$\abs{\Fl{\rho\bm{\upsilon}}}$ at $t=10$ $s$ with $\Fl{\varepsilon}=0.1$ $kg/(N \cdot m)$]
{\includegraphics[width=0.49\textwidth]{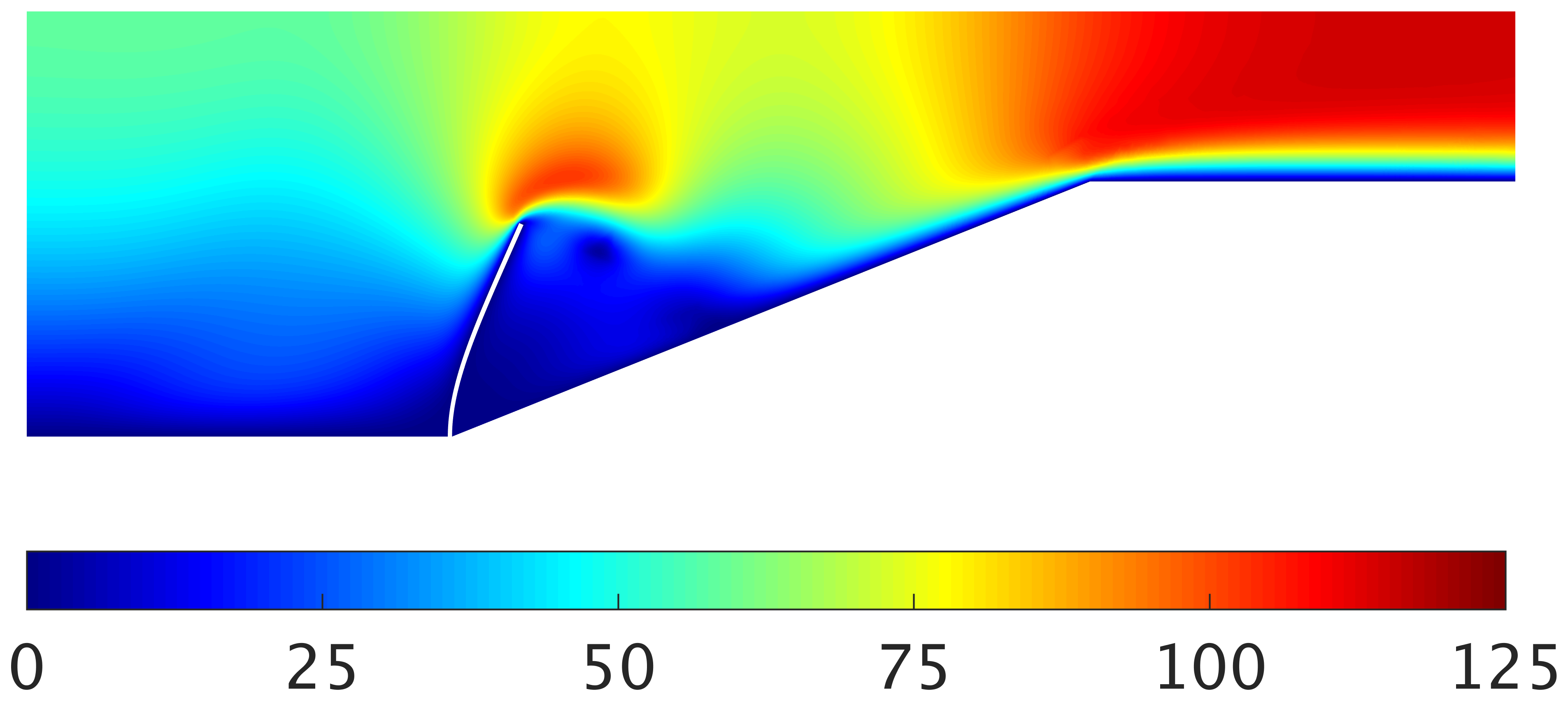}} \\
\subfigure[$\Fl{\rho}$ at $t=25$ $s$ with $\Fl{\varepsilon}=0.1$ $kg/(N \cdot m)$]
{\includegraphics[width=0.49\textwidth]{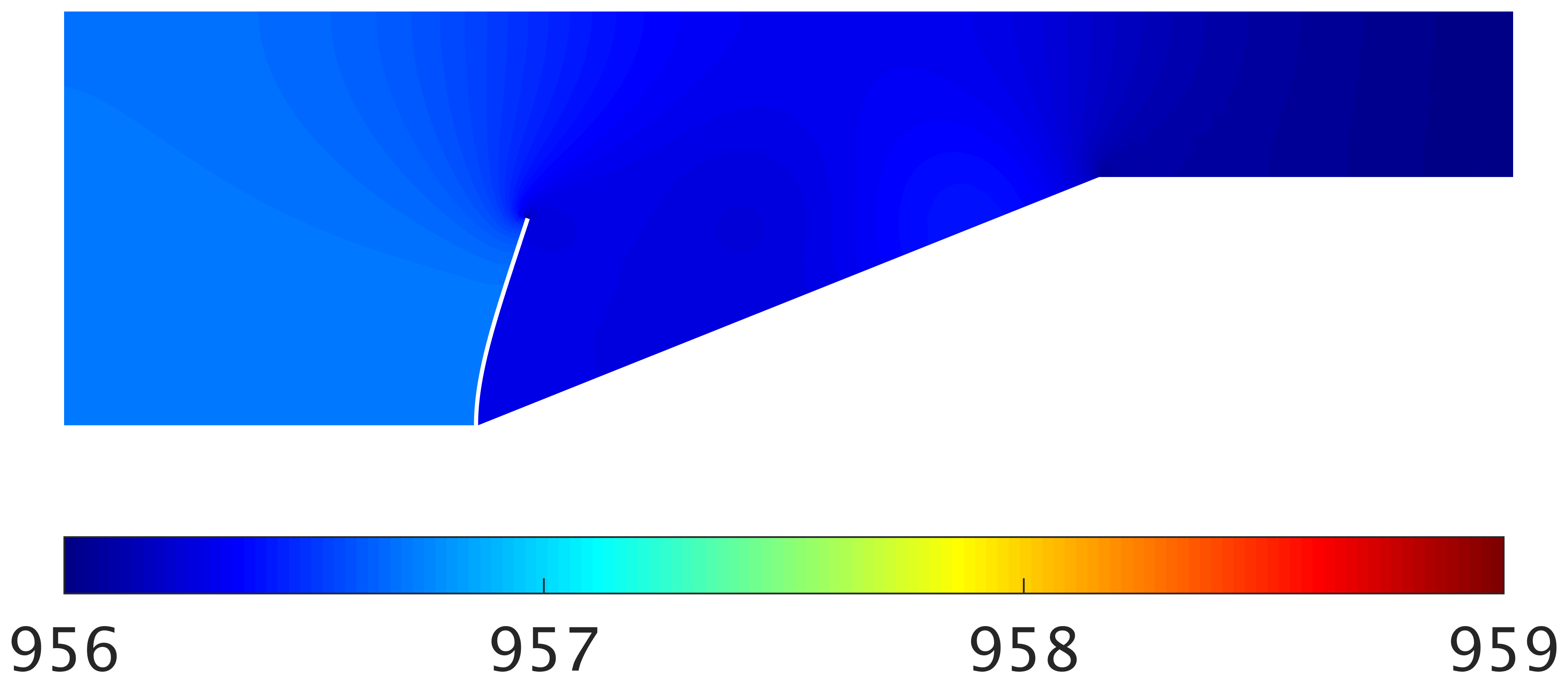}}
\subfigure[$\abs{\Fl{\rho\bm{\upsilon}}}$ at $t=25$ $s$ with $\Fl{\varepsilon}=0.1$ $kg/(N \cdot m)$]
{\includegraphics[width=0.49\textwidth]{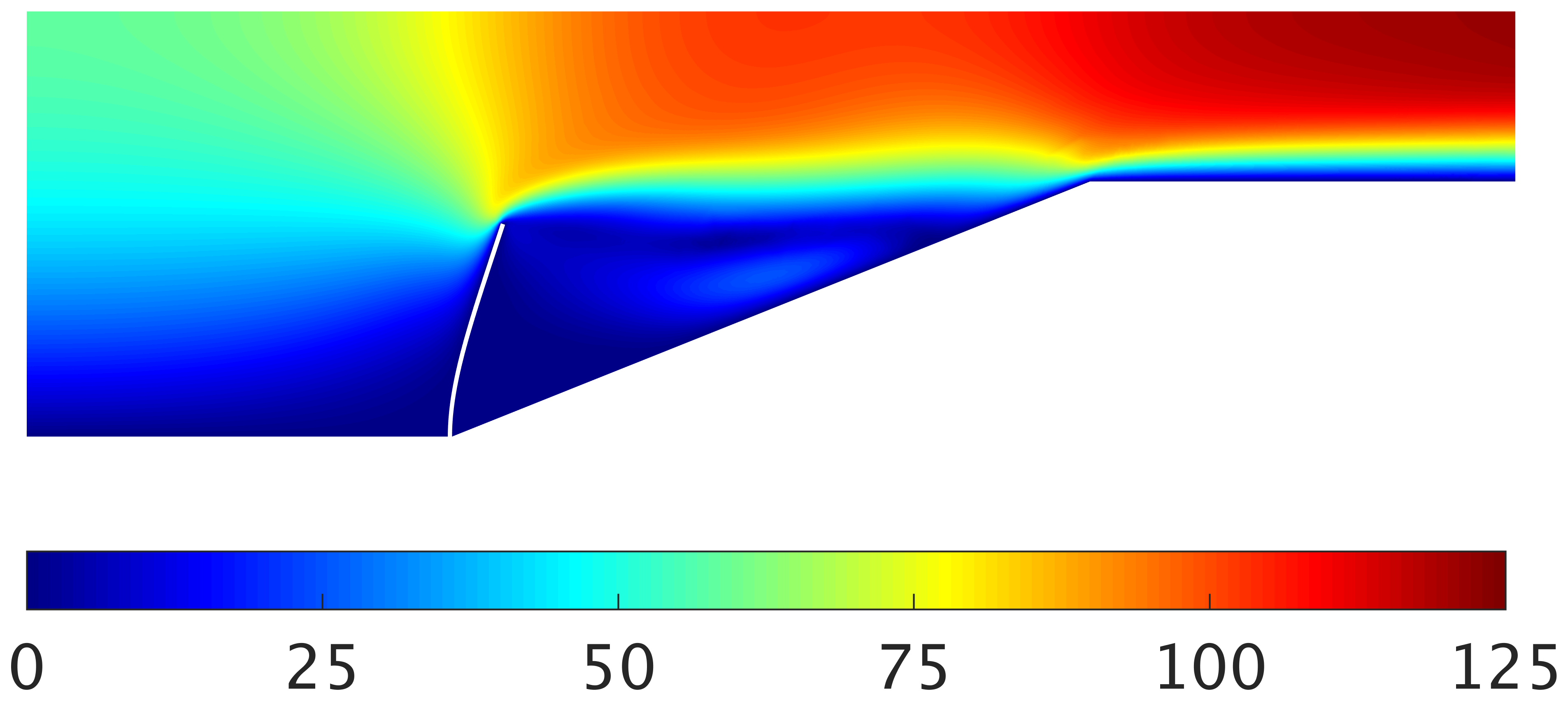}}
\caption{Approximation of the density and the momentum field of the channel with flexible wall at different time instants with $\Fl{\varepsilon}=0.1$ $kg/(N \cdot m)$.}
\label{fig:solutionchannelflexiblewall}
\end{figure}

The approximation of the density and the momentum field obtained with the monolithic Nitsche-based coupling with $\Fl{\varepsilon}=0.1$ $kg/(N \cdot m)$ is shown in Figure \ref{fig:solutionchannelflexiblewall} at different time instants.

The average value of the dynamic relaxation parameter $\omega_{avg}$ (evaluated over all the coupling iterations) and the average number of coupling iterations $i_{avg}$ (evaluated over all the time steps) needed by the Dirichlet--Neumann algorithm are summarized in Table \ref{tab:relaxanditerationschannelflexiblewall}.
\begin{table}
\centering
\begin{tabular}{ccccccc}
\hline
                         & \multicolumn{2}{c}{$\Delta t=0.1$ $s$} & \multicolumn{2}{c}{$\Delta t=0.01$ $s$} & \multicolumn{2}{c}{$\Delta t=0.001$ $s$} \\
                           \cline{2-3}                              \cline{4-5}                               \cline{6-7}
                         & $\omega_{avg}$ & $i_{avg}$             & $\omega_{avg}$ & $i_{avg}$              & $\omega_{avg}$ & $i_{avg}$               \\
\hline
Incompressible           & 0.17           & 27.5                  & --             & --                     & --             & --                      \\
$\Fl{\varepsilon}=0.001$ $kg/(N \cdot m)$ & 0.19           & 22.5                  & 0.20           & 34.0                   & 0.45           & 16.1                    \\
$\Fl{\varepsilon}=0.01$ $kg/(N \cdot m)$  & 0.20           & 22.5                  & 0.29           & 22.6                   & 0.64           & 8.2                     \\
$\Fl{\varepsilon}=0.1$ $kg/(N \cdot m)$   & 0.23           & 18.4                  & 0.47           & 11.4                   & 0.86           & 4.6                     \\
\hline
\end{tabular}
\caption{Average value of the dynamic relaxation parameter (evaluated over all the coupling iterations) and average number of coupling iterations (evaluated over all the time steps) needed by the Dirichlet--Neumann algorithm for the channel with flexible wall.}
\label{tab:relaxanditerationschannelflexiblewall}
\end{table}

When using the fully incompressible solver, a large number of coupling iterations is required to satisfy the convergence criterion \eqref{eqn:fsiconvergencecheckDN} for the largest time step size considered.
Moreover, for smaller time step sizes, the coupling fails to converge.
As expected from the analysis in \cite{Laspina2020a}, the weakly compressible formulation for the fluid field instead leads to a smaller number of coupling iterations and to a larger value of the dynamic relaxation parameter, thanks to the reduction of the maximal eigenvalue of the so-called added mass operator.
As also shown in \cite{Laspina2020a}, these beneficial effects are proportional to the compressibility coefficient and more pronounced for small time step sizes.
On the one hand, the monolithic scheme, not requiring sub-iterations of the single-field problems, is observed to outperform the partitioned one for small $\Fl{\varepsilon}$ and large $\Delta t$.
On the other hand, the partitioned scheme becomes very competitive for larger $\Fl{\varepsilon}$ and smaller $\Delta t$.
However, since the two algorithms have been implemented on distinct platforms, a quantitative comparison of the computational cost cannot be fairly assessed in the context of the present work and it constitutes a topic of future research.

\subsection{3D channel with flexible wall}
\label{sec:3dchannelflexiblewall}

\begin{figure}
\begin{center}
\begin{tikzpicture}
\def \fac{2.4}
\def \Linuns{0.5}
\def \LFuns{3.0}
\def \HFuns{0.5}
\def \BFuns{1.0}
\def \TSuns{0.05}
\def \HSuns{0.4}
\def \BSuns{0.6}
\def \umaxuns{0.4}
\def \dsizeuns{0.1}
\def \dtextuns{0.2}
\def \dbcuns{0.02}
\def \dviewuns{0.8}
\def \arrows{7}
\def \nconst{3}
\def \dconstuns{0.01}

\def \Lin{\Linuns*\fac}
\def \LF{\LFuns*\fac}
\def \HF{\HFuns*\fac}
\def \BF{\BFuns*\fac}
\def \TS{\TSuns*\fac}
\def \HS{\HSuns*\fac}
\def \BS{\BSuns*\fac}
\def \umax{\umaxuns*\fac}
\def \dview{\dviewuns*\fac}
\def \dsize{\dsizeuns*\fac}
\def \dtext{\dtextuns*\fac}
\def \dbc{\dbcuns*\fac}
\def \dconst{\dconstuns*\fac}

\draw[fill=lightgray,draw=black,thick]
(0,0)     -- node[above=\dbc]                   {$\Fl{\rho\bm{\upsilon}}=\mathbf{0}$}
(\LF,0)   -- node[left=\dbc]                    {$\Fl{\rho}=\Fl{\rho_0}$}
(\LF,\HF) -- node[below=\dbc]                   {$\Fl{\rho\bm{\upsilon}}=\mathbf{0}$}
(0,\HF)   -- node[above=\dbc,sloped,rotate=180] {$\Fl{\rho\bm{\upsilon}}={\Fl{\rho\bm{\upsilon}}}_{in}$}
cycle;

\draw[fill=gray,draw=black,thin]
(\Lin,0)       --
(\Lin+\TS,0)   --
(\Lin+\TS,\HS) --
(\Lin,\HS)     --
cycle;

\draw[fill=white,draw=black,thick,domain={0}:{\HF},samples=100] plot[smooth] ({4*(\umax)*(\x/\fac)*(1-(\x/\fac)/(\HF/\fac))},{\x}) --
(0,\HF) --
cycle;
\foreach \i in {2,...,\numexpr(\arrows)-1}
{
  \def \xi{0}
  \def \yi{(\i-1)*(\HF)/(\arrows-1)}
  \def \xf{4*(\umax)*(\yi/\fac)*(1-(\yi/\fac)/(\HF/\fac))}
  \def \yf{(\i-1)*(\HF)/(\arrows-1)}
  \draw [->,draw=black,thick] ({\xi},{\yi}) -- ({\xf},{\yf});
}

\draw [<->,draw=black,thin] ({0},{-\dsize}) -- node[below=\dtext] {$0.5$} ({\Lin},{-\dsize});

\draw [<->,draw=black,thin] ({0},{\HF+\dsize}) -- node[above=\dtext] {$3.0$} ({\LF},{\HF+\dsize});

\draw [<->,draw=black,thin] ({\LF+\dsize},{0}) -- node[right=\dtext] {$0.5$} ({\LF+\dsize},{\HF});


\draw [<->,draw=black,thin] ({\Lin},{\HS+\dsize/2}) -- node[left=\dtext] {$0.05$} ({\Lin+\TS},{\HS+\dsize/2});

\draw [<->,draw=black,thin] ({\Lin+\TS+\dsize/2},{0}) -- node[right=\dtext] {$0.4$} ({\Lin+\TS+\dsize/2},{\HS});


\draw[fill=lightgray,draw=black,thick]
(\LF+\dview,0)       --
(\LF+\dview+\BF,0)   --
(\LF+\dview+\BF,\HF) --
(\LF+\dview,\HF)     --
cycle;

\draw[fill=gray,draw=black,thin]
(\LF+\dview+\BF/2-\BS/2,0)       --
(\LF+\dview+\BF/2-\BS/2+\BS,0)   --
(\LF+\dview+\BF/2-\BS/2+\BS,\HS) --
(\LF+\dview+\BF/2-\BS/2,\HS)     --
cycle;



\draw [<->,draw=black,thin] ({\LF+\dview+\BF+\dsize},{0}) -- node[right=\dtext] {$0.5$} ({\LF+\dview+\BF+\dsize},{\HF});

\draw [<->,draw=black,thin] ({\LF+\dview},{\HF+\dsize}) -- node[above=\dtext] {$1.0$} ({\LF+\dview+\BF},{\HF+\dsize});


\draw [<->,draw=black,thin] ({\LF+\dview-\dsize},{0}) -- node[left=\dtext] {$0.4$} ({\LF+\dview-\dsize},{\HS});

\draw [<->,draw=black,thin] ({\LF+\dview+\BF/2-\BS/2},{-\dsize}) -- node[below=\dtext] {$0.6$} ({\LF+\dview+\BF/2-\BS/2+\BS},{-\dsize});


\draw[fill=lightgray,draw=black,thick]
(0,-\dview-\BF)   -- node[above=\dbc]                   {$\Fl{\rho\bm{\upsilon}}=\mathbf{0}$}
(\LF,-\dview-\BF) -- node[left=\dbc]                    {$\Fl{\rho}=\Fl{\rho_0}$}
(\LF,-\dview)     -- node[below=\dbc]                   {$\Fl{\rho\bm{\upsilon}}=\mathbf{0}$}
(0,-\dview)       -- node[above=\dbc,sloped,rotate=180] {$\Fl{\rho\bm{\upsilon}}={\Fl{\rho\bm{\upsilon}}}_{in}$}
cycle;

\draw[fill=gray,draw=black,thin]
(\Lin,-\dview-\BF/2-\BS/2)         --
(\Lin+\TS,-\dview-\BF/2-\BS/2)     --
(\Lin+\TS,-\dview-\BF/2-\BS/2+\BS) --
(\Lin,-\dview-\BF/2-\BS/2+\BS)     --
cycle;

\draw[fill=white,draw=black,thick,domain={0}:{\BF},samples=100] plot[smooth] ({2*(\umax)*(\x/\fac)*(1-(\x/\fac)/(\BF/\fac))},{\x-\dview-\BF}) --
(0,-\dview) --
cycle;
\foreach \i in {2,...,\numexpr(2*\arrows)-2*1}
{
  \def \xi{0}
  \def \yi{-\dview-\BF+(\i-1)*(\BF)/(2*\arrows-2*1)}
  \def \xf{2*(\umax)*((\yi+\dview+\BF)/\fac)*(1-((\yi+\dview+\BF)/\fac)/(\BF/\fac))}
  \def \yf{-\dview-\BF+(\i-1)*(\BF)/(2*\arrows-2*1)}
  \draw [->,draw=black,thick] ({\xi},{\yi}) -- ({\xf},{\yf});
}

\draw [<->,draw=black,thin] ({0},{-\dview-\BF-\dsize}) -- node[below=\dtext] {$0.5$} ({\Lin},{-\dview-\BF-\dsize});

\draw [<->,draw=black,thin] ({0},{-\dview+\dsize}) -- node[above=\dtext] {$3.0$} ({\LF},{-\dview+\dsize});


\draw [<->,draw=black,thin] ({\LF+\dsize},{-\dview-\BF}) -- node[right=\dtext] {$1.0$} ({\LF+\dsize},{-\dview});

\draw [<->,draw=black,thin] ({\Lin},{-\dview-\BF/2-\BS/2+\BS+\dsize/2}) -- node[left=\dtext] {$0.05$} ({\Lin+\TS},{-\dview-\BF/2-\BS/2+\BS+\dsize/2});


\draw [<->,draw=black,thin] ({\Lin+\TS+\dsize/2},{-\dview-\BF/2-\BS/2}) -- node[right=\dtext] {$0.6$} ({\Lin+\TS+\dsize/2},{-\dview-\BF/2-\BS/2+\BS});

\end{tikzpicture}
\end{center}
\caption{Geometry and boundary conditions of the 3D channel with flexible wall.}
\label{fig:3dchannelflexiblewallgeometrybcs}
\end{figure}
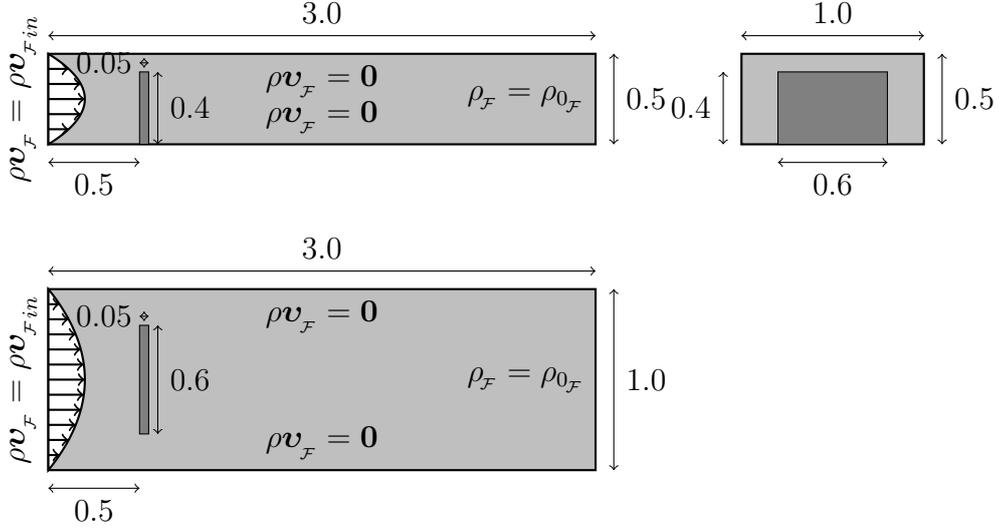

The fifth numerical example considers an elastic wall embedded in a channel flow.
This example was set up by \cite{Gerstenberger2010} and the goal of this study is to show the advantages of the weakly compressible fluid formulation for the solution of FSI problems in a three dimensional setting.
The geometry and the boundary conditions of the problem are depicted in Figure \ref{fig:3dchannelflexiblewallgeometrybcs}.

The dynamic viscosity of the fluid is $\Fl{\mu}=0.01$ and the reference density $\Fl{\rho_0}=1$ evaluated at the reference pressure $\Fl{p_0}=0$.
Three different orders of magnitude are considered for the compressibility coefficient, i.e., $\Fl{\varepsilon}=[0.001,0.01,0.1]$.
The structure is modelled as a Neo--Hookean material with density $\St{\rho}=1$, Young's modulus $\St{E}=500$ and Poisson's ratio $\St{\nu}=0$.
The following parabolic momentum profile is imposed at the inlet:
\begin{equation}
\begin{split}
&{\Fl{\rho\upsilon}}_x(0,y,z,t)=
\left\{
\begin{aligned}
&\left(1-16y^2\right)\left(1-4z^2\right)\dfrac{1-\cos\left(\pi t/5\right)}{2}\Fl{\rho_0}\Fl{\bar{\upsilon}}
\quad&&\text{if }t\leq 5, \\
&\left(1-16y^2\right)\left(1-4z^2\right)\Fl{\rho_0}\Fl{\bar{\upsilon}}
\quad&&\text{if }t  >  5,
\end{aligned}
\right. \\
&{\Fl{\rho\upsilon}}_y(0,y,z,t)=0, \\
&{\Fl{\rho\upsilon}}_z(0,y,z,t)=0,
\end{split}
\label{eqn:3dinflowchannelflexiblewall}
\end{equation}
with $\Fl{\bar{\upsilon}}=0.1$, resulting in a Reynolds number of about $5$.
No-slip boundary conditions are applied on the top and bottom walls as well as on the lateral walls.
At the channel exit, the fluid density is imposed equal to its reference value.
The HDG fluid discretization contains $25664$ hexahedral elements, while the CG structural discretization contains $1536$ hexahedral elements.
The degree of approximation considered is $k=1$ for both subdomains.
The time interval studied is $10$ $s$ and the temporal integration is performed by means of the BDF2 method.
Three different orders of magnitude of the time step size are considered, i.e., $\Delta t=[0.1,0.01,0.001]$.
The stabilization parameters are $\tau_\rho=1/\Fl{\varepsilon}$ and $\tau_{\rho\upsilon}=1$.

The problem is solved with the partitioned Dirichlet--Neumann coupling presented in section \ref{sec:partitioneddirichletneumannalgorithm}, with convergence tolerance $\eta=10^{-7}$.

\begin{figure}
\centering
{\includegraphics[width=0.49\textwidth]{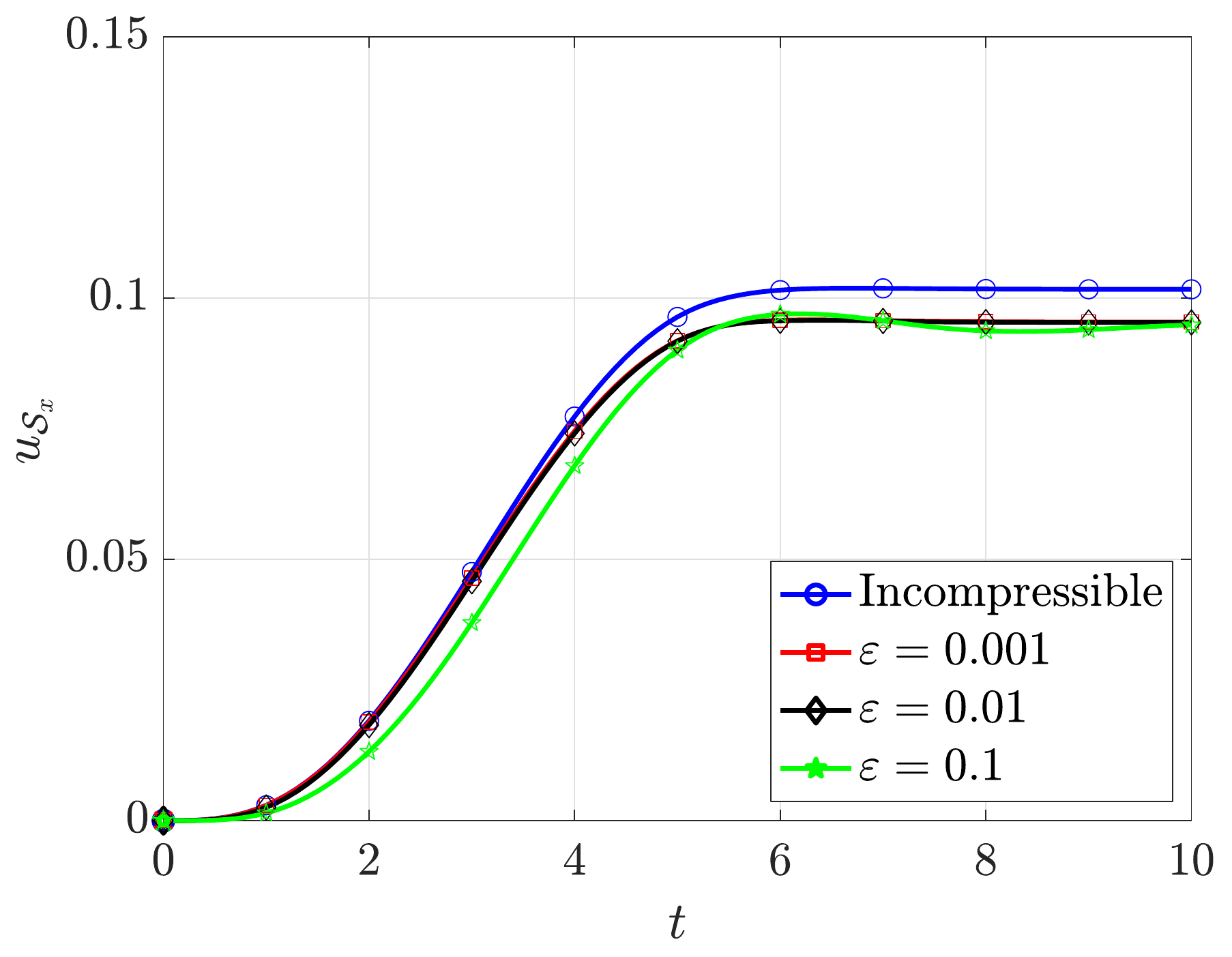}}
\caption{$x$-displacement component of the center top point of the flexible wall with different values of the compressibility coefficient computed with the partitioned Dirichlet--Neumann coupling with $\Delta t=0.1$.}
\label{fig:displacement3dchannelflexiblewall}
\end{figure}

The $x$ component of the displacement of the center top point of the wall is shown in Figure \ref{fig:displacement3dchannelflexiblewall} for the different compressibility coefficients considered.
The final displacement differs from the one obtained with a fully incompressible solver with the same settings for the coupling algorithm by about $6\%$.

\begin{figure}
\centering
\subfigure[$\left|\Fl{\bm{\upsilon}}\right|$ at $t=10$ with $\Fl{\varepsilon}=0.01$]
{\includegraphics[width=0.70\textwidth]{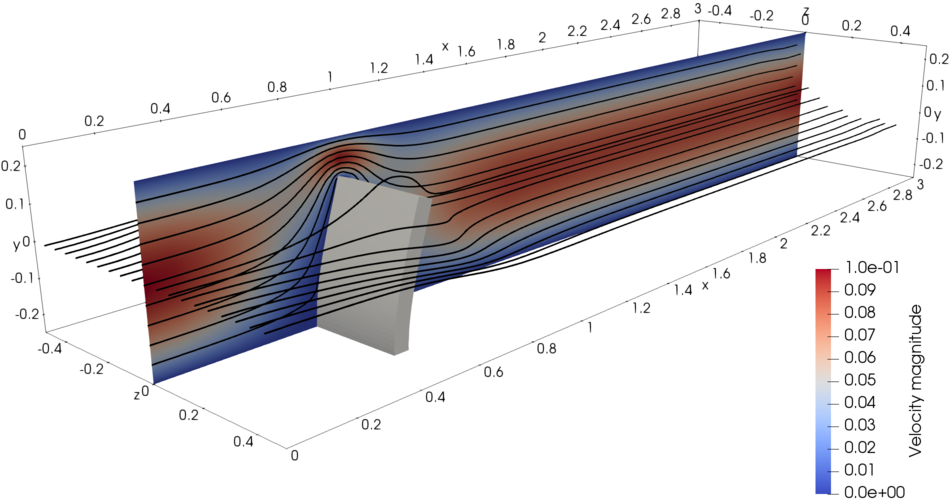}} \\
\subfigure[$\Fl{p}$ at $t=10$ with $\Fl{\varepsilon}=0.01$]
{\includegraphics[width=0.70\textwidth]{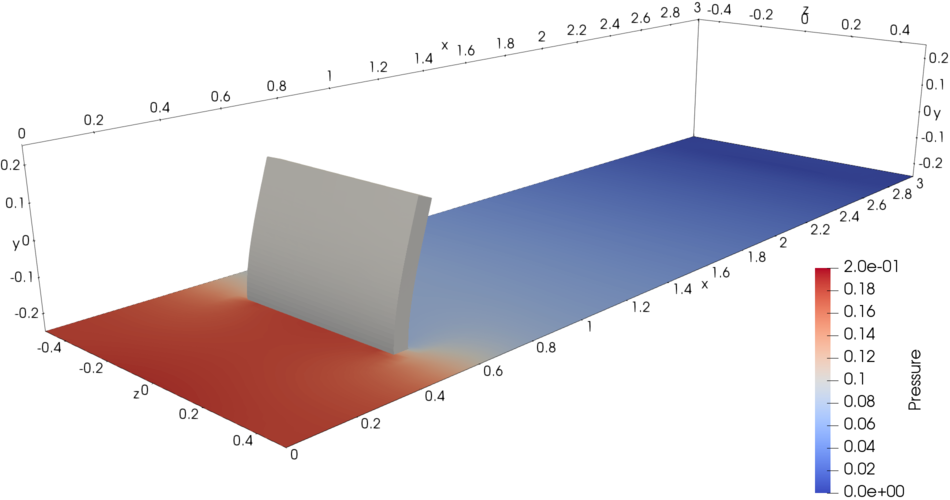}} \\
\subfigure[$\left|\St{\mathbf{u}}\right|$ at $t=10$ with $\Fl{\varepsilon}=0.01$]
{\includegraphics[width=0.70\textwidth]{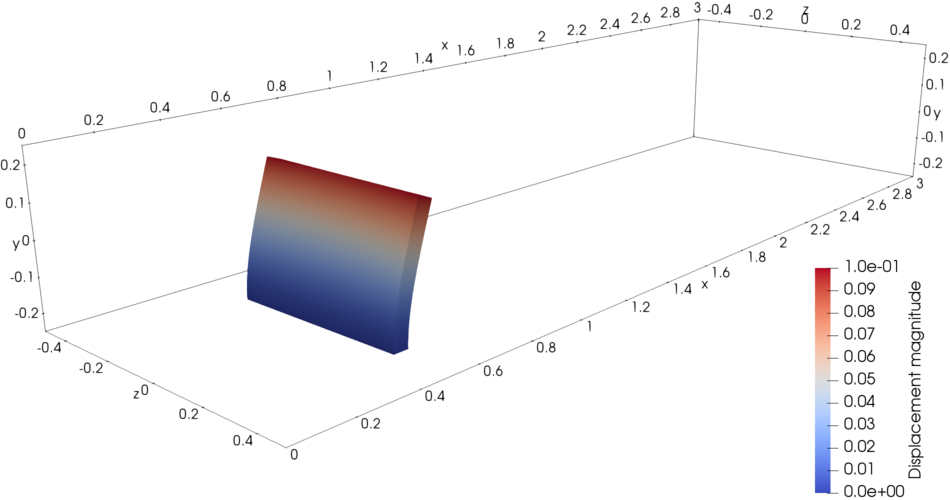}}
\caption{Approximation of the fluid velocity and pressure field and the structural displacement of the 3D channel with flexible wall at $t=10$ with $\Fl{\varepsilon}=0.01$.}
\label{fig:solution3dchannelflexiblewall}
\end{figure}

The approximation of the fluid velocity and pressure field and the structural displacement obtained with $\Fl{\varepsilon}=0.01$ and $\Delta t=0.1$ is shown in Figure \ref{fig:solution3dchannelflexiblewall} for the final simulation time.

The average value of the dynamic relaxation parameter $\omega_{avg}$ (evaluated over all the coupling iterations) and the average number of coupling iterations $i_{avg}$ (evaluated over all the time steps) needed by the partitioned coupling algorithm are summarized in Table \ref{tab:relaxanditerations3dchannelflexiblewall}.
The response obtained with a higher polynomial degree ($k=2$) in the fluid field with $\Delta t=0.1$ is very close to the one obtained with linear elements and the resulting average value of the relaxation parameter and the average number of coupling iterations are $\omega_{avg}=[0.36,0.39,0.44]$ and $i_{avg}=[11.6,8.1,6.5]$ for $\Fl{\varepsilon}=[0.001,0.01,0.1]$.
Considering the required computational effort, the influence of the time step size on the robustness of the partitioned scheme is analyzed here only for the linear case.
\begin{table}
\centering
\begin{tabular}{ccccccccc}
\hline
                         & \multicolumn{2}{c}{$\Delta t=0.1$} & \multicolumn{2}{c}{$\Delta t=0.01$} & \multicolumn{2}{c}{$\Delta t=0.001$} \\
                           \cline{2-3}                          \cline{4-5}                           \cline{6-7}
                         & $\omega_{avg}$ & $i_{avg}$         & $\omega_{avg}$ & $i_{avg}$          & $\omega_{avg}$ & $i_{avg}$           \\
\hline
Incompressible           & 0.33           & 10.9              & --             & --                 & --             & --                  \\
$\Fl{\varepsilon}=0.001$ & 0.37           & 10.1              & 0.43           & 7.7                & 0.54           & 2.2                 \\
$\Fl{\varepsilon}=0.01$  & 0.39           & 8.7               & 0.57           & 4.2                & 0.74           & 0.8                 \\
$\Fl{\varepsilon}=0.1$   & 0.44           & 6.2               & 0.78           & 2.4                & 1.00           & 0.4                 \\
\hline
\end{tabular}
\caption{Average value of the dynamic relaxation parameter (evaluated over all the coupling iterations) and average number of coupling iterations (evaluated over all the time steps) needed by the Dirichlet--Neumann algorithm for the 3D channel with flexible wall.}
\label{tab:relaxanditerations3dchannelflexiblewall}
\end{table}

Similarly to the two dimensional problem in section \ref{sec:channelflexiblewall}, the incompressible solver needs a relatively high number of coupling iterations for large $\Delta t$ and it fails to reach convergence of the partitioned algorithm for smaller time step sizes.
On the other hand, the weakly compressible solver is always able to converge, exhibiting a decreasing number of coupling iterations when decreasing $\Delta t$.
For $\Fl{\varepsilon}=0.1$ and $\Delta t=0.001$, no relaxation is performed at all (always $\omega=1$) and almost no coupling iterations are needed to meet the convergence criterion.

\section{Conclusion}
\label{sec:conclusion}

A hybrid HDG-CG formulation for the solution of fluid-structure interaction problems has been proposed.
Special emphasis has been devoted to the derivation and validation of a novel hybridizable discontinuous Galerkin approach for weakly compressible flows with the Arbitrary Lagrangian--Eulerian description.
A partitioned Dirichlet--Neumann scheme has been revisited for the hybrid discretization and a minimally-intrusive monolithic Nitsche-based coupling has been proposed for the coupling of the HDG and the CG subproblems, by exploiting the definition of the numerical flux and the trace of the solution to impose the coupling conditions.
The numerical examples demonstrate the convergence of the HDG and CG primal and mixed variables with order $k+1$ and the superconvergence of the fluid velocity with order $k+2$, through an inexpensive element-by-element postprocessing.
Moreover, the advantages of introducing a weak compressibility in the fluid have been confirmed on two and three dimensional FSI problems.
In particular, the constraints that the incompressibility poses on the fluid-structure coupling are alleviated by the proposed formulation, resulting in a more robust and efficient FSI solver.
The investigation of the computational efficiency of the proposed schemes in a high-performance framework is subject of future work.

\section*{Acknowledgments}
\label{sec:acknowledgments}

This work was supported by the European Education, Audiovisual and Culture Executive Agency (EACEA) under the Erasmus Mundus Joint Doctorate {\lq\lq{Simulation in Engineering and Entrepreneurship Development}\rq\rq} (SEED), FPA 2013-0043.

\bibliographystyle{abbrv}
\bibliography{bibliography}

\begin{thebibliography}{10}

\bibitem{Bassi2006}
F.~Bassi, A.~Crivellini, D.~{Di Pietro}, and S.~Rebay.
\newblock An artificial compressibility flux for the discontinuous galerkin
  solution of the incompressible navier--stokes equations.
\newblock {\em Journal of Computational Physics}, 218(2):794--815, 2006.

\bibitem{Bassi2007}
F.~Bassi, A.~Crivellini, D.~A. {Di Pietro}, and S.~Rebay.
\newblock An implicit high-order discontinuous galerkin method for steady and
  unsteady incompressible flows.
\newblock {\em Computers \& Fluids}, 36(10):1529--1546, 2007.
\newblock Special Issue Dedicated to Professor Michele Napolitano on the
  Occasion of his 60th Birthday.

\bibitem{Causin2005}
P.~Causin, J.~Gerbeau, and F.~Nobile.
\newblock Added-mass effect in the design of partitioned algorithms for
  fluid--structure problems.
\newblock {\em Computer Methods in Applied Mechanics and Engineering},
  194(42):4506 -- 4527, 2005.

\bibitem{Chung1993}
J.~Chung and G.~M. Hulbert.
\newblock A time integration algorithm for structural dynamics with improved
  numerical dissipation: The generalized-α method.
\newblock {\em Journal of Applied Mechanics}, 60(2):371--375, 1993.

\bibitem{Cockburn2008}
B.~Cockburn, B.~Dong, and J.~Guzm{\'a}n.
\newblock A superconvergent ldg-hybridizable galerkin method for second-order
  elliptic problems.
\newblock {\em Mathematics of Computation}, 77(264):1887--1916, 2008.

\bibitem{Cockburn2009a}
B.~Cockburn, J.~Gopalakrishnan, and R.~Lazarov.
\newblock Unified hybridization of discontinuous galerkin, mixed, and
  continuous galerkin methods for second order elliptic problems.
\newblock {\em SIAM Journal on Numerical Analysis}, 47(2):1319--1365, 2009.

\bibitem{Cockburn2012}
B.~Cockburn and K.~Shi.
\newblock {Superconvergent HDG methods for linear elasticity with weakly
  symmetric stresses}.
\newblock {\em IMA Journal of Numerical Analysis}, 33(3):747--770, 10 2012.

\bibitem{Degroote2009}
J.~Degroote, K.-J. Bathe, and J.~Vierendeels.
\newblock Performance of a new partitioned procedure versus a monolithic
  procedure in fluid-structure interaction.
\newblock {\em Comput. Struct.}, 87(11-12):793--801, June 2009.

\bibitem{Epshteyn2007}
Y.~Epshteyn and B.~Rivi{\`e}re.
\newblock Estimation of penalty parameters for symmetric interior penalty
  galerkin methods.
\newblock {\em Journal of Computational and Applied Mathematics},
  206(2):843--872, 2007.

\bibitem{Etienne2012}
S.~{\'E}tienne, A.~Garon, and D.~Pelletier.
\newblock Some manufactured solutions for verification of fluid-structure
  interaction codes.
\newblock {\em Computers \& Structures}, 106-107:56 -- 67, 2012.

\bibitem{Farhat2006}
C.~Farhat, K.~G. van~der Zee, and P.~Geuzaine.
\newblock Provably second-order time-accurate loosely-coupled solution
  algorithms for transient nonlinear computational aeroelasticity.
\newblock {\em Computer Methods in Applied Mechanics and Engineering},
  195(17):1973 -- 2001, 2006.
\newblock Fluid-Structure Interaction.

\bibitem{Fehn2019}
N.~Fehn, W.~A. Wall, and M.~Kronbichler.
\newblock A matrix-free high-order discontinuous galerkin compressible
  navier-stokes solver: A performance comparison of compressible and
  incompressible formulations for turbulent incompressible flows.
\newblock {\em International Journal for Numerical Methods in Fluids},
  89(3):71--102, 2019.

\bibitem{Fish2007}
J.~Fish and T.~Belytschko.
\newblock {\em A First Course in Finite Elements}.
\newblock John Wiley \& Sons, 2007.

\bibitem{Forster2007}
C.~F{\"o}rster, W.~A. Wall, and E.~Ramm.
\newblock Artificial added mass instabilities in sequential staggered coupling
  of nonlinear structures and incompressible viscous flows.
\newblock {\em Computer Methods in Applied Mechanics and Engineering},
  196(7):1278 -- 1293, 2007.

\bibitem{Froehle2014}
B.~Froehle and P.-O. Persson.
\newblock A high-order discontinuous galerkin method for fluid--structure
  interaction with efficient implicit--explicit time stepping.
\newblock {\em Journal of Computational Physics}, 272:455 -- 470, 2014.

\bibitem{Gerstenberger2010}
A.~Gerstenberger and W.~A. Wall.
\newblock An embedded dirichlet formulation for 3d continua.
\newblock {\em International Journal for Numerical Methods in Engineering},
  82(5):537--563, 2010.

\bibitem{Giacomini2018}
M.~Giacomini, A.~Karkoulias, R.~Sevilla, and A.~Huerta.
\newblock A superconvergent hdg method for stokes flow with strongly enforced
  symmetry of the stress tensor.
\newblock {\em Journal of Scientific Computing}, 77(3):1679--1702, Dec 2018.

\bibitem{Giacomini2020}
M.~Giacomini, R.~Sevilla, and A.~Huerta.
\newblock {\em Tutorial on Hybridizable Discontinuous Galerkin (HDG)
  Formulation for Incompressible Flow Problems}, pages 163--201.
\newblock Springer International Publishing, Cham, 2020.

\bibitem{Giorgiani2014}
G.~Giorgiani, S.~Fern{\'a}ndez-M{\'e}ndez, and A.~Huerta.
\newblock Hybridizable discontinuous galerkin with degree adaptivity for the
  incompressible navier--stokes equations.
\newblock {\em Computers \& Fluids}, 98:196--208, 2014.

\bibitem{Giorgiani2013}
G.~Giorgio, F.-M. Sonia, and H.~Antonio.
\newblock Hybridizable discontinuous galerkin p-adaptivity for wave propagation
  problems.
\newblock {\em International Journal for Numerical Methods in Fluids},
  72(12):1244--1262, 2018/07/05 2013.

\bibitem{Griebel2003}
M.~Griebel and M.~A. Schweitzer.
\newblock {\em A Particle-Partition of Unity Method Part V: Boundary
  Conditions}, pages 519--542.
\newblock Springer Berlin Heidelberg, Berlin, Heidelberg, 2003.

\bibitem{Hansbo2005}
P.~Hansbo.
\newblock Nitsche's method for interface problems in computational mechanics.
\newblock {\em GAMM-Mitteilungen}, 28(2):183--206, 2020/07/24 2005.

\bibitem{Heil2004}
M.~Heil.
\newblock An efficient solver for the fully coupled solution of
  large-displacement fluid--structure interaction problems.
\newblock {\em Computer Methods in Applied Mechanics and Engineering}, 193(1):1
  -- 23, 2004.

\bibitem{Housiadas2016}
K.~D. Housiadas and G.~C. Georgiou.
\newblock New analytical solutions for weakly compressible newtonian poiseuille
  flows with pressure-dependent viscosity.
\newblock {\em International Journal of Engineering Science}, 107:13--27, 2016.

\bibitem{Huang2019}
D.~Huang, P.-O. Persson, and M.~Zahr.
\newblock High-order, linearly stable, partitioned solvers for general
  multiphysics problems based on implicit--explicit runge--kutta schemes.
\newblock {\em Computer Methods in Applied Mechanics and Engineering},
  346:674--706, 2019.

\bibitem{Jansen2000}
K.~E. Jansen, C.~H. Whiting, and G.~M. Hulbert.
\newblock A generalized-α method for integrating the filtered navier--stokes
  equations with a stabilized finite element method.
\newblock {\em Computer Methods in Applied Mechanics and Engineering},
  190(3):305 -- 319, 2000.

\bibitem{Donea2004}
D.~Jean, H.~Antonio, P.~J.-Ph., and R.-F. A.
\newblock {\em Arbitrary Lagrangian--Eulerian Methods}, chapter~14.
\newblock American Cancer Society, 2004.

\bibitem{Kirby2012}
R.~M. Kirby, S.~J. Sherwin, and B.~Cockburn.
\newblock To cg or to hdg: A comparative study.
\newblock {\em Journal of Scientific Computing}, 51(1):183--212, 2012.

\bibitem{Krank2017}
B.~Krank, N.~Fehn, W.~A. Wall, and M.~Kronbichler.
\newblock A high-order semi-explicit discontinuous galerkin solver for 3d
  incompressible flow with application to dns and les of turbulent channel
  flow.
\newblock {\em Journal of Computational Physics}, 348:634 -- 659, 2017.

\bibitem{Kuttler2006}
U.~K{\"u}ttler, C.~F{\"o}rster, and W.~A. Wall.
\newblock A solution for the incompressibility dilemma in partitioned
  fluid--structure interaction with pure dirichlet fluid domains.
\newblock {\em Computational Mechanics}, 38(4):417--429, Sep 2006.

\bibitem{Kuttler2008}
U.~K{\"u}ttler and W.~A. Wall.
\newblock Fixed-point fluid--structure interaction solvers with dynamic
  relaxation.
\newblock {\em Computational Mechanics}, 43(1):61--72, Dec 2008.

\bibitem{Laspina2020a}
A.~La~Spina, C.~F{\"o}rster, M.~Kronbichler, and W.~A. Wall.
\newblock On the role of (weak) compressibility for fluid-structure interaction
  solvers.
\newblock {\em International Journal for Numerical Methods in Fluids},
  92(2):129--147, 2020.

\bibitem{Laspina2020b}
A.~La~Spina, M.~Giacomini, and A.~Huerta.
\newblock Hybrid coupling of cg and hdg discretizations based on nitsche's
  method.
\newblock {\em Computational Mechanics}, 65(2):311--330, Feb 2020.

\bibitem{Mayr2015}
M.~Mayr, T.~Kl{\"o}ppel, W.~A. Wall, and M.~W. Gee.
\newblock A temporal consistent monolithic approach to fluid-structure
  interaction enabling single field predictors.
\newblock {\em SIAM J. Scientific Computing}, 37, 2015.

\bibitem{Mok2001}
D.~Mok and W.~Wall.
\newblock Partitioned analysis schemes for the transient interaction of
  incompressible flows and nonlinear flexible structures.
\newblock {\em Trends in Computational Structural Mechanics}, 01 2001.

\bibitem{Montlaur2018}
A.~Montlaur, S.~Fernandez-Mendez, and A.~Huerta.
\newblock Discontinuous galerkin methods for the stokes equations using
  divergence-free approximations.
\newblock {\em International Journal for Numerical Methods in Fluids},
  57(9):1071--1092, 2008.

\bibitem{Nguyen2009}
N.~C. Nguyen, J.~Peraire, and B.~Cockburn.
\newblock An implicit high-order hybridizable discontinuous galerkin method for
  linear convection--diffusion equations.
\newblock {\em Journal of Computational Physics}, 228(9):3232--3254, 2009.

\bibitem{Nguyen2009a}
N.~C. Nguyen, J.~Peraire, and B.~Cockburn.
\newblock An implicit high-order hybridizable discontinuous galerkin method for
  nonlinear convection--diffusion equations.
\newblock {\em Journal of Computational Physics}, 228(23):8841--8855, 2009.

\bibitem{Nguyen2010}
N.~C. Nguyen, J.~Peraire, and B.~Cockburn.
\newblock A hybridizable discontinuous galerkin method for stokes flow.
\newblock {\em Computer Methods in Applied Mechanics and Engineering},
  199(9):582--597, 2010.

\bibitem{Nguyen2011}
N.~C. Nguyen, J.~Peraire, and B.~Cockburn.
\newblock An implicit high-order hybridizable discontinuous galerkin method for
  the incompressible navier--stokes equations.
\newblock {\em Journal of Computational Physics}, 230(4):1147--1170, 2011.

\bibitem{Paipuri2019}
M.~Paipuri, C.~Tiago, and S.~Fern{\'a}ndez-M{\'e}ndez.
\newblock Coupling of continuous and hybridizable discontinuous galerkin
  methods: Application to conjugate heat transfer problem.
\newblock {\em Journal of Scientific Computing}, 78(1):321--350, Jan 2019.

\bibitem{Peraire2010}
J.~Peraire, N.~Nguyen, and B.~Cockburn.
\newblock {\em A Hybridizable Discontinuous Galerkin Method for the
  Compressible Euler and Navier-Stokes Equations}.
\newblock American Institute of Aeronautics and Astronautics, 2018/07/05 2010.

\bibitem{Persson2009}
P.-O. Persson, J.~Bonet, and J.~Peraire.
\newblock Discontinuous galerkin solution of the navier--stokes equations on
  deformable domains.
\newblock {\em Computer Methods in Applied Mechanics and Engineering},
  198(17):1585 -- 1595, 2009.

\bibitem{Schott2018}
B.~Schott, C.~Ager, and W.~A. Wall.
\newblock A monolithic approach to fluid-structure interaction based on a
  hybrid eulerian-ale fluid domain decomposition involving cut elements.
\newblock {\em CoRR}, abs/1808.00343, 2018.

\bibitem{Sevilla2018}
R.~Sevilla, M.~Giacomini, A.~Karkoulias, and A.~Huerta.
\newblock A superconvergent hybridisable discontinuous galerkin method for
  linear elasticity.
\newblock {\em International Journal for Numerical Methods in Engineering},
  116(2):91--116, 2018.

\bibitem{Sevilla2018a}
R.~Sevilla and A.~Huerta.
\newblock Hdg-nefem with degree adaptivity for stokes flows.
\newblock {\em Journal of Scientific Computing}, 77(3):1953--1980, 2018.

\bibitem{Sheldon2016}
J.~P. Sheldon, S.~T. Miller, and J.~S. Pitt.
\newblock A hybridizable discontinuous galerkin method for modeling
  fluid--structure interaction.
\newblock {\em Journal of Computational Physics}, 326:91 -- 114, 2016.

\bibitem{Sheldon2018}
J.~P. Sheldon, S.~T. Miller, and J.~S. Pitt.
\newblock An improved formulation for hybridizable discontinuous galerkin
  fluid-structure interaction modeling with reduced computational expense.
\newblock {\em Communications in Computational Physics}, 24(5), 6 2018.

\bibitem{Soon2009}
S.~C. Soon, B.~Cockburn, and H.~K. Stolarski.
\newblock A hybridizable discontinuous galerkin method for linear elasticity.
\newblock {\em International Journal for Numerical Methods in Engineering},
  80(8):1058--1092, 2018/07/05 2009.

\bibitem{Taliadorou2008}
E.~Taliadorou, G.~C. Georgiou, and E.~Mitsoulis.
\newblock Numerical simulation of the extrusion of strongly compressible
  newtonian liquids.
\newblock {\em Rheologica Acta}, 47(1):49--62, Jan 2008.

\bibitem{Venerus2006}
D.~C. Venerus.
\newblock Laminar capillary flow of compressible viscous fluids.
\newblock {\em Journal of Fluid Mechanics}, 555:59--80, 2006.

\bibitem{Vinay2006}
G.~Vinay, A.~Wachs, and J.-F. Agassant.
\newblock Numerical simulation of weakly compressible bingham flows: The
  restart of pipeline flows of waxy crude oils.
\newblock {\em Journal of Non-Newtonian Fluid Mechanics}, 136(2):93 -- 105,
  2006.

\bibitem{Wall2006}
W.~A. Wall, A.~Gerstenberger, P.~Gamnitzer, C.~F{\"o}rster, and E.~Ramm.
\newblock Large deformation fluid-structure interaction -- advances in ale
  methods and new fixed grid approaches.
\newblock In H.-J. Bungartz and M.~Sch{\"a}fer, editors, {\em Fluid-Structure
  Interaction}, pages 195--232, Berlin, Heidelberg, 2006. Springer Berlin
  Heidelberg.

\bibitem{Zienkiewicz2000}
O.~Zienkiewicz and R.~Taylor.
\newblock {\em The Finite Element Method, Volume 2: Solid Mechanic}.
\newblock 01 2000.

\end{thebibliography}

\end{document}